\documentclass[%
 reprint,
superscriptaddress,
nofootinbib,
 amsmath,amssymb,
 aps,
pra,
]{revtex4-2}
\usepackage{physics}
\usepackage{bbold}
\usepackage{latexsym}
\usepackage{graphicx}
\usepackage{amsthm}
\usepackage{lipsum}
\usepackage[english]{babel}
\usepackage{slashed}
\usepackage{appendix}
\usepackage{multirow}
\usepackage{mathtools}
\usepackage{mathrsfs}
\usepackage{amsbsy}
\newcommand{\sam}[1]{{\color{blue} #1}}

\usepackage{ragged2e}
\usepackage{comment}
\newcommand{\M}{\mathcal{M}}
\usepackage{xcolor}
\usepackage{tikz}
\usepackage{subcaption}
\usepackage{footnotehyper}
\usepackage{hyperref}
\usepackage[utf8]{inputenc}
\usepackage[T1]{fontenc}

\tikzset{every picture/.style={line width=0.75pt}} 

\newcommand{\bh}{\mathcal{B}(\hi)}
\newcommand{\Bor}{\text{Bor}}

\newcommand{\hi}{\mathcal{H}}

\newcommand{\U}{\mathcal{U}}
\newcommand{\V}{\mathcal{V}}
\def\state{\mathscr{D}(\mathcal{H})}

\theoremstyle{plain}

\newcommand{\inner}[2]{\langle #1, #2 \rangle}

\usepackage[bottom]{footmisc}

\makeatletter
\def\l@subsubsection#1#2{}
\newcommand*{\StopTOCentries}{%
  \let\l@section\@gobbletwo
  \let\l@subsection\@gobbletwo
  \let\l@subsubsection\@gobbletwo}
\makeatother
\makeatother

\makesavenoteenv{align}

\begin{document}

\title{A relativistic gravitisation of quantum theory?}
\date{\today}
\author{Samuel Fedida}
\affiliation{Centre for Quantum Information and Foundations, DAMTP, Centre for Mathematical Sciences, University of Cambridge, Wilberforce Road, Cambridge CB3 0WA, UK}
\author{Adrian Kent}
\affiliation{Centre for Quantum Information and Foundations, DAMTP, Centre for Mathematical Sciences, University of Cambridge, Wilberforce Road, Cambridge CB3 0WA, UK}
\affiliation{Perimeter Institute for Theoretical
	Physics, 31 Caroline Street North, Waterloo, ON N2L 2Y5, Canada.}

\bigskip

\begin{abstract}    
    We develop a model, that we call local semiclassical gravity, based on (i) the distributional semiclassical Einstein field equations sourced by the local quantum state, i.e. the reduced density matrix of the local system updated by collapses (only) in its past light cone; (ii) a modified measurement rule that postulates an additional source that is singular on the future light cone of collapse events.
     
    We propose this as the basis of a relativistically causal generally covariant gravitisation of relativistic quantum theory in $1+3$ spacetime dimensions.  We derive the weak-field limit which exhibits the explicitly causal dynamical structure of the theory, and recover the Schrödinger-Newton system in the nonrelativistic limit.     
    This proposal can thus be understood as a causal completion of the Schrödinger-Newton equation, a nonlinear extension to quantum theory.
    
    We analyze relativistic quantum measurements in this nonlinear setup, showing that these do not lead to superluminal signalling if quantum states update (only) in the future light-cone of the measurement region.  We give explicit examples of quantum measurements on a spatial superposition and on entangled subsystems showing how, in the weak field limit, discontinuities in the expectation value of the stress-energy tensor can be consistent with distributional solutions to the (linearised) semiclassical Einstein field equations.
    
    We also highlight that the theory allows non-standard measurements via its nonlinearities.   These effective ``readout devices'' can approximately determine a quantum state without affecting it through the state's influence on the classical gravitational field
    and hence can implement approximate quantum state cloning.   Nonetheless, the theory does not allow superluminal signalling, although it does violate the 
    quantum no-signalling principle. We further show that the classical gravitational field is generally not locally causal in the sense of Bell.
    
    We present several paths to test the model against low-energy quantum gravity in tabletop experiments. We discuss extensions to the relativistic gravitisation of quantum field theory and other outstanding problems and open questions.
\end{abstract}

\maketitle

\onecolumngrid

\vspace{22pt}


\tableofcontents

\newpage

\twocolumngrid

\section{Introduction}

The unification of gravity with relativistic quantum theory is one of the major open problems of contemporary physics. Significant theoretical and experimental efforts go into probing the interface between gravitational and quantum phenomena -- yet, the essence of spacetime itself is still debated. Does it arise as an emergent structure from deeper quantum mechanical origins; is it fundamentally a classical arena which interacts with quantum particles and fields in a nontrivial way; or is it something else entirely?

There are several possibilities for coupling the Einstein field equations (EFEs) to quantum field theory (QFT) through the stress-energy tensor $\hat{T}_{ab}$ of the other quantum fields. One possibility is, of course, quantum (Einstein) gravity, for which the Einstein tensor is raised to a quantum operator:
\begin{equation}
    \label{eqn: QEG}
    \hat{G}_{ab}(x) = \kappa \hat{T}_{ab}(x) \, ,
\end{equation}
where $\kappa = \frac{8\pi G}{c^4}$ and we ignore the cosmological constant for simplicity. This (and related quantum gravitational models) faces many issues, both at the conceptual and technical levels, including the existence of a Hilbert space of suitable Lorentzian metrics \cite{isham_canonical_1993,wald_proposal_1993}, renormalisability \cite{t_hooft_one-loop_1974, goroff_quantum_1985}, the black hole information paradox \cite{hawking_breakdown_1976,almheiri_black_2013,akil_quantum_2025}, the tension between Einstein's equivalence principle and quantum theory \cite{penrose_gravitization_2014,giacomini_einsteins_2020,giacomini_quantum_2022}, the problem of time \cite{isham_canonical_1993,wald_proposal_1993, kuchar_time_2011}, the quantum hole argument \cite{schmelzer_quantum_2009,gomes_hole_2023,kabel_quantum_2025}, the notion of events and indefinite causality \cite{oreshkov_quantum_2012,hamette_what_2024,de_la_hamette_indefinite_2025, fedida_knot_2025}, amongst other challenges.

Another attempt at coupling the Einstein field equations to QFT is M{\o}ller-Rosenfeld semiclassical gravity \cite{moller_theories_1962,rosenfeld_quantization_1963}, for which the Einstein tensor remains classical, but the operator-valued right-hand side requires to be a c-number, leading to
\begin{equation}
    \label{eqn:MR SCG}
    G_{ab}(x) = \kappa \expval{\hat{T}_{ab}(x)}{\Psi[\Sigma]}
\end{equation}
whenever the \emph{whole} system on the spacelike Cauchy hypersurface $\Sigma \ni x$ is in state $\Psi[\Sigma]$. In the nonrelativistic limit, this framework, when coupled to the Tomonaga-Schwinger equation,  reduces to the Schrödinger-Newton equation \cite{bahrami_schrodingernewton_2014}
\begin{multline}
    \label{eqn:SNE}
    i \hbar \frac{\partial}{\partial t}\psi(t,\vec{x}) = \Bigg[-\frac{\hbar^2}{2m} \nabla^2 + V \\ - G m^2 \iiint_{\mathbb{R}^3} \frac{\abs{\psi(t,\vec{y})}^2}{\abs{\vec{x}-\vec{y}}} d^3 \vec{y}\Bigg] \psi(t,\vec{x}) \, ,
\end{multline}
i.e. yields a nonlinear modification to standard quantum theory. Nonlinear extensions to the Schrödinger equation have been considered from a foundational point of view, e.g. see \cite{mielnik_generalized_1974,bialynicki-birula_nonlinear_1976}. If nature is described by such nonlinear models, this would have not only important foundational repercussions but could also have major technological impact on computational and communication methods \cite{kent_quantum_2021}.

Other schemes to combine gravity with (relativistic) quantum fields exist, notably by Diósi \cite{diosi_gravitation_1984,diosi_universal_1987}, Penrose \cite{penrose_gravitys_1996}, Tilloy \cite{tilloy_sourcing_2016,tilloy_general_2024} and Oppenheim \cite{oppenheim_postquantum_2023,oppenheim_path_2023}. These typically take the form of dynamical objective collapse models to couple quantum fields with classical gravity through a noisy Schrödinger equation. 

\subsection{Some issues of earlier versions of semiclassical gravity}

Some of these stochastic objective-collapse models have been developed in light of the many objections to a semiclassical approach to gravity which have been raised over the years \cite{page_indirect_1981,hossenfelder_real_2012,tilloy_binding_2018,tilloy_does_2019,maudlin_status_2020,grossardt_three_2022}. We consider in this paper a version of Eqn.~\eqref{eqn:MR SCG} in which $G_{ab}$ is sourced by the expectation value of the (quantum) stress-energy tensor with respect to the local quantum state. The right-hand side of Eqn.~\eqref{eqn:MR SCG} should not be sourced by the \emph{global} quantum state $\Psi[\Sigma]$ (nor by a foliation-dependent reduced state): the equation is instead written for a specific alternative choice of the state that was not considered by M{\o}ller and Rosenfeld. We start by reviewing the objections against mean-field approaches and responding to some of them.

\subsubsection{The Page-Geilker experiment}

\begin{figure}
    \centering
    \begin{tikzpicture}[x=0.75pt,y=0.75pt,yscale=-1,xscale=1]

\draw   (198.67,58.33) -- (279.33,58.33) -- (279.33,98.33) -- (198.67,98.33) -- cycle ;
\draw    (279.33,78) .. controls (281.01,76.35) and (282.68,76.36) .. (284.33,78.03) .. controls (285.99,79.7) and (287.66,79.71) .. (289.33,78.05) .. controls (291,76.4) and (292.67,76.41) .. (294.33,78.08) .. controls (295.99,79.75) and (297.66,79.76) .. (299.33,78.1) .. controls (301,76.45) and (302.67,76.46) .. (304.33,78.13) .. controls (305.99,79.8) and (307.66,79.81) .. (309.33,78.15) .. controls (311,76.5) and (312.67,76.51) .. (314.33,78.18) .. controls (315.99,79.85) and (317.66,79.86) .. (319.33,78.2) .. controls (321,76.55) and (322.67,76.56) .. (324.33,78.23) .. controls (325.99,79.9) and (327.66,79.91) .. (329.33,78.25) .. controls (331,76.6) and (332.67,76.61) .. (334.33,78.28) .. controls (335.99,79.95) and (337.66,79.96) .. (339.33,78.3) .. controls (341,76.65) and (342.67,76.66) .. (344.33,78.33) -- (347.67,78.34) -- (355.67,78.38) ;
\draw [shift={(358.67,78.4)}, rotate = 180.29] [fill={rgb, 255:red, 0; green, 0; blue, 0 }  ][line width=0.08]  [draw opacity=0] (8.93,-4.29) -- (0,0) -- (8.93,4.29) -- cycle    ;
\draw   (358.67,58.33) -- (500,58.33) -- (500,98.33) -- (358.67,98.33) -- cycle ;
\draw    (441.87,99.07) -- (441.72,103.41) -- (441.44,111.4) ;
\draw [shift={(441.33,114.4)}, rotate = 271.99] [fill={rgb, 255:red, 0; green, 0; blue, 0 }  ][line width=0.08]  [draw opacity=0] (8.93,-4.29) -- (0,0) -- (8.93,4.29) -- cycle    ;
\draw   (386,115) -- (501.87,115) -- (501.87,155) -- (386,155) -- cycle ;
\draw   (200,116) -- (342.67,116) -- (342.67,226.4) -- (200,226.4) -- cycle ;
\draw    (268.67,115.73) -- (235.33,141.07) ;
\draw    (268.67,115.73) -- (298,141.07) ;
\draw    (235.33,141.07) -- (235.33,181.07) ;
\draw    (298,141.07) -- (298,181.07) ;
\draw    (386,134.4) .. controls (384.33,136.07) and (382.67,136.07) .. (381,134.4) .. controls (379.33,132.73) and (377.67,132.73) .. (376,134.4) .. controls (374.33,136.07) and (372.67,136.07) .. (371,134.4) .. controls (369.33,132.73) and (367.67,132.73) .. (366,134.4) .. controls (364.33,136.07) and (362.67,136.07) .. (361,134.4) .. controls (359.33,132.73) and (357.67,132.73) .. (356,134.4) -- (353.67,134.4) -- (345.67,134.4) ;
\draw [shift={(342.67,134.4)}, rotate = 360] [fill={rgb, 255:red, 0; green, 0; blue, 0 }  ][line width=0.08]  [draw opacity=0] (8.93,-4.29) -- (0,0) -- (8.93,4.29) -- cycle    ;
\draw  [dash pattern={on 0.84pt off 2.51pt}] (224.8,191.6) .. controls (224.8,185.78) and (229.52,181.07) .. (235.33,181.07) .. controls (241.15,181.07) and (245.87,185.78) .. (245.87,191.6) .. controls (245.87,197.42) and (241.15,202.13) .. (235.33,202.13) .. controls (229.52,202.13) and (224.8,197.42) .. (224.8,191.6) -- cycle ;
\draw  [dash pattern={on 0.84pt off 2.51pt}] (287.47,191.6) .. controls (287.47,185.78) and (292.18,181.07) .. (298,181.07) .. controls (303.82,181.07) and (308.53,185.78) .. (308.53,191.6) .. controls (308.53,197.42) and (303.82,202.13) .. (298,202.13) .. controls (292.18,202.13) and (287.47,197.42) .. (287.47,191.6) -- cycle ;
\draw    (344,181.33) .. controls (345.74,179.74) and (347.41,179.81) .. (349,181.55) .. controls (350.59,183.29) and (352.25,183.36) .. (353.99,181.77) .. controls (355.72,180.18) and (357.39,180.25) .. (358.99,181.98) .. controls (360.58,183.72) and (362.24,183.79) .. (363.98,182.2) .. controls (365.72,180.61) and (367.39,180.68) .. (368.98,182.42) -- (373.01,182.59) -- (381,182.94) ;
\draw [shift={(384,183.07)}, rotate = 182.48] [fill={rgb, 255:red, 0; green, 0; blue, 0 }  ][line width=0.08]  [draw opacity=0] (8.93,-4.29) -- (0,0) -- (8.93,4.29) -- cycle    ;
\draw   (383.33,162.33) -- (503.2,162.33) -- (503.2,226.4) -- (383.33,226.4) -- cycle ;

\draw (202.67,74.67) node [anchor=north west][inner sep=0.75pt]  [font=\scriptsize] [align=left] {Geiger counter};
\draw (366.67,73.33) node [anchor=north west][inner sep=0.75pt]  [font=\scriptsize] [align=left] {Reading by Page \& Geilker};
\draw (390,122) node [anchor=north west][inner sep=0.75pt]  [font=\scriptsize] [align=center] {Larger masses moved\\depending on count};
\draw (387.33,170.67) node [anchor=north west][inner sep=0.75pt]  [font=\scriptsize] [align=center] {Displacement of smaller\\masses observed to \\depend on larger \\masses' locations};

\end{tikzpicture}
    \caption{\justifying Setup of the Page-Geilker experiment. A Geiger counter generates some ``ticks'' with a certain probability. After some time, Page and Geilker read the Geiger counter, and decide whether to place a large mass on the left or the right of a torsion pendulum depending on the number of those ticks. Consequently, they observed that, contrary to a no-collapse approach to semiclassical gravity, the smaller masses depend on the larger masses' location instead of depending on the average gravitational field, thus falsifying Everettian semiclassical gravity.}
    \label{fig:Page-Geilker}
\end{figure}
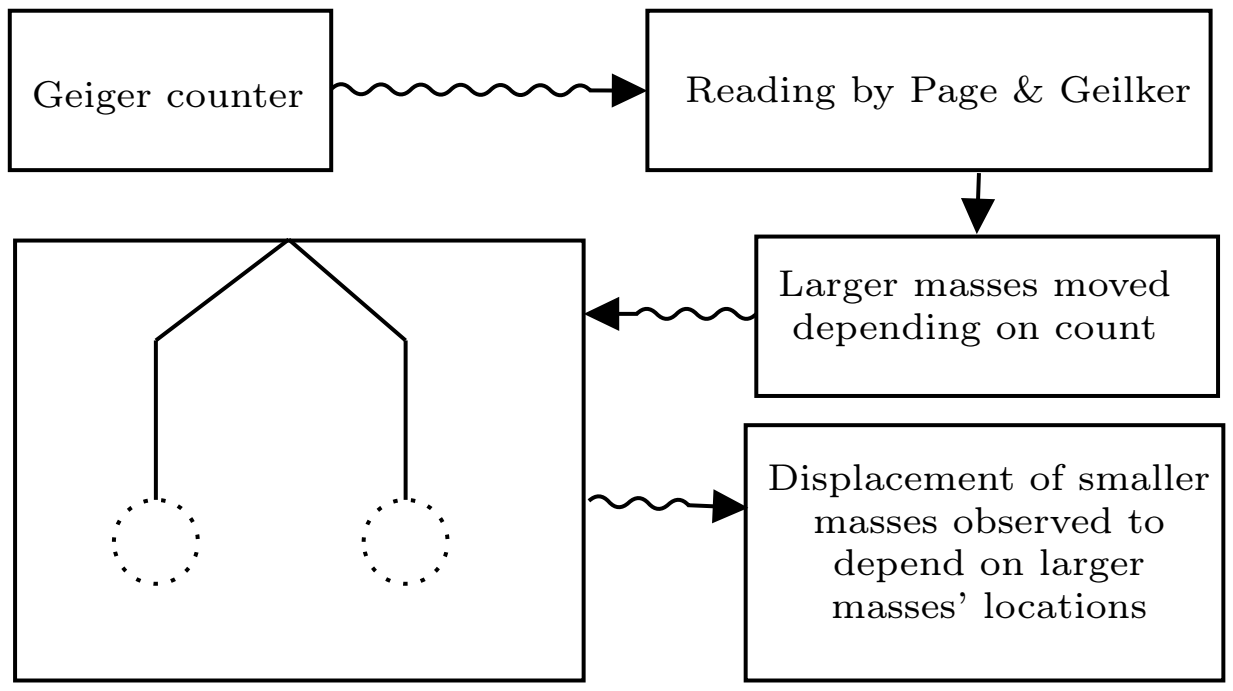

The original interpretation of Eqns.~\eqref{eqn:MR SCG} and \eqref{eqn:SNE} was that they imply a continuous evolution law for the quantum state, with no measurement postulate or intrinsic collapse hypothesis -- essentially an Everettian picture, though with a nonlinear evolution.    
It is not clear how to interpret this in the context of cosmology, where the expectation value is over a state associated with multiple different classical spacetimes.   If this makes sense at all, it seems to suggest that the gravitational fields we observe should be sourced by matter in many macroscopically different cosmologies, which suggests the gravitational field in (for example) the Solar System should be almost uncorrelated with the configuration of Sun and planets.   Even on the extremely unlikely hypothesis that the correct theory of cosmology is dominated by quantum states whose matter distribution is very similar to that of the heavenly bodies we observe, Page and Geilker showed \cite{page_indirect_1981} that the semiclassical equations \eqref{eqn:MR SCG} are falsified experimentally. The setup of their experiment is given in Fig.~\ref{fig:Page-Geilker}: the experimenters read out the outcome of a Geiger counter, decided whether to place a large mass at the left or right of a torsion pendulum, and carried out a Cavendish experiment to determine whether the gravitational field was sourced by the average position or by each individual ball position.   Unsurprisingly, the latter was experimentally verified, thus falsifying this version of no-collapse semiclassical gravity. 

This leaves the possibility of ``one-world'' theories in which semiclassical gravity is defined with respect to a quantum state that collapses upon quantum measurements or follows an intrinsic collapse law.   These are not immediately refutable by astronomical observation or the Page-Geilker experiment.  This line of argument has been discussed in, for example, Refs.~\cite{tilloy_sourcing_2016,tilloy_binding_2018, kent_testing_2021,grossardt_three_2022}. 
However, like ``many-worlds'' semiclassical gravity, ``one-world'' semiclassical gravity gives us only a formal set of equations and an intuition about their supposed implications in experiments or cosmology.  Work still needs to be done to decide whether these produce a well-defined theory at all, let alone one aligned with the intuitions.    

\subsubsection{Quantum measurements} 

An important issue about semiclassical Einstein field equations which, to our knowledge, has not been resolved previously in the literature, is that quantum measurements appear inconsistent with Eqn.~\eqref{eqn:MR SCG}. Indeed, while away from quantum measurements we have $\nabla^a \expval{\hat{T}^\mathrm{ren}_{ab}}{\Psi[\Sigma]} = 0$, quantum measurements typically change the value of $\expval{\hat{T}^\mathrm{ren}_{ab}}{\Psi[\Sigma]}$ discontinuously, meaning $\nabla^a \expval{\hat{T}^\mathrm{ren}_{ab}}{\Psi[\Sigma]} \neq 0$ and so the field equations do not hold. This issue is well-known \cite{diez-tejedor_towards_2012,hossenfelder_real_2012,maudlin_status_2020,juarez-aubry_initial_2023,mucino_fully_2025}.   For example, Page and Geilker argued that \cite{page_indirect_1981}
\begin{quote}
    ``[...] our theoretical arguments show that the semiclassical Einstein equations \eqref{eqn:MR SCG} are mathematically inconsistent if the matter wave function collapses arbitrarily during a measurement, and our experimental results show that these equations are inconsistent with nature (to a high confidence level) if the wave function does not collapse.''
\end{quote}
A way of resolving this issue, leading to the possibility of a consistent version of semiclassical gravity with quantum state collapse, in the context of ``standard'' relativistic quantum measurements, will be proposed in Sec. \ref{sec:measurements and divergence}.\footnote{We note that if quantum state collapses arise dynamically from the nonlinear equations, this may not be an issue in the first place. There have been claims \cite{diosi_gravitation_1984,penrose_gravitys_1996,penrose_quantum_1998,penrose_gravitization_2014} that the Schrödinger-Newton equation, which can be recovered as a nonrelativistic limit of M{\o}ller-Rosenfeld semiclassical gravity (and, as it turns out, the version of semiclassical gravity we discuss in this paper), can at least partly explain the collapse of the wavefunction. Such claims have been contested in \cite{bahrami_schrodingernewton_2014}.
We do not find these counterarguments absolutely definitive and are not aware of a completely conclusive refutation. 
However, the claim relies on the Born rule effectively emerging from nonlinear (and perhaps chaotic) dynamics and initial conditions, which would seem surprising (though there exist claims that this might happen in other models of nonlinear quantum mechanics \cite{geszti_nonlinear_2019}), and we are not aware even of a convincing plausibility argument as to why this should be true in general.  The burden of proof clearly lies with proponents.}

\subsubsection{Thermodynamics and information} It is sometimes argued \cite{terno_classical-quantum_2024} that models of nonlinear quantum mechanics of the form discussed here (e.g. the Schrödinger-Newton system, which is the nonrelativistic limit of M{\o}ller-Rosenfeld semiclassical gravity \cite{bahrami_schrodingernewton_2014}) are thermodynamically unstable, in that they typically violate the second law of thermodynamics. These claims rely on three arguments, first discussed by von Neumann \cite{neumann_mathematical_1955}, Peres \cite{peres_quantum_1993} and Hänggi and Wehner \cite{hanggi_violation_2013}, respectively. 

Broadly speaking, von Neumann's and Peres' arguments are that the (in principle) perfect distinguishability of non-orthogonal states in such nonlinear extensions of quantum theory allows one to violate the second law of quantum thermodynamics. However, the notion of thermodynamical entropy changes in such nonlinear theories, so that it is not possible to extract work from this perfect distinguishability in semiclassical gravity \cite{fedida_thermodynamics_2026}.

While norms are preserved under wavefunction evolution, inner products are not \cite{mielnik_mobility_1979,kibble_is_1981}. This, as well as the existence of readout devices, points towards violations of the uncertainty principle.

The argument by Hänggi and Wehner is based on the violation of the uncertainty principle whereby, under certain assumptions, it implies a violation of a generalised second law for the theory.  Crucially, most of the assumptions for this claim fail to hold in semiclassical gravity, and a detailed treatment of the setup shows no net work extraction from the thermodynamical cycle. A rebuttal to these claims in the context of readout devices and semiclassical gravity was given in more detail in \cite{fedida_thermodynamics_2026}.


\subsubsection{Faster-than-quantum computation}
    Another argument raised against semiclassical gravity in the literature has to do with the computing power allowed by readout devices: black boxes which are not present in quantum theory but which can be constructed in semiclassical gravity as we review below. In particular, it was argued in \cite{fox_semiclassical_2026} that semiclassical gravity can efficiently solve NP-hard problems. This violates the Physical Extended Church-Turing Thesis (PECTT) \cite{deutsch_quantum_1985}, which conjectures that no physical procedure should be able to decide an NP-complete problem in polynomially many steps. 

    The fact that semiclassical gravity can efficiently compute NP-hard problems was already explored in \cite{kent_quantum_2021,kent2025fundamental}, and is a specific example of an older argument by Abrams and Lloyd \cite{abrams_nonlinear_1998} that nonlinear quantum mechanics can be exploited to get polynomial time solutions for NP-complete and \#P problems.
    It is certainly true that not all post-classical theories of computation can efficiently perform such tasks. For example, quantum theory is not believed to be able to solve NP-complete problems in polynomial time \cite{aaronson_space_2014}. 
    The authors of \cite{fox_semiclassical_2026} view the violation of the PECTT as an argument against semiclassical gravity, and more generally ``regard [it] as evidence for the quantization of gravity''. We disagree for several reasons.   
    
    First, we see no reason to reject any physical theory on the grounds of its computational power, even if it violates the PECTT.   We agree of course that it is interesting to try to characterise the information-theoretic properties of quantum theory and other physical theories, and to investigate axiomatisations of these theories that include information-theoretic principles.  But quantum theory and general relativity are considered (within their domains) outstandingly successful scientific theories because they explain empirical data successfully and parsimoniously.   Loosely speaking, they provide enormous data compression, in the sense that a few postulates explain and predict a wealth of empirical data.   These postulates have not, historically, included postulates about computational complexity.    If it were possible to reformulate quantum theory and/or general relativity equally or more elegantly by introducing computational complexity postulates -- which we believe remains unproven -- this would be interesting, but it would not change the fact that the theories are outstandingly successful {\em as currently formulated}.     
    
    We find it natural to see the computational power of a physical theory as something to be derived from the theory and the computational power of the universe as something to be determined empirically.    If it were the case (which, as we argue below, remains questionable) that a realistic semiclassical gravity theory would allow us to decide NP-complete problems in polynomially many steps, our main conclusion would be that there is extremely strong motivation to test for semiclassical gravity, as the technological consequences of living in a world governed by such a theory could be revolutionary \cite{kent_fundamental_2025}.  In short, there is nothing intrinsically problematic (nor arguably surprising) in the possibility that theories beyond classical and quantum physics could alter the complexity hierarchy or even collapse it dramatically. 
    In contrast, for example, a version of quantum theory that violated the no-superluminal signalling principle would be inconsistent with the existing formulations of special and general relativity, so that taking the former seriously as a fundamental theory would require, at the very least, revising our formulation of another very successful fundamental theory.    This is why theories that allow superluminal signalling are generally disfavoured (although not dogmatically rejected).     

    Second, the true computational power of semiclassical gravity remains to be understood \cite{fedida_mixture_2025}.  It depends \textit{inter alia} on how the resources required to measure gravitational fields scale with precision, allowing for background noise -- including background gravitational fields from all sources -- and device errors and limitations. If semiclassical gravity were correct, resolving this issue would require a combination of theoretical and empirical analyses, including a detailed understanding of the quantum state of matter and the gravitational field over extended regions.     
    
    Third, it is important to stress here that there are already compelling arguments from cosmology and terrestrial experiment \cite{page_indirect_1981} against semiclassical gravity as a universal theory with deterministic evolution laws for quantum matter and the gravitational field.     Analyses of computational power can thus only be relevant to versions of semiclassical gravity that involve extra hypotheses, most pertinently some form of collapse or measurement postulate.  The above issues of precision, noise and background then very much depend on the details and on the (as yet unknown) fundamental underlying theory.    
    
    We also note there are interesting lines of thought (e.g. \cite{diosi_gravitation_1984,tilloy_binding_2018,tilloy_does_2019,oppenheim_postquantum_2023}) about alternatives to quantum gravity that do not involve any type of semiclassical gravity theory in the usual sense of a theory in which the semiclassical Einstein equations hold, which is the focus of the analysis of \cite{fox_semiclassical_2026}.   Thus, even if the premises and analysis of \cite{fox_semiclassical_2026} were accepted at face value, they would not be compelling evidence for the quantization of gravity (although in a Bayesian sense they would constitute some evidence, since they would reduce credence in an alternative).

\subsubsection{Statistical interpretation}

Another peculiarity of models based on Eqn.~\eqref{eqn:MR SCG} is that the state of a subsystem is not fully described by the partial trace of the global quantum state of the joint system. In Tilloy's words \cite{tilloy_binding_2018},
\begin{quote}
    ``The essence of the difficulty [...] is that reduced states cannot be trusted, \emph{even approximately}. To make predictions, Alice needs to know \emph{precisely} what Bob did and more generally everything that happened to whatever shared a common past with her. Reduced states (i.e. typically the only objects one can consider in practice) have no statistical interpretation. Extracting predictions from the theory is a daunting if not impossible task.''
\end{quote}

A similar argument was also expressed in \cite{tilloy_does_2019}. The idea is the following: in a nonlinear theory, such as the Schrödinger-Newton system, in which the nonlinearity arises from the \emph{global} state, the evolution of subsystems becomes dependent on the entire system, the information of which may not be accessible to subsystems. This makes reduced states carry incomplete information for making predictions.

The response to this argument in the framework we propose is that the nonlinearities arise from the \emph{local states} rather than the \emph{global states} \cite{kent_nonlinearity_2005}. We will expand on the construction of local state nonlinearity in Sec. \ref{sec:local quantum states}.

The state space of the theory is not simply that of quantum theory anymore. The \emph{state} of a subsystem is not fully described by the partial trace of the global quantum state on a given spacelike Cauchy hypersurface. 
Models extending quantum theory in this way were introduced in \cite{kent_nonlinearity_2005} and discussed further in \cite{kent_simple_2018,kent_quantum_2021,kent_testing_2021-1,fedida_mixture_2025,fedida_thermodynamics_2026,kent_measurement_2025}.    
Observers localised in spacetime do not have access to the reduced states of the global quantum state (and thus do not have access to the global state information), but rather have access to the local quantum state that characterises the locally accessible information.  A similar point has recently been made under the name of the ``polyperspective formalism'' \cite{polo-gomez_state_2025}, which we will review in Sec. \ref{sec:mmts}.  

These concerns are ultimately linked with violations of the Mixture Equivalence Principle (MEP) \cite{fedida_mixture_2025}, whereby proper (statistical) and improper (subsystems) mixtures with the same density matrix yield different measurement statistics. This is an interesting feature of such theories, but
does not {\it per se} make them inconsistent nor intrinsically problematic.

\subsubsection{Superluminal signalling} 

It has been argued that  nonlinear modifications of quantum theory that are not stochastic allow faster than light signalling \cite{gisin_stochastic_1989}, essentially because (in our language) they violate the MEP.

The idea is the following: suppose one has access to two proper mixtures $\rho_1(t) = \sum_{i=1}^N p^{(1)}_i \ketbra{\psi_i}$ and $\rho_2(t) = \sum_{i=1}^N p^{(2)}_i \ketbra{\varphi_i}$, $\sum_{i=1}^N p_i^{(1)} = \sum_{i=1}^N p_i^{(2)} = 1$, such that, at time $t=t_0$, the mixtures are equal:
\begin{equation}
    \rho_1(t_0) = \rho_2(t_0) \, .
\end{equation}
Then, in general for nonlinear Schrödinger dynamics \cite{gisin_stochastic_1989},
\begin{equation}
    \rho_1(t_1) \neq \rho_2(t_1)
\end{equation}
for $t_1 > t_0$. That is, the MEP \cite{fedida_mixture_2025} is violated in nonlinear extensions of the Schrödinger equation. These density matrices are then distinguishable, and Gisin showed \cite{gisin_stochastic_1989} that any two mixtures corresponding to the same density operator can be prepared at a distance thanks to EPR-like correlations, and thus naively defined nonlinear Schrödinger dynamics allow signalling.

However, with an appropriate treatment of quantum measurements and nonlinearities, this argument fails, as was highlighted in \cite{kent_nonlinearity_2005,kent_quantum_2021}. A key point is that one \emph{expects} signalling at a distance to be possible in a nonrelativistic limit of a causal theory ($c \to \infty$). So, a more fundamental theory may be explicitly relativistically causal yet lead to a nonlinear Schrödinger equation in the nonrelativistic limit. In the treatment we advocate, nonlinearities arise from \emph{local quantum states} \cite{kent_nonlinearity_2005}: states understood to be ``on'' the past light cone at a point. Light cones ``flatten'' in the nonrelativistic limit, so if a \emph{global} quantum state evolution depends on \emph{local quantum states} there can be nonlinearities in the quantum state in the nonrelativistic limit of a relativistically causal model.    This construction will be reviewed in Sec. \ref{sec:local quantum states}.

Finally, in a controversial paper whose arguments and conclusions have been refuted \cite{huggett_why_2001,mattingly_why_2006,albers_measurement_2008,kent_simple_2018}, Eppley and Hannah \cite{eppley_necessity_1977} wrongly claimed (amongst other things) that if gravity was not quantum, then there would exist ways to communicate faster-than-light. Ref.~\cite{kent_simple_2018} uses the ideas just described to simply refute those claims. 

\subsubsection{Arbitrariness of semiclassical gravity} Another commonly raised point against semiclassical gravity is the arbitrariness of the assumptions of the theory. Indeed, why should one postulate that classical gravity couples to quantum fields through some expectation value? For example, Tilloy argued \cite{tilloy_binding_2018} that a semiclassical gravity of the form of Møller and Rosenfeld is arbitrary:
\begin{quote}
    ``In my opinion, trying to redeem it is hopeless and unnecessary: after all, the prescription of Møller and Rosenfeld (making expectation values gravitate) is completely arbitrary.''
\end{quote}
One could however argue for the contrary: the Banach space isomorphism $\mathcal{T}(\mathcal{H}) \cong \mathcal{B}(\mathcal{H})_*$ makes expectation values canonical (e.g. basis independent) for recovering a $c$-number from a $q$-number\footnote{That is, every trace-class operator (and, in particular, every quantum state) $\chi \in \mathcal{T}(\hi)$ defines a unique continuous linear functional $\omega_\chi : \bh \to \mathbb{C}$ that maps an observable $A \in \bh$ to an ``expectation value'' (strictly speaking, it is only interpreted as such for quantum states): $\omega_\chi(A) = \Tr[\chi A]$. This is basis independent. Note however that the quantum stress-energy tensor may typically be unbounded. In such cases one may consider the duality between $O^*$-algebras of unbounded operators \cite{schmudgen_spaces_1990} and continuous linear functionals (algebraic states) on them.}. In fact, our prescription is arguably even more natural than that of Møller and Rosenfeld: they could be criticised for a ``category error'' in assuming that the \emph{local} gravitational field backreacts from the \emph{global state} $\Psi[\Sigma]$ (or a foliation-dependent reduced state), rather than asserting that it backreacts from the \emph{local state} as we do.

Other approaches coupling classical gravity to quantum fields through stochastic feedback, such as the models of gravitationally-induced collapse first proposed by Diósi \cite{diosi_universal_1987,diosi_gravitation_1984}, Penrose \cite{penrose_gravitys_1996} and Tilloy \cite{tilloy_sourcing_2016,tilloy_binding_2018,tilloy_general_2024}, and later by Oppenheim \cite{oppenheim_postquantum_2023,oppenheim_covariant_2026,oppenheim_path_2023}, are independently interesting.   Arguably, they are more convoluted, in that they require constraints on the noise from its construction \cite{oppenheim_constraints_2022}, even before any empirical test \cite{oppenheim_gravitationally_2023}, as well as introducing additional fundamental parameters, which arguably makes the coupling of quantum fields with gravity less ``minimal''.   That said, there are doubtless arguments and counter-arguments in both directions.   
Our aim in this paper is to motivate and develop a new approach, not to provide a critique of others. 

In the context of Algebraic QFT (AQFT), it has also been recently claimed \cite{dorau_quantum_2026} that, at least in certain regimes, some thermodynamical arguments\footnote{These are based on quantum thermodynamics
rather than the generalised thermodynamics discussed in Ref. \cite{fedida_thermodynamics_2026}.} imply the semiclassical Einstein field equations.  As we understand it, the claim is that these hold with no
$\mathcal{O}(\hbar)$ corrections (i.e. not as a semiclassical limit of quantum gravity), although perhaps with higher gravitational (e.g.~$R^2$) corrections.   If valid, this would be a quantum field theoretic analogue of Jacobson's derivation of the classical Einstein field equations \cite{jacobson_thermodynamics_1995}. Although linking algebraic quantum field theoretic considerations of semiclassical gravity with the version of semiclassical gravity presented in this paper is an open problem, this opens the interesting possibility of recovering semiclassical Einstein field equations from deeper principles, rather than being an ``arbitrary'' choice made \textit{ad hoc}.

Sourcing the local gravitational field at a point through the local quantum state at that point is arguably very natural, respects causality and is parameter-free in its simplest form.
Moreover, table-top experiments can test the theory against perturbative quantum gravity and other post-quantum approaches to gravitational quantum physics, as we shall further explore in Sec. \ref{sec:falsifiability}. The possible technological consequences that could open up should such a theory be (even approximately) correct, e.g. through computational speed-ups via readout devices \cite{kent_quantum_2021,kent_fundamental_2025}, also makes it an important avenue to explore from the perspectives of applied physics, novel astronomical methods, potential novel forms of computation, potential novel attacks on quantum cryptography, and possibly fundamental questions about the future development of intelligence.

This paper presents a framework for semiclassical gravity that addresses several standard objections. A number of questions remain open, see Sec.~\ref{sec:conclusion}.


The theory we develop involves some ideas related to those of Ref.~\cite{mucino_fully_2025} but has important differences.  Several open conceptual and technical questions remain to be answered in the approach of \cite{mucino_fully_2025}; these include: 
\begin{enumerate}
    \item providing an explicit realistic example of a suitable quantum field theoretic version of dynamical objective collapse models. In the model considered in the present paper, we instead take a ``Copenhagenist'' view on quantum measurements and do not need to introduce additional parameters for wavefunction collapse.  (We agree with the authors of Ref. \cite{mucino_fully_2025} that there is strong motivation for combining explicit dynamical collapse models with semiclassical gravity, and this also motivates our work.   Our point is that our techniques also apply even to the seemingly more technically difficult project of combining Copenhagen quantum measurements with semiclassical gravity.)

    \item discussing the integrability of the relativistic quantum field equations in the presence of nonlinearities. We show that the Tomonaga-Schwinger equation still satisfies the integrability condition despite nonlinearities. This is nontrivial. It would be important to understand whether such conditions also hold in relativistic versions of objective-collapse models.
    
    \item a conceptual and technical analysis of the ``glued spacetimes'' they introduce: a series of $C^2$ spacetimes ``glued'' together along the future light cones of the quantum measurements. The universe would then be such a ``gluing'' of spacetimes, which poses a number of issues, including 
    \begin{enumerate}
        \item how to understand clocks, proper times, causality, and relative localisations across different $C^2$ spacetimes.
        \item a precise understanding of the initial value problem in this nonstandard geometric setting.
    \end{enumerate}
    In our model, we manage to accommodate quantum measurements by working in a single $C^{0,1}_{\mathrm{loc}}$ spacetime, which is arguably conceptually and technically much simpler, and in fact follows directly from the fact that the right-hand side of the Einstein field equations can be distributional since quantum fields are. Arguably, one should expect the consideration of distributional solutions to the semiclassical Einstein field equations to also be present in an objective collapse model setting: quantum fields are operator-valued distributions even away from quantum measurements, so if one rejects distributional solutions to the field equations on the basis of unphysicality one should also discuss the domain of definition of the field equations away from quantum measurements. 
    
    \item an understanding of the well-posedness of the initial value problem assumed in Ref.~\cite{mucino_fully_2025} (that the induced metrics and characteristic data match on both sides of the collapse hypersurface). 
    We will retrieve in Sec.~\ref{sec:measurements and divergence} a locally Lipschitz metric on the whole spacetime, whereas the model outlined in Ref.~\cite{mucino_fully_2025} is based on the rejection of the continuity of the metric across the collapse hypersurface.
    
    For this reason, we expect they should end up with a distributional geometry based on the Geroch-Traschen class of metrics \cite{geroch_strings_1987} (see Sec.~\ref{sec:distributional geometry}) for which the Barrabès-Israel formalism can still apply, but which presents many exotic behaviours. For example, as reviewed in Sec.~\ref{sec:C0 causal theory}, even $C^0$ spacetimes which are not Lipschitz can be bubbling and need not have the push-up property, which is an important feature for the characteristic initial value problem (reviewed in Sec.~\ref{sec:measurements and divergence}) to be well-posed.
    
    It is still an open question, both in \cite{mucino_fully_2025} and in our work, to understand whether the induced metrics are continuous across the collapse hypersurface for the full nonlinear semiclassical Einstein field equations.  However, we show in the present paper that it holds in the linearised limit in our model.
    
    \item We discuss both quantum and non-quantum measurements, showing how causality is maintained explicitly. The authors of \cite{mucino_fully_2025} do not discuss non-quantum measurements, yet claim their theory is ``fully self-consistent'' without considering this key point.  Unless one is very careful with the prescription for non-quantum measurements, this leads to superluminal signalling \cite{gisin_weinbergs_1990}.
\end{enumerate}
In summary, we here propose a new theory, which shares similarities with \cite{mucino_fully_2025} but also has new features and represents an independent proposal.

\subsection{Organisation of the paper}

The paper is structured as follows. In Sec.~\ref{sec:preliminaries}, we recall standard notions of distributional and $C^0$ Lorentzian geometry, as well as the basic theory of operators on Hilbert spaces and Hilbert bundles.

In Sec.~\ref{sec:local quantum states}, we review the notion of local quantum states in spacetime and build the appropriate Hilbert bundle from which such states arise.

In Sec.~\ref{sec:semiclassical equations}, we make use of these local states to define the semiclassical field equations, as well as the evolution equations of the global wavefunctionals. We discuss the distributional nature of these equations and review the integrability condition in this nonlinear context.

In Sec.~\ref{sec:MEP}, we review the mixture equivalence principle and discuss the difference of treatment between proper and improper mixtures in this semiclassical theory. We show that proper mixtures give rise to statistical ensembles of spacetimes, while improper mixtures source the local gravitational field in a single spacetime.

In Sec.~\ref{sec:limits} we discuss some of the limits of the theory, notably the weak gravitational field limit, which displays the explicitly causal and retarded features of the gravitational field. We show that the Schrödinger-Newton system is recovered in the nonrelativistic limit.

In Sec.~\ref{sec:mmts} we explore different aspects of measurements in this semiclassical model. We first analyse the different possibilities for quantum state collapse, and argue that within this semiclassical theory the only sensible approach to collapse is to consider future light-cone updates. 

In Sec.~\ref{sec:measurements and divergence}, we show that a collapse with future light-cone updates need not violate the distributional Einstein field equations by exploring the Barrabès-Israel null-shell formalism, which allows one to join two patches of spacetime with different stress-energy tensors across a null hypersurface (the future light cone of a collapse point).  To implement this, we postulate an additional source term, singular on the future light cone of a collapse point.

In Sec.~\ref{sec:examples}, we provide explicit examples for the computation of the gravitational field following quantum measurements.   We first discuss the case of a single mass in a spatial superposition whose location is measured, effectively collapsing it to one component of the superposition.  
We then consider separate measurements of two masses in a bipartite entangled state in position space,
separately localized in spacelike separated regions.

In Sec.~\ref{sec:relativistic readout devices}, we examine how classical gravitational field measurements are an instance of an exotic extension of quantum theory, defining a version of so-called ``readout devices'' \cite{kent_causal_2005,kent_quantum_2021,kent_fundamental_2025,fedida_mixture_2025,fedida_thermodynamics_2026}, which are hypothetical devices that can consistently be introduced as extensions of quantum theory without allowing superluminal signalling.   These devices provide classical information about a quantum state without altering it, are more general than cloning devices; yet are consistent with relativistic causality.

In Sec.~\ref{sec:nsp}, we explore different versions and variations of the no-superluminal signalling principle in quantum theory and the quantum no-signalling principle and extensions to more general theories, including the semiclassical model presented in this paper. We also highlight that the gravitational field is generally Bell nonlocal.


In Sec.~\ref{sec:falsifiability}, we explore different tests of the theory. We show that violations of the mixture equivalence principle can lead to experimental signatures; we highlight that the power spectrum of gravitational waves can have features distinct from those of quantum gravity; we review the predictions of our semiclassical model in the context of gravity mediated entanglement experiments; and finish by arguing that easier proposals testing our model can be carried out in a tabletop experiment that is implementable in the nearer future.

We finish in Sec.~\ref{sec:conclusion} by arguing that the model we develop in this paper appears a serious contender for a gravitational theory with quantum fields, while pointing out important open questions that need to be addressed.

\section{Preliminaries and notation}

\label{sec:preliminaries}

We first review some notions of differential geometry and of functional analysis, and clarify the notation used in the rest of the paper.

\subsection{Lorentzian geometry}

In this paper we work with $1+3$-dimensional (classical) Lorentzian spacetimes in $(-,+,+,+)$ signature. We write $d\mu_g$ or $\sqrt{\abs{g}} d^dx$ for the volume form on $\mathcal{M}$ of a spacetime $(\M,g)$. Throughout this paper, $\M$ denotes a connected, oriented, $C^\infty$ differentiable (smooth) manifold.

\subsubsection{Distributions}

\label{sec:distributions}

We now review the basics of the theory of distributions. We write $\mathbb{F} = \mathbb{R}$ or $\mathbb{C}$.

We denote by $\mathcal{D}(\mathcal{M},\mathbb{F})\equiv C^\infty_c(\mathcal{M},\mathbb{F})$ compactly supported $\mathbb{F}$-valued smooth functions on $\mathcal{M}$; if the context is clear $\mathbb{F}$ will be omitted. We write $\mathcal{D}'(\mathcal{M},\mathbb{F})$ for the space of $\mathbb{F}$-valued distributions, i.e. the space of continuous linear maps $\mathcal{D}(\mathcal{M},\mathbb{F}) \to \mathbb{F}$. If $\mathcal{M} = \mathbb{R}^4$, we write $\mathscr{S}(\mathbb{R}^4,\mathbb{F})$ for the space of Schwartz functions: smooth functions whose derivatives of all orders are rapidly decreasing at infinity, with $\mathcal{D}(\mathbb{R}^4,\mathbb{F}) \subset \mathscr{S}(\mathbb{R}^4,\mathbb{F})$ and thus $\mathscr{S}'(\mathbb{R}^4,\mathbb{F}) \subset \mathcal{D}'(\mathbb{R}^4,\mathbb{F})$.

Distributions allow for the differentiation of functions whose derivatives do not exist in the standard sense. If $F \in \mathcal{D}'(\mathcal{M},\mathbb{F})$, then the \emph{distributional derivative} is defined as
\begin{equation}
    (\partial_\mu F)(f) := -F(\partial_\mu f) \qquad \forall f \in \mathcal{D}(\mathcal{M},\mathbb{F}).
\end{equation}

For an open $\U \subset \mathcal{M}$ and $p \in \mathbb{N}$, the space $L^p_\mathrm{loc}(\U,\mathbb{F})$ is defined as
\begin{multline}
    L^p_\mathrm{loc}(\U,\mathbb{F}) := \Bigg\{f : \U \to \mathbb{F} \text{ measurable} \mid \\\norm{f}_p \equiv \left(\int_K \abs{f(x)}^p d\mu_g(x)\right)^{1/p} < \infty \\ \forall \text{ compact } K \subset \U  \Bigg\} / \sim_{\mu_g \text{ a.e.}}.
\end{multline}
For example, the space $L^2_{\mathrm{loc}}(\U,\mathbb{F})$ is defined as the space of locally square-integrable $\mu_g$-measurable functions, here taken modulo equality $\mu_g$ almost everywhere:
\begin{multline}
    L^2_\mathrm{loc}(\U,\mathbb{F}) := \Bigg\{f : \U \to \mathbb{F} \text{ measurable} \mid \\\int_K \abs{f(x)}^2 d\mu_g(x) < \infty \\ \forall \text{ compact } K \subset \U  \Bigg\} / \sim_{\mu_g \text{ a.e.}}.
\end{multline}
Similarly, the space $L^\infty_\mathrm{loc}(\U,\mathbb{F})$ is defined as the space of essentially bounded $\mu_g$-measurable functions, again taken modulo equality $\mu_g$ almost everywhere:
\begin{multline}
    L^\infty_\mathrm{loc}(\U,\mathbb{F}) := \Bigg\{f : \U \to \mathbb{F} \text{ measurable}\mid \\ \mathrm{ess \; sup }_{x \in K} \abs{f(x)} < \infty \\ \forall \text{ compact } K \subset \U  \Bigg\} / \sim_{\mu_g \text{ a.e.}}.
\end{multline}
An important example of an $L^\infty_\mathrm{loc}$ function is the Heaviside $\Theta$ function
\begin{equation}
    \Theta(y) = \begin{cases}
        1 \qquad \text{if } y \geq 0, \\
        0 \qquad \text{otherwise}.
    \end{cases}
\end{equation}
Note that the value $\Theta(0)$ changes in different conventions, but this is unimportant for $L^\infty_\mathrm{loc}$ functions: a single point has measure zero. The distributional derivative of a Heaviside $\Theta$ function is a Dirac delta:
\begin{equation}
    \Theta'(x) = \delta(x),
\end{equation}
defined distributionally as
\begin{equation}
    \delta_x(f) = f(x), \qquad f \in \mathcal{D}(\mathcal{M}).
\end{equation}

For $p \in \mathbb{N} \cup \{\infty\}$ and $k \in \mathbb{N}$, the \emph{Sobolev spaces} are defined as
\begin{multline}
    W^{k,p}_\mathrm{loc}(\U,\mathbb{F}) := \{f \in L^p_\mathrm{loc}(\U,\mathbb{F}) \mid \\f^{(j)} \in L^p_\mathrm{loc}(\U,\mathbb{F}) \; \forall j \leq k \}.
\end{multline}
For $p=2$, these spaces are written $W^{k,2}_\mathrm{loc} \equiv H^k_\mathrm{loc}$:
\begin{equation}
    H^k_\mathrm{loc}(\U,\mathbb{F}) := \left\{f \in L^2_\mathrm{loc}(\U,\mathbb{F}) \mid f^{(j)} \in L^2_\mathrm{loc}(\U,\mathbb{F}) \; \forall j \leq k \right\}.
\end{equation}
$H^k_\mathrm{loc}(\U,\mathbb{F})$ is the space of functions for which every distributional derivative to order $k$ is locally square-integrable. 
Taking $2k$ distributional derivatives of functions in $H^k_\mathrm{loc}(\U,\mathbb{F})$ yields a distribution in a space denoted $H^{-k}_{\mathrm{loc}}(\U,\mathbb{F})$.

Let $h$ be a smooth complete Riemannian metric on $\mathcal{M}$. A function $f : \mathcal{M} \to \mathbb{R}$ is \emph{locally Lipschitz continuous} if for every point $p \in \M$ there exists a neighbourhood $\U \subset \M$, there exists a $C_\U \in \mathbb{R}_+$ such that for any $x,y \in \U$,
\begin{equation}
    \abs{f(x)-f(y)} \leq C_\U d_h(x,y).
\end{equation}
The definition of a locally Lipschitz function is independent of the choice of Riemannian metric $h$. The space of locally Lipschitz functions is denoted $C^{0,1}_\mathrm{loc}$. Every locally Lipschitz continuous function is $L^\infty_\mathrm{loc} \cap H^1_\mathrm{loc}$, and in particular $C^{0,1}_\mathrm{loc} = W^{1,\infty}_\mathrm{loc}$ \cite[Theorem 4.5]{evans_measure_2015}: the distributional derivative of a locally Lipschitz function is $L^\infty_\mathrm{loc}$.

\subsubsection{$C^0$ causal theory}

\label{sec:C0 causal theory}

\begin{figure}
    \centering
    \begin{tikzpicture}[x=0.75pt,y=0.75pt,yscale=-1,xscale=1]

\draw    (275.73,251.6) .. controls (303.07,219.6) and (277.07,160.93) .. (261.33,137) ;
\draw    (261.33,137) -- (261.07,102.93) ;
\draw    (259.07,55.6) .. controls (251.07,64.93) and (245.73,96.93) .. (261.07,102.93) ;
\draw    (384.4,256.27) .. controls (335.73,240.93) and (326.4,160.27) .. (376.4,120.27) ;
\draw  [dash pattern={on 4.5pt off 4.5pt}]  (376.4,120.27) -- (313.73,174.27) ;
\draw  [dash pattern={on 4.5pt off 4.5pt}]  (376.4,120.27) -- (435.07,172.93) ;
\draw    (376.4,120.27) -- (375.73,46.27) ;

\end{tikzpicture}
    \caption{\justifying Both of these curves are causal curves in two-dimensional Minkowski space. The curve on the left is a timelike curve. The curve on the right is a timelike almost everywhere curve. It is not a timelike curve under our definition \cite{ling_aspects_2020v2} since it approaches a null vector at its break point, so the curve’s timelike tangents approach null closely enough near the breakpoint that uniform timelikeness fails.\cite{ling_aspects_2020v2}}
    \label{fig:timelike curve c0}
\end{figure}
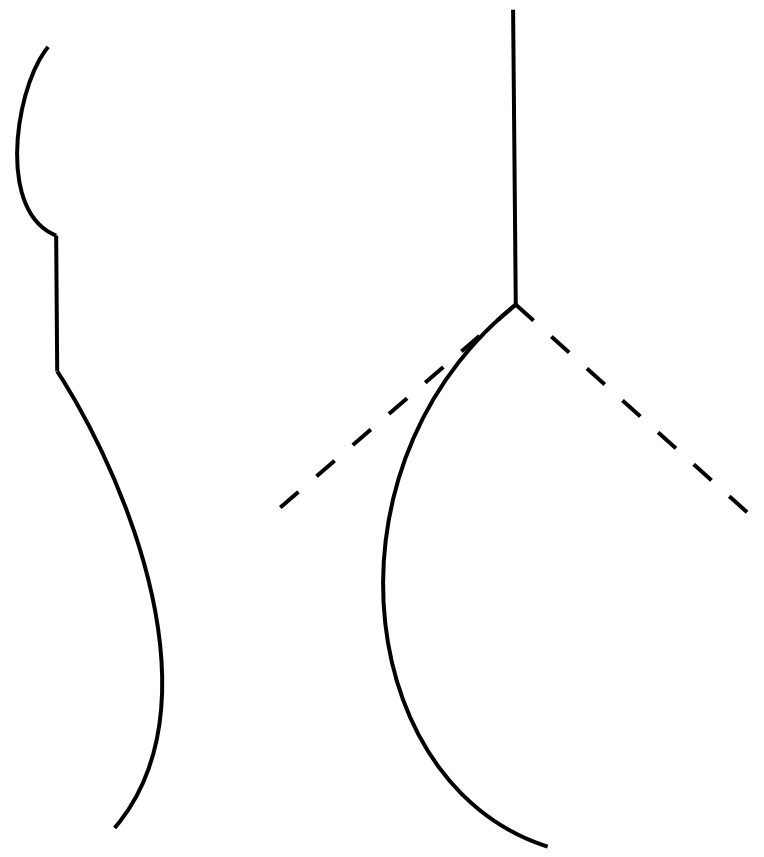

\begin{figure}
    \centering
    \begin{tikzpicture}[x=0.75pt,y=0.75pt,yscale=-1,xscale=1]

\draw  [fill={rgb, 255:red, 0; green, 0; blue, 0 }  ,fill opacity=1 ] (213.7,195.03) .. controls (213.7,192.67) and (215.72,190.75) .. (218.21,190.75) .. controls (220.7,190.75) and (222.71,192.67) .. (222.71,195.03) .. controls (222.71,197.39) and (220.7,199.3) .. (218.21,199.3) .. controls (215.72,199.3) and (213.7,197.39) .. (213.7,195.03) -- cycle ;
\draw [fill={rgb, 255:red, 155; green, 155; blue, 155 }  ,fill opacity=0.38 ] [dash pattern={on 4.5pt off 4.5pt}]  (363.19,58.4) .. controls (356.18,79.3) and (218.46,193.93) .. (218.21,195.03) .. controls (217.96,196.12) and (335.15,168.57) .. (367.2,122.04) ;
\draw    (409.2,109.07) .. controls (401.64,90.17) and (388.13,89.13) .. (371.59,90.14) ;
\draw [shift={(368.67,90.33)}, rotate = 355.94] [fill={rgb, 255:red, 0; green, 0; blue, 0 }  ][line width=0.08]  [draw opacity=0] (8.93,-4.29) -- (0,0) -- (8.93,4.29) -- cycle    ;
\draw    (321.87,203.73) .. controls (316.83,193.65) and (315.36,185.95) .. (305.2,173.27) ;
\draw [shift={(303.33,171)}, rotate = 49.85] [fill={rgb, 255:red, 0; green, 0; blue, 0 }  ][line width=0.08]  [draw opacity=0] (8.93,-4.29) -- (0,0) -- (8.93,4.29) -- cycle    ;
\draw    (260.53,95.73) .. controls (262.41,107.64) and (273.72,115.42) .. (286.14,123.51) ;
\draw [shift={(288.53,125.07)}, rotate = 213.02] [fill={rgb, 255:red, 0; green, 0; blue, 0 }  ][line width=0.08]  [draw opacity=0] (8.93,-4.29) -- (0,0) -- (8.93,4.29) -- cycle    ;

\draw (201.33,198.4) node [anchor=north west][inner sep=0.75pt]  [font=\scriptsize]  {$p$};
\draw (380.53,113.8) node [anchor=north west][inner sep=0.75pt]  [font=\scriptsize]  {$\mathscr{B}^{+}( p)$};
\draw (241.33,75.73) node [anchor=north west][inner sep=0.75pt]  [font=\scriptsize]  {$\partial I^{+}( p)$};
\draw (309.33,209.07) node [anchor=north west][inner sep=0.75pt]  [font=\scriptsize]  {$\partial J^{+}( p)$};

\end{tikzpicture}
    \caption{\justifying A possible bubbling set in a $C^0$ spacetime. Here the shaded region represents the open set $\mathscr{B}^+(p)$. \cite{ling_aspects_2020v2}}
    \label{fig:bubbling}
\end{figure}
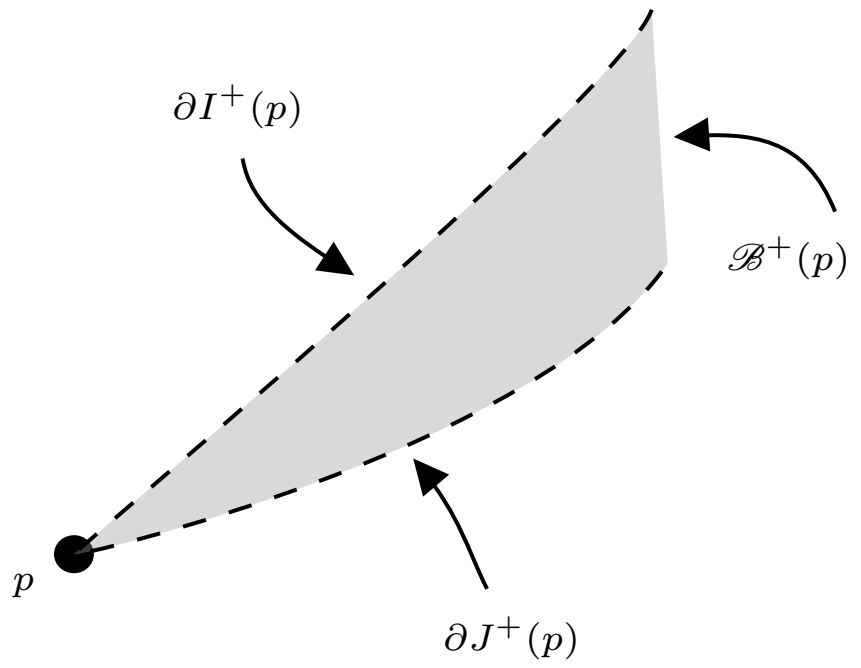

The theory of $C^0$ Lorentzian spacetimes presents interesting and arguably exotic phenomena that are not found in $C^2$ spacetimes. We here follow the introduction to $C^0$ causal theory of Ref.~\cite{ling_aspects_2020v2}, see Refs.~\cite{samann_global_2016,grant_future_2020,garcia-heveling_causality_2021,garcia-heveling_causal_2022} for other recent developments. Figures \ref{fig:timelike curve c0} and \ref{fig:bubbling} are taken from Figures 1 and 3 in Ref.~\cite{ling_aspects_2020v2}. 

A $C^0$ metric on a smooth manifold $\M$ is a nondegenerate ($g(X,Y)=0$ for all $Y \in T\M$ implies $X=0$) symmetric ($g(X,Y) = g(Y,X)$) tensor $g : T\M \times T\M \to \mathbb{R}$ with constant signature whose components $g_{\mu\nu} = g(\partial_\mu,\partial_\nu)$ in any coordinate system are $C^0$ functions. 

The tangent and cotangent spaces at $p \in \M$ are denoted by $T_p \M$ and $T_p^*\M$, respectively, and the corresponding bundles by $T\M = \bigcup_{p \in \M} T_p \M$ and $T^*\M := \bigcup_{p \in \M} T_p^*\M$. 

A nonzero vector $X \in T_p\M$ is \emph{timelike} if $g(X,X) < 0$, \emph{null} if $g(X,X)=0$, \emph{spacelike} if $g(X,X)>0$. A nonzero vector is \emph{causal} if it is either timelike or null. A causal vector $Y \in T_p\M$ is \emph{future-directed} if $g(X,Y) < 0$ and \emph{past-directed} if $g(X,Y) > 0$.

A Lorentzian manifold $(\M,g)$ is \emph{time-oriented} if there is a $C^1$ timelike vector field $X$ on $\M$. A $C^0$ spacetime $(\M,g)$ is a time-oriented Lorentzian manifold with a $C^0$ metric such that $\M$ is connected, Hausdorff and second-countable.

Let $I \subset \mathbb{R}$ be an interval and $h$ be a smooth complete Riemannian metric on $\M$. Almost everywhere statements here are made with respect to the Lebesgue measure on $I$. A \emph{locally Lipschitz curve} $\gamma : I \to \M$ is a continuous function such that for any compact $K\subset I$, there exists a $C_K \in \mathbb{R}_+$ such that for any $a,b \in K$,
\begin{equation}
    d_h(\gamma(a),\gamma(b)) \leq C_K \abs{b-a}.
\end{equation}
The definition of a locally Lipschitz curve is independent of the choice of Riemannian metric $h$. If $\gamma$ is locally Lipschitz, then the components $\gamma^\mu = x^\mu \circ \gamma$ in any coordinate system are differentiable almost everywhere and $(\gamma^\mu)' \in L^\infty_\mathrm{loc}$.

A \emph{causal curve} is a locally Lipschitz curve $\gamma : I \to \M$ such that $\gamma'$ is future-directed causal almost everywhere. A \emph{timelike almost everywhere curve} is a causal curve such that $\gamma'$ is future-directed timelike almost everywhere. A curve is 
{\emph{uniformly timelike in the given parametrisation}}
if it is a causal curve $\gamma : I \to \M$ such that $g(\gamma',\gamma') < -\epsilon$ almost everywhere for some $\epsilon>0$. We follow \cite{ling_aspects_2020v2} in referring to this as a \emph{timelike curve} (although the terminology is arguably less precise).  See Fig.~\ref{fig:timelike curve c0} \cite{ling_aspects_2020v2}.

Given a set $S \subset \M$ contained in an open neighbourhood $\U \subset \M$, we define the \emph{causal future} and \emph{timelike future} of $S$ within $\U$ as
\begin{multline}
    J^+(S,\U) := \{p \mid \exists \text{ causal } \gamma : [a,b] \to \U \\ \text{ s.t. } \gamma(a) \in S, \, \gamma(b)=p\} \cup S
\end{multline}
and
\begin{multline}
    I^+(S,\U) := \{p \mid \exists \text{ timelike } \gamma : [a,b] \to \U \\ \text{ s.t. } \gamma(a) \in S, \, \gamma(b)=p\}
\end{multline}
The \emph{causal past} $J^-(S,\U)$ and \emph{timelike past} $I^-(S,\U)$ are defined time-dually. If $U=\M$, we write $J^\pm(S)$ and $I^\pm(S)$, respectively. If $S=\{p\}$, we write $J^\pm(p,\U)$ and $I^\pm(p,\U)$, respectively. If we wish to emphasize the Lorentzian metric $g$ being used, we write $J^\pm_g$ and $I_g^\pm$, respectively.

$I^\pm(p)$ are open sets. However, if $I^+$ had been defined via timelike almost everywhere curves, then it would not necessarily be open \cite{grant_future_2020}. $J^\pm(p)$ is closed for all $p \in \M$ if $(\M,g)$ is globally hyperbolic. 

A causal curve $\gamma : [a,b) \to \M$ is \emph{future extendible} if there is a causal curve $\tilde{\gamma} : [a,b] \to \M$ with $\tilde{\gamma}(t) = \gamma(t)$ for all $t \in [a,b)$. Otherwise $\gamma$ is \emph{future inextendible}. Time-dualizing gives \emph{past inextendible} causal curves. 
$\gamma$ is \emph{inextendible} if it is both future and past inextendible.

A Cauchy hypersurface for $(\M,g)$ is a set $S \subset \M$ such that every inextendible causal curve intersects $S$ exactly once.  A set $S \subset \M$ is \emph{achronal} if $I^+(S) \cap S = \varnothing$. A Cauchy hypersurface is a $C^0$ achronal closed hypersurface. Global hyperbolicity is equivalent to the existence of a Cauchy hypersurface \cite{samann_global_2016}. There are no closed causal curves in a globally hyperbolic spacetime.

We denote by $B^\pm(S)$ the future (respectively past) boundary of $S$, i.e. \begin{equation}B^{\pm}(S) = \{p \in \overline{S} \mid J^\pm(p) \cap \overline{S} = \{p\}\} \, .\end{equation}
We write $\partial S$ for the topological boundary of $S$, and $\mathrm{int}[S]$ for its interior.

In a $C^0$ spacetime, it may be that $I^+(p) \subsetneq \mathrm{int}[J^+(p)]$. Given a set $S$ within a neighbourhood $\U$, we define the \emph{future bubbling set of $S$ within $\U$} as the open set
\begin{equation}
    \mathscr{B}^+(S,\U) = \mathrm{int}[J^+(S,\U)]\smallsetminus \overline{I^+(S,\U)}.
\end{equation}
Past bubbling sets are defined time-dually. When $\U = \M$, we simply write $\mathscr B^\pm(S)$ respectively. When $S=\{p\}$, we simply write $\mathscr B^\pm(p,\U)$, respectively.

A $C^0$ spacetime is called \emph{causally plain} if $\mathscr B^+(p) = \varnothing$ for all $p \in \M$. It is said to satisfy the \emph{push-up property} if $I^+(J^+(p)) = I^+(p)$ for all $p \in \M$. If $(\M,g)$ satisfies the push-up property, then $(\M,g)$ is causally plain. $C^2$ spacetimes are causally plain. In a $C^0$ spacetime that is not causally plain, there is no canonical notion of ``light cone'', since $\partial I^+(p)$ need not be $\partial J^+(p)$, as is shown in Fig.~\ref{fig:bubbling} \cite{ling_aspects_2020v2}. In the following, we write $\Lambda^\pm(p) := \partial J^\pm(p)$. 

A \emph{partial Cauchy hypersurface} is a closed achronal set which is a topological hypersurface \cite{bar_initial_2015}. Every Cauchy hypersurface is a partial Cauchy hypersurface. A \emph{characteristic} partial Cauchy hypersurface is a lightlike partial Cauchy hypersurface. Light cones are (non smooth) characteristic partial Cauchy hypersurfaces.

A $C^0$ spacetime is Lipschitz, or $C^{0,1}_\mathrm{loc}$, if the components of the metric $g_{\mu\nu}$ in any coordinate system are locally Lipschitz functions. If $(\M,g)$ is $C^{0,1}_\mathrm{loc}$, then $(\M,g)$ satisfies the push-up property and hence is causally plain. In the present paper, we will focus on Lipschitz spacetimes: we will show that continuous metrics are necessary to work with because of quantum measurements yet Lipschitz metrics can be recovered in the patching of different spacetime regions across light-cones.

\subsubsection{Distributional geometry}

\label{sec:distributional geometry}

We now use basic notions of distribution theory to apply it to differential geometry. Every statement made in Sec.~\ref{sec:distributions} is at the level of scalar functions for conciseness; it can be generalised to tensors and to sections of vector bundles. See also \cite{geroch_strings_1987,steinbauer_use_2006,lefloch_definition_2007,bar_initial_2015} for the theory of generalised functions and distributions in General Relativity.

A \emph{distributional metric} $g$ on $\M$ is a symmetric and non-degenerate $(0,2)$-tensor distribution on $\M$. A distributional metric is said to be of \emph{Geroch-Traschen class} \cite{geroch_strings_1987}, and is said to be a \emph{GT-metric}, if $g,g^{-1} \in H^1_\mathrm{loc} \cap L^\infty_\mathrm{loc}$. This is sufficient to define the curvature tensors found in General Relativity, now in a distributional way, see \cite{geroch_strings_1987,lefloch_definition_2007}.

Heuristically, this can be seen as follows. A (smeared) Einstein-Hilbert action of the form
\begin{equation}
    S[f] = \frac{1}{2\kappa}\int_{\mathcal{M}} R(x) f(x) \sqrt{\abs{\det g}} \, d^4x, \qquad f \in \mathcal{D}(\M),
\end{equation}
is only well-defined analytically if $\sqrt{\abs{\det g}} R \in \mathcal{D}'(\M)$. But $R = g^{ab} R_{ab} \sim  g^{-1} \Gamma \Gamma + g^{-1}\partial \Gamma$, with $\Gamma \sim g^{-1}\partial g$. If $g$ is of Geroch-Traschen class then $\sqrt{|\det g|} g^{-1} \in H^1_\mathrm{loc} \cap L^\infty_\mathrm{loc}$, $\partial g \in L^2_\mathrm{loc}$ so $\Gamma \sim g^{-1} \partial g \in L^2_\mathrm{loc}$ so $\Gamma \Gamma \in L^1_\mathrm{loc}$ so $\sqrt{|\det g|} g^{-1} \Gamma \Gamma \in L^1_\mathrm{loc} \hookrightarrow \mathcal{D}'(\M)$. Moreover, $\partial \Gamma \in H^{-1}_\mathrm{loc}$ so $\sqrt{|\det g|} g^{-1} \partial \Gamma \in \mathcal{D}'(\M)$, i.e. $R \in \mathcal{D}'(\M)$. Hence, $S[f]$ is indeed well-defined for $f \in \mathcal{D}(\M)$.

For distributional metrics that are not $C^0$, the notion of distance between points breaks down. Instead, only the notion of distance between regions is meaningful. This is arguably a very operational view on geometry, but one we will not expand on further here. 

Although $C^0 \subset L^\infty_\mathrm{loc}$, a $C^0$ function need not be in $H^1_\mathrm{loc}$. Hence, not all $C^0$ metrics are GT-metrics. However, $C^{0,1}_\mathrm{loc} \subsetneq L^\infty_\mathrm{loc} \cap H^1_\mathrm{loc} \cap C^0$ and $g \in C^{0,1}_\mathrm{loc} \Rightarrow g^{-1} \in C^{0,1}_\mathrm{loc}$, i.e. $C^{0,1}_\mathrm{loc}$ metrics are continuous GT-metrics and so both distributional and $C^0$ geometry hold on Lipschitz spacetimes.

For a GT-metric $g_{\mu\nu} = \eta_{\mu\nu} +h_{\mu\nu}$ on a Minkowski background, define
\begin{equation}
    h := \eta^{\mu\nu} h_{\mu\nu}, \quad \overline{h}_{\mu\nu} = h_{\mu\nu}-\frac{1}{2}\eta_{\mu\nu} h,
\end{equation}
and we write
\begin{equation}
    C_\nu[h] := \nabla^\mu \overline{h}_{\mu\nu}
\end{equation}
where $\nabla$ is the Levi-Civita connection. The linearised Einstein tensor for $h$ is
\begin{multline}
    \label{eqn:linearised Einstein tensor}
    G_{\mu\nu}^{(1)}[h] = -\frac{1}{2}(\Box \overline{h}_{\mu\nu} - \nabla_\mu C_\nu[h] - \nabla_\nu C_{\mu}[h] \\ + \eta_{\mu\nu} \nabla^\rho C_\rho[h]).
\end{multline}
This expression is valid whether or not $h$ satisfies the \emph{Lorenz condition}
\begin{equation}
    C_\nu[h] = \nabla^\mu \overline{h}_{\mu\nu} = 0.
\end{equation}

\subsubsection{The Chabauty-Fell topology}

\label{subsec:topologies}

We wish to discuss the convergence of \emph{hypersurfaces} in the spacetime manifold $\M$, for which a standard choice is the Chabauty-Fell topology \cite{chabauty_limite_1950,fell_hausdorff_1962}. The spacetime is equipped with its ordinary manifold topology $(\M,\tau_\M)$. $\tau_\M$ is a topology that provides a notion of convergence for \emph{points} in $\M$, and it is in this topology on $\M$ that we say that a subset of $\M$ is ``open'', ``closed'' and ``compact'' in $\M$. We wish to provide a topology for \emph{closed subsets of $\M$}: spacelike Cauchy hypersurfaces and light cones of a globally hyperbolic $C^{0,1}_\mathrm{loc}$ spacetime are closed. Here, we examine the closed-set space
\begin{equation}
    \mathrm{CL}(\M) := \{O \subseteq \M \mid O\text{ is closed}\} \subset 2^\M.
\end{equation}
For an open set $\U \subseteq \M$ and a compact set $K \subseteq \M$, define
\begin{equation}
    \mathbf{H}_\U := \{O \in \mathrm{CL}(\M) \mid O \cap \U \neq \varnothing\}
\end{equation}
and
\begin{equation}
    \mathbf{M}_K := \{O \in \mathrm{CL}(\M) \mid O \cap K = \varnothing\}.
\end{equation}
The Chabauty-Fell topology $\tau_\mathrm{CF}$ on $\mathrm{CL}(\M)$ is the topology generated by the subbasis
\begin{equation}
    \{ \mathbf{H}_\U \mid \U \subseteq \M \text{ open}\} \cup \{\mathbf{M}_K \mid K \subseteq \M \text{ compact}\}.
\end{equation}
In particular, for an arbitrary sequence $(F_n) \in \mathrm{CL}(\M)$,
\begin{equation}
    F_n \longrightarrow F \qquad \text{ in } \tau_\mathrm{CF}
\end{equation}
if and only if $\forall n \geq N$, $\forall \U \subseteq \M$ open and $\forall K \subseteq \M$ compact,
\begin{equation}
    \begin{cases}
        F \cap U \neq \varnothing \Longrightarrow F_n \cap \U \neq \varnothing \text{ eventually,} \\
        F \cap K = \varnothing \Longrightarrow F_n \cap K = \varnothing \text{ eventually.}
    \end{cases}
\end{equation}
Equivalently \cite{paulin_geometrie_2007v2,harpe_spaces_2008}, $F_n \stackrel{\tau_\mathrm{CF}}{\to} F$ if and only if the two following conditions holds:
\begin{enumerate}
    \item for all $x \in F$, there exists for all sufficiently large $i$ a point $x_i \in F_i$ such that $x_i \to x$ in $(\M,\tau_\M)$,
    \item for all strictly increasing sequences $(i_n)_{n\geq1}$ and for all sequences $(x_{i_n})_{n \geq 1}$ such that $x_{i_n} \in F_{i_n}$ and $x_{i_n} \to x$ in $(\M,\tau_\M)$, we have $x \in F$.
\end{enumerate}
This will prove to be useful for understanding the convergence of spacelike Cauchy hypersurfaces to a past light cone.

\subsection{Operators on Hilbert spaces}

Every Hilbert space under consideration in this paper is complex and separable. The space of bounded operators on a separable Hilbert space $\hi$ will be denoted $\bh$, while the linear operators on a dense domain $\mathcal{D} \subset \hi$ will be denoted $\mathcal{L}(\mathcal{D},\hi)$. Trace-class operators are written $\mathcal{T}(\hi)$, while quantum states (positive trace-class operators with trace $1$) live in the convex subset $\state$ of $\mathcal{T}(\hi)$. $\mathcal{T}(\hi)$ is the predual of $\bh$, written $\mathcal{T}(\hi) \cong \bh_*$. For $\chi \in \mathcal{T}(\hi)$, the mapping $\omega_\chi : \bh \to \mathbb{C}$, $\omega_\chi(A) = \Tr[\chi A]$, maps observables to expectation values in a basis independent way. 

\subsubsection{Topologies}
\label{sec:topologies}

There are several classes of topologies to use for infinite-dimensional Hilbert spaces. We review here two types of topologies: operational ones based on expectation values, and ontic ones based on norms.

The \emph{ultraweak topology on $\bh$} is the topology of convergence in expectation value for bounded operators. If $(A_n)$ is a sequence of bounded operators in $\bh$, we say that $A_n$ converges ultraweakly to $A \in \bh$ if and only if $\Tr[A_n \chi] \to \Tr[A\chi]$ in $\mathbb{C}$ for all $\chi \in \mathcal{T}(\hi)$. Equivalently, $A_n \stackrel{\mathrm{uw}}{\to} A$ if and only if $\Tr[A_n \rho] \to \Tr[A \rho]$ in $\mathbb{C}$ for all $\rho \in \state$. That is, convergence means convergence in expectation values over all states. More generally, as we shall see in Sec.~\ref{sec:Hilbert bundles}, the ultraweak topology concerns von Neumann algebras and their preduals.

The predual topology on $\mathcal{T}(\hi)$, sometimes called the \emph{operational topology} \cite{carette_operational_2025}, is similar but now applies to trace-class operators (and thus to states). A sequence $(\chi_n)$ of trace-class operators converges to $\chi \in \mathcal{T}(\hi)$ if and only if $\Tr[\chi_n A] \to \Tr[\chi A]$ in $\mathbb{C}$ for all $A \in \bh$. That is, convergence means convergence in expectation values over all operators. 

On the other hand, one may look at the \emph{norm topology on $\bh$}. A sequence $(A_n)$ of bounded operators in $\bh$ converge in norm to $A \in \bh$ if and only if $\norm{A_n - A} \to 0$ in $\mathbb{R}$. This is stronger than the ultraweak topology on $\bh$.

Likewise, the \emph{trace-norm topology on $\mathcal{T}(\hi)$} is stronger than the operational topology on $\mathcal{T}(\hi)$. A sequence $(\chi_n)$ of trace-class operators converges to $\chi \in \mathcal{T}(\hi)$ if and only if $\norm{\chi_n - \chi}_1 \to 0$ in $\mathbb{C}$, where $\norm{\chi}_1 = \Tr[\abs{\chi}]$.

\subsubsection{Hilbert bundles}

\label{sec:Hilbert bundles}

A Hilbert bundle \cite{dixmier_champs_1963,dixmier_von_1981,bos_continuous_2007} is best understood as a measurable field of (separable complex) Hilbert spaces, that we take for simplicity over the spacetime manifold $\M$ 
\begin{equation}
    \mathscr{H} = \bigsqcup_{x \in \mathcal{M}} \hi_x \stackrel{\pi}{\longrightarrow} \mathcal{M}
\end{equation}
with $\pi(\xi) = x$ when $\xi \in \hi_x$. That is, every $x \in \mathcal{M}$ labels its own Hilbert space $\hi_x$. We allow these Hilbert spaces to vary measurably with $x$. We want to specify what it means for a field
\begin{equation}
    x \longmapsto \psi(x) \in \hi_x
\end{equation}
to vary measurably when the vectors $\psi(x)$ live in different Hilbert spaces. Importantly, vectors in different fibres cannot yet be compared: if $\xi(x) \in \hi_x$, $\eta(y) \in \hi_y$ with $x \neq y$, then there is no intrinsic inner product $\braket{\xi(x)}{\eta(y)}$ nor any addition $\xi(x) + \eta(y)$ defined. Thus, we require a set of conditions for $\mathscr{H}$ to be a Hilbert bundle and start making sense of different $\psi(x)$'s.

\paragraph*{1. Every fibre is a Hilbert space. } We require that for all $x \in \mathcal{M}$, the space $\pi^{-1}(x) = \hi_x$ is a Hilbert space.

\paragraph*{2. There exists a countable family of measurable reference sections. } For $i \in \mathbb{N}$, the maps
\begin{equation}
    \psi_i : \mathcal{M} \to \mathscr{H}, \qquad \pi(\psi_i(x)) = x,
\end{equation}
are sections of the bundle, with $\psi_i(x) \in \hi_x$. We want a countable set of such sections to be such that their span is dense in the Hilbert spaces:
\begin{equation}
    \overline{\mathrm{span}\left\{\psi_i(x)\right\}^{\infty}_{i=1}} = \hi_x \qquad \forall x \in \mathcal{M}.
\end{equation}

\paragraph*{3. The overlaps of the reference sections are measurable. } For every $i,j \in \mathbb{N}$, we require 
\begin{equation}
    x \longmapsto \braket{\psi_i(x)}{\psi_j(x)}_{\hi_x}
\end{equation}
to be a Borel function.

\paragraph*{4. Measurability is specified by the reference sections. } Let
\begin{equation}
    \phi : \mathcal{M} \to \mathscr{H}
\end{equation}
be an arbitrary map, and define $b(x) := \pi(\phi(x))$. Then $\phi(x) \in \hi_{b(x)}$. We require that $\phi$ be Borel if and only if
\begin{enumerate}
    \item the fibre label $b(x)$ is Borel, and
    \item every inner product
    \begin{equation}
        x \longmapsto \braket{\phi(x)}{\psi_i(b(x))}_{\hi_{b(x)}}
    \end{equation}
    is Borel.
\end{enumerate}
If $\phi$ is a section i.e. $b(x)=\pi(\phi(x)) = x$, this simplifies to
\begin{equation}
    \phi \text{ is Borel} \Longleftrightarrow x \longmapsto \braket{\phi(x)}{\psi_i(x)}_{\hi_x} \text{ is Borel } \forall i \in \mathbb{N}.
\end{equation}
These four conditions define a \emph{Hilbert bundle over $\mathcal{M}$}. 

Hilbert bundles are not Hilbert spaces, but can be used to construct Hilbert spaces as follows. Let $d\mu_g := \sqrt{\abs{g}} \, d^dx$ be the volume measure of the spacetime $(\mathcal{M},g)$. The direct integral Hilbert space of a Hilbert bundle
\begin{equation}
    \mathscr{H} = \bigsqcup_{x \in \M} \hi_x \stackrel{\pi}{\longrightarrow} \M
\end{equation}
is defined as \cite{dixmier_champs_1963}
\begin{equation}
    \hi := \int^\oplus_\mathcal{M} \hi_x \, d\mu_g(x).
\end{equation}
Such direct integral Hilbert spaces arise naturally from the representation theory of groupoids \cite{dixmier_von_1981,bos_continuous_2007,renault_c-algebra_1980,fedida_groupoidal_2026}, i.e. by looking at the notion of local symmetries in general Lorentzian spacetimes which need not have global symmetries.

Operator fields associated with $\mathscr{H}$ live in
\begin{widetext}
    \begin{equation}
        \mathcal{A}(\mathscr{H}) := \int^\oplus_\M \mathcal{B}(\hi_x) d\mu_g(x)= \left\{A = \int^\oplus_\M A(x) \, d\mu_g(x) \mid x \mapsto A(x) \text{ measurable}, \norm{A}_{\mathcal{A}(\mathscr{H})} := \mathrm{ess \; sup}_x \norm{A(x)}_{\mathcal{B}(\hi_x)} < \infty \right\}.
    \end{equation}
\end{widetext}
This is a type $I$ von Neumann algebra \cite{takesaki_tensor_1979} with a canonical semifinite normal trace. Its predual is
\begin{widetext}
    \begin{equation}
        \mathcal{A}(\mathscr{H})_* := \left\{\chi = \int^\oplus_\M \chi(x) \, d\mu_g(x) \mid x \mapsto \chi(x) \in \mathcal{T}(\hi_x) \text{ measurable}, \norm{\chi}_1 := \int_\M \norm{\chi(x)}_1 \, d\mu_g(x) < \infty\right\}
    \end{equation}
\end{widetext}
with trace norm $\norm{\chi}_1 = \int_\M \Tr[\abs{\chi(x)}] \, d\mu_g(x)$, and
\begin{equation}
    \inner{A}{\chi} := \Tr[A \chi] = \int_\M \Tr[A(x) \chi(x)] d\mu_g(x).
\end{equation}
This induces the ultraweak topology $\sigma(\mathcal{A}(\mathscr{H}),\mathcal{A}(\mathscr{H})_*)$ on $\mathcal{A}(\mathscr{H})$.

\section{Local quantum states}

\label{sec:local quantum states}

The idea that if one lies at a point $x$, one must describe the quantum state with respect to some local density operator lying on the past light-cone of $x$ (in some suitably-defined sense, the past light-cone $\Lambda^-(x)$ not being a spacelike Cauchy hypersurface) has been discussed and (re)discovered in various contexts, see e.g. \cite{ghirardi_local_2000,kent_nonlinearity_2005,bedingham_relativistic_2011,polo-gomez_state_2025}.

\subsection{The limiting sequence picture}

\label{sec:local quantum states Kent}

In standard ``physicists'" QFT, wavefunctionals lie on spacelike Cauchy hypersurfaces, evolve unitarily away from quantum measurements and through some form of Lüders' rule during quantum measurements. If one wishes to capture the information about the unitary evolution and quantum measurements undertaken by the quantum system until some point $x$, one should look at the wavefunctional on spacelike Cauchy hypersurfaces which lie ``close'' to the past causal cone of $x$ and which contains $x$, and trace out the parts of the wavefunctional away from $x$. This is the essence of the definition of the local quantum state in \cite{kent_nonlinearity_2005}, which we now review.

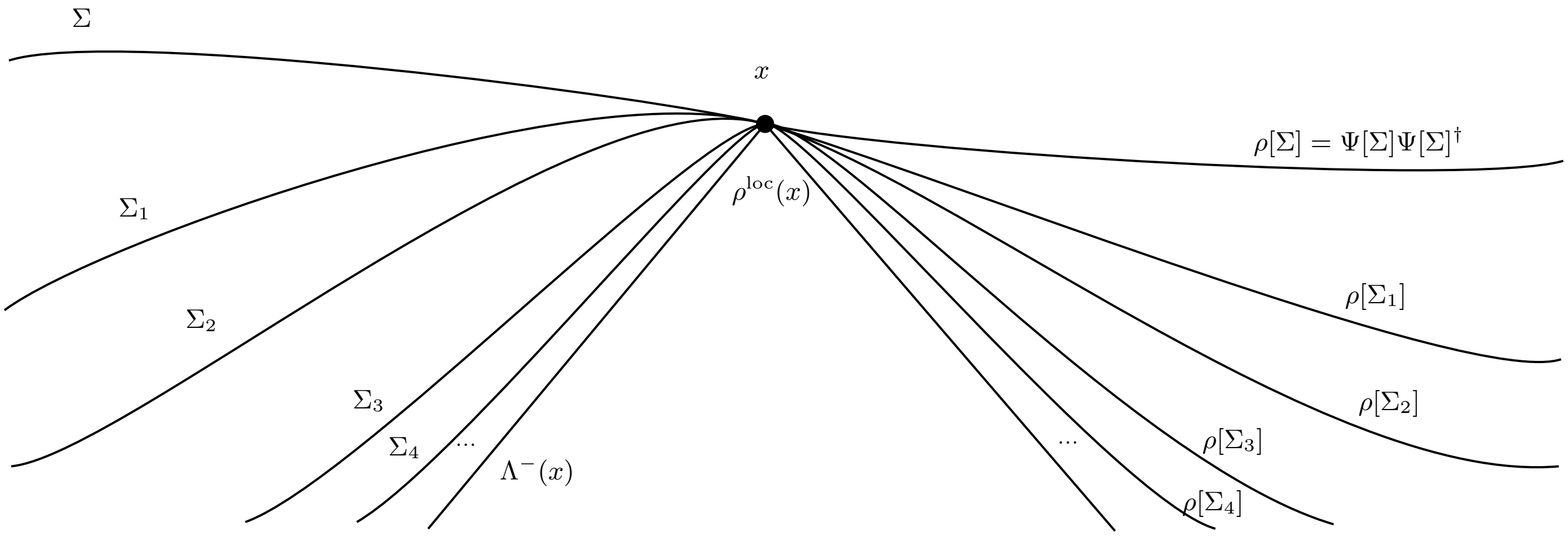
\begin{figure*}
    \begin{tikzpicture}[x=0.75pt,y=0.75pt,yscale=-1,xscale=1]

\draw  [fill={rgb, 255:red, 0; green, 0; blue, 0 }  ,fill opacity=1 ] (337.9,118.55) .. controls (337.9,116.59) and (339.49,115) .. (341.45,115) .. controls (343.41,115) and (345,116.59) .. (345,118.55) .. controls (345,120.51) and (343.41,122.1) .. (341.45,122.1) .. controls (339.49,122.1) and (337.9,120.51) .. (337.9,118.55) -- cycle ;
\draw    (2.36,90.1) .. controls (49.36,75.1) and (291.54,106) .. (341.45,118.55) .. controls (391.36,131.1) and (658.36,147.1) .. (699.36,135.1) ;
\draw    (341.45,118.55) -- (190.36,300.1) ;
\draw    (341.45,118.55) -- (498.36,301.1) ;
\draw    (0.36,202.1) .. controls (40.36,172.1) and (258.36,94.1) .. (341.45,118.55) .. controls (424.54,143) and (658.36,237.1) .. (698.36,224.1) ;
\draw    (3.36,272.1) .. controls (55.36,267.1) and (258.36,94.1) .. (341.45,118.55) .. controls (424.54,143) and (593.36,281.1) .. (697.36,272.1) ;
\draw    (108.36,297.1) .. controls (158.36,279.1) and (310.54,120) .. (341.45,118.55) .. controls (372.36,117.1) and (505.36,271.1) .. (596.36,298.1) ;
\draw    (158.36,297.1) .. controls (204.36,270.1) and (323.54,120) .. (341.45,118.55) .. controls (359.36,117.1) and (498.36,287.1) .. (543.36,300.1) ;

\draw (201,260.4) node [anchor=north west][inner sep=0.75pt]  [font=\scriptsize]  {$...$};
\draw (471,259.4) node [anchor=north west][inner sep=0.75pt]  [font=\scriptsize]  {$...$};
\draw (29,65.4) node [anchor=north west][inner sep=0.75pt]  [font=\small]  {$\Sigma $};
\draw (50,150.4) node [anchor=north west][inner sep=0.75pt]  [font=\small]  {$\Sigma _{1}$};
\draw (80,200.4) node [anchor=north west][inner sep=0.75pt]  [font=\small]  {$\Sigma _{2}$};
\draw (155,236.4) node [anchor=north west][inner sep=0.75pt]  [font=\small]  {$\Sigma _{3}$};
\draw (171,257.4) node [anchor=north west][inner sep=0.75pt]  [font=\small]  {$\Sigma _{4}$};
\draw (221,266.4) node [anchor=north west][inner sep=0.75pt]  [font=\small]  {$\Lambda ^{-}( x)$};
\draw (335,91.4) node [anchor=north west][inner sep=0.75pt]  [font=\small]  {$x$};
\draw (559,118.4) node [anchor=north west][inner sep=0.75pt]  [font=\small]  {$\rho [ \Sigma ] =\Psi [ \Sigma ] \Psi [ \Sigma ]^{\dagger }$};
\draw (600,188.4) node [anchor=north west][inner sep=0.75pt]  [font=\small]  {$\rho [ \Sigma _{1}]$};
\draw (606,236.4) node [anchor=north west][inner sep=0.75pt]  [font=\small]  {$\rho [ \Sigma _{2}]$};
\draw (536,253.4) node [anchor=north west][inner sep=0.75pt]  [font=\small]  {$\rho [ \Sigma _{3}]$};
\draw (527,281.4) node [anchor=north west][inner sep=0.75pt]  [font=\small]  {$\rho [ \Sigma _{4}]$};
\draw (325,140.4) node [anchor=north west][inner sep=0.75pt]  [font=\small]  {$\rho^{\mathrm{loc}}( x)$};

\end{tikzpicture}
\caption{\justifying Limiting sequence picture of the local quantum state \cite{kent_nonlinearity_2005}. Here, each $\rho[\Sigma_n]$ is the quantum state on $\Sigma_n$, and $\rho_n(x) = \Tr_{\Sigma_n \smallsetminus \{x\}}(\rho[\Sigma_n])$ is the reduced state at $x \in \Sigma_n$. The local quantum state is then the reduced state at $x$ in the limit $n \to \infty$ where $x \in \Sigma_n$ for all $n$ and $\Sigma_n \to \Lambda^-(x)$.}
\label{fig:local quantum state kent}
\end{figure*}

Consider the following construction, also depicted in Fig. \ref{fig:local quantum state kent}. Let $\{\Sigma_n\}_{n \in \mathbb{N}}$ be a sequence of spacelike Cauchy hypersurfaces such that
\begin{itemize}
    \item $\forall n \in \mathbb{N}$, $x \in \Sigma_n$,
    \item $\Sigma_n \stackrel{n \to \infty}{\to} \Lambda^-(x)$ in $(\mathrm{CL}(\M),\tau_\mathrm{CF})$.
\end{itemize}

Convergence in the Chabauty-Fell topology is natural here: it says that
\begin{enumerate}
    \item Every open set $\U$ intersecting $\Lambda^-(x)$ eventually meets $\Sigma_n$:
    \begin{equation}
        \U \cap \Lambda^-(x) \neq \varnothing \Longrightarrow \U \cap \Sigma_n \neq \varnothing
    \end{equation}
    for all sufficiently large $n$.
    \item Every compact set $K$ disjoint from $\Lambda^-(x)$ is eventually disjoint from $\Sigma_n$:
    \begin{equation}
        K \cap \Lambda^-(x) = \varnothing \Longrightarrow K \cap \Sigma_n = \varnothing
    \end{equation}
    for all sufficiently large $n$.
\end{enumerate}
Equivalently, 
\begin{enumerate}
    \item $\forall p \in \Lambda^-(x)$, $\exists p_n \in \Sigma_n$ such that $p_n \to p$,
    \item For $n_j \uparrow \infty$, $p_j \in \Sigma_{n_j}$, $p_j \to p \Rightarrow p \in \Lambda^-(x)$.
\end{enumerate}
The first condition states that for every point $p$ on the past light cone, there exists a sequence of points on the spacelike Cauchy hypersurfaces that converges to $p$. The second states that no sequence of points on the sequence $(\Sigma_n)$ can converge to a point away from the past light cone.

We construct spacelike Cauchy hypersurfaces converging in the Chabauty-Fell topology to the past light cone at a point in $1+3$-dimensional Minkowski spacetime in Appendix \ref{app:limit sequences Minkowski}. We discuss the convergence of metrics in spacetime in Appendix \ref{app:convergence metrics}.

We now construct the local quantum state. Let $\rho_n(x) := \Tr_{\Sigma_n \smallsetminus \{x\}}[\Psi[\Sigma_n] \Psi[\Sigma_n]^\dagger]$. Note that in relativistic QFT, we cannot generally write $\hi_{\Sigma_n} \cong \hi_x \otimes \hi_{\Sigma_n \smallsetminus \{x\}}$. This construction of local quantum states is thus to be understood in a relativistic quantum theoretic setting (e.g. where the number of particles is fixed and particles approximately decompose as independent subsystems) but not yet in a relativistic quantum \emph{field} theoretic one.

Then the local quantum state is taken to be \cite{kent_nonlinearity_2005,kent_simple_2018}
\begin{equation}
    \rho^\mathrm{loc}(x) = \lim_{n \to \infty} \rho_n(x)
\end{equation}
in an appropriate topology of $\state$ (or of the space $\mathcal{T}(\hi)$ spanned by the states). 

We 
do not fix a precise choice of topology on $\state$ for this convergence here, but outline some natural choices and their meaning.  One is the operational \cite{carette_operational_2025} $\sigma(\mathcal{T}(\hi),\bh)$ topology on $\mathcal{T}(\hi)$.  Since local quantum states will source the expectation value of the stress-energy tensor in the semiclassical Einstein field equations, this choice is appropriate if one wants gravity to back-react from (quantumly) operationally accessible resources. Another is the trace-norm topology. This is stronger than the operational topology, and would be an appropriate choice if gravity is sensitive to the ontology of states underlying (quantumly) operationally indistinguishable equivalence classes of states. Note that on $\state$ these two topologies coincide \cite{dellantonio_limits_1967}.  It is possible that $\rho^\mathrm{loc}$ may be an algebraic state but not a normal state in some topologies, so it may in principle define different functionals in different topologies.  
We do not expect this to be an issue for the models considered in this paper, which are based on relativistic quantum theory with fixed numbers of particles, but it may need to be addressed in extending the models to relativistic quantum field theory.

Another question is whether and how expectation values of unbounded operators can be evaluated with respect to the local quantum state. Relevant observables (and in particular quantum fields and quantum stress-energies) may only be defined on dense domains $D$ of $\hi$. Let $D^\times$ be the space of continuous antilinear functionals (kets) on $D$ so that $D \subset \hi \subset D^\times$ defines a rigged Hilbert space \cite{de_la_madrid_role_2005}. A pointwise quantum field can be defined as a mapping $\mathcal{M} \to \mathcal{L}(D,D^\times)$ \cite{rehberg_quantum_1986,wollenberg_quantum_1986, antoine_physical_2002}. It remains to be understood under which conditions $\rho^\mathrm{loc}(x)$ is a density operator representative of a linear map $\omega_x :\mathcal{L}(D,D^\times) \to \mathbb{C}$ such that $\omega_x(A) \geq 0$ whenever $\expval{A}{\psi} \geq 0$ for every $\psi \in D$ and whose restriction to bounded operators satisfies $\omega_x(B) = \Tr_\hi[\rho^\mathrm{loc}(x) B]$ for all $B \in \bh$.\footnote{More generally, we can look at rigged Hilbert spaces $D_x \subset \hi_x \subset D_x^\times$ for $x \in \mathcal{M}$ (see Sec.~\ref{sec:polystates hilbert bundle} for a Hilbert bundle picture of local quantum states) and ask when $\rho^\mathrm{loc}(x)$ is a density operator representative of a linear map $\omega_x :\mathcal{L}(D_x,D_x^\times) \to \mathbb{C}$ for $x \in \mathcal{M}$ such that $\omega_x(A_x) \geq 0$ whenever $\expval{A_x}{\psi_x} \geq 0$ for every $\psi_x \in D_x$ and whose restriction to bounded operators satisfies $\omega_x(B_x) = \Tr_{\hi_x}[\rho^\mathrm{loc}(x) B_x]$ for all $B_x \in \mathcal{B}(\hi_x)$. In such a case, we can define $D_x := \bigcap_{\mu,\nu} \mathrm{Dom}(\hat{T}_{\mu\nu}(x))$ instead of taking an intersection over all $x \in \mathcal{M}$.} In particular, we are interested in the case
\begin{equation}
    D := \bigcap_{x \in \mathcal{M}}\bigcap_{\mu,\nu} \mathrm{Dom}(\hat{T}_{\mu\nu}(x)),
\end{equation}
so that $\expval{\hat{T}_{\mu\nu}(x)}_{\rho^\mathrm{loc}(x)} < \infty$ for all $x \in \mathcal{M}$. We leave verification for the local quantum states considered here to future work.

Regardless of the topology considered and of these important technical considerations, the construction of local quantum states makes the Lorentz invariant nature of $\rho^\mathrm{loc}$ explicit. The physical interpretation of the local quantum state at $x$ is that it is the quantum state that a maximally informed observer at $x$ should use: it captures all the information about the dynamics and measurements in its past light cone.  This is transparent in a model of quantum theory in Minkowski space.
We will argue that it continues to hold in semiclassical models where the gravitational field is sourced by the expectation value of the stress-energy of quantum fields, using retarded sourcing.

\subsection{The Hilbert bundle picture}

\label{sec:polystates hilbert bundle}

Another perspective on the local quantum state, which mirrors more closely detector-based models of quantum measurements \cite{polo-gomez_state_2025}, is as follows. To describe what it means in mathematical terms to be the state of a maximally informed observer at $x$, we need to consider Hilbert spaces at individual spacetime points and accessible operator algebras at spacetime points. A useful formalism for this is the notion of Hilbert bundles \cite{dixmier_von_1981,bos_continuous_2007}, reviewed in Sec.~\ref{sec:Hilbert bundles}.

Consider once again the Hilbert bundle
\begin{equation}
    \mathscr{H} = \bigsqcup_{x \in \M} \hi_x \stackrel{\pi}{\longrightarrow} \M.
\end{equation}
The construction of $\mathcal{A}(\mathscr{H})$ describes operator fields for a single system described by the bundle $\mathscr{H}$. To describe the composition of systems in this formalism, there are two possibilities: one either composes operator algebras locally (fibre-wise), or globally. We adopt the latter here; for an exposition of the former, see \cite{fedida_groupoidal_2026}.

Let us start with $n=2$. To make a new von Neumann algebra from two $\mathcal{A}(\mathscr{H})$'s based on the same Hilbert bundle $\mathscr{H}$, we need to close the \emph{algebraic} tensor product $\odot$ ultraweakly, i.e. use the spatial tensor product $\bar{\otimes}$:\footnote{Equivalently, use von Neumann's bicommutant theorem \cite{takesaki_theory_1979} and take the double commutant of the algebraic tensor product of the two $\mathcal{A}(\mathscr{H})$'s.}
\begin{widetext}
    \begin{equation}
        \mathcal{A}^{(2)}(\mathscr{H}) := \mathcal{A}(\mathscr{H}) \bar{\otimes} \mathcal{A}(\mathscr{H})
    = \overline{\mathcal{A}(\mathscr{H}) \odot \mathcal{A}(\mathscr{H})}^{\mathrm{uw}}  =\overline{\left\{\sum_{i=1}^n A_i \otimes B_i \mid n \in \mathbb{N}, A_i, B_i \in \mathcal{A}(\mathscr{H})\right\}}^{\mathrm{uw}}
    \end{equation}
\end{widetext}
More generally, for $n \in \mathbb{N}$, we write
\begin{equation}
    \mathcal{A}^{(n)}(\mathscr{H}) := \overline{\mathcal{A}(\mathscr{H})^{\odot n}}^{\mathrm{uw}}.
\end{equation}

This describes the operator fields of $n$ systems that can fibre-wise be localised over the same or different spacetime points. Indeed, since there exists a canonical unitary
\begin{equation}
    U_n : \hi^{(n)} := \bigotimes^n \hi^{(1)}  \longrightarrow \int^\oplus_{\mathcal{M}^n} \left(\bigotimes_{i=1}^n \hi_{x_i}\right) \prod_{i=1}^n d\mu_g(x_i)
\end{equation}
given by\footnote{This is easily seen to be a unitary: it is straightforwardly linear and isometric, with a dense range in $\int^\oplus_{\mathcal{M}^n} \left(\bigotimes_{i=1}^n \hi_{x_i}\right) \prod_{i=1}^n d\mu_g(x_i)$ i.e. it is a surjective linear isometry and thus a unitary operator.}
\begin{equation}
    [U_n(\xi_1 \otimes \cdots \otimes \xi_n)](x_1,\cdots,x_n) = \xi_1(x_1) \otimes \cdots \otimes \xi_n(x_n),
\end{equation}
we have
\begin{equation}
    \hi^{(n)}  \cong \int_{\mathcal{M}^n}^{\oplus} \left(\bigotimes_{i=1}^n \hi_{x_i}\right) \prod_{i=1}^n d\mu_g(x_i) \, .
\end{equation}
Moreover,  if $A_1,\cdots,A_n \in \mathcal{A}(\mathscr{H})$, then
\begin{multline}
    U_n(A_1 \otimes \cdots \otimes A_n)U_n^* \\ = \int_{\mathcal{M}^n}^\oplus (A_1(x_1) \otimes \cdots \otimes A_n(x_n)) \prod_{i=1}^n d\mu_g(x_i).
\end{multline}
Thus,
\begin{equation}
    U_n \mathcal{A}^{(n)}(\mathscr{H}) U_n^* = \int_{\mathcal{M}^n}^{\oplus} \mathcal{B}\left(\bigotimes_{i=1}^n \hi_{x_i}\right) \prod_{i=1}^n d\mu_g(x_i),
\end{equation}
i.e.
\begin{equation}
    \mathcal{A}^{(n)}(\mathscr{H}) \cong \int_{\mathcal{M}^n}^{\oplus} \mathcal{B}\left(\bigotimes_{i=1}^n \hi_{x_i}\right) \prod_{i=1}^n d\mu_g(x_i).
\end{equation}
In general, $\mathcal{A}^{(n)}(\mathscr{H}) \subsetneq \mathcal{B}(\hi^{(n)})$. This is important for the following.

For the total Hilbert space that can accommodate a potential infinity of systems that can be localised over the same or different spacetime points, we then write
\begin{equation}
    \hi_{\mathcal{M}} = \bigoplus_{n=0}^\infty \hi^{(n)}.
\end{equation}
We include the $n=0$ vacuum sector, where $\hi^{(0)} = \mathbb{C}$. As for a Fock space construction, there exists operators in $\mathcal{B}(\hi_{\mathcal{M}})$ that can transfer between sectors (the algebra is not diagonal). On the other hand, consider
\begin{align}
    \mathcal{A}(\mathscr{H})_{\mathrm{phys}} &= \bigoplus_{n=0}^\infty \mathcal{A}^{(n)}(\mathscr{H})\\ &\cong \bigoplus_{n=0}^\infty\int_{\mathcal{M}^n}^{\oplus} \mathcal{B}\left(\bigotimes_{i=1}^n \hi_{x_i}\right) \prod_{i=1}^n d\mu_g(x_i)
\end{align}
This is now diagonal and so induces superselection sectors. \emph{Polystates} \cite{polo-gomez_state_2025} $\rho$ are then direct sums of positive trace-class operators with trace 1 on each of the sectors\footnote{Since decomposable operator fields are typically defined up to equality almost everywhere, this decomposition should really be understood $\mu_g$-almost everywhere.}:
\begin{equation}
    \rho = 1^{(0)} \oplus \rho^{(1)}(x_1) \oplus \rho^{(1)}(x_2) \oplus \cdots \oplus \rho^{(2)}(x_1,x_2) \oplus \cdots
\end{equation}
Note that polystates are \emph{not} quantum states: they do not have trace $1$, though each sector component does. We can recast the definition of the local quantum state \cite{kent_nonlinearity_2005} in the language of polystates by defining \cite{polo-gomez_state_2025} 
\begin{equation}
    \label{eqn:local state update}
    \rho^{\mathrm{loc}}(x) := \frac{\Psi_{J^-(x)}(\rho)(x)}{\Tr[\Psi_{J^-(x)}(\rho)(x)]}\in \mathscr{D}(\hi_x)
\end{equation}
where, for any spacetime region $\mathcal{S} \subset \mathcal{M}$,
\begin{equation}
    \label{eqn:update channel}
    \Psi_\mathcal{S} : \mathcal{A}(\mathscr{H})_{\mathrm{phys}} \to \mathcal{A}(\mathscr{H})_{\mathrm{phys}}
\end{equation}
is a map that encodes all the transformations undergone by all the systems at every point $x \in \mathcal{S} \subset \mathcal{M}$, including time evolutions and measurements. This construction is precisely that of the local state, where $\Psi_{J^-(x)}$ is the function mapping the global state $\Psi$ to the local quantum state using a sequence of Cauchy hypersurfaces converging to $\Lambda^-(x)$. We will use both the original definition of the local quantum state and the definition in the language of the polystate formalism in the context of quantum measurements in Sec. \ref{sec:mmts}.

\subsection{Local quantum states in quantum field theory?}

One may (reasonably) think that the construction of $\rho^\mathrm{loc}$ in Sec.~\ref{sec:local quantum states Kent} typically might not lead to a \emph{normal algebraic state} (an algebraic state $\bh \to \mathbb{C}$ that can be represented as the trace over a density operator). Moreover, it may be that the Hilbert space $\hi_{\Sigma_n}$ cannot be decomposed as $\hi_x \otimes \hi_{\Sigma_n \smallsetminus \{x\}}$.

Likewise, the construction of local quantum states in Sec.~\ref{sec:polystates hilbert bundle} involve type $I$ von Neumann algebras. It is well known \cite{yngvason_role_2005} that local algebras in AQFT are generically type $III$ von Neumann algebras however. Hence, such local quantum states do not describe states of the accessible local algebras found in AQFT.

AQFT is based on Haag-Kastler nets, defined as follows \cite{haag_algebraic_1964}. Let $\mathcal{K}(\M)$ be the collection of nonempty, relatively compact (the topological closure of the subset is compact in $\M$), causally convex open subsets of $\M$. A \emph{Haag-Kastler net} is a family of unital $C^*$-algebras
\begin{equation}
    O \longmapsto \mathfrak{A}(O,g), \qquad O \in \mathcal{K}(\M),
\end{equation}
with injective unital inclusion morphisms. An AQFT in curved spacetimes in its most basic form satisfies \emph{isotony}
\begin{equation}
    O_1 \subset O_2 \Longrightarrow \mathfrak{A}(O_1,g) \subset \mathfrak{A}(O_2,g),
\end{equation}
and \emph{Einstein causality}
\begin{equation}
    O_1 \perp_g O_2 \Longrightarrow \comm{\mathfrak{A}(O_1,g)}{\mathfrak{A}(O_2,g)} = \{0\}.
\end{equation}
Other assumptions are often made, such as covariance under isometries, the time-slice property, the split property, Haag duality and others \cite{horuzhy_axiomatic_1990,halvorson_algebraic_2007,fewster_algebraic_2015}. \emph{Algebraic states} are continuous linear functionals $\omega$ on $\mathfrak{A}(\mathcal{M},g)$ such that $\omega(A^* A) \geq 0$ for all $A \in \mathfrak{A}(\mathcal{M},g)$ and $\omega(\mathbb{1}) = 1$. The restriction of algebraic states to local algebras is written $\omega_O := \omega|_{\mathfrak{A}(O,g)}$.

One can then fix Hilbert space representations $(\pi_\omega,\hi_\omega,\Omega_\omega)$ of these abstract nets of $C^*$-algebras to get von Neumann subalgebras of $\bh$:
\begin{equation}
    \mathcal{N}_\omega(O,g) = \pi_\omega(\mathfrak{A}(O,g))''.
\end{equation}
We write $\mathcal{N}_\omega(\M,g) = \pi_\omega(\mathfrak{A}(\M,g))''$. Once a representation is fixed, the normal states are elements $\varphi$ of $\mathcal{N}_\omega(\M,g)_*$; on local algebras these are $\omega_O =: \varphi_O \circ \pi_\omega$ with $\varphi_O \in \mathcal{N}_\omega(O,g)_*$, such that $\omega_O(A) = \varphi_O(\pi_\omega(A))$ for $A \in \mathfrak{A}(O,g)$. 

Defining local quantum states in AQFT is nontrivial because the axioms of AQFT refer to regions rather than points.   Algebras at specific points are trivial, so cannot be used to define ``the local quantum state at $x \in \M$''.  More precisely, under certain assumptions (unique cyclic vacuum, weak additivity, Einstein causality, covariance and the spectrum condition),
\begin{equation}
    \mathcal{N}_\omega(x,g) := \bigcap_{x \in \mathcal{O}} \mathcal{N}_\omega(O,g) = \mathbb{C} 1.
\end{equation}
This is Wightman's theorem \cite[Theorem 3.2]{wightman_theorie_1964}, see also \cite[Proposition 1.3.18]{horuzhy_axiomatic_1990} and \cite[Theorem 2]{fedida_no-go_2026}. Hence, we cannot simply talk about the states of algebras of bounded observables at a point. 

Instead, we suggest to use germs of states, defined in\footnote{Haag and Ojima attributed the idea to John E. Roberts in a private communication around 1985.} \cite{haag_problem_1996}, for the definition of local quantum states in AQFT. Indeed, one may declare the local quantum state at $x \in \M$ to be the equivalence class of algebraic states that all give the same expectation values when evaluated on locally accessible observables in local neighbourhoods of $x$. To make this precise, we define the equivalence relation $\sim_x$, given by
\begin{multline}
    \omega \sim_x \omega' \Longleftrightarrow \exists V \in \mathcal{K}(\M) \\ \text{ s.t. } V \text{ is a neighbourhood of } x \;  \& \;
    \omega_V = \omega'_V.
\end{multline}
We write elements (germs) of the induced equivalence class of states as $[\rho^\mathrm{loc}](x)$. Note that on any $W \subset V$ in $\mathcal{K}(\M)$, two representatives of the same equivalence class also coincide. This is because of isotony: if $W \subset V$ then $\forall A \in \mathfrak{A}(W,g)$, $A \in \mathfrak{A}(V,g)$ i.e. if $\omega_V(B) = \omega'_V(B)$ for all $B \in \mathfrak{A}(V,g)$ then $\omega_W(A) \equiv \omega_V|_W(A) = \omega_V'|_W(A) \equiv \omega'_W(A)$ for all $A \in \mathfrak{A}(W,g)$.

This meaningfully captures the expectation values of observables accessible at (a neighbourhood of) $x$. In the Fewster-Verch picture of quantum measurements in Algebraic QFT \cite{fewster_measurement_2024}, expectation values of states over local algebras are insensitive to nonselective quantum measurements undertaken over spacelike separations: what they call the ``principle of blissful ignorance'' \cite{fewster_lectures_2025}. Yet $[\rho^\mathrm{loc}](x)$ captures the information available in the past light cone of $x$. That is, a maximally informed observer at (a neighbourhood of) $x$ would use such a local quantum state.

Moreover, one may wish to check that the local quantum state is indeed a Hadamard state, i.e. one that has ``good'' ultraviolet behaviour \cite{fewster_algebraic_2015}. Whether we should pick by hand a Hadamard representative (when it exists) of $[\rho^\mathrm{loc}](x)$, or whether these representatives of a same germ are all either (perhaps locally) Hadamard or not Hadamard, is an interesting open question. Such considerations should be treated in a future work implementing such ideas in the context of AQFT. In this paper we take the simplifying assumption that we can
work with local quantum states that have finite stress-energy expectation value.

\section{Local semiclassical equations}

\label{sec:semiclassical equations}

Given a foliation $\{\Sigma_\tau\}_{\tau \in \mathbb{R}}$ of $\mathcal{M}$ into spacelike Cauchy hypersurfaces $\Sigma_\tau$, we postulate that\footnote{One could instead postulate a more general hypersurface evolution mapping Hilbert spaces $\hi_\Sigma \to \hi_{\Sigma'}$ which need not be unitarily equivalent. The Tomonaga-Schwinger evolution of quantum fields can be understood as some sort of interaction picture for this kind of hypersurface evolution \cite{lienert_borns_2020}, which need not be appropriate in general \cite{helfer_stress_1996,torre_functional_1999}. We take this simplifying assumption as it is commonly done in the theoretical physics literature, and helps to highlight where the nonlinearity in the wavefunctional evolution arises.} the states in this foliation evolve as
\begin{equation}
    \label{eqn:modified TS eqn GR}
    i \hbar c \frac{\delta}{\delta \Sigma_\tau(x)}\Psi[\Sigma_\tau] = \widehat{\mathscr{H}}[g[\rho^\mathrm{loc}]](x) \Psi[\Sigma_\tau]
\end{equation}
where $\widehat{\mathscr{H}}$ is the kinetic Hamiltonian density of the quantum fields in the theory. The dynamics of the gravitational field \emph{away from quantum measurements} are postulated to be given by \begin{equation}
    \label{eqn:retarded grav field GR}
    G_{ab}[\rho^\mathrm{loc}](x) + \Lambda g_{ab}[\rho^\mathrm{loc}](x) = \kappa T_{ab}^\mathrm{loc}(x)
\end{equation}
where, in a relativistic quantum theoretic setting, 
\begin{equation}
    T_{ab}^\mathrm{loc}(x) := \expval{\hat{T}_{ab}(x)}_{\rho^\mathrm{loc}(x)} = \Tr[\rho^{\mathrm{loc}}(x) \hat{T}_{ab}(x)] \, .
\end{equation}
Here, we assume that $T_{ab}^\mathrm{loc}$ is such that $T_{ab}^\mathrm{loc} \in C^\infty$ and $\nabla^a T_{ab}^\mathrm{loc} = 0$ away from quantum measurements (see Sec.~\ref{sec:measurements and divergence} for the treatment of the semiclassical equations in the vicinity of quantum measurements).

A relativistic quantum field theoretic setting requires a more careful treatment. Indeed, as mentioned previously, $\rho^\mathrm{loc}(x)$ is not typically a normal state of a local algebra. Moreover, $\hat{T}_{ab}(x)$ is typically not an operator (not even an unbounded one; rather it is a quadratic form \cite{rehberg_quantum_1986,wollenberg_quantum_1986}) so taking its trace with respect to a quantum state is meaningless. Instead, $\hat{T}_{ab}$ is an \emph{operator-valued distribution} $\mathcal{D}(\mathcal{M}) \to \mathcal{L}(D,\mathcal{H})$ in general, though we can turn its expectation value with respect to a Hadamard state into a smooth real-valued function as follows.

Let $\varphi$ be a smooth real scalar classical field propagating in a $C^\infty$ globally hyperbolic spacetime $(\M,g)$ with equation of motion $P\varphi = 0$ where
\begin{equation}
    P = -\nabla^\mu \nabla_\mu + m^2 + \xi R(x) + V(x),
\end{equation}
where $\xi \in \mathbb{R}$ is a constant, $R$ is the scalar curvature, $m \geq 0$ is the mass of the field and $V \in C^\infty(\mathcal{M},\mathbb{R})$. Then the (for now classical) symmetric stress-energy tensor, obtained by taking the variational derivative with respect to the metric of the matter action for $\varphi$ reads \cite{wald_general_1984,moretti_comments_2003}
\begin{widetext}
    \begin{multline}
        T_{\mu\nu}(x) = \nabla_\mu \varphi(x) \nabla_\nu \varphi(x) - \frac{1}{2}g_{\mu\nu}(x) \left(\nabla_\gamma \varphi(x) \nabla^\gamma \varphi(x) + \varphi^2(x) (m^2 + V(x)) \right) \\ + \xi \left[\left(R_{\mu\nu}(x) - \frac{1}{2}g_{\mu\nu}(x) R(x)\right)+g_{\mu\nu}(x) \nabla_\gamma \nabla^\gamma - \nabla_\mu\nabla_\nu \right]\varphi^2(x).
    \end{multline}
\end{widetext}
Since $P \varphi = 0$, classically we may re-define
\begin{equation}
    T_{\mu\nu}^\mathrm{ren}(x) := T_{\mu\nu}(x) + \frac{1}{3} g_{\mu\nu}(x) \varphi(x) P \varphi(x).
\end{equation}
Quantumly ($T_{\mu\nu}^\mathrm{ren} \mapsto \hat{T}_{\mu\nu}^\mathrm{ren}$) this is used for point-splitting renormalisation (see \cite{moretti_comments_2003}), and preserves the idea that quantum observables must reduce to the classical observables in the classical limit. One can then show \cite[Theorem 2.1]{moretti_comments_2003} that if $\omega$ is a Hadamard state, the map
\begin{equation}
    \label{eqn:Cinfty expval Tab}
    x \longmapsto \expval{\hat{T}_{\mu\nu}^\mathrm{ren}(x)}_\omega
\end{equation}
is $C^\infty$ (as a regular distribution), and so is $x \mapsto \expval{\hat{\varphi}^2(x)}_\omega$. Moreover \cite[Theorem 2.1]{moretti_comments_2003}, 
\begin{equation}
    \label{eqn:contracted bianchi renormalised Tab}
    \nabla^\mu \expval{\hat{T}_{\mu\nu}^\mathrm{ren}(x)}_\omega = -\frac{1}{2}\expval{\hat{\varphi}^2(x)}_\omega \nabla_\nu V(x).
\end{equation}
Hence, if $V=0$, the renormalised quantum stress-energy tensor is conserved. Thus, provided $[\rho^\mathrm{loc}](x)$ is Hadamard, we should instead write
\begin{equation}
    T^\mathrm{loc}_{ab}(x) = \expval{\hat{T}^\mathrm{ren}_{ab}(x)}_{[\rho^\mathrm{loc}](x)} = [\rho^\mathrm{loc}](x)\left(\hat{T}^\mathrm{ren}_{ab}(x)\right).
\end{equation}
This raises several open questions concerning the properties of such a $T_{ab}^\mathrm{loc}$, including:
\begin{enumerate}
    \item whether $[\rho^\mathrm{loc}](x)$ (or some representative of this equivalence class) is Hadamard,
    \item whether the ``Hadamardness'' (or some other appropriate regularity property) of (some representative of) $[\rho^\mathrm{loc}](x)$ is sufficient to show that $T^\mathrm{loc}_{ab} \in C^\infty$ away from quantum measurements (note that Eqn.~\eqref{eqn:Cinfty expval Tab} holds for a \emph{fixed Hadamard} $\omega$, not necessarily when $\omega$ is replaced by the local quantum state map $x \mapsto [\rho^\mathrm{loc}](x)$),
    \item whether the above are sufficient to show that the contracted Bianchi identities are satisfied (again, Eqn.~\eqref{eqn:contracted bianchi renormalised Tab} holds for a \emph{fixed Hadamard} $\omega$, not necessarily when $\omega$ is replaced by $x \mapsto [\rho^\mathrm{loc}](x)$).
\end{enumerate}
These questions are important for a future extension of this formalism to the gravitisation of (algebraic) quantum field theory. For the following, we take as a working assumption that, away from quantum measurements, $T_{ab}^\mathrm{loc}$ is $C^\infty$ and $\nabla^a T_{ab}^\mathrm{loc} = 0$.\footnote{Let $\V \subset \mathcal{M}$ be an open region disjoint from any collapse hypersurface of a quantum measurement, on which the metric $g$ is smooth. A sufficient and arguably reasonable condition for both (i) $T_{ab}^\mathrm{loc}$ to be $C^\infty$ and (ii) $\nabla^a T_{ab}^\mathrm{loc} = 0$ to hold in $\V$ is that, for every $x \in \V$, $\exists$ a neighbourhood $\U$ of $x$ in $\V$ and a Hadamard state $\omega_\U$ such that
\begin{equation}
    T^\mathrm{loc}_{ab}(y)
=
\omega_\U\left(\hat{T}^\mathrm{ren}_{ab}(y)\right)
\qquad\text{for all }y\in \U.
\end{equation}} Thus, the semiclassical Einstein field equations \eqref{eqn:retarded grav field GR} sourced by the local quantum state are meaningful, at least away from quantum measurements. We focus our discussion on whether Eqn.~\eqref{eqn:retarded grav field GR} can be meaningfully interpreted in the presence of quantum measurements.

We now see that, unlike M{\o}ller-Rosenfeld semiclassical gravity, the way gravity is sourced in this model is covariant: we do not source the gravitational field through the \emph{global} expectation value $\expval{\hat{T}_{ab}(x)}{\Psi[\Sigma_\tau]}$ where $x \in \Sigma_\tau$, nor through $\expval{\hat{T}_{ab}(x)}_{\rho^{\Sigma_\tau}(x)}$ where $\rho^{\Sigma_\tau}(x) := \Tr_{\Sigma_\tau \smallsetminus \{x\}}[\Psi[\Sigma_\tau] \Psi[\Sigma_\tau]^\dagger]$, which would be foliation dependent (and raises other issues, including superluminal signalling and thus potential causal paradoxes). 

Since we source the classical gravitational field through the expectation value of the quantum stress-energy tensor evaluated on the local quantum state, we call the version of semiclassical gravity discussed in this paper \emph{local semiclassical gravity}.


\subsection{Distributional equations and solutions}

\label{sec:distributional EFEs}

It is sometimes overlooked that discontinuous, or continuous but not differentiable, functions can be (distributional) solutions to partial differential equations and can be physically
meaningful. As a simple example, consider the wave equation in $1+1$ dimensions:
\begin{equation}
    \label{eqn:wave 1+1d}
    (\partial_t^2 - \partial_x^2) f = 0
\end{equation}
Then the Heaviside step function $\Theta$, which is a locally integrable discontinuous function given by
\begin{equation}
    f(t,x) = \Theta(t-x)= \begin{cases}
        1 \quad \text{ if } t \geq x \\
        0 \quad \text{ otherwise}
    \end{cases}
\end{equation}
is a solution to the wave equation \eqref{eqn:wave 1+1d} when understood as a distribution in $\mathcal{D}(\mathbb{R}^2)'$. The lesson from this is that one can work with discontinuous functions (and, indeed, continuous but not differentiable functions) and still satisfy differential equations at the level of distributions. This is commonly used to model phenomena like shocks and sudden pressure changes across some boundary.

We would argue that the Einstein field equations \eqref{eqn:retarded grav field GR} as well as the Tomonaga-Schwinger equation \eqref{eqn:modified TS eqn GR} should be understood \emph{distributionally} for (at least) two reasons:

\begin{enumerate}
    \item Quantum fields (such as $\hat{T}_{ab}$ as well as $\widehat{\mathscr{H}}$) cannot generally be understood as operator-valued functions $\mathcal{M} \to \mathcal{L}(D,\hi)$ (where $D$ is a dense subset of $\hi$ for unbounded quantum fields) \cite{halvorson_algebraic_2007,fedida_no-go_2026}. Rather, quantum fields are generally to be understood as operator-valued \emph{distributions} $\mathcal{D}(\mathcal{M}) \to \mathcal{L}(D,\hi)$, as is for example axiomatised in Wightman QFT \cite{wightman_fields_1965}.\footnote{In Wightman's approach, quantum fields in Minkowski spacetime are operator-valued \emph{tempered} distributions $\mathscr{S}(\mathbb{R}^4) \to \mathcal{L}(D,D)$ that preserve the dense domain $D$. This difference is not important for the point we make here however.} The value of a quantum field at specific spacetime points is sometimes ill-defined, and so it may not always possible to assign a pointwise backreaction in the Einstein field equations.
    
    \item In some standard models of quantum measurement, the quantum states are updated so that $T_{ab}^\mathrm{loc}$ jumps discontinuously. Thus, although $T_{ab}^\mathrm{loc}$ is $C^\infty$ and $\nabla^a T_{ab}^\mathrm{loc} = 0$ away from quantum measurements, this is typically not the case in the vicinity of the future light cone of the localisation of quantum measurements. As we will see in Sec. \ref{sec:measurements and divergence}, this need not cause inconsistency with the Einstein field equations as long as the latter are treated distributionally.
\end{enumerate}

We should thus generally look at \emph{distributional solutions} to the Einstein field equations and to the Tomonaga-Schwinger equation, i.e. solutions to these partial differential equations that are distributions and thus not differentiable everywhere.  Specifically, the matter source and the Einstein tensor may be discontinuous and indeed distributional.  We require the metrics to be continuous (and, in particular, we will be able to require local Lipschitz continuity), since we require a well-defined classical Lorentzian geometry. This will be crucial in the treatment of quantum measurements, as we will show in Sec. \ref{sec:mmts}.
We thus write (heuristically)\footnote{This notation with an integral kernel is indeed heuristic: we should really be thinking of these distributions more generally without assuming their regularity (i.e. the fact that there exists a smearing-function-independent pointwise-defined function being smeared under the integral).}
\begin{widetext}
\begin{equation}
\begin{dcases}
    \int_{\mathcal{M}} f^{ab}(x) \left(G_{ab}[\rho^\mathrm{loc}](x) + \Lambda g_{ab}[\rho^\mathrm{loc}](x) - \kappa T_{ab}^\mathrm{loc}(x)\right) \sqrt{\abs{g[\rho^\mathrm{loc}](x)}} d^dx = 0 \\
    \int_{\mathcal{M}} f(x) \left(i \hbar c \dfrac{\delta}{\delta \Sigma_\tau(x)}\Psi[\Sigma_\tau] - \widehat{\mathscr{H}}[g[\rho^\mathrm{loc}]](x) \Psi[\Sigma_\tau] \right) \sqrt{\abs{g[\rho^\mathrm{loc}](x)}} d^dx = 0
\end{dcases} \qquad 
\end{equation}
\end{widetext}
for all ``suitable'' ($\mathcal{D}(\M)$) functions $f^{ab}$ and $f$, \emph{away from quantum measurements}, with some modification to the stress-energy in the vicinity of quantum state collapses, as we will see in Sec.~\ref{sec:measurements and divergence}. 

\subsection{Integrability}

\label{sec:integrability}

We note here Tomonaga-Schwinger dynamics should be
foliation-independent and hence require integrability. That is, that the evolution from one leaf of the chosen foliation to another leaf is independent of the path used to evolve the state:
\begin{equation}
    \comm{\frac{\delta}{\delta \Sigma_\tau(x)}}{\frac{\delta}{\delta \Sigma_\tau(y)}} \Psi[\Sigma_\tau] = 0 \, .
\end{equation}
Within the context of the standard Tomonaga-Schwinger equation, this corresponds to the requirement of (distributional) microcausality for the kinetic Hamiltonian density of the quantum fields in the theory:
\begin{equation}
    \comm{\widehat{\mathscr{H}}(x)}{\widehat{\mathscr{H}}(y)} = 0 \qquad \forall x \perp y \, .
\end{equation}
In \cite{hsu_relativistic_2026}, it was argued that nonlinear quantum theories typically violate the integrability condition of the Tomonaga-Schwinger equation. This analysis was rebutted in \cite{oppenheim_nonlinear_2025,diosi_comment_2026}, which showed that the criterion for integrability is the locality of the Hamiltonian density.  In our case, the Hamiltonian density is indeed local. 

A direct application of the product rule implies that the integrability condition for the semiclassical Tomonaga-Schwinger equation \eqref{eqn:modified TS eqn GR} becomes
\begin{widetext}
    \begin{equation}
        \left(\comm{\widehat{\mathscr{H}}[g[\rho^\mathrm{loc}]](x)}{\widehat{\mathscr{H}}[g[\rho^\mathrm{loc}]](y)} + i \hbar c \left(\frac{\delta}{\delta \Sigma_\tau(y)} \widehat{\mathscr{H}}[g[\rho^\mathrm{loc}]](x) - \frac{\delta}{\delta \Sigma_\tau(x)}\widehat{\mathscr{H}}[g[\rho^\mathrm{loc}]](y)\right)\right) \Psi[\Sigma_\tau] = 0.
    \end{equation}
\end{widetext}
with $x \perp y$. Note however that the source data at any spacetime point $z \in J^-(x)$ is a functional of $\rho^\mathrm{loc}(z)$ (local expectation values of local stress-energy), which is not affected by a local deformation of the global state $\Psi[\Sigma]$ at $y \perp x$, and we also have $y \notin J^-(z)$, thus $\rho^\mathrm{loc}(z)$ is unchanged by $\frac{\delta}{\delta \Sigma_\tau(y)}$, and thus so is $g[\rho^\mathrm{loc}](x)$ and hence so is $\widehat{\mathscr{H}}[g[\rho^\mathrm{loc}]](x)$, i.e.
\begin{equation}
    \frac{\delta}{\delta \Sigma_\tau(y)} \widehat{\mathscr{H}}[g[\rho^\mathrm{loc}]](x) = 0 \qquad \forall x \perp y \, .
\end{equation}
Thus, the integrability condition reduces to the ``standard'' one
\begin{equation}
    \comm{\widehat{\mathscr{H}}[g[\rho^\mathrm{loc}]](x)}{\widehat{\mathscr{H}}[g[\rho^\mathrm{loc}]](y)} = 0 \, ,
\end{equation}
which is satisfied if $\widehat{\mathscr{H}}[g[\rho^\mathrm{loc}]]$ is a microcausal (in the distributional sense) Hamiltonian density, which is the standard criterion in QFT. We highlight in Sec.~\ref{sec:relativistic readout devices} that quantum measurements preserve the causal property of the metric, 
i.e. that the evolution is also integrable in the presence of quantum measurements, 
at least in the weak-field regime.

\section{The mixture equivalence principle}

\label{sec:MEP}

In quantum theory, there are two ontologically distinct types of mixtures: improper mixtures, which arise as a description of a subsystem obtained from tracing out a degree of freedom of a multipartite system, and proper mixtures, which arise from (classical) statistical considerations of an ensemble of states \cite{fedida_mixture_2025}. In standard quantum mechanics, proper and improper mixtures described by the same density operator $\rho \in \state$ are empirically indistinguishable since quantum states capture all the physically accessible information obtainable from acting only on the relevant subsystem. That is, if a proper mixed state $\rho_\mathrm{proper}$ and an improper mixed state $\rho_\mathrm{improper}$ are such that $\rho_\mathrm{proper} = \rho_\mathrm{improper}$, then within standard quantum theory no experiment can distinguish both cases.

This idea is encapsulated in the \emph{mixture equivalence principle} (MEP) \cite{fedida_mixture_2025}\footnote{As well as in the closely related \emph{purification principle} \cite{galley_any_2018} and \emph{weak mixture equivalence principle} \cite{fedida_mixture_2025}.}.
\begin{quote}
    \textbf{MEP: } Let $\{\ket{\psi_i}\}_i \in \mathcal{H_S}$ be an ensemble of pure states
with associated probabilities $\{p_i\}_i$, $\sum_i p_i = 1$, where $\hi_\mathcal{S}$ is the Hilbert space representing a system $\mathcal{S}$, and let $\rho_\mathrm{proper} = \sum_{i} p_i \ketbra{\psi_i}$ be the corresponding proper mixed state. Let $\ket{\Psi} \in \hi_\mathcal{S} \otimes \hi_\mathcal{A}$ be a state in the Hilbert space representing the combined system $\mathcal{S} + \mathcal{A}$, and let $\rho_\mathrm{improper} = \Tr_{\hi_A}[\ketbra{\Psi}]$ be the corresponding improper mixed state of $\mathcal{S}$. Suppose $\rho_\mathrm{proper} = \rho_\mathrm{improper}$. Then, no experiment on $\mathcal{S}$ can distinguish these two cases.
\end{quote}
By transitivity, the MEP also implies that no experiment can distinguish any pair of proper mixtures, or any pair of improper mixtures, represented by the same density matrix. A much weaker statement is the so-called \emph{gravitational weak mixture equivalence principle} (GWMEP) \cite{fedida_mixture_2025}.
\begin{quote}
\textbf{GWMEP: } There exists an ensemble of experimentally distinguishable pure states $\{\ket{\psi_i}\}_i \in \hi_\mathcal{S}$ with associated probabilities $\{p_i\}_i$, $\sum_i p_i = 1$ such that $\rho_\mathrm{proper} = \sum_{i} p_i \ketbra{\psi_i}$ is the corresponding proper mixed state, and a state $\ket{\Psi} \in \hi_\mathcal{S} \otimes \hi_\mathcal{A}$ such that $\rho_\mathrm{improper} = \Tr_{\hi_A}[\ketbra{\Psi}]$ is the corresponding improper mixed state of $\mathcal{S}$, such that $\rho_\mathrm{proper} = \rho_\mathrm{improper}$ and such that no measurable gravitational effect of $\mathcal{S}$ can distinguish these two cases.
\end{quote}
Standard quantum gravity models are unitary and do not violate the GWMEP.
However, our model (and mean-field models more generally) violate it, as was shown in \cite{fedida_mixture_2025}. 

It is easy to see that our theory violates these mixture equivalence principles.
First, note that in many physical situations $\rho^\mathrm{loc}$ 
is an improper mixture that is not well approximated by a pure state.  While improper mixtures source (within the expectation value of the stress-energy tensor) the gravitational field of a single spacetime, proper mixtures are associated with a statistical ensemble of spacetimes. For example, given a statistical ensemble describing a gas  in a localised box, the associated dynamics of the matter and of the spacetime follow the statistics of the relevant ensemble. Thus, we end up with an ensemble of spacetimes:
\begin{align*}
    \rho_\mathrm{proper}^\mathrm{loc}(x) &=\{p_i,\rho_i^\mathrm{loc}(x)\} \\
    &\downarrow \\
    \{p_i&,g_i[\rho^\mathrm{loc}_i](x)\} \\
    &\downarrow \\
    \{p_i&,(\mathcal{M}_i,g_i)\} \, .
\end{align*}
On the other hand, for an improper (local) mixture $\rho_\mathrm{improper}^\mathrm{loc}(x)$ (which could have the same density matrix as in the proper mixture described above), only a single spacetime is sourced:
\begin{equation*}
    \rho_\mathrm{improper}^\mathrm{loc}(x) \rightarrow g[\rho^\mathrm{loc}_\mathrm{improper}](x) \rightarrow (\mathcal{M},g) \, .
\end{equation*}
Hence the GWMEP, and hence the MEP, is violated.    Violating the MEP has been used to argue that some deterministic nonlinear extensions of quantum mechanics lead to superluminal signalling \cite{gisin_stochastic_1989}.
However, an MEP-violating theory (such as the present model) need not allow superluminal signalling \cite{kent_nonlinearity_2005,fedida_mixture_2025}.  
Conversely, unitary quantum mechanics in Galilean spacetime allows superluminal signalling but does not violate the MEP.  That is, the two notions are independent.


\section{Limits of the theory}

\label{sec:limits}

Note that we will use ``physicists'" tools and rigour here, ignoring from now on the fact that equal-time canonical commutation relations are generally ill-defined for pointwise quantum fields \cite{halvorson_algebraic_2007,fedida_no-go_2026}. Mathematically minded readers are invited to consider the following discussion distributionally.

\subsection{Weak Gravitational Field Limit}

\label{sec:weak field limit}

From now on, for simplicity, we work at scales where $\Lambda$ is negligible, and on an approximately Minkowski background. We write the metric as $g_{\mu \nu} = \eta_{\mu \nu} + h_{\mu \nu}$, with $\abs{h_{\mu \nu}} \ll 1$. The linearised field equations in Lorenz gauge reduce to
\begin{align}
    \label{eqn:linearised EFEs}
        \Box \overline{h}_{\mu \nu} &= -2 \kappa T_{\mu \nu}^{\text{loc}}, \\
        \label{eqn:Lorenz condition}
        \partial^\mu \overline{h}_{\mu\nu} &= 0 \,, 
\end{align}
where $\overline{h}_{\mu\nu} = h_{\mu\nu} - \frac{1}{2}\eta_{\mu\nu} \eta^{\rho\sigma} h_{\rho\sigma}$ denotes the trace-reversal of $h$.
The retarded Green's function for $\Box$ in flat spacetime is
\begin{equation}
    \label{eqn:retarded Green function}
    G_\text{ret}(x-x') = \frac{\delta(t'-t+\abs{\vec{x}-\vec{x'}}/c)}{4\pi \abs{\vec{x}-\vec{x}'}} \Theta(t-t') 
\end{equation}
where $\Theta$ is the Heaviside theta function, so that
\begin{equation}
    \Box G_\text{ret}(x-x') = -\delta^{(4)}(x-x'),
\end{equation}
and so
\begin{equation}
    \label{eqn: weak field gravitational field}
    \bar{h}_{\mu \nu}(t,\vec{x}) = \frac{4G}{c^4} \iiint_{\mathbb{R}^3}  \frac{T_{\mu \nu}^\mathrm{loc}(t_\text{ret},\vec{x}')}{\abs{\vec{x}-\vec{x}'}} d^3 \vec{x}'
\end{equation}
where $t_\text{ret} = t - \frac{\abs{\vec{x}-\vec{x}'}}{c}$. This is explicitly relativistically causal since $\rho^\mathrm{loc}$ is itself retarded. That is, ignoring quantum measurements for now, one cannot signal faster than light through the gravitational field by (i) any gravitational field measurement, (ii) gravitational dynamics, and (iii) quantum field dynamics. Likewise, there is no gravitational backreaction on the quantum matter sector that is not retarded.  The nonlinearity in the quantum state thus only depends on data in the past light cone.
As we will see in Sec. \ref{sec:mmts}, in our treatment quantum measurements update the wavefunction in the future light cone of the measurement events, so also cannot lead to superluminal signalling.

\subsection{Nonrelativistic Limit}

Let us now show that Eqns.~\eqref{eqn:modified TS eqn GR} and \eqref{eqn:retarded grav field GR} reduce to the standard Schrödinger-Newton system in the non-relativistic limit on a ${1+3}$-dimensional Minkowski background. First, we consider the passage from the Tomonaga-Schwinger picture to the Schrödinger picture in Minkowski spacetime. Choose a Minkowski foliation into constant-time slices $\Sigma_t = \{(t,\vec{x}) : \vec{x} \in \mathbb{R}^3\}$ with unit normal $n^\mu = (1,0,0,0)$. Advancing $\Sigma_t$ by $dt$ yields
\begin{equation}
    \frac{d}{dt} \Psi[\Sigma_t] = \iiint_{\mathbb{R}^3} c \frac{\delta}{\delta \Sigma_t(t,\vec{x})} \Psi[\Sigma_t] \, d^3\vec{x} \, .
\end{equation}
Integrating eqn. \eqref{eqn:modified TS eqn GR} over $\mathbb{R}^3$ thus gives
\begin{equation}
    i \hbar \frac{d}{dt} \Psi[\Sigma_t] = \iiint_{\mathbb{R}^3} \widehat{\mathscr{H}}[g[\rho^\mathrm{loc}]](x) \Psi[\Sigma_t] \, d^3\vec{x} \, .
\end{equation}
Define $\ket{\psi(t)} := \Psi[\Sigma_t]$ and
\begin{equation}
    \hat{H}[\rho^\mathrm{loc}](t) := \iiint_{\mathbb{R}^{3}} \widehat{\mathscr{H}}[g[\rho^\mathrm{loc}]](x) \, d^3\vec{x} \, .
\end{equation}
Then
\begin{equation}
    \label{eqn: NL TDSE}
    i \hbar \partial_t \ket{\psi(t)} = \hat{H}[\rho^\mathrm{loc}](t) \ket{\psi(t)}
\end{equation}
is a (nonlinear) Schrödinger equation. We identify
\begin{equation}
    g_{00} = -\left(1 + \frac{2\Phi}{c^2}\right), \quad g_{ij} = \left(1 - \frac{2\Phi}{c^2}\right) \delta_{ij}
\end{equation}
and
\begin{equation}
    g^{00} = -\left(1 - \frac{2 \Phi}{c^2}\right), \quad g^{ij} = \left(1 + \frac{2\Phi}{c^2}\right) \delta^{ij}
\end{equation}
with $\sqrt{-g} \approx 1-\frac{2\Phi}{c^2}$, so that
\begin{equation}
    h_{00} = -\frac{2\Phi}{c^2}, \quad h_{ij} = -\frac{2\Phi}{c^2} \delta_{ij}, \quad h = -h_{00} + \delta^{ij} h_{ij} = -\frac{4\Phi}{c^2}
\end{equation}
so
\begin{equation}
    \bar{h}_{00} = -\frac{4\Phi}{c^2}, \qquad \overline{h}_{ij} \approx 0 \, .
\end{equation}
To leading order in $\frac{1}{c^2}$, since $\Box = -\frac{1}{c^2}\partial_t^2 + \nabla^2$, and $\hat{T}_{00} = \hat{T}^{00}=\hat{\varrho}_m c^2$ where $\hat{\varrho}_m(\vec{x}) = m \hat{\psi}^\dagger(\vec{x})\hat{\psi}(\vec{x})$ is the mass density of the quantum fields (we here assume only one kind of particle is present for simplicity), and other components of the energy-momentum tensor are subleading in $c$, we have
\begin{equation}
    \nabla^2 \Phi(t,\vec{x}) = 4 \pi G m \Tr[\hat{\psi}^\dagger(\vec{x}) \hat{\psi}(\vec{x}) \rho^\mathrm{loc}(t,\vec{x})],
\end{equation}
with $\Phi(t,\vec{x}) \to 0$ as $\abs{\vec{x}} \to \infty$ as boundary conditions. In this limit, the nonrelativistic gravitational interaction Hamiltonian is \cite{bahrami_schrodingernewton_2014}
\begin{align}
    \hat{H}_\text{grav}[\rho^\mathrm{loc}](t) &= -\frac{1}{2} \iiint_{\mathbb{R}^3} h_{\mu \nu}(t,\vec{x}) \hat{T}^{\mu \nu}(\vec{x}) d^3 \vec{x} \\ &\approx \iiint_{\mathbb{R}^3} \frac{\Phi[\rho^\mathrm{loc}](t,\vec{x})}{c^2} \hat{T}_{00}(\vec{x}) d^3 \vec{x} \\
    &= \iiint_{\mathbb{R}^3} \Phi[\rho^\mathrm{loc}](t,\vec{x}) m \hat{\psi}^\dagger(\vec{x}) \hat{\psi}(\vec{x}) d^3 \vec{x}
\end{align}
while the kinetic Hamiltonian is
\begin{equation}
    \hat{H}_\text{kin} = \iiint_{\mathbb{R}^3} \frac{\hbar^2}{2m} \nabla \hat{\psi}^\dagger(\vec{x}) \cdot \nabla \hat{\psi}(\vec{x}) d^3 \vec{x}
\end{equation}
i.e. the full nonrelativistic Hamiltonian is
\begin{multline}
    \hat{H}[\rho^\mathrm{loc}](t) = \iiint_{\mathbb{R}^3} \Big[\frac{\hbar^2}{2m} \nabla \hat{\psi}^\dagger(\vec{x}) \cdot \nabla \hat{\psi}(\vec{x}) \\ + m \Phi[\rho^\mathrm{loc}](t,\vec{x}) \hat{\psi}^\dagger(\vec{x}) \hat{\psi}(\vec{x})\Big] d^3 \vec{x} \, .
\end{multline}
The passage from the Schrödinger to the Heisenberg picture cannot be done as in the unitary theory, at least not in a state-independent way. We thus work in Schrödinger picture, for which quantum fields are time-independent. 
We take the number of particles to be constant. Thus, we define the $N$-particle position ket as
\begin{equation}
    \ket{\vec{x}_1,\cdots,\vec{x}_N} := \frac{1}{\sqrt{N!}} \hat{\psi}^\dagger(\vec{x}_1) \cdots \hat{\psi}^\dagger(\vec{x}_N) \ket{0}
\end{equation}
so that the $N$-particle wavefunction is
\begin{equation}
    \Psi_N(t,\vec{x}_1,\cdots,\vec{x}_N) := \braket{\vec{x}_1,\cdots,\vec{x}_N}{\psi(t)} \, .
\end{equation}
For the kinetic part of the Hamiltonian, we can use $\comm{\hat{\psi}(\vec{x})}{\hat{\psi}^\dagger(\vec{y})} = \delta^3(\vec{x}-\vec{y})$ and integration by parts to find
\begin{multline}
    \bra{\vec{x}_1,\cdots,\vec{x}_N} \hat{H}_\text{kin} \ket{\psi(t)} \\ = \left(- \sum_{k=1}^N \frac{\hbar^2}{2m} \nabla^2_{\vec{x}_k}\right) \Psi_N(t,\vec{x}_1,\cdots,\vec{x}_N) \, .
\end{multline}
For the gravitational term, we also use the commutation relations to find
\begin{multline}
    \bra{\vec{x}_1,\cdots,\vec{x}_N} \hat{H}_\text{grav}[\rho^\mathrm{loc}](t) \ket{\psi(t)} \\ = \left(\sum_{k=1}^N m \Phi[\rho^\mathrm{loc}](t,\vec{x}_k)\right) \Psi_N(t,\vec{x}_1,\cdots,\vec{x}_N) \, .
\end{multline}
Thus,
\begin{align}
    i \hbar &\partial_t \Psi_N(t,\vec{x}_1,\cdots,\vec{x}_N) = \bra{\vec{x}_1,\cdots,\vec{x}_N} \hat{H}[\rho^\mathrm{loc}](t) \ket{\psi(t)} \\
    &= \left(\sum_{k=1}^N\left(- \frac{\hbar^2}{2m} \nabla^2_{\vec{x}_k} + m \Phi[\rho^\mathrm{loc}](t,\vec{x}_k) \right)\right) \nonumber \\ & \qquad \qquad \qquad \qquad \qquad \qquad \Psi_N(t,\vec{x}_1,\cdots,\vec{x}_N) \, .
\end{align}
Now note that for a pure $1$-particle state, $\Tr[\hat{\psi}^\dagger(\vec{x}) \hat{\psi}(\vec{x})\rho^{\mathrm{loc}}(t,\vec{x})]$ becomes $\abs{\psi(t,\vec{x})}^2$ since the light-cones flatten, and $\expval{\hat{\varrho}_m}_{\rho^{\mathrm{loc}}(t,\vec{x})} = m \abs{\psi(t,\vec{x})}^2$ is the mass density for a single particle, so

\begin{equation}
    \label{eqn:SNE 1 particle}
\begin{cases}
    i \hbar \partial_t \Psi_1(t,\vec{x}) = \left(- \frac{\hbar^2}{2m}\nabla^2 + m\Phi[\Psi_1](t,\vec{x})\right)\Psi_1(t,\vec{x}) \, , \\
    \nabla^2 \Phi(t,\vec{x}) = 4\pi G m \abs{\Psi_1(t,\vec{x})}^2 \, .
\end{cases}
\end{equation}

This is precisely the Schrödinger-Newton system for a single particle. 
The discussion for the many-particle case is similar. Again, in the nonrelativistic limit, the boundary of the past light-cone flattens onto an equal-time slice. Thus, the local state prescription yields a $1$-particle reduced state on $\Sigma_t$, tracing out over the other particles' configurations:
\begin{multline}
    \Tr[\hat{\psi}^\dagger(\vec{x}) \hat{\psi}(\vec{x}) \rho^\mathrm{loc}(t,\vec{x})] = \\ N \idotsint_{\mathbb{R}^{3(N-1)}} \abs{\Psi_N(t,\vec{x},\vec{x}_2,\cdots,\vec{x}_N)}^2 d^3\vec{x}_2 \cdots d^3 \vec{x}_N \, .
\end{multline}
Thus,
\begin{multline}
    \label{eqn:N-body Poisson}
    \nabla^2 \Phi(t,\vec{x}) = 4 \pi G mN \\ \idotsint_{\mathbb{R}^{3(N-1)}} \abs{\Psi_N(t,\vec{x},\vec{x}_2,\cdots,\vec{x}_N)}^2 d^3 \vec{x}_2 \cdots d^3 \vec{x}_N \, .
\end{multline}
The wavefunction dynamics obeys
\begin{multline}
    \label{eqn:N-body SNE}
    i \hbar \partial_t \Psi_N(t,\vec{x}_1,\cdots,\vec{x}_N) \\ = \left(\sum_{k=1}^N\left(- \frac{\hbar^2}{2m} \nabla^2_{\vec{x}_k} + m \Phi[\rho^\mathrm{loc}](t,\vec{x}_k) \right)\right) \\  \Psi_N(t,\vec{x}_1,\cdots,\vec{x}_N) \, .
\end{multline}
Eqns.~ \eqref{eqn:N-body Poisson} and \eqref{eqn:N-body SNE} define the Schrödinger-Newton system for $N$ particles. The model considered  in this paper at the level of Eqns.~\eqref{eqn:modified TS eqn GR} and \eqref{eqn:retarded grav field GR} can thus be seen as a relativistically causal completion of the Schrödinger-Newton model -- up to the consideration of quantum measurements, which we now examine.

\section{Quantum measurements and quantum state updates}

\label{sec:mmts}

Although the dynamics of the gravitational field and the wavefunctional are relativistically causal away from measurements, one must consider the latter to get a full story of 1) the causal nature of the theory and 2) the available operations within the theory. We first treat quantum measurements before examining measurements of the classical gravitational field in Sec.\ref{sec:relativistic readout devices}.

We seek a postulate defining how measurements induce wavefunction updates that is consistent with relativistic causality. We will examine three possibilities: past light-cone updates à la Hellwig-Kraus \cite{hellwig_pure_1969,hellwig_formal_1970,kraus_general_1971}, instantaneous updates à la Aharonov-Albert \cite{aharonov_states_1980,aharonov_can_1981,aharonov_is_1984,breuer_state_1999,breuer_measurements_2007}, and future light-cone updates in the sense of detector-based models \cite{polo-gomez_detector-based_2022,perche_particle_2024,pranzini_detector-based_2025,polo-gomez_state_2025}.\footnote{
A radically different possibility would be to treat quantum measurements as a causal physical process, leading to some form of causal quantum theory \cite{kent_causal_2005,kent_testing_2018}. In such models, Bell correlations are not generally maintained over spacelike separations. 
Despite this, causal versions of quantum theory can be logically
consistent and some versions are consistent with experiments to date.
This is because of the collapse locality loophole \cite{kent_causal_2005,kent_stronger_2020}, which arises because of the ambiguity in quantum theory as to when and where a definite measurement outcome actually occurs. 
In models in which collapse depends on displacement of significant masses, or on conscious observation, the measurement outcomes in standard Bell experiments need not arise in the spacelike separated regions where they are normally assumed to arise.  
Instead a joint measurement outcome may arise in a single region in their joint causal future. 
Whether a given experiment thus
reports data that appear consistent with Bell correlations 
depends on the details of the experiment and on the specific collapse model considered \cite{kent_testing_2018,salart_spacelike_2008}. We shall not further expand on this possibility in this paper.}

An alternative interesting approach would be to consider algebraic methods to model quantum measurements in QFT \cite{fewster_quantum_2020,fewster_measurement_2024}. It is known \cite{fewster_hadamard_2025} that non-selective Fewster-Verch quantum measurements preserve the Hadamard property; the case of selective measurements is not known. This is important because if $[\rho^\mathrm{loc}](x)$ is Hadamard prior to a Fewster-Verch quantum measurement then it would also be after the collapse hypersurface. Hence, one would potentially be guaranteed to have $T_{ab}^{\mathrm{loc},\pm} \in C^\infty$ and $\nabla^a T_{ab}^{\mathrm{loc},\pm} = 0$ on both sides of the collapse hypersurface. Such an approach would however be better adapted in a model of semiclassical gravity that is embedded within the AQFT framework, which we leave to future work.

\subsection{Past light-cone updates}

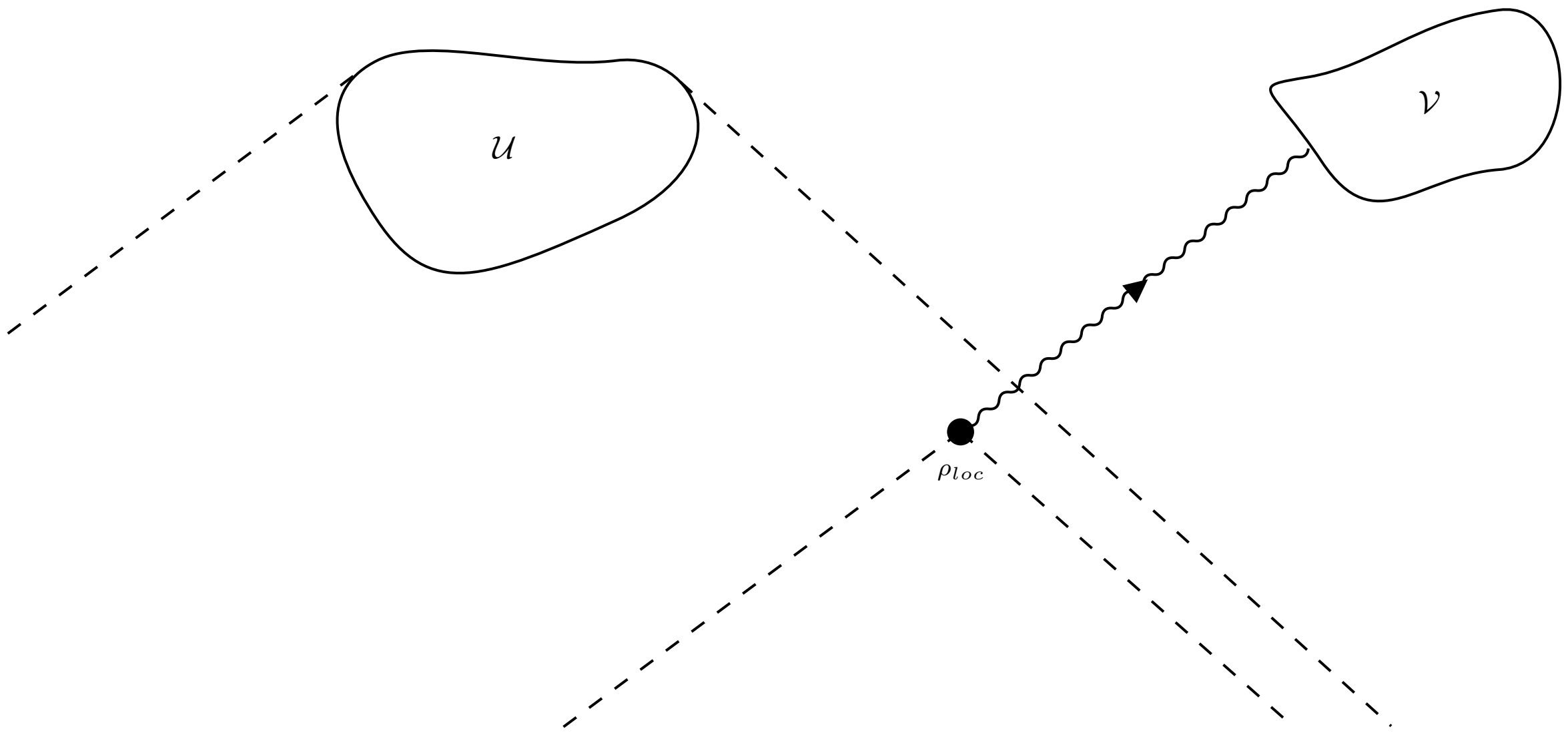
\begin{figure*}
    \begin{tikzpicture}[x=0.75pt,y=0.75pt,yscale=-1,xscale=1]

\draw   (140,48) .. controls (160,38) and (199.64,51.82) .. (230,48) .. controls (260.36,44.18) and (280.36,85.18) .. (230,108) .. controls (179.64,130.82) and (160,138) .. (140,108) .. controls (120,78) and (120,58) .. (140,48) -- cycle ;
\draw   (490.36,54.18) .. controls (514.36,50.18) and (531.64,32.82) .. (562,29) .. controls (592.36,25.18) and (592.64,86.82) .. (562,89) .. controls (531.36,91.18) and (515.36,116.18) .. (495.36,86.18) .. controls (475.36,56.18) and (466.36,58.18) .. (490.36,54.18) -- cycle ;
\draw  [dash pattern={on 4.5pt off 4.5pt}]  (254,56) -- (521.36,298.18) ;
\draw  [dash pattern={on 4.5pt off 4.5pt}]  (131,53.82) -- (0.36,151.18) ;
\draw  [fill={rgb, 255:red, 0; green, 0; blue, 0 }  ,fill opacity=1 ] (354.82,187.59) .. controls (354.82,185.06) and (356.88,183) .. (359.41,183) .. controls (361.94,183) and (364,185.06) .. (364,187.59) .. controls (364,190.12) and (361.94,192.18) .. (359.41,192.18) .. controls (356.88,192.18) and (354.82,190.12) .. (354.82,187.59) -- cycle ;
\draw  [dash pattern={on 4.5pt off 4.5pt}]  (359.41,187.59) -- (205.2,302.08) ;
\draw  [dash pattern={on 4.5pt off 4.5pt}]  (359.41,187.59) -- (484.2,298.08) ;
\draw    (490.2,81.08) .. controls (489.96,83.43) and (488.67,84.48) .. (486.32,84.24) .. controls (483.98,84) and (482.69,85.05) .. (482.45,87.39) .. controls (482.21,89.74) and (480.92,90.79) .. (478.57,90.55) .. controls (476.22,90.31) and (474.93,91.36) .. (474.69,93.71) .. controls (474.45,96.06) and (473.16,97.11) .. (470.81,96.87) .. controls (468.47,96.63) and (467.18,97.68) .. (466.94,100.02) .. controls (466.7,102.37) and (465.41,103.42) .. (463.06,103.18) .. controls (460.71,102.94) and (459.42,103.99) .. (459.18,106.34) .. controls (458.95,108.69) and (457.66,109.74) .. (455.31,109.5) .. controls (452.96,109.25) and (451.67,110.3) .. (451.43,112.65) .. controls (451.19,115) and (449.9,116.05) .. (447.55,115.81) .. controls (445.2,115.57) and (443.91,116.62) .. (443.68,118.97) .. controls (443.44,121.32) and (442.15,122.37) .. (439.8,122.12) .. controls (437.45,121.88) and (436.16,122.93) .. (435.92,125.28) .. controls (435.68,127.63) and (434.39,128.68) .. (432.04,128.44) .. controls (429.69,128.2) and (428.4,129.25) .. (428.17,131.6) .. controls (427.93,133.95) and (426.64,135) .. (424.29,134.75) .. controls (421.94,134.51) and (420.65,135.56) .. (420.41,137.91) .. controls (420.18,140.26) and (418.89,141.31) .. (416.54,141.07) .. controls (414.19,140.83) and (412.9,141.88) .. (412.66,144.23) .. controls (412.42,146.58) and (411.13,147.63) .. (408.78,147.38) .. controls (406.43,147.14) and (405.14,148.19) .. (404.91,150.54) .. controls (404.67,152.89) and (403.38,153.94) .. (401.03,153.7) .. controls (398.68,153.46) and (397.39,154.51) .. (397.15,156.86) .. controls (396.91,159.21) and (395.62,160.26) .. (393.27,160.01) .. controls (390.92,159.77) and (389.63,160.82) .. (389.4,163.17) .. controls (389.16,165.52) and (387.87,166.57) .. (385.52,166.33) .. controls (383.17,166.08) and (381.88,167.13) .. (381.64,169.48) .. controls (381.41,171.83) and (380.12,172.88) .. (377.77,172.64) .. controls (375.42,172.4) and (374.13,173.45) .. (373.89,175.8) .. controls (373.65,178.15) and (372.36,179.2) .. (370.01,178.96) .. controls (367.66,178.71) and (366.37,179.76) .. (366.13,182.11) .. controls (365.9,184.46) and (364.61,185.51) .. (362.26,185.27) -- (359.41,187.59) -- (359.41,187.59) ;
\draw [shift={(429.85,130.23)}, rotate = 140.84] [fill={rgb, 255:red, 0; green, 0; blue, 0 }  ][line width=0.08]  [draw opacity=0] (8.93,-4.29) -- (0,0) -- (8.93,4.29) -- cycle    ;

\draw (349,198.4) node [anchor=north west][inner sep=0.75pt]  [font=\scriptsize]  {$\rho _{loc}$};
\draw (182,75) node [anchor=north west][inner sep=0.75pt]   [align=left] {$\U$};
\draw (530,58) node [anchor=north west][inner sep=0.75pt]   [align=left] {$\V$};

\end{tikzpicture}
    \caption{\justifying Hellwig-Kraus past cone updates inducing both retrocausation and superluminal signalling in the context of local semiclassical gravity: the local state at $x \in J^-(\U) \cap J^-(\V)$ updates; this change of local state changes the local expectation value of the stress-energy tensor, and so the gravitational field backreacts in the past of $\V$, so an observer in $\V$ can, in principle, determine whether a quantum measurement has been undertaken in $\U$ through a Cavendish experiment.}
    \label{fig:Past light cone problems}
\end{figure*}

Past light-cone collapse was first examined by Hellwig and Kraus \cite{hellwig_pure_1969}. In this approach, upon a measurement, given by measurement operators\footnote{Let $X$ be a topological space and $\Bor(X)$ the Borel $\sigma$-algebra on $X$. A \emph{measurement (Kraus) operator} \cite{fedida_einstein_2026} is a map $\mathsf{M} : \Bor(X) \to \bh$ such that
\begin{equation}
    \mathsf{M}\Bigg(\bigcup_{n=1}^\infty \Delta_n \Bigg)^\dagger \mathsf{M}\Bigg(\bigcup_{n=1}^\infty \Delta_n \Bigg) = \sum_{n=1}^\infty \mathsf{M}(\Delta_n)^\dagger \mathsf{M}(\Delta_n)
\end{equation}
where $\{\Delta_n\}_{n}$ is a countable family of disjoint sets in $\Bor(X)$ and the series converges in the some chosen (e.g. weak-*) topology of $\mathcal{B}(\hi)$, and such that
\begin{equation}
    \mathsf{M}(X)^\dagger \mathsf{M}(X) = \mathbb{1}_{\bh} \, .
\end{equation} Roughly speaking, it is the square-root of a positive operator-valued measure, and $\mathsf{M}(\Delta)$ represents a quantum operation in $\Delta$.} $\{\mathsf{M}_k : \Bor(\mathcal{M}) \to \bh\}_{k=1}^{N \in \mathbb{N}}$, $\sum_{k=1}^N \mathsf{M}_k(\U)^\dagger \mathsf{M}_k(\U) = \mathbb{1}_{\bh}$, in a region $\U$, the wavefunctional on the past light cone of $B^+(\U)$ becomes
\begin{equation}
    \lim_{n \to \infty} \left(\Psi[\Sigma_n^{\U_+}] - \frac{\mathsf{M}_k(\U) K[\Psi]_{\Sigma_n^{\U_-}}^{\Sigma_n^{\U_+}} \Psi[\Sigma_n^{\U_-}]}{\norm{\mathsf{M}_k(\U) K[\Psi]_{\Sigma_n^{\U_-}}^{\Sigma_n^{\U_+}} \Psi[\Sigma_n^{\U_-}]} } \right) = 0
\end{equation}
with probability
\begin{equation}
    p(k) = \lim_{n \to \infty} \norm{\mathsf{M}_k(\U) K[\Psi]_{\Sigma_n^{\U_-}}^{\Sigma_n^{\U_+}} \Psi[\Sigma_n^{\U_-}]}^2
\end{equation}
where
\begin{itemize}
    \item $\{\Sigma_n^{\U_{\pm}}\}_{n \in \mathbb{N}}$ are two sequences of spacelike Cauchy hypersurfaces of $\mathcal{M}$ such that \begin{itemize}
        \item $\forall n \in \mathbb{N}, \Sigma_n^{\U_\pm} \cap J^-(B^\pm(\U)) = B^\pm(\U)$, respectively, and
        \item $\Sigma_n^{\U_{\pm}} \to \partial J^-(B^\pm(\U))$, respectively.
    \end{itemize}
    \item $K[\Psi] : \hi \to \hi$ is the propagator of the wavefunctional time evolution of Eqn. \eqref{eqn:modified TS eqn GR}.
\end{itemize}

Hellwig-Kraus collapse is, at least formally, retrocausal. It also gives rise to Sorkin's impossible measurement paradox \cite{sorkin_impossible_1993}  and has also been criticised on other grounds, see e.g. \cite{aharonov_can_1981}. 

These issues are already present in quantum theory in Minkowski spacetime, and are aggravated if semiclassical gravity is introduced. 
Indeed, if $p$ and $q$ are spacelike separated, and Alice measures at $q$ then collapse happens all along $\Lambda^-(q)$ which overlaps with $J^-(p)$, so the gravitational field at $p$, though retarded, depends on Alice's decision to measure.  It follows that Bob can tell whether Alice measured or not. Thus this past light-cone prescription is not relativistically causal once (classical) gravitational measurements are permitted. This is shown pictorially in Fig. \ref{fig:Past light cone problems}. We thus need to consider other options for relativistic state update for local semiclassical gravity.

\subsection{Instantaneous update}

Another collapse postulate of relativistic quantum theory could be used, for example ``instantaneous'' collapse \cite{aharonov_states_1980,aharonov_can_1981,aharonov_is_1984,breuer_relativistic_1998,breuer_state_1999,breuer_measurements_2007,fedida_einstein_2026,polo-gomez_state_2025} for which the covariance of the state histories breaks (though whether this is really an issue is unclear in standard quantum theory where the evolution is unitary outside measurement regions). In this approach, upon a measurement, given by measurement operators $\{\mathsf{M}_k : \Bor(\mathcal{M}) \to \bh\}_{k=1}^{N \in \mathbb{N}}$, $\sum_{k=1}^N \mathsf{M}_k(\U)^\dagger \mathsf{M}_k(\U) = \mathbb{1}_{\bh}$, in a region $\U$, the wavefunctional evolves as
\begin{multline}
    \label{eqn:instantaneous mmts}
    \lim_{\epsilon \to 0^+} \Bigg(\Psi[\tau^{-1}(t_f + \epsilon)] \\ - \frac{\mathsf{M}_k(\U) K[\Psi]_{\tau^{-1}(t_i-\epsilon)}^{\tau^{-1}(t_f+\epsilon)} \Psi[\tau^{-1}(t_i - \epsilon)]}{\norm{\mathsf{M}_k(\U) K[\Psi]_{\tau^{-1}(t_i-\epsilon)}^{\tau^{-1}(t_f+\epsilon)} \Psi[\tau^{-1}(t_i - \epsilon)]}} \Bigg) = 0
\end{multline}
with probability
\begin{equation}
    p(k) = \lim_{\epsilon \to 0} \norm{\mathsf{M}_k(\U) K[\Psi]_{\tau^{-1}(t_i-\epsilon)}^{\tau^{-1}(t_f+\epsilon)} \Psi[\tau^{-1}(t_i - \epsilon)]}^2
\end{equation}
where
\begin{itemize}
    \item $\Sigma_i$ and $\Sigma_f$ are any disjoint pair of acausal Cauchy hypersurfaces such that $B^-(\U) \subset \Sigma_i$ and $B^+(\U) \subset \Sigma_f$\footnote{These exist, see e.g. \cite{fedida_einstein_2026}.}.
    \item $\tau$ is any Cauchy time function such that $\tau^{-1}(t_i) = \Sigma_i$ and $\tau^{-1}(t_f) = \Sigma_f$, $t_i < t_f$,
    \item $K[\Psi] : \hi \to \hi$ is the generator of the wavefunctional time evolution of eqn. \eqref{eqn:modified TS eqn GR}.
\end{itemize}

For such an instantaneous collapse rule, Einstein causality (measurements undertaken at spacelike-separated locations commute) is only known to hold when the dynamics are unitary outside measurement regions \cite{fedida_einstein_2026}. 

Nonlinear dynamics with an instantaneous collapse rule imply a violation of Einstein causality and generically allow superluminal signalling even at the level of wavefunctional dynamics following an argument similar to that of Gisin \cite{gisin_weinbergs_1990}.   The non-covariant state histories also lead to ambiguities in the definition of $\rho^\mathrm{loc}$.

\begin{figure*}
    \begin{tikzpicture}[x=0.75pt,y=0.75pt,yscale=-1,xscale=1]

\draw   (139.64,38.82) .. controls (159.64,28.82) and (199.28,42.64) .. (229.64,38.82) .. controls (260,35) and (280,76) .. (229.64,98.82) .. controls (179.28,121.64) and (159.64,128.82) .. (139.64,98.82) .. controls (119.64,68.82) and (119.64,48.82) .. (139.64,38.82) -- cycle ;
\draw   (485,30) .. controls (509,26) and (526.28,8.64) .. (556.64,4.82) .. controls (587,1) and (587.28,62.64) .. (556.64,64.82) .. controls (526,67) and (510,92) .. (490,62) .. controls (470,32) and (461,34) .. (485,30) -- cycle ;
\draw  [fill={rgb, 255:red, 0; green, 0; blue, 0 }  ,fill opacity=1 ] (514.12,110.41) .. controls (514.12,107.88) and (515.98,105.82) .. (518.27,105.82) .. controls (520.56,105.82) and (522.42,107.88) .. (522.42,110.41) .. controls (522.42,112.94) and (520.56,115) .. (518.27,115) .. controls (515.98,115) and (514.12,112.94) .. (514.12,110.41) -- cycle ;
\draw  [dash pattern={on 4.5pt off 4.5pt}]  (518.27,110.41) -- (342,252) ;
\draw  [dash pattern={on 4.5pt off 4.5pt}]  (518.27,110.41) -- (659.2,251) ;
\draw    (229.64,38.82) .. controls (290.92,32.74) and (405.92,133.74) .. (694.92,110.74) ;
\draw    (2.92,60.74) .. controls (56.92,63.74) and (89.92,59.74) .. (139.64,38.82) ;
\draw    (518.92,76.74) .. controls (520.55,78.44) and (520.51,80.11) .. (518.81,81.74) .. controls (517.11,83.37) and (517.07,85.04) .. (518.7,86.74) .. controls (520.33,88.45) and (520.29,90.11) .. (518.58,91.74) .. controls (516.88,93.37) and (516.84,95.03) .. (518.47,96.73) .. controls (520.1,98.43) and (520.06,100.1) .. (518.36,101.73) -- (518.27,105.82) -- (518.27,105.82) ;
\draw [shift={(518.74,84.78)}, rotate = 91.29] [fill={rgb, 255:red, 0; green, 0; blue, 0 }  ][line width=0.08]  [draw opacity=0] (8.93,-4.29) -- (0,0) -- (8.93,4.29) -- cycle    ;

\draw (507.99,121.22) node [anchor=north west][inner sep=0.75pt]  [font=\scriptsize]  {$\rho _{loc}$};
\draw (181.64,65.82) node [anchor=north west][inner sep=0.75pt]   [align=left] {$\U$};
\draw (350,50) node [anchor=north west][inner sep=0.75pt]   [align=left] {$\Sigma_t$};
\draw (524.64,33.82) node [anchor=north west][inner sep=0.75pt]   [align=left] {$\V$};

\end{tikzpicture}
\caption{\justifying Aharonov-Albert instantaneous updates can induce superluminal signalling in the context of the local semiclassical gravitational theory presented in this paper.  The local quantum state at $x \in \Sigma_t$ updates in the causal past of $\V$ where collapse occurs on $\Sigma_t$.   This change of local state changes the local expectation value of the stress-energy tensor, and so the gravitational field backreacts in the past of $\V$.
Hence an observer in $\V$ can, in principle, determine whether a quantum measurement has been undertaken in $\U$ through a Cavendish experiment.}
\label{fig:Side cone problems}
\end{figure*}
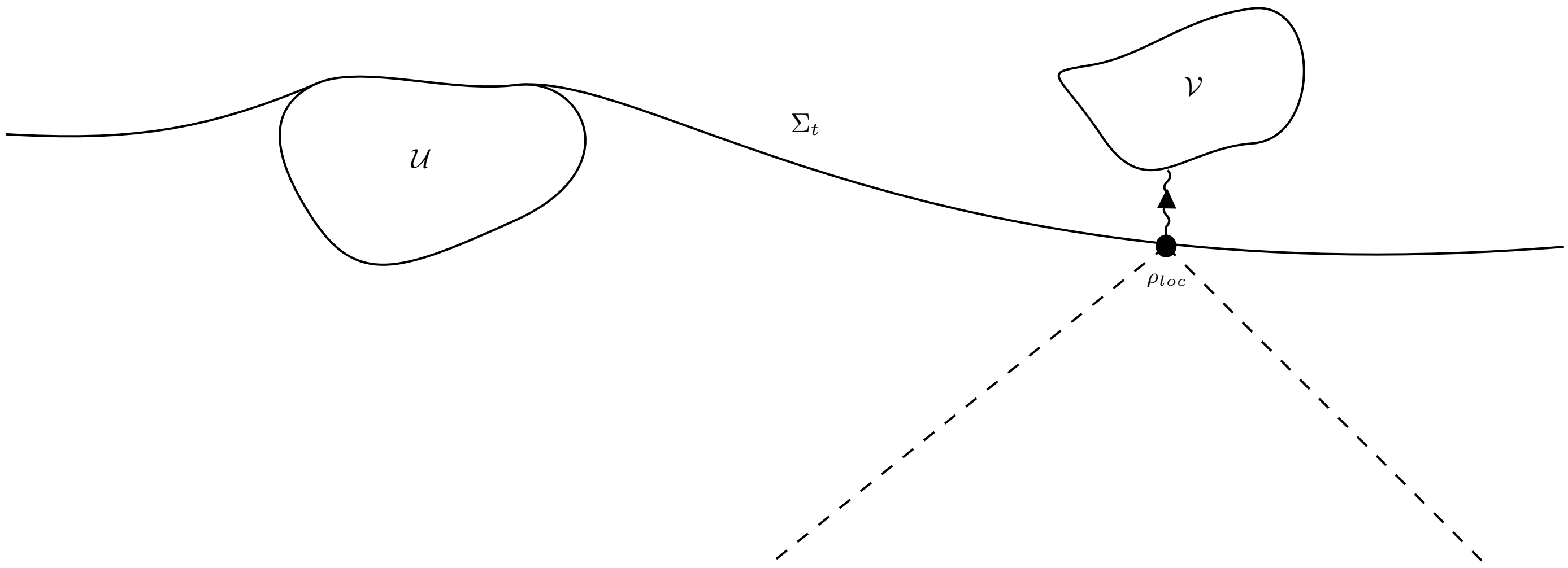

Further superluminal signalling issues arise because the collapse hypersurface may lie in the causal future of 
one experimenter, in a region $\U$, and the causal past of another, in $\V$, where $\U$ and $\V$ are spacelike separated. This is shown pictorially in Fig. \ref{fig:Side cone problems}. The foliation-dependent
backreaction in the causal past of $\V$ also makes the gravitational field equations ill-defined.

This leaves one natural option: future light-cone collapse.

\subsection{Future light-cone updates}

\label{sec:polyperspective}

The remaining natural option is that collapses update the state on the future-light cone. Here, we consider detector-based models \cite{polo-gomez_detector-based_2022,perche_particle_2024,pranzini_detector-based_2025,polo-gomez_state_2025} of quantum measurements.  These model quantum detectors explicitly. For example, one can consider a spatially smeared Unruh-DeWitt detector \cite{unruh_notes_1976} coupled to a real scalar field, for which, in the interaction picture, the interaction Hamiltonian is
\begin{equation}
    \hat{H}_I(t) = \lambda\chi(t) \hat{\mu}(t) \int_{\Sigma_t} F(\vec{x}) \hat{\phi}(t,\vec{x}) \sqrt{h(t,\vec{x})} d\vec{x} 
\end{equation}
where $\lambda$ is the coupling strength between the detector and the quantum field $\hat{\phi}$, $\chi(t)$ is the switching function controlling the time-dependence of the coupling,
\begin{equation}
    \hat{\mu}(t) = \ketbra{g}{e} e^{-i \Omega t/\hbar} + \ketbra{e}{g} e^{i \Omega t/\hbar}
\end{equation}
is the monopole moment of the detector which has two levels $\ket{g}$ (ground) and $\ket{e}$ (excited) with an energy gap $\Omega$, $h$ is the induced $3$-metric on the leaf $\Sigma_t$, and $F$ is the spatial smearing function that models the localisation of the detector. If the interaction between the quantum field and the detector yields negligible gravitational effects,\footnote{This may or may not be a suitable approximation in a real-life situation. It would be interesting to study detector-based models with nonlinear semiclassical backreactions.} then the updated state of the field is known to lie in the future causal cone of the interaction region.

In this approach, one needs to distinguish global quantum states from local quantum states, 
as the readout device formalism \cite{kent_nonlinearity_2005,kent_quantum_2021,fedida_mixture_2025,kent_measurement_2025} makes clear.  This has also been more recently been discussed using the terminology of ``polystates'' and ``polyperspectives'' \cite{polo-gomez_state_2025}, which we will use below. Recalling Eqns.~\eqref{eqn:local state update} and \eqref{eqn:update channel}, the prescription for quantum measurements (forgetting about free evolution for now) is then the following: if a selective quantum measurement is conducted in $\mathcal{S} \subset \mathcal{M}$ in which a collapse event occurs at $y \in \mathcal{S}$ with outcome $a$ and associated projection operator $P_a$, then
\begin{equation}
    \rho^\mathrm{loc}(x) = \begin{cases}
        \rho^\mathrm{loc}(x) \;& \text{ if } \; y \notin J^-(x), \\
        \left(\frac{P_a \rho^\mathrm{loc} P_a}{\Tr[P_a \rho^\mathrm{loc} P_a]}\right)(x) \;& \text{ if } \; y \in J^-(x).
    \end{cases}
\end{equation} 

More generally, $\rho^\mathrm{loc}(x)$ is given by Eqn.~\eqref{eqn:local state update}. Consider, for example, the case of a Bell state \cite{polo-gomez_state_2025}
\begin{equation}
    \ket{\Psi^-} = \frac{1}{\sqrt{2}}\left(\ket{0_A}\ket{1_B} - \ket{1_A} \ket{0_B}\right) \, .
\end{equation}
Suppose Alice measures in $\sigma_z$ at $\tau^A = \tau^A_\text{collapse}$, while Bob measures $\sigma_{\vec{n}}$ at $\tau^B = \tau^B_\text{collapse}$. Let $a$ and $b$ be the outcomes that Alice and Bob obtain in their respective measurements, and let $\ket{a}$ and $\ket{b}$ (and $\ket{-a}$ and $\ket{-b}$) the corresponding eigenstates (and their orthonormal counterparts) of $\sigma_z$ and $\sigma_{\vec{n}}$. Then the polystate for this bipartite case evolves as
\begin{widetext}
    \begin{align}
        \rho_{AB}(\tau^A,\tau^B) = \begin{cases}
            \frac{1}{2} \mathbb{1}_A \oplus \frac{1}{2}\mathbb{1}_B \oplus \ket{\Psi^-}\bra{\Psi^-}  \qquad \qquad \qquad  & \text{ if } \tau^A < \tau^A_\text{collapse}, \tau^B < \tau^B_\text{collapse}, \\
            \ket{a_A}\bra{a_A} \oplus \frac{1}{2}\mathbb{1}_B \oplus (\ketbra{a_A}{a_A} \otimes \ketbra{-a_B}{-a_B}) & \text{ if } \tau^A > \tau^A_\text{collapse}, \tau^B < \tau^B_\text{collapse}, \\
            \frac{1}{2} \mathbb{1}_A \oplus \ketbra{b_B}{b_B} \oplus (\ketbra{-b_A}{-b_A} \otimes \ketbra{b_B}{b_B}) & \text{ if } \tau^A < \tau^A_\text{collapse}, \tau^B > \tau^B_\text{collapse}, \\
            \ketbra{a_A}{a_A} \oplus \ketbra{b_B}{b_B} \oplus (\ketbra{a_A}{a_A} \otimes \ketbra{b_B}{b_B}) & \text{ if } \tau^A > \tau^A_\text{collapse}, \tau^B > \tau^B_\text{collapse}.
        \end{cases}
    \end{align}
\end{widetext}
This can be shown to preserve Bell correlations even at spacelike separations, from the difference between expectation values of $\Tr[\rho_{AB} \mathcal{O} \otimes \mathbb{1}]$ (expectation values of joint observables) and $\Tr[\rho_A \mathcal{O}]$ (expectation values of individual observables), where $\mathcal{O} \in \mathcal{L}(\hi_\mathcal{A})$. To see this, let $\theta_{AB}$ be the angle between $\vec{n}$ and the $z$ axis. Then the average of the product of outcomes, which is an \emph{a priori} expectation value computed prior to the experiment, is \cite{polo-gomez_state_2025}
\begin{align}
    \expval{ab} &= \Tr[(\sigma_z \otimes \sigma_{\vec{n}}) \rho_{AB}(\tau^A < \tau^A_\text{collapse}, \tau^B < \tau^B_\text{collapse})] \\
    &= \expval{\sigma_z \otimes \sigma_{\vec{n}}}{\Psi^-} \\
    &= - \cos(\theta_{AB})
\end{align}
as in standard quantum theory. Likewise, the joint \emph{a priori} probability distribution is given by \cite{polo-gomez_state_2025}
\begin{align}
    \mathbb{P}(a,b) &= \Tr\Big((\ketbra{a_A} \otimes \ketbra{b_B}) \nonumber \\ & \qquad\rho_{AB}(\tau^A < \tau^A_\text{collapse}, \tau^B < \tau^B_\text{collapse}) \Big) \\
    &= \frac{1}{2} \left(\delta_{ab} \sin^2(\theta_{AB}/2) + (1-\delta_{ab}) \cos^2(\theta_{AB}/2)\right) \, .
\end{align}
The marginal probabilities are \cite{polo-gomez_state_2025}
\begin{align}
    \mathbb{P}(a) &= \Tr[\ketbra{a_A} \rho_{AB}(\tau^A<\tau^A_\text{collapse},\tau^B)] \nonumber \\ &= \expval{\frac{1}{2}\mathbb{1}_A}{a_A} = \frac{1}{2} \\
    \mathbb{P}(b) &= \Tr[\ketbra{b_B} \rho_{AB}(\tau^A,\tau^B<\tau^B_\text{collapse})] \nonumber \\  &= \expval{\frac{1}{2}\mathbb{1}_B}{b_B} = \frac{1}{2} \, .
\end{align}
These marginal probabilities involve only expectation values of individual observables. The marginal probabilities conditioned on the outcome of the other experimenter involve non-local information from the other experimenter's outcome, i.e. use expectation values of joint observables, and are given by \cite{polo-gomez_state_2025}
\begin{align}
    \mathbb{P}(a \mid b) &= \Tr[(\ketbra{a_A} \otimes \mathbb{1}_B) \rho_{AB}(\tau^A<\tau^A_\text{collapse},\tau^B)] \nonumber \\
    &= \abs{\braket{-b_B}{a_A}}^2 \nonumber \\
    &= \delta_{ab} \sin^2(\theta_{AB}/2) + (1-\delta_{ab}) \cos^2(\theta_{AB}/2) \\
    \mathbb{P}(b \mid a) &= \Tr[(\mathbb{1}_A \otimes \ketbra{b_B}) \rho_{AB}(\tau^A,\tau^B<\tau^B_\text{collapse})] \nonumber \\
    &= \abs{\braket{-a_A}{b_B}}^2 \nonumber \\
    &= \delta_{ab} \sin^2(\theta_{AB}/2) + (1-\delta_{ab}) \cos^2(\theta_{AB}/2) \, .
\end{align}
In summary, individual observables are used in marginal probabilities while joint observables are used in conditional probabilities \cite{polo-gomez_state_2025}.

More generally, consider a quantum measurement conducted in some compact region $K \subset \mathcal{M}$ in which a collapse event occurs at $y \in K$. If the pre-measurement local state is $\rho^\mathrm{loc}$ while the post-measurement local state is $\tilde{\rho}^\mathrm{loc}$, then the expectation value of the stress-energy tensor is given by\footnote{In the setting of AQFT, this would be given by $T_{ab}^\mathrm{loc}(x) = \expval{\hat{T}_{ab}^\mathrm{ren}(x)}_{[\rho^\mathrm{loc}](x)}$ in $\mathcal{M} \smallsetminus J^+(y)$ and $T_{ab}^\mathrm{loc}(x) = \expval{\hat{T}_{ab}^\mathrm{ren}(x)}_{[\tilde{\rho}^\mathrm{loc}](x)}$ in $I^+(y)$.}
\begin{equation}
    \label{eqn:update Tabloc}
    T_{ab}^\mathrm{loc}(x) = \begin{cases}
        \Tr[\rho^\mathrm{loc}(x)\hat{T}_{ab}(x)] \;  \text{ if } x \in \mathcal{M} \smallsetminus J^+(y), \\
        \Tr[\tilde{\rho}^\mathrm{loc}(x) \hat{T}_{ab}(x)] \; \text{ if } x \in I^+(y).
    \end{cases}
\end{equation}
One must then determine what happens on $\Lambda^+(y)$, where $T_{a b}^\mathrm{loc}$ is discontinuous.

\section{The divergence of the local stress-energy}

\label{sec:measurements and divergence}

In any purely metric, diffeomorphism-invariant theory with minimal matter coupling, one must have $\nabla^a T_{ab} = 0$ (in some distributional sense when thin shells are present). However, in the vicinity of quantum measurements it seems that $\nabla^a T_{ab}^\mathrm{loc}(x) \neq 0$, while $\nabla^a G_{ab}(x)=0$, so the EFEs are not satisfied. 

As we have seen in Sec.~\ref{sec:semiclassical equations}, under certain conditions, the expectation value of the quantum stress-energy tensor of matter is conserved under free evolution \cite{wald_back_1977,wald_axiomatic_1978,moretti_comments_2003,hollands_conservation_2005, dappiaggi_stable_2008,verch_local_2012,pinamonti_initial_2011} in semiclassical gravity. However, it is not evident that this can hold true everywhere if we assume objective quantum state collapse. For example, if the state were in a superposition of two localised wavepackets, then a quantum measurement collapsing the quantum state to a single localised wavepacket would generate a nonzero divergence. \textit{Prima facie} this is problematic: it seems to imply that the gravitational equations are only satisfied away from quantum measurements and that around quantum measurements the theory is not well-defined.

This issue is well-known \cite{page_indirect_1981,diez-tejedor_towards_2012,hossenfelder_real_2012,maudlin_status_2020,juarez-aubry_initial_2023,mucino_fully_2025} and seen as a serious roadblock to a fully consistent approach to semiclassical gravity. One response is to consider modifications of the Einstein field equations that recover General Relativity in some limit. Indeed, one might argue that the theory presented up to this point should be understood as an effective field theory that may need to be completed by some generalised classical gravitational theory. For example, $f(R,Q,T)$ gravity (which has higher-order Ricci terms, non-zero metricity and/or torsion) with non-minimal coupling can have $\nabla^a T_{ab} \neq 0$ in general.

Another related response is to accept that semiclassical field equations hold only ``sufficiently far away'' from quantum measurements, and that one should suitably modify the field equations in the vicinity of measurements. 

Yet another approach is that of the so-called \emph{semiclassical self-consistent configuration} model \cite{diez-tejedor_towards_2012,mucino_fully_2025}, which works within an objective collapse formalism and ``glues'' metrics along the future light cone $\Lambda^+(x)$ of the measurement location $x$. Their approach is to make the semiclassical field equations hold in patches. It is motivated by the initial-value formulation of General Relativity \cite{juarez-aubry_initial_2023,juarez-aubry_recent_2026}. In this approach, though the induced metric on $\Lambda^+(x)$ is continuous, the full metric in general is not, i.e. the physics does not take place within a single classical Lorentzian spacetime. Refs.~\cite{diez-tejedor_towards_2012,mucino_fully_2025} argue that by ensuring that the induced metric is in fact continuous, a suitable prescription can be provided to determine the post-measurement metric.

We here argue for a different approach using the Barrabès-Israel null-shell formalism \cite{barrabes_thin_1991,israel_singular_1966,poisson_reformulation_2002,poisson_hypersurfaces_2004} of classical General Relativity, which we now review.   This approach avoids having to work with several spacetimes. The idea is that at an interface between two different energy-momentum tensors, the metric need only be $C^0 \cap H^1_\mathrm{loc}$ across the collapse hypersurface.    This allows a discontinuity in $\partial g_{ab}$, but no direct conflict with standard Lorentzian geometry given that $g_{ab} \in C^0$ (in particular, we will show that $g_{ab} \in C^{0,1} \subset C^0 \cap H^1_\mathrm{loc}$).   It allows the possibility that, while the pointwise Einstein field equations do not generally have solutions given by functions over $\Lambda^+(x)$, the distributional Einstein field equations do. As we argued above, the Einstein field equations should be understood distributionally.

Consider a system–probe interaction confined to a bounded connected region $K$. In the future light-cone detector-based approach to quantum measurements (see Sec. \ref{sec:polyperspective}), $T_{ab}^\mathrm{loc}$ agrees with its pre-measurement value outside $J^+(K)$, as given in Eqn. \eqref{eqn:update Tabloc}. With the local semiclassical Einstein equations \eqref{eqn:retarded grav field GR}, $g$ therefore agrees with the pre-measurement metric away from $J^+(K)$. 

Across $\Lambda^+(K)$ however, the source term does not have a vanishing divergence.   This is addressed by the Barrabès-Israel formalism, which we review in Appendix \ref{app:BI}. In general, $K$, being the measurement region, contains collapse events, being elements of a set of points $\{x, y, ... \}$ in $K$. Across the future light cone of each of these points, the stress-energy does not have a vanishing divergence. Let us thus examine what happens at each individual point. For simplicity, we suppose that only one collapse event lies in $\mathcal{M}$. The generalisation to timelike-separated chains of collapse events follows straightforwardly. The generalisation to spacelike-separated collapse events requires more care and will be covered in Sec. \ref{subsec:mmt Bell state}.

We assume here the most singular case, i.e. collapses happening at points, rather than continuous forms of collapse \cite{pearle_combining_1989,ghirardi_markov_1990} occurring over regions. We will show that, applying this to the collapse of a superposed delocalized state, in the weak field limit, there are consistent distributional solutions of the (linearised) Einstein field equations.

The Barrabès-Israel formalism \cite{barrabes_thin_1991,israel_singular_1966,poisson_reformulation_2002,poisson_hypersurfaces_2004} is algorithmic, depending on non-trivial conditions, in the following sense.  We have inputs:
\begin{enumerate}
    \item[(a)] a sufficiently regular $T_{ab}^-$ in $\mathcal{M} \smallsetminus J^+(x)$,
    \item[(b)] a strong solution $g^-$ in $\mathcal{M} \smallsetminus J^+(x)$ to the EFE sourced by $T_{ab}^-$ with some chosen initial conditions (e.g. an asymptotically flat metric).
\end{enumerate}
These determine the induced metric $\sigma_{AB}$ on $\Lambda^+(x)$. 
We have further inputs: 
\begin{enumerate}
    \item[(c)] a sufficiently regular $T_{ab}^+$ in $I^+(x)$,
    \item[(d)] a strong solution $g^+$ in $I^+(x)$ to the EFE sourced by $T_{ab}^+$ with initial conditions given by $g^+|_{\Lambda^+(x)} = g^-|_{\Lambda^+(x)}$ on $\Lambda^+(x)$. That is, the pullback metrics from $g^+$ and $g^-$ match on $\Lambda^+(x)$. 
\end{enumerate}
This last implies the existence of a global $C^{0,1}$ metric $g_{ab} = g_{ab}^+ \Theta(u) + g_{ab}^- (1-\Theta(u))$ (where $\Lambda^+(x) = \{u=0\}$, $u>0$ on $I^+(x)$, $u<0$ on $\mathcal{M} \smallsetminus J^+(x)$) on $\mathcal{M}$ such that $g|_{I^+(x)} = g^+$ and $g|_{\mathcal{M} \smallsetminus J^+(x)} = g^-$. This result was formalised by Clarke and Dray \cite{clarke_junction_1987} and later made more precise by Mars and Senovilla \cite{mars_geometry_1993}. It was more recently shown \cite[Theorem A.4]{manzano_generalizing_2025} that for two general $C^\infty$ spacetimes $(\M^-,g^-)$ and $(\M^+,g^+)$ whose induced metrics match on a smooth $\partial \M^- = \partial \M^+  =: \mathcal{N}$ \footnote{In fact, the statement need not assume that $\partial \M^- = \partial \M^+$, but more generally that there exists a smooth diffeomorphism $\phi : \partial \M^- \to \partial \M^+$ such that $\phi$ is a smooth isometry, i.e. $\phi_*(g^-|_{\partial \M^-}) = g^+|_{\partial \M^+}$, as well as some other related assumptions, see \cite[Definition A.3]{manzano_generalizing_2025}.}, there exists a locally Lipschitz $g$ on $\M = \M^- \cup \mathcal{N} \cup \M^+$ such that $g|_{\M^\pm} = g^\pm$. Note that this result applies to a smooth $\mathcal{N}$, the future light cone is only smooth if the vertex $x$ is removed, i.e. we need a separate check in a neighbourhood of $x$.

This results in a distributional Einstein tensor $G_{ab}[g]$ with both a regular part and a singular part. This needs to be matched by a stress-energy with both a regular part and a singular part for the EFEs to hold. The regular stress-energy is given by $T_{ab}^\text{reg} = T_{ab}^+ \chi_{I^+(x)} + T_{ab}^- \chi_{\mathcal{M} \smallsetminus J^+(x)}$, where for any region $R \subset \mathcal{M}$, \begin{equation}
    \chi_R(x) = \begin{cases}
    1 \text{ if x} \in R \\
    0 \text{ otherwise}
\end{cases}
\end{equation} is the indicator function.

The output of the Barrabès-Israel formalism is the singular stress-energy $T^\mathrm{sing}_{ab}$ supported on $\Lambda^+(x)$ required for the distributional EFE of which $g$ is a distributional solution on $\mathcal{M}$ to hold. The full source of the distributional EFE is then $T^\mathrm{tot}_{ab} = T_{ab}^+ \chi_{I^+(x)} + T_{ab}^- \chi_{\mathcal{M} \smallsetminus J^+(x)} + T_{ab}^\mathrm{sing}$.

To determine the applicability of the Barrabès-Israel formalism, one needs to determine $\sigma_{AB}$ on $\Lambda^+(x)$ from (a) and (b), combine these data with (c), and determine whether a $g^+$ satisfying (d) exists\footnote{Should nature operate according to the framework proposed in this work, the existence of a valid $g^+$ solution is indispensable; without it, the dynamical evolution of spacetime would terminate abruptly.} (and optionally, though perhaps desirably, whether it is unique up to diffeomorphisms). Steps (a) and (b), as well as steps (c) and (d), are instances of the (characteristic) initial value problem of semiclassical gravity \cite{juarez-aubry_initial_2023,juarez-aubry_recent_2026}.\footnote{In \cite[Conjecture 3.10]{juarez-aubry_initial_2023}, the authors hypothesize that what they call $\hbar^0$ semiclassical gravity, which is an algebraic quantum field theoretic relative of the model presented in the current paper, has a well-posed initial-value problem.}  For the general nonlinear problem, finding (d) (if it exists) is hard already in classical general relativity: see \cite{choquet-bruhat_cauchy_2011,chrusciel_existence_2012,chrusciel_many_2012} for some results in that direction.

The full nonlinear problem is much more difficult than the linearised problem examined in Sec.\ref{sec:weak field solutions} since the problem of the existence and uniqueness (up to gauge) of solutions becomes a fixed-point problem. Indeed, in curved backreacting spacetimes, $\hat{T}^+ = \hat{T}^+[g]$ depends on the metric, so it is not sufficient to fix a $T^+$ in $\M^+$ and check for the existence of a $g^+$ in $\M^+$ satisfying the Einstein field equations. Rather, one needs to check the existence and uniqueness (up to gauge) of a ``suitable'' pair $(T^+[g],g^+)$ in $\M^+$ given the updated $\rho^\mathrm{loc}$, such that $T^+$ arises as the expectation value of a quantum stress-energy tensor $\hat{T}^+[g]$ with background metric $g$ sourced by the updated local quantum state $\rho^\mathrm{loc}$, and $g^+$ satisfies the full Einstein field equations sourced by $T^+[g] = \expval{\hat{T}^+[g]}_{\rho^\mathrm{loc}}$ in $\M^+$. We do not attempt to solve this difficult question in this paper.

The nature of $T^\mathrm{sing}_{ab}$ is different to that of $T_{ab}^\mathrm{loc}$. The latter arises from local quantum states evaluating the quantum stress-energy tensor, the former from the effect of quantum state collapses in local semiclassical gravity.  We find it natural to interpret this distributional term 
as completing the definition of the measurement postulate in spacetime.   
It is perhaps worth stressing that the introduction of $T_{ab}^\mathrm{sing}$ is qualitatively and conceptually distinct from the non-conservation of energy effect implied by most dynamical collapse models.

We highlight that this ``collapse stress-energy'' is smaller in magnitude than the matter's stress-energy, and the analysis of Appendix ~\ref{app:BI} suggests it only interacts gravitationally\footnote{See Sec.~\ref{sec:conclusion} for further discussion.}. It is thus no surprise that experiments to date have not been able to probe its effect. One could, in principle, superpose macroscopic lumps of matter in the position basis and examine gravitational effects resulting from inducing a collapse, though this is far out of reach of current technology.

To sum up, the gravitational field equations are distributional, singular on the light cone of collapse events, and are given by
\begin{equation}
    G_{ab}[\rho^\mathrm{loc};f] + \Lambda g_{ab}[\rho^\mathrm{loc};f] = \kappa T_{ab}^{\text{tot}}[\rho^\mathrm{loc};f]
\end{equation}
for all ``suitable'' ($\mathcal{D}(\M)$) functions $f$. The Bianchi identities hold distributionally \cite{mars_geometry_1993}, and the singular contributions on the left-hand side of the equation are compensated by the singular contributions in $T_{ab}^\text{tot}$. 

We can test whether the approach may be viable by analysing the linearised problem in the weak field limit, that we take for simplicity on a Minkowski background $\mathbb{M}$. Recall that we are working in a model in which wave function collapses occur at (or rather emanate from) discrete space-time points.  
It is worth stressing that this is consistent with a picture in which quantum measurements -- by which we mean interactions between one quantum system and another probe/apparatus system -- are continuous processes taking place over a space-time region.  
Of course, there are natural models in which collapses (not just measurement processes) are continuous processes taking place over a region (for example Continuous Spontaneous Localization models \cite{ghirardi_unified_1986, pearle_combining_1989, ghirardi_markov_1990, bassi_dynamical_2003}).  
For such models, the (very much analogous) Israel formalism for spacelike and/or timelike transitions \cite{israel_singular_1966,barrabes_thin_1991} may be used.\footnote{Depending on what quantities are continuous and/or differentiable, the Barrabès-Israel formalism may also play a role.}
We hypothesize collapses as discrete events here not because we think it is necessarily more natural, but because it necessarily creates discontinuities and so poses the sharpest challenge to the idea that semiclassical models can be constructed with a single well-defined space-time.
Our point is that even this challenge can be met, at least in the weak field limit.

\subsection{Locally Lipschitz solutions to the linearised semiclassical Einstein field equations}

\label{sec:weak field solutions}

For the following, we work on a Minkowski background $\mathbb{M}$ and take $x \in \mathbb{M}$, without loss of generality $x=(0,\vec{0})$ in some chart. Suppose that $h^- \in C^\infty$ in $\M^- \equiv \mathbb{M} \smallsetminus J^+(0)$, $T^- \in C^\infty$ in $\M^-$ is such that $\partial^\mu T^-_{\mu\nu}=0$ in $\M^-$, and that $h^-$ satisfies the linearised Einstein field equations sourced by $T^-$, given some suitable initial condition for $h^-$. We also suppose that $T^+ \in C^\infty$ in $\M^+ \equiv I^+(0)$ such that $\partial^\mu T_{\mu\nu}^+ = 0$ in $\M^+$, and we assume that $T^+_{\mu\nu}$ extends smoothly to a neighbourhood of $\Lambda^+(0)$.

We assume that $h^-$ has a smooth continuation $\underline{h}^-$ through $\partial \M^- \equiv \Lambda^+(0) =: \mathcal{N}$ to an open neighbourhood of $J^+(0)$, agreeing with $h^-$ on $\mathcal{M}^-$. Define
\begin{equation}
    \underline{T}_{\mu\nu}^{-} := \kappa^{-1} G_{\mu\nu}^{(1)}[\underline{h}^-]
\end{equation}
which satisfies $\underline{T}^- = T^-$ on $\M^-$. Note that in general, $\underline{h}^- \neq 0$ in $\M^+$.  See Whitney's extension theorem \cite{whitney_analytic_1934,seeley_extension_1964v2} for the existence of a smooth continuation.\footnote{In principle, $\underline{h}^-$ need not be nondegenerate (e.g. $\underline{h}^-=0$ is consistent) because $\eta + \underline{h}^-$ would then be nondegenerate. Instead, one needs to ensure that $\eta + \underline{h}^-$ is nondegenerate. A sufficient condition for this is that the continuation remains a weak field perturbation, and in particular that $\norm{\underline{h}^-(x)}_{\mathrm{op}} < 1$ for all $x \in \M^+$ in matrix operator norm. Then if $(\eta + \underline{h}^-(x)) v = 0$ for some vector $v$ and some $x \in \M^+$ (i.e. the metric induced is degenerate), then $v=-\eta^{-1} \underline{h}^-(x) v$ so since $\norm{\eta^{-1}}_{\mathrm{op}} = 1$, we have $\norm{v} \leq \norm{\underline{h}^-(x) v}< \norm{v}$ which is a contradiction. In the following, we assume that we remain in a weak field regime across quantum state collapses.}

For example, if there have been no other collapses in the timelike past, we can take $\underline{h}^-$ to be given in a retarded way as in \eqref{eqn: weak field gravitational field} and interpret this $\underline{h}^-$ counterfactually: it would be the metric $h^-$ in $\M^+$ had there not been a quantum measurement being undertaken.

The goal is to find an $h^+ \in C^\infty$ in $\M^+$ such that $h^+$ satisfies the linearised Einstein field equations
\begin{equation}
    G^{(1)}[h^+] = \kappa T^+ \quad \text{ in } \M^+
\end{equation}
with initial condition $h^+|_{\mathcal{N}} = h^-|_{\mathcal{N}}$. This will allow us to build Barrabès-Israel distributional stress-energy $T^\mathrm{sing}$ on $\mathcal{N}$ and a locally Lipschitz solution $h = h^+ \Theta(u) + h^- (1-\Theta(u))$ on $\mathbb{M} = \M^- \cup \mathcal{N} \cup \M^+$ such that
\begin{equation}
    G^{(1)}[h] = \kappa (T^- \chi_{\M^-} + T^+ \chi_{\M^+} + T^\mathrm{sing}) \quad \text{ on } \mathbb{M},
\end{equation} 
such that $h|_{\M^-} = h^-$ and $h|_{\M^+} = h^+$. Note that $\chi|_{\M^+} = \Theta(u)$ and $\chi|_{\M^-} = 1-\Theta(u)$ distributionally.

The idea behind the ensuing construction is the following. If one simply tried to retrieve a global retarded solution to the linearised Einstein field equations, with an asymptotically flat metric at past infinity and sourced by $T^+ \chi_{\M^+} + T^- \chi_{\M^-}$, of the form
\begin{multline}
    \label{eqn:naive linearised metric}
    \overline{h}_{\mu\nu}^{\mathrm{ret}}(y) = 2 \kappa \int_\mathbb{M} G_\mathrm{ret}(y,y') (T^-(y') \chi_{\M^-}(y') \\ + T^+(y') \chi_{\M^+}(y')) d^4y',
\end{multline}
then although $\overline{h}^{\mathrm{ret}}_{\mu\nu}|_{\M^-} = \overline{h}_{\mu\nu}^-$, $\overline{h}^{\mathrm{ret}}_{\mu\nu}|_{\M^+}$ would be a solution to \eqref{eqn:linearised EFEs} in $\M^+$ but would \emph{not} be in Lorenz gauge (i.e. would not satisfy \eqref{eqn:Lorenz condition} in $\M^+$), hence it would \emph{not} be a solution to the linearised Einstein field equations \cite{gralla_rigorous_2008,gralla_rigorous_2011}.

Instead, let $u=ct-r=0$ define the future light cone and write in $\M^+$ the difference source
\begin{equation}
    \Delta T := T^+ - \underline{T}^-.
\end{equation}
The linearised Bianchi identity $\partial^\mu G_{\mu\nu}^{(1)}[\underline{h}^-] = 0$ and the conservation of $T^+$ imply
\begin{equation}
    \partial^\mu \Delta T_{\mu\nu} = 0 \quad \text{ in }\M^+.
\end{equation}
Moreover,
\begin{equation}
    \Delta T_{\mu\nu}|_{\mathcal{N}} = (T^+_{\mu\nu} - T_{\mu\nu}^-)|_{\mathcal{N}}.
\end{equation}
We seek a $\mathfrak{h}^+_{\mu\nu}$ such that
\begin{equation}
    G^{(1)}[\mathfrak{h}^+] = \kappa \Delta T \quad \text{ in } \M^+, \qquad \mathfrak{h}^+|_{\mathcal{N}} = 0.
\end{equation}
Once such a $\mathfrak{h}^+$ has been found, we can define
\begin{equation}
    h^+ := \underline{h}^- + \mathfrak{h}^+
\end{equation}
Linearity then gives
\begin{equation}
    G^{(1)}[h^+] = G^{(1)}[\underline{h}^-] + \kappa \Delta T = \kappa T^+ \quad \text{ in } \M^+,
\end{equation}
while $\mathfrak{h}^+|_{\mathcal{N}} = 0$ gives
\begin{equation}
    h^+|_{\mathcal{N}} = \underline{h}^-|_{\mathcal{N}} = h^-|_{\mathcal{N}}.
\end{equation}

We now construct this correction, which contains two terms: 
\begin{equation}
    \mathfrak{h}^+ := \mathfrak{H} + H
\end{equation}
where $\mathfrak{H}$ is a field fully supported in a one-sided neighbourhood of $\mathcal{N}_\mathrm{reg} := \mathcal{N} \smallsetminus \{0\}$ inside $\M^+$ that vanishes on $\mathcal{N}$ with a prescribed first transverse derivative there, and $H$ is a retarded Lorenz-gauge field which solves the remaining bulk equation. Indeed, if one immediately attempted to retrieve a retarded solution to the linearised Einstein field equations in Lorenz gauge with the source $\Theta(u) \Delta T_{\mu\nu}$, then
\begin{equation}
    \partial^\mu (\Theta(u) \Delta T_{\mu\nu}) = \Theta(u) \underbrace{\partial^\mu \Delta T_{\mu\nu}}_{=0} + \delta(u) n^\mu \Delta T_{\mu\nu},
\end{equation}
where $n:= du$, so unless
\begin{equation}
    \label{eqn:condition no distribution}
    n^\mu \Delta T_{\mu\nu}|_{\mathcal{N}}=0,
\end{equation}
this source is not distributionally conserved. It therefore cannot be the right-hand side of a global linearised Einstein field equation, because the left-hand side obeys the Bianchi identity. Hence, we will here look for a $\mathfrak{H}$ that satisfies
\begin{equation}
    \label{eqn:linearised EFE mathfrak H}
    n^\mu G^{(1)}_{\mu\nu}[\mathfrak{H}]|_{\mathcal{N}} = \kappa n^\mu \Delta T_{\mu\nu}|_{\mathcal{N}}, \quad \mathfrak{H}|_\mathcal{N} = 0.
\end{equation}
The source for $H$ will then be
\begin{equation}
    \label{eqn:H source}
    \tau_{\mu\nu} := \Theta(u)(\Delta T_{\mu\nu} - \kappa^{-1}G^{(1)}_{\mu\nu}[\mathfrak{H}]),
\end{equation}
with
\begin{equation}
    \partial^\mu \tau_{\mu\nu} = \delta(u) n^\mu ( \Delta T_{\mu\nu} -\kappa^{-1} G_{\mu\nu}^{(1)}[\mathfrak{H}])|_{\mathcal{N}}
\end{equation}
which vanishes under Eqn.~\eqref{eqn:linearised EFE mathfrak H}. Note that $H|_{\mathcal{N}}=0$ since for $y \in \mathcal{N}$, $J^-(y) \cap J^+(0)$ has measure $0$. Hence, the retarded field of a source supported in $\overline{\M^+}$ vanishes on $\mathcal{N}$, and only then does $\mathfrak{h}^+|_{\mathcal{N}} = (\mathfrak{H}+H)|_{\mathcal{N}}=\mathfrak{H}|_{\mathcal{N}} =0$. 

Note however that $T_{\mu\nu}^\mathrm{loc} = T_{\mu\nu}^{\mathrm{loc},-} \chi_{\M^-} + T_{\mu\nu}^{\mathrm{loc},+} \chi_{\M^+}$ can be discontinuous across the collapse hypersurface whilst satisfying Eqn.~\eqref{eqn:condition no distribution}: the latter is weaker than continuity. If Eqn.~\eqref{eqn:condition no distribution} is satisfied, no Barrabès-Israel distributional stress-energy is introduced on the collapse hypersurface, and in such cases the global metric can potentially have higher regularity (though the Weyl tensor may still be distributional while the Ricci tensor remains bounded). This condition might be satisfied in some models of quantum measurements, though it is not in the (highly singular) examples considered in Sec.~\ref{sec:examples} with discrete wavefunction collapses.

The computation of a future metric, its associated global metric and Barrabès-Israel stress-energy is given in Appendix \ref{app:future metric}. We have that
\begin{align}
    h^+ &= \underline{h}^- + \mathfrak{H} + H \quad \text{ in } \M^+, \\
    h &= \underline{h}^- + \Theta(u) \mathfrak{H} +H \quad \text{ in } \mathbb{M}.
\end{align}
Explicitly, 
\begin{widetext}
    \begin{align}
        \label{eqn:sol linearised EFE}
        h^+_{\mu\nu}(ct,\vec{x}) &= \underline{h}^-_{\mu\nu}(ct,\vec{x}) + \mathfrak{H}^{(\epsilon)}_{\mu\nu}(ct,\vec{x}) + H_{\mu\nu}(ct,\vec{x}), \\
        \mathfrak{H}^{(\epsilon)}_{\mu\nu}(u,v,\theta) &= u \chi_\epsilon(u/v) \left[\mathfrak{b}_{vv}(v,\theta)\delta^v_{\mu} \delta^v_{\nu} + 2\mathfrak{b}_{vA}(v,\theta) \delta^v_{(\mu} \delta^A_{\nu)} + \frac{\Omega^{CD} \mathfrak{b}_{CD}(v,\theta)}{2} \Omega_{AB} \delta^A_{(\mu} \delta^B_{\nu)} \right], \\
        \chi_\epsilon(z)&=\frac{e(\epsilon-z)}{e(\epsilon-z)+e\left(z-\frac{\epsilon}{2}\right)}, \qquad e(s)=\begin{cases}\exp(-1/s),&s>0,\\0,&s\leq0,\end{cases} \\
        \label{eqn:characteristic bvv bvA}
        \mathfrak{b}_{vv}(v,\theta) &= \frac{\kappa v}{2} \Delta T_{vv}(0,v,\theta), \qquad \qquad
        \mathfrak{b}_{vA}(v,\theta) = \frac{\kappa}{v^2} \int_0^v s^2 \left[\Delta T_{vA}(0,s,\theta) + \frac{s}{2} D_A \Delta T_{vv}(0,s,\theta)\right] ds, \\
        \Omega^{AB} \mathfrak{b}_{AB}(v,\theta) &= \int_0^v \left[\Omega^{AB} D_B \mathfrak{b}_{vA}(s,\theta) + \frac{\kappa s^2}{2} (\Delta T_{vv}+\Delta T_{uv})(0,s,\theta) \right] ds, \\
        \overline{H}_{\mu\nu}(ct,\vec{x}) = \frac{\kappa}{2\pi}& \iiint_{\mathbb{R}^3} \frac{\tau_{\mu\nu}(ct-\abs{\vec{x}-\vec{y}},\vec{y})}{\abs{\vec{x}-\vec{y}}} d^3 \vec{y}, \qquad   
        \tau_{\mu\nu}(u,v,\theta) = \Theta(u)\left(\Delta T_{\mu\nu}(u,v,\theta) - \kappa^{-1}G^{(1)}_{\mu\nu}[\mathfrak{H}](u,v,\theta)\right),   
    \end{align}
\end{widetext}
where $0<\epsilon<1$, $D$ is the covariant derivative for the unit-sphere metric $\Omega$ and $\theta^A=\vartheta,\phi$, $A=1,2$, $u=ct-\abs{\vec{x}}$ and $v=ct+\abs{\vec{x}}$, and one can use the standard Jacobian formulae to go from $(u,v,\theta)$ null coordinates to Cartesian coordinates. Here,
\begin{equation}
    \chi_\epsilon(z) = \begin{cases}
        1 \qquad & z \leq \epsilon/2, \\
        \text{a smooth transition} \\ \text{ between 1 and 0} \qquad & \epsilon/2<z<\epsilon, \\
        0 & z \geq \epsilon.
    \end{cases}
\end{equation}

If $\overline{h}^-_{\mu\nu}$ is a retarded metric,
\begin{equation}
    \underline{\overline{h}}^-_{\mu\nu}(ct,\vec{x}) = \frac{\kappa}{2\pi} \iiint_{\mathbb{R}^3} \frac{T_{\mu\nu}^\mathrm{loc,-}(ct-\abs{\vec{x}-\vec{y}},\vec{y})}{\abs{\vec{x}-\vec{y}}} d^3 \vec{y},
\end{equation}
where $T_{\mu\nu}^\mathrm{loc,-}$ is extended to all of $\mathbb{M}$. 

Further note that different choices of $0<\epsilon<1$ give a valid $\mathfrak{H}$. This $\epsilon$ parameter is dimensionless. Thus, different values of $\epsilon$ can (but need not to, e.g. if $\Delta T=0$ then $\mathfrak{H}$ vanishes for every $\epsilon$) produce different $h^+$ in the specified coordinates. However, the freedom in $\epsilon$ does not establish physically inequivalent solutions. Indeed, whenever they differ, these solutions are related by a linearised gauge transformation; their distributional linearised Riemann tensor and shell stress-energies agree (i.e. they are operationally indistinguishable), as we now show.

Let $0<\epsilon_1<\epsilon_2<1$ and define
\begin{align}
    Q_{\mu\nu} &:= \Theta(u)\left(\mathfrak{H}^{(\epsilon_2)}-\mathfrak{H}^{(\epsilon_1)}\right), \\
    \delta h &:= h^{(\epsilon_2)} - h^{(\epsilon_1)} = Q + H^{(\epsilon_2)}-H^{(\epsilon_1)}.
\end{align}
Note that since $\delta h$ is independent of $\underline{h}^-$, we do not need to assume for the following that the common continuation is retarded.

Both cutoffs equal $1$ whenever $0 \leq \frac{u}{v} \leq \frac{\epsilon_1}{2}$. Their difference therefore vanishes on a neighbourhood of every point of the regular cone $\mathcal{N}_\mathrm{reg}$. In particular, changing $\epsilon$ does not change the value of $\mathfrak{H}^{(\epsilon)}$ on the cone, nor its first transverse derivative there. Thus, the singular stress-energies match.

Hence,
\begin{equation}
    G^{(1)}[Q] = \Theta(u) G^{(1)}[\mathfrak{H}^{(\epsilon_2)}-\mathfrak{H}^{(\epsilon_1)}].
\end{equation}
It follows that
\begin{multline}
    \tau^{(\epsilon_2)} - \tau^{(\epsilon_1)} = -\kappa^{-1} \Theta(u) G^{(1)}[\mathfrak{H}^{(\epsilon_2)}-\mathfrak{H}^{(\epsilon_1)}]\\  = -\kappa^{-1} G^{(1)}[Q],
\end{multline}
and so
\begin{equation}
    \overline{H}^{(\epsilon_2)} - \overline{H}^{(\epsilon_1)} = -2G_\mathrm{ret}*G^{(1)}[Q].
\end{equation}
Thus,
\begin{equation}
    \overline{\delta h}_{\mu\nu} = \overline{Q}_{\mu\nu} - 2 G_\mathrm{ret} * G^{(1)}_{\mu\nu}[Q].
\end{equation}
We have
\begin{multline}
    2 G^{(1)}_{\mu\nu}[Q] = - \Box \overline{Q}_{\mu\nu} + \partial_\mu \partial^\rho \overline{Q}_{\rho \nu}+\partial_\nu \partial^\rho \overline{Q}_{\rho\mu}\\-\eta_{\mu\nu} \partial^\rho \partial^\sigma \overline{Q}_{\rho \sigma}
\end{multline}
But \cite[Lemma 4.1]{bar_green-hyperbolic_2015}
\begin{equation}
    G_{\mathrm{ret}} * \Box \overline{Q} = - \overline{Q},
\end{equation}
where we use the sign convention $\Box G_{\mathrm{ret}} = -\delta^{(4)}$. Hence, the $\overline{Q}$ term cancels and we get
\begin{equation}
    \overline{\delta h}_{\mu\nu} = -G_\mathrm{ret} * (\partial_\mu \partial^\rho \overline{Q}_{\rho\nu} + \partial_\nu \partial^\rho \overline{Q}_{\rho\mu} - \eta_{\mu\nu} \partial^\rho \partial^\sigma \overline{Q}_{\rho\sigma}).
\end{equation}
Let
\begin{equation}
    \xi_\mu := -G_\mathrm{ret} * (\partial^\rho \overline{Q}_{\rho\mu}).
\end{equation}
Then
\begin{equation}
    \overline{\delta h}_{\mu\nu} = \partial_\mu \xi_\nu + \partial_\nu \xi_\mu - \eta_{\mu\nu} \partial^\rho \xi_\rho,
\end{equation}
and so
\begin{multline}
    h_{\mu\nu}^{(\epsilon_2)} - h_{\mu\nu}^{(\epsilon_1)} = \delta h_{\mu\nu} = \overline{\delta h}_{\mu\nu} - \frac{1}{2}\eta_{\mu\nu} \eta^{\rho\sigma}\overline{\delta h}_{\rho\sigma} \\ = \partial_\mu \xi_\nu  + \partial_\nu \xi_\mu - \eta_{\mu\nu} \partial^\rho \xi_\rho - \frac{1}{2} \eta_{\mu\nu}(- 2 \partial^\rho \xi_\rho) \\ = \partial_\mu \xi_\nu + \partial_\nu \xi_\mu,
\end{multline}
i.e. they differ by a linearised gauge transformation. More generally, we show in Sec.~\ref{sec:uniqueness future metric} that the solution is actually unique up to local gauge transformations.

We thus have a suitable $h^-$ solution of the linearised EFE for $T^-$ in $\mathcal{M} \smallsetminus J^+(x)$ and a suitable $h^+$ solution of the linearised EFE for $T^+$ in $I^+(x)$ with appropriate matching conditions on $\Lambda^+(x)$. We leave for future work the verification of the applicability of the Barrabès-Israel formalism in the full post-Newtonian expansion. In the full nonlinear problem, the analogous task is to construct $g^+$ and $g^-$ with matching data on $\Lambda^+(x)$.  When this is possible, the Barrabès-Israel formalism then gives us an explicit distributional stress-energy on $\Lambda^+(x)$ such that $g_{ab} = g^+_{ab} \Theta(u) + g_{ab}^- (1-\Theta(u))$ satisfies the distributional Einstein field equations. The construction of such a distributional stress-energy in the full nonlinear case is reviewed in Appendix \ref{app:BI}. 

The construction of the distributional stress-energy in the linearised regime described above is given in Appendix \ref{app:BI linearised}. Explicitly,
\begin{align}
    T^{\mathrm{sing}}_{uu}
   &=-\frac{\Omega^{AB} \mathfrak{b}_{AB}(v,\theta)}{2\kappa r^2}\delta(u),\\
 T^{\mathrm{sing}}_{uA}=T^{\mathrm{sing}}_{Au}  
   &=-\frac{\mathfrak b_{vA}(v,\theta)}\kappa\delta(u),\\
 T^{\mathrm{sing}}_{AB}
   &=-\frac{2r^2}{\kappa}\Omega_{AB}
                  \mathfrak b_{vv}(v,\theta)\delta(u),\\
 T^{\mathrm{sing}}_{v\nu}&=0.
\end{align}

\subsection{Late time linearised metrics}

\label{subsec:late time}

It is also interesting to understand the properties of $h^+$ at late times, i.e., long after the quantum measurement takes place. We show here that if $\overline{h}^-$ (or more precisely its smooth continuation $\overline{\underline{h}}^-$ on the whole of $\mathbb{M}$) is a retarded potential, then $\overline{h}^+$ eventually becomes a retarded potential sourced by $\underline{T}^-_{\mu\nu} + \tau_{\mu\nu}$.

We start with
\begin{equation}
    \underline{\overline{h}}^- = 2 \kappa G_\mathrm{ret} * \underline{T}^-.
\end{equation}
The future field is
\begin{equation}
    \overline{h}^+ = \underline{\overline{h}}^- + \overline{\mathfrak{H}} + \overline{H}
\end{equation}
where
\begin{align}
    \overline{H} &= 2 \kappa G_{\mathrm{ret}} * \tau, \\
    \tau &= \Theta(u) \left(\Delta T - \kappa^{-1} G^{(1)}[\mathfrak{H}]\right).
\end{align}
The field $\mathfrak{H}$ disappears after a finite time at every fixed spatial position. Indeed, the smooth cutoff $\chi_\epsilon$ satisfies
\begin{equation}
    \chi_\epsilon(z)=0 \qquad \text{ for } z \geq \epsilon, \qquad 0<\epsilon<1.
\end{equation}
Since
\begin{equation}
    \mathfrak{H} = u \chi_\epsilon(u/v) \cdot (\text{finite terms}),
\end{equation}
it vanishes wherever $\dfrac{ct-r}{ct+r} > \epsilon$, i.e. when
\begin{equation}
    ct > \frac{1+\epsilon}{1-\epsilon} r \approx r(1+2\epsilon) + \mathcal{O}(\epsilon^2).
\end{equation}
For such $(ct,\vec{x})$, we get
\begin{align}
    \overline{h}^+ &= \underline{\overline{h}}^- + \overline{H} \\
    &= 2 \kappa G_\mathrm{ret} * \underline{T}^- + 2 \kappa G_\mathrm{ret} * \tau \\
    &= 2 \kappa G_\mathrm{ret} * (\underline{T}^- + \tau),
\end{align}
hence
\begin{equation}
    \overline{h}_{\mu\nu}^+(t,\vec{x}) = \frac{4G}{c^4} \iiint_{\mathbb{R}^3} \frac{(\underline{T}^-_{\mu\nu} + \tau_{\mu\nu})(t-\abs{\vec{x}-\vec{y}}/c,\vec{y})}{\abs{\vec{x}-\vec{y}}} d^3 \vec{y}
\end{equation}
for $ct > \dfrac{1+\epsilon}{1-\epsilon} r$. Note for such times, $\underline{T}^- + \tau = T^+$. However, this does not immediately imply that $\overline{h}_{\mu\nu}^+$ is necessarily sourced solely by $T^+$ at late times, since as a retarded potential it depends on the values of the source at earlier times. In particular, since
\begin{equation}
    \underline{T}_{\mu\nu}^- + \tau_{\mu\nu} = \chi_{\M^+} T^+ + \chi_{\M^-} T^- - \Theta(u) \kappa^{-1} G^{(1)}[\mathfrak{H}]
\end{equation}
distributionally, where we have interchanged Heaviside theta functions and indicator functions since they are distributionally equal when multiplied by smooth functions such as $T^\pm_{\mu\nu}$, we have that for $ct > \dfrac{1+\epsilon}{1-\epsilon} r$,
\begin{align}
    \label{eqn:retarded potential delta}
    \overline{h}^+_{\mu\nu}(t,\vec{x}) \stackrel{\eqref{eqn:naive linearised metric}}{=} \overline{h}^\mathrm{ret}_{\mu\nu}(t,\vec{x}) + \delta_{\mu\nu}(t,\vec{x})
\end{align}
where
\begin{multline}
    \delta_{\mu\nu}(t,\vec{x}) = \frac{-1}{2\pi} \iiint_{\mathbb{R}^3} \frac{\Theta(ct- \abs{\vec{x}-\vec{y}}-\abs{\vec{y}}) }{\abs{\vec{x}-\vec{y}}} \\ \cdot G_{\mu\nu}^{(1)}[\mathfrak{H}]\left(t-\frac{\abs{\vec{x}-\vec{y}}}{c},\vec{y}\right) d^3 \vec{y}.
\end{multline}
One may therefore source $\overline{h}^+$ by $T^+$ in such a retarded way at late times only should the correction $\delta$ from $G^{(1)}[\mathfrak{H}]$ be subleading to the rest of the source. $\overline{h}^\mathrm{ret}_{\mu\nu}$ is thus an approximation to $\overline{h}^+_{\mu\nu}$ for $ct > \dfrac{1+\epsilon}{1-\epsilon}r$. Note that for fixed $(ct,\vec{x}) \in I^+(0)$, we can let $\epsilon < \frac{ct-\abs{\vec{x}}}{ct+\abs{\vec{x}}}$ to have \eqref{eqn:retarded potential delta} hold locally around $(ct,\vec{x})$ up to linearised gauge equivalence.

\subsection{Uniqueness of the future linearised metric solution}

\label{sec:uniqueness future metric}

We now prove that the solution given in Eqn.~\eqref{eqn:sol linearised EFE} of the linearised Einstein field equations with the prescribed data ($T^- \in C^\infty(\M^-)$, $T^+ \in C^\infty(\M^+)$, $h^- \in C^\infty(\M^-)$ solution to the linearised Einstein field equations in $\M^-$ sourced by $T^-$) is \emph{unique} up to gauge transformations. More precisely, if $h^{(1)},h^{(2)} \in C^{0,1}_\mathrm{loc}(\mathbb{M})$ are two global solutions for this characteristic initial value problem that are regular at the cone, then we show that there is a covector field
\begin{equation}
    \label{eqn:properties xi}
    \xi_a \in C^{1,1}_{\mathrm{loc}}(\mathbb{M}), \qquad \operatorname{supp} \xi \subseteq J^+(0),
\end{equation}
such that
\begin{equation}
    h^{(1)}_{\mu\nu} - h^{(2)}_{\mu\nu} = \partial_\mu \xi_\nu + \partial_\nu \xi_\mu.
\end{equation}
Moreover, $\xi_a|_{\mathcal{N}}=\partial_b \xi_a|_\mathcal{N} = 0$ and $T^\mathrm{sing}[h^{(1)}] = T^\mathrm{sing}[h^{(2)}]$ and $R_{\mu\nu\rho\sigma}[h^{(1)}] = R_{\mu\nu\rho\sigma}[h^{(2)}]$. The reason we want $\xi \in C^{1,1}_\mathrm{loc}$ is that one derivative makes it $C^{0,1}_\mathrm{loc}$ i.e. preserves the regularity of the metric.

We can reduce this problem as follows. For any $q=h^{(1)}-h^{(2)}$, we have that 
\begin{equation}
    \label{eqn:conditions q}
    \begin{cases}
        q \in C^{0,1}_\mathrm{loc}(\mathbb{M}), \\
        \operatorname{supp} q \subseteq J^+(0), \\
        G^{(1)}[q] = 0 \text{ in } I^+(0) \cup \M^-.
    \end{cases}
\end{equation}
That is, $\operatorname{supp} G^{(1)}[q] \subseteq \mathcal{N}$. We now prove that every $q$ satisfying \eqref{eqn:conditions q} obeys $G^{(1)}[q]=0$ on all of $\mathbb{M}$ and has the form $q_{\mu\nu}=\partial_\mu \xi_\nu + \partial_\nu \xi_\mu$ with $\xi$ as in \eqref{eqn:properties xi}.

By linearity, $G^{(1)}[q] = \kappa (T^{\mathrm{sing}}[h^{(1)}]-T^{\mathrm{sing}}[h^{(2)}])$, $u=ct-r$, $r=\abs{\vec{x}}$. Since $q$ is locally Lipschitz, $G^{(1)}[q]$ has a bounded surface density on $\mathcal{N}$ i.e. can be written as $G^{(1)}[q]=2\kappa S[q] \delta(u)$, and so the Cartesian components of $S[q]$ are essentially bounded on every finite cone segment, including uniformly towards the vertex, and there is no additional vertex-supported distribution.

Since the two solutions agree outside the cone,
\begin{equation}
    q=0 \quad \text{ in } \M^- \Rightarrow q|_\mathcal{N}=0.
\end{equation}
Differentiating this in any direction tanget to the cone gives zero. Thus the first derivatives can jump only in the transverse direction:
\begin{equation}
    [\partial_\alpha q_{\mu\nu}] = n_\alpha b_{\mu\nu}
\end{equation}
for some $b_{\mu\nu}$. Differentiating once more produces a delta function:
\begin{equation}
    (\partial_\alpha \partial_\beta q_{\mu\nu})_{\mathrm{sing}} = n_\alpha n_\beta b_{\mu\nu} \delta(u).
\end{equation}
Substituting this into $G^{(1)}[q]=2\kappa S[q] \delta(u)$ gives $S$ the explicit form
\begin{multline}
    2\kappa S_{\mu\nu} = \frac{1}{2} (n_\mu n^\alpha b_{\alpha \nu} + n_\nu n^\alpha b_{\alpha \mu} \\ - n_\mu n_\nu \eta^{\rho\sigma}b_{\rho\sigma} - \eta_{\mu\nu} n^\alpha n^\beta b_{\alpha \beta}),
\end{multline}
where we used that the hypersurface is null $\eta^{\alpha\beta} n_\alpha n_\beta = 0$. Two immediate consequences of this are that $n^\mu S_{\mu\nu} = 0$, i.e. that $S^{u\nu} = 0$ for every $\nu$; and that taking both indices angular so that $n_A = n_B = 0$ gives
\begin{equation}
    2 \kappa S_{AB} = -\frac{1}{2} r^2 \Omega_{AB} n^\alpha n^\beta b_{\alpha \beta}.
\end{equation}
Hence, the angular block is proportional to the spherical metric. Together, this gives
\begin{equation}
    \begin{cases}        
    S^{u\nu} = 0, \\
    S^{vv} = K, \\
    S^{vA} = B^A, \\
    S^{AB} = \frac{p}{r^2} \Omega^{AB}
    \end{cases}
\end{equation}
for some $K$, $B^A$, $p$. One can extend the coefficients off $u=0$ while keeping $S^{u\nu}=0$, and so the distributional Bianchi identity $\nabla_\mu G^{(1),\mu\nu}[q]=\nabla_\mu (2 \kappa S^{\mu\nu}[q] \delta(u))=0$ becomes
\begin{equation}
    \begin{cases}
        \frac{2p}{r}=0, \qquad &\nu=u, \\
        \partial_v B^A + \frac{4}{v} B^A + r^{-2} D^A p =0, &\nu=A, \\
        \partial_v K + \frac{2}{v} K + D_A B^A - \frac{4}{v}p=0, & \nu = v.
    \end{cases}
\end{equation}
Every coefficient is evaluated at $u=0$ where $r=v/2$. The first equation forces $p=0$. The second equation gives
\begin{equation}
    \partial_v (v^4 B^A)=0 \Rightarrow B^A(v,\theta) = v^{-4} J^A(\theta)
\end{equation}
for some $J^A(\theta)$. However, we require the Cartesian components of $S$ to be uniformly bounded for small $v>0$, which gives $J=0$ i.e. $B=0$ (see Appendix \ref{app:vertex regularity} for more details on the interplay between constants of integration and uniform boundedness at $x=0$). Likewise, the third equation gives
\begin{equation}
    \partial_v(v^2 K) = 0 \Rightarrow K(v,\theta) = v^{-2} a(\theta),
\end{equation}
and by uniform boundedness at the $x=0$ once again gives $a(\theta)=K=0$. Thus,
\begin{equation}
    G^{(1)}[q] = 0 \qquad \text{ over } \mathbb{M}.
\end{equation}
This establishes shell uniqueness:
\begin{equation}
    G^{(1)}[q] = \kappa (T^\mathrm{sing}[h^{(1)}]-T^{\mathrm{sing}}[h^{(2)}]) = 0.
\end{equation}

Define
\begin{equation}
    \xi_\mu = - G_\mathrm{ret} * (\partial^\rho \overline{q}_{\rho \mu}).
\end{equation}
Since $\operatorname{supp}(G_\mathrm{ret}*f)\subseteq J^+(\operatorname{supp} f)$, this gives $\xi_\mu$ support in $J^+(0)$ \cite[Lemma 4.1]{bar_green-hyperbolic_2015}. Moreover, $\Box (G_\mathrm{ret}*f)=-f=G_\mathrm{ret}*(\Box f)$, so since
\begin{equation}
    \Box \overline{q}_{\mu\nu} = \partial_\mu \partial^\rho \overline{q}_{\rho \nu} + \partial_\nu \partial^\rho \overline{q}_{\rho \mu} - \eta_{\mu\nu}\partial^\rho \partial^\sigma \overline{q}_{\sigma \rho},
\end{equation}
we can apply $G_{\mathrm{ret}}$ on both sides, which gives
\begin{equation}
    \overline{q}_{\mu\nu} = \partial_\mu \xi_\nu + \partial_\nu \xi_\mu - \eta_{\mu\nu} \partial^\rho \xi_\rho.
\end{equation}
Since $\eta^{\mu\nu} \overline{q}_{\mu\nu} = -2\partial^\rho \xi_\rho$, we find
\begin{equation}
    h^{(1)}_{\mu\nu}-h^{(2)}_{\mu\nu}=q_{\mu\nu} = \partial_\mu \xi_\nu + \partial_\nu \xi_\mu.
\end{equation}
Since $h^{(1)}$ and $h^{(2)}$ are both in $C^{0,1}_\mathrm{loc}$, this makes $\xi \in C^{1,1}_{\mathrm{loc}}$ as required.

Since
\begin{multline}
    R^{(1)}_{\alpha\beta\gamma\delta}[q]
 =\tfrac12( \partial_\gamma\partial_\beta q_{\alpha\delta}
 +\partial_\delta\partial_\alpha q_{\beta\gamma}
 \\-\partial_\delta\partial_\beta q_{\alpha\gamma}
 -\partial_\gamma\partial_\alpha q_{\beta\delta}),
\end{multline}
substitution of $q_{\mu\nu}=\partial_\mu\xi_\nu + \partial_\nu \xi_\mu$ cancels every term, also for distributions, i.e.
\begin{equation}
    R^{(1)}_{\mu\nu\rho\sigma}[h^{(1)}] = R^{(1)}_{\mu\nu\rho\sigma}[h^{(2)}].
\end{equation}
Thus both the Ricci (and thus Einstein) as well as the Weyl curvatures are uniquely fixed up to gauge.

Note that metric uniqueness (up to gauge equivalence) and shell uniqueness are equivalent here. Indeed, since $h^{(1)}_{\mu\nu}-h^{(2)}_{\mu\nu}=\partial_\mu \xi_\nu + \partial_\nu \xi_\mu$ and $G^{(1)}[\partial_\mu \xi_\nu + \partial_\nu \xi_\mu] = 0$, we have that
\begin{multline}
    G^{(1)}[h^{(1)}-h^{(2)}]=0=\kappa (T^{\mathrm{sing}}[h^{(1)}]-T^{\mathrm{sing}}[h^{(2)}]) \\ \Rightarrow T^{\mathrm{sing}}[h^{(1)}] = T^{\mathrm{sing}}[h^{(2)}].
\end{multline}
Conversely, if $T^{\mathrm{sing}}[h^{(1)}] = T^{\mathrm{sing}}[h^{(2)}]$ then $G^{(1)}[h^{(1)}-h^{(2)}]=G^{(1)}[h^{(1)}]-G^{(1)}[h^{(2)}]=0$ i.e.
\begin{equation}
    G^{(1)}[h^{(1)}]=G^{(1)}[h^{(2)}],
\end{equation}
and the above applies to show gauge equivalence.

\subsection{The collapse hypersurface's mass}

\label{subsec:shell mass}

The introduction of a distributional stress-energy induces a notion of mass on the collapse hypersurface. We now analyse its properties. We first briefly review the Barrabès-Israel construction of the shell stress-energy, outlined in more detail in Appendix \ref{app:BI}. 

Consider the coordinates $y^a = (\lambda,\theta^A)$ on $\mathcal{N} \equiv \Lambda^+(0)$, where $a=1,2,3$ and $A = 2,3$, which are taken to be the same on both sides of $\mathcal{N}$. Take $\lambda$ to be an arbitrary parameter on the null generators of the hypersurface, and use $\theta^A$ to label the generators: to each generator we assign two coordinate $\theta^A$. Changing $\lambda$ while keeping $\theta^A$ constant produces a displacement along a single generator; changing $\theta^A$ while keeping $\lambda$ constant produces a displacement across generators. Hence, $\mathcal{N}$ as seen from $\mathcal{M}^\pm$ is described by a set of parametric relations $x^\alpha_\pm(y^a)$, which can be used to define tangent vectors $e^\alpha_{\pm a} = \partial x^\alpha_\pm/\partial y^a$ on each side of the hypersurface. The null vector
\begin{equation}
    k_\pm^\alpha = \left(\frac{\partial x_\pm^\alpha}{\partial \lambda}\right)_{\theta^A} \equiv e^\alpha_{\pm \lambda}
\end{equation}
is proportional to the null generator, while the two spacelike vectors
\begin{equation}
    e^\alpha_{\pm A} = \left(\frac{\partial x_\pm^\alpha}{\partial \theta^A}\right)_\lambda
\end{equation}
are orthogonal to the null generator.  By construction,
\begin{equation}
    k_{\pm \alpha } k^{\alpha}_\pm = 0 = k_{\pm \alpha} e^{\alpha}_{\pm A} \, .
\end{equation}
The remaining inner products, that we write
\begin{equation}
    \sigma_{AB}^\pm(\lambda,\theta^C) = g_{\alpha \beta}^{\pm} e^{\alpha}_{\pm A} e^\beta_{\pm B}
\end{equation}
do not vanish, though we assume by construction that they are the same on both sides of $\mathcal{N}$:
\begin{equation}
    [\sigma_{AB}] \equiv \sigma^+_{AB}|_{\mathcal{N}} - \sigma^-_{AB}|_{\mathcal{N}} = 0.
\end{equation}
This is satisfied if, in particular, the global metric is continuous. It can be shown \cite{poisson_reformulation_2002,poisson_hypersurfaces_2004} (see Appendix \ref{app:BI}) that the surface stress-energy flux on either sides of $\mathcal{N}$ is given by
\begin{align}
    S^{\alpha \beta}_\pm &= \mu k^\alpha_\pm k^\beta_\pm + j^A(k^\alpha_\pm e^\beta_{\pm A} + e^{\alpha}_{\pm A} k^\beta_\pm) + p \sigma^{AB} e^\alpha_{\pm A} e^{\beta}_{\pm B}, \\    T^{\mathrm{sing}}_{\mu\nu}&=2S_{\mu\nu}\delta(u).
\end{align}
The first term represents a flow of matter along the null generators of the hypersurface, i.e. $\mu$ represents a mass density; the second term represents a flow of matter in the directions transverse to the generators, i.e. $j^A$ represents a current density; the last term expresses the fact that $p$ is an isotropic pressure.

Write $\omega_i := \frac{x_i}{r}$. Using \eqref{eq:BIc-S} and \eqref{eq:BIc-normalized-delta} gives the Cartesian stress-energy components
\begin{align}
 T^{\mathrm{sing}}_{00}(ct,\vec x)
 &=\frac\mu2\delta(ct-r),\\
 T^{\mathrm{sing}}_{0i}(ct,\vec x)
 &=-\left(\frac\mu2\omega_i+rj^A\partial_A\omega_i\right)\delta(ct-r),\\
 T^{\mathrm{sing}}_{ij}(ct,\vec x)
 &=\Big[
    \frac\mu2\omega_i\omega_j
    +rj^A(\omega_i\partial_A\omega_j+(\partial_A\omega_i)\omega_j) \nonumber \\
    &\qquad +2p(\delta_{ij}-\omega_i\omega_j)
   \Big]\delta(ct-r).
 \label{eq:BIc-cartesian}
\end{align}
$\mu$, $p$ and $j^A$ are given by \eqref{eq:BIc-BI-relations}:
\begin{align}
 \mu(2r,\theta)&=-\frac{\Omega^{AB} \mathfrak{b}_{AB}(2r,\theta)}{\kappa r^2},\\
 j^A(2r,\theta)&=\frac{\Omega^{AB}\mathfrak b_{vB}(2r,\theta)}{\kappa r^2},\\
 \qquad p(2r,\theta)&=-\frac{\mathfrak b_{vv}(2r,\theta)}\kappa.
\end{align}

The surface coefficients above are evaluated at $(v,\theta)=(2r,\theta) = (2ct,\theta)$. For $t>0$, the energy is concentrated on the sphere $r=ct$. But $\Sigma_t \cap \Lambda^+(0) = S^2$. It follows that the mass on the shell is
\begin{align}
 M_{\mathrm{shell}}(t)c^2
 &:=\int_{\mathbb R^3}T^{\mathrm{sing}}_{00}(ct,\vec x)\,d^3\vec x
 \\
 &=\frac{(ct)^2}{2}\iint_{S^2}\mu(2ct,\theta)\,d\Omega .
 \label{eq:BIc-energy}
\end{align}
The energy per area on this time slice is $\mu/2$. Dividing \eqref{eq:BIc-energy} by $c^2$ gives a mass, whereas \eqref{eq:BIc-energy} itself is an energy. 

Note that
\begin{equation}
    \iint_{S^2} D^A \mathfrak{b}_{vA}(s,\theta) d\Omega = 0
\end{equation}
as it is a divergence on a compact surface without boundary. Also note that from Eqn.~\eqref{eq:BIc-mu-source-intermediate},
\begin{align}
    \mu(v,\theta)
 &=-\frac4{\kappa v^2}\int_0^v
 \Big[D^A\mathfrak b_{vA}(s,\theta) \nonumber \\
       &+\frac{\kappa s^2}{2}
           (\Delta T_{vv}+\Delta T_{uv})(0,s,\theta)
           \Big]ds.
\end{align}
Thus,
\begin{align}
     M_{\mathrm{shell}}(t)c^2 &= -\frac{4}{\kappa (2ct)^2} \frac{(ct)^2}{2} \frac{\kappa}{2} \nonumber \\ \cdot  \int_0^{2ct} &s^2 \iint_{S^2} (\Delta T_{vv} + \Delta T_{uv})(0,s,\theta) d\Omega ds \\
     = -\frac{1}{4}\int_0^{2ct} &s^2 \iint_{S^2} (\Delta T_{vv} + \Delta T_{uv})(0,s,\theta) d\Omega ds.
\end{align}
Hence, since $\frac{d(2ct)}{d(ct)} = 2$,
\begin{multline}
 \frac d{d(ct)}\big(M_{\mathrm{shell}}(t)c^2\big)
 \\=-\frac12(2ct)^2\iint_{S^2}
          (\Delta T_{vv}+\Delta T_{uv})(0,2ct,\theta)\,d\Omega.
\end{multline}
The null-to-Cartesian conversion \eqref{eq:vc-source-components} gives
\begin{multline}
    \label{eqn:null to cartesian}
 (\Delta T_{vv}+\Delta T_{uv})(0,2ct,\theta)
 \\=\frac12\left(\Delta T_{00}
                +\omega^i\Delta T_{0i}\right)
                                  (ct,ct\vec\omega).
\end{multline}
Consequently,
\begin{multline}
 \frac d{d(ct)}\big(M_{\mathrm{shell}}(t)c^2\big)
 \\ =-(ct)^2\iint_{S^2}
 \left(\Delta T_{00}
             +\omega^i\Delta T_{0i}\right)
                                  (ct,ct\vec\omega)\,d\Omega.
 \label{eq:BIc-energy-flux}
\end{multline}

We now compare with the regular source. Note that
\begin{multline}
    \frac{1}{\epsilon} \int_{ct < \abs{\vec{x}}<ct+\epsilon} \Delta T_{00} (ct+\epsilon,\vec{x}) d^3 \vec{x} \\
    = \frac{1}{\epsilon} \iint_{S^2}\int_{ct}^{ct+\epsilon} r^2 \Delta T_{00} (ct+\epsilon,r \vec{\omega}) dr d\Omega.
\end{multline}
Make the substitution $r=ct+\epsilon s$, $0 \leq s \leq 1$, $dr = \epsilon ds$. Then
\begin{multline}
    \frac{1}{\epsilon} \int_{ct < \abs{\vec{x}}<ct+\epsilon} \Delta T_{00} (ct+\epsilon,\vec{x}) d^3 \vec{x} \\
    = \iint_{S^2} \int_0^1 (ct+\epsilon s)^2 \Delta T_{00} (ct+\epsilon,(ct+\epsilon s) \vec{\omega}) ds d\Omega.
\end{multline}
As $\epsilon \downarrow 0$, $(ct + \epsilon,(ct+\epsilon s) \vec{\omega}) \to (ct, ct \vec{\omega})$ and $(ct+\epsilon s)^2 \Delta T_{00} (ct+\epsilon,(ct+\epsilon s)\vec{\omega}) \to (ct)^2 \Delta T_{00} (ct,ct\vec{\omega})$ uniformly, so
\begin{multline}
    \lim_{\epsilon \downarrow 0} \frac{1}{\epsilon} \int_{ct < \abs{\vec{x}}<ct+\epsilon} \Delta T_{00} (ct+\epsilon,\vec{x}) d^3 \vec{x} \\ =(ct)^2 \iint_{S^2} \Delta T_{00}(ct,ct \vec{\omega}) d\Omega.
\end{multline}
Thus,
\begin{align}
    \frac{d}{d(ct)} \int_{\abs{\vec{x}} < ct}& \Delta T_{00}(ct,\vec{x}) d^3 \vec{x} \nonumber \\ &= (ct)^2 \iint_{S^2} \Delta T_{00}(ct,ct \vec{\omega}) d\Omega \nonumber \\
    &+ \iiint_{\abs{\vec{x}}<ct} \partial_{0} \Delta T_{00}(ct,\vec{x}) d^3 \vec{x}.
\end{align}

Conservation with a lower time index means $-\partial_0\Delta T^{\mathrm{Cart}}_{00} +\partial_i\Delta T^{\mathrm{Cart}}_{i0}=0$. The divergence theorem implies that
\begin{equation}
    \iiint_{\abs{\vec{x}}<ct} \partial_{0} \Delta T_{00}(ct,\vec{x}) d^3 \vec{x} = (ct)^2\iint_{S^2}\omega^i\Delta T_{0i}\,d\Omega.
\end{equation}
Thus,
\begin{equation}
    \label{eqn:diff eqn mass}
    \frac{d}{d(ct)} \int_{\abs{\vec{x}} < ct} \Delta T_{00}(ct,\vec{x}) d^3 \vec{x} = -\frac{d}{d(ct)}\big(M_{\mathrm{shell}}(t)c^2\big).
\end{equation}
Since $\underline{T}^-$ is smooth in a neighbourhood of the vertex, $\Delta T$ is bounded near the vertex, so there exists a $C>0$ such that
\begin{equation}
    \abs{\Delta T_{00}(ct,\vec{x})} \leq C
\end{equation}
for all sufficiently small t and for $\abs{\vec{x}} < ct$. Thus
\begin{align}
    \abs{\int_{\abs{\vec{x}} < ct} \Delta T_{00}(ct,\vec{x}) d^3 \vec{x}} &\leq \int_{\abs{\vec{x}} < ct} \abs{\Delta T_{00}(ct,\vec{x})} d^3 \vec{x} \\
    &\leq C \int_{\abs{\vec{x}} < ct} d^3 \vec{x} \\
    &= C \frac{4\pi}{3}(ct)^3 \stackrel{t \downarrow 0}{\longrightarrow} 0.
\end{align}
Thus,
\begin{equation}
    \lim_{t \downarrow 0} \int_{\abs{\vec{x}} < ct} \Delta T_{00}(ct,\vec{x}) d^3 \vec{x} = 0.
\end{equation}
Moreover, from Eqn.~\eqref{eqn:null to cartesian} and following the same argument, there exists $C >0$ such that
\begin{equation}
    \abs{(\Delta T_{vv} + \Delta T_{uv})(0,s,\theta)} \leq C
\end{equation}
for all sufficiently small $s>0$ and all angles. Thus,
\begin{align}
    \abs{M_\mathrm{shell}(t) c^2} &\leq \frac{1}{4} \int_0^{2ct} s^2 \iint_{S^2}\abs{(\Delta T_{vv} + \Delta T_{uv})(0,s,\theta)} d\Omega ds \\
    &\leq \frac{C}{4} \int_0^{2ct} s^2 \iint_{S^2} d\Omega ds \\
    &= \frac{\pi C}{3} (2ct)^3 \stackrel{t \to 0}{\longrightarrow} 0.
\end{align}
Thus,
\begin{equation}
    \lim_{t \downarrow0} M_\mathrm{shell}(t) c^2 = 0
\end{equation}
Thus, the differential equation \eqref{eqn:diff eqn mass} can be integrated with these boundary conditions to get
\begin{equation}
    \label{eqn:shell mass}
    M_\mathrm{shell}(t) c^2 = -\int_{\abs{\vec{x}} < ct} \Delta T_{00}(ct,\vec{x}) d^3 \vec{x}.
\end{equation}

This naturally leads to mass conservation across the collapse hypersurface: $M_{\mathrm{shell}}(t) c^2$ counteracts the jump in energy density across the collapse hypersurface. Note that \emph{this shell mass can be negative}. This is linked to violations of energy conditions, expanded upon in Sec.~\ref{sec:examples}.

We further check explicitly in Appendix \ref{app:distributional conservation} that the total stress-energy is indeed consequently distributionally conserved, and importantly that the shell stress-energy defines a meaningful distribution in a neighbourhood of the vertex, where $\Lambda^+(0)$ is not smooth. We also show that the Einstein tensor does not contain any additional term concentrated only at the vertex, and that stress-energy conservation does not fail by a term concentrated only at the vertex.

\subsection{Timelike-separated wavefunction collapses}

This procedure can be generalised to timelike separated quantum measurements. Indeed, since $h^+ \in C^\infty$ and $T^+ \in C^\infty$, one can apply the same construction patchwise. For example, if there are two successive quantum measurements with wavefunction collapses emanating from $x$ and $y \in I^+(x)$ respectively, then divide $\mathbb{M}$ into three patches: let $\M^{(1)} := \mathbb{M} \smallsetminus J^+(x)$, $\M^{(2)} := I^+(x) \cap (\mathbb{M} \smallsetminus J^+(y))$ and $\M^{(3)} := I^+(y)$, so that $\mathbb{M} = \M^{(1)} \cup \Lambda^+(x) \cup \M^{(2)} \cup \Lambda^+(y) \cup \M^{(3)}$. Suppose the stress-energies in $\M^{(1)}$, $\M^{(2)}$ and $\M^{(3)}$ are $T^{(1)}$, $T^{(2)}$ and $T^{(3)}$, respectively, and that the metric on $\M^{(1)}$ that satisfies the linearised Einstein field equations sourced by $T^{(1)}$ is $h^{(1)}$. Then 
\begin{enumerate}
    \item use the construction above to build $h^{(2)}_{\mu\nu} \in C^\infty(\M^{(2)})$ from $h^{(1)}$, $T^{(1)}$ and $T^{(2)}$, with a metric $h^{(12)}$ locally Lipschitz across $\Lambda^+(x)$ and patchwise smooth that satisfies the distributional linearised Einstein field equations in $\M^{(1)} \cup \Lambda^+(x) \cup \M^{(2)}$ sourced by $T^{(1)}$, $T^{(2)}$ and the singular Barrabès-Israel source $T^\mathrm{sing,(12)}$ on $\Lambda^+(x)$ with initial condition given by $h^{(1)}$,
    \item use the construction above to build $h^{(3)}_{\mu\nu} \in C^\infty(\M^{(3)})$ from $h^{(2)}$, $T^{(2)}$ and $T^{(3)}$, with a metric $h^{(23)}$ locally Lipschitz across $\Lambda^+(y)$ and patchwise smooth that satifies the distributional linearised Einstein field equations in $\M^{(2)} \cup \Lambda^+(y) \cup \M^{(3)}$ sourced by $T^{(2)}$, $T^{(3)}$ and the singular Barrabès-Israel source $T^\mathrm{sing,(23)}$ on $\Lambda^+(y)$ with initial condition given by $h^{(2)}|_{\Lambda^+(y)}$,
    \item Since $h^{(23)}|_{\M^{(2)}} = h^{(2)} = h^{(12)}|_{\M^{(2)}}$ and the sources are retarded in the linearised Einstein field equations, we can patch $h^{(12)}$ and $h^{(23)}$ to get a locally Lipschitz metric $h^{(123)}$ on $\mathbb{M}$ that is patchwise smooth and recovers each individual metric in the relevant regions, and satisfies the linearised Einstein field equations sourced by $T^{(1)}$, $T^{(2)}$, $T^{(3)}$ and the two Barrabès-Israel singular terms $T^\mathrm{sing,(12)}$ and $T^\mathrm{sing,(23)}$ on $\Lambda^+(x)$ and $\Lambda^+(y)$ respectively, with initial condition given by $h^{(1)}$.
the \end{enumerate}
The issue of spacelike-separated quantum measurements requires separate scrutiny however, since there are now crossing shells and different solutions that need to match. The validity of the formalism in those situations relies heavily on the fact that the local stress-energies are built from local quantum states which do not update until the collapse hypersurface reaches their spacetime localisation. We show that a consistent picture can be drawn for spacelike-separated measurements of (possibly entangled) qudits in Sec.~\ref{subsec:mmt Bell state}.

\section{Examples of wavefunction collapses in local semiclassical gravity}

\label{sec:examples}

We now provide two explicit examples of the procedure described above: that of a qubit in superposition undergoing a single quantum measurement, and that of two entangled qubits undergoing quantum measurements over spacelike separations. We provide additional details of the computations undertaken in Appendix \ref{app:computations qubit in superposition}.

\subsection{A qubit in superposition}

\label{subsec:mmt qubit in superposition}

Consider the simple example whereby a stationary quantum particle of mass $m$ lies in a superposition of two spatially separated wavepackets at $t=t_0$ of the form
\begin{equation}
    \label{eqn:qubit}
    \Psi(t,\vec{x}) = a_1 \psi_1(t,\vec{x}) + a_2 \psi_2(t,\vec{x}),
\end{equation}
where $\abs{a_1}^2 + \abs{a_2}^2 = 1$, with $\psi_i(t,\cdot) \in \mathscr{S}(\mathbb{R}^3)$ having negligible overlap and each being $L^2$-normalised $\norm{\psi_i(t,\cdot)}_2 = 1$.\footnote{It is worth emphasizing that the discussion thus does not apply to the unphysical idealised case of ``pointlike'' wavefunctions perfectly localised as a $\delta$ distribution.   These 1) cannot be $L^2$ normalised and 2) do not live in Hilbert space (but rather in a rigged Hilbert space \cite{de_la_madrid_role_2005}).} 

\subsubsection{The past gravitational field}

In this regime, we work in the weak field limit on a Minkowski background in some foliation $\{\Sigma_t\}_{t \in \mathbb{R}}$. The mass density of the particle around $t=t_0$ (prior but suitably close to the quantum measurement) is
\begin{align}
    \varrho_m^-(t,\vec{x}) &= m \abs{\Psi(t,\vec{x})}^2 \\
    &\approx m(\abs{a_1}^2 \abs{\psi_1(t,\vec{x})}^2 + \abs{a_2}^2 \abs{\psi_2(t,\vec{x})}^2).
\end{align}
We have
\begin{equation}
    \iiint_{\mathbb{R}^3} \varrho_m^- d^3 \vec{x} = m(\abs{a_1}^2 + \abs{a_2}^2) = m.
\end{equation}

We also assume that the wavepackets are roughly stationary in this foliation during the light-travel times under consideration prior but close to the subsequent quantum measurement, and so
\begin{multline}
    T_{\mu\nu}^{\mathrm{loc},-}(t,\vec{x})  \approx c^2 \varrho_{m}^-(t,\vec{x}) \delta^0_\mu \delta^0_\nu \\
    \approx mc^2(\abs{a_1}^2 \abs{\psi_1(t,\vec{x})}^2 + \abs{a_2}^2 \abs{\psi_2(t,\vec{x})}^2) \delta^0_\mu \delta^0_\nu.
\end{multline}
The assumption that these are stationary implies that
\begin{equation}
    \partial^\mu T_{\mu\nu}^{\mathrm{loc},-} = 0
\end{equation}
exactly on the Minkowski background. The ensuing discussion described intervals short compared to wavepacket motion or spreading. Late-time statements within this stationary model will require the relevant light-travel times to remain short compared with those dynamical times.

For example, the pre-measurement wavepackets can be sharply peaked normalised Gaussians of the form
\begin{align}
    \label{eqn:psi gaussian}
    \psi_i(\vec{x}) &= (2\pi \sigma_i^2)^{-3/4} \exp\left(-\frac{\abs{\vec{x}-\vec{x}_i}^2}{4\sigma_i^2}\right), \\
    \vec{x}_1 &= \vec{0}, \qquad \abs{\vec{x}_2} = L,
\end{align}
where $\sigma_i$ is some standard deviation and $L$ is the distance between the peaks of both wavepackets; since we assume these are sharply peaked, we take $\sigma_i \ll L$. These are $L^2$-normalised wavepackets $\iiint_{\mathbb{R}^3} \abs{\psi_i(\vec{x})}^2 d^3 \vec{x} = 1$, with exponentially small overlap:
\begin{multline}
    \iiint_{\mathbb{R}^3} \psi_1(\vec{x})^* \psi_2(\vec{x}) d^3 \vec{x} \\ = \left(\frac{2 \sigma_1 \sigma_2}{\sigma_1^2 + \sigma_2^2}\right)^{3/2} \exp\left(-\frac{L^2}{4(\sigma_1^2 + \sigma_2^2)}\right).
\end{multline}
Not only are we neglecting here wavepacket spreading, we are also neglecting the gravitational self-interaction of the wavefunction over the timescales considered. We show in Appendix \ref{app:interval stationary approximation} that this is a good approximation over timescales
\begin{equation}
    \tau \ll \min_{i=1,2} \min\left(\frac{2m \sigma_i^2}{\hbar},\sqrt{\frac{2\sigma_i L^2}{Gm}}\right)
\end{equation}
In the following, we will thus consider the wavefunction and its associated gravitational field in $\M^-$ in the time-interval $t_0 \in (-\tau,0)$. Although the past gravitational field is retarded (as in Eqn.\eqref{eqn:retarded past gravitational field}), we will assume (in the goal of providing explicit analytic expressions) that ``just before'' the collapse point $x = (0,\vec{0})$ the contribution to the gravitational field is dominated by the stationary Gaussian wavefunction in this time interval. In general this is not necessary, though otherwise it becomes harder to provide exact analytic expressions and one should instead turn to numerical methods.

At point $x = (0,\vec{0})$ Alice performs a position measurement which yields outcome ``$i$'' where $i$ is one of $1,2$. Write $\bar{i}$ for the other outcome. We make further assumptions here. First, we assume that the wavefunction collapse affects  the Born probability amplitudes but does not significantly change the $\sigma_i$'s, i.e. does not sharpen or widen the peak of the selected wavepacket.\footnote{Roughly speaking, this requires that the quantum measurement resolves distances of the order of the $\sigma_i$ or coarser.} We also assume that after the measurement, the selected wavepacket is roughly stationary in this foliation during the light-travel times under consideration (for analytic expressions in the Gaussian case, we again require $t \in (0,\tau)$, though qualitatively we expect the results to be similar). Thus, in $I^+(0)$,
\begin{equation}
    \varrho_{m}^{+,(i)}(t,\vec{x}) \approx m \abs{\psi_i(t,\vec{x})}^2
\end{equation}
and the mass density difference is
\begin{multline}
    \Delta \varrho_m^{(i)} := \varrho_m^{+,(i)}-\varrho_m^- \approx m \abs{a_{\bar{i}}}^2(\abs{\psi_i}^2 - \abs{\psi_{\bar{i}}}^2) \\ = m \sum_{j=1}^2 (\delta_{ij}-\abs{a_j}^2) \abs{\psi_j(\vec{x})}^2 \, .
\end{multline}
In particular, in $I^+(0)$,
\begin{align}
    T_{\mu\nu}^{\mathrm{loc},+,(i)}(t,\vec{x}) &\approx c^2 \varrho_{m}^{+,i}(t,\vec{x}) \delta_{\mu}^0 \delta_{\nu}^0 \\
    \partial^\mu T_{\mu\nu}^{\mathrm{loc},+,(i)} &= 0 \qquad \text{ in } I^+(0).
\end{align}
This is saying that we assume that $T_{0j}^{\mathrm{loc},+,(i)}$ and $T_{jk}^{\mathrm{loc},+,(i)}$ are subleading in $v/c$ in this regime. Moreover,
\begin{equation}
    \label{eqn:difference stress energy}
    \Delta T_{\mu\nu}^{(i)} \approx c^2 \Delta \varrho_m^{(i)} \delta^0_\mu \delta^0_\nu.
\end{equation}

These $\Delta \varrho_m^{(i)}$ are smooth and rapidly decaying if the wavepackets are: we have assumed the wavepackets are Schwartz, and since the product of Schwartz functions is a Schwartz function, $\Delta \varrho_m^{(i)}(t,\cdot)$ is Schwartz, and thus so are $T_{\mu\nu}^{\mathrm{loc},-}$, $T_{\mu\nu}^{\mathrm{loc},+,(i)}$ and $\Delta T_{\mu\nu}^{(i)}$ (in particular they are $C^\infty$ and are conserved in their respective regions).

In the weak field limit, the gravitational field in the past of $\Lambda^+(x)$ is given by
\begin{equation}
    \label{eqn:retarded past gravitational field}
    \Phi^-(t,\vec{x}) = -G \iiint_{\mathbb{R}^3} \frac{\varrho_{m}^{-}(t_\text{ret},\vec{x}')}{\abs{\vec{x}-\vec{x}'}} d^3 \vec{x}'
\end{equation}
where $t_\text{ret} = t-\frac{\abs{\vec{x} - \vec{x}'}}{c}$. For example, if the $\psi_i$ are given as in \eqref{eqn:psi gaussian}, we have (see Appendix \ref{app:semiclassical newtonian potential})
\begin{equation}
    \Phi^-(t,\vec{x}) \approx - Gm \sum_{j=1}^2 \frac{\abs{a_j}^2}{\abs{\vec{x}-\vec{x}_j}} \mathrm{erf}\left(\frac{\abs{\vec{x}-\vec{x}_j}}{\sqrt{2} \sigma_j}\right),
\end{equation}
where $\mathrm{erf}(x) = \frac{2}{\sqrt{\pi}} \int_0^x e^{-z^2} dz$ is the error function. That is, just before the collapse takes place, the linearised metric in Lorenz gauge in Cartesian coordinates is
\begin{multline}
    ds_-^2 = -\left(1- \frac{2Gm}{c^2} \sum_{j=1}^2 \frac{\abs{a_j}^2}{\abs{\vec{x}-\vec{x}_j}} \mathrm{erf}\left(\frac{\abs{\vec{x}-\vec{x}_j}}{\sqrt{2} \sigma_j}\right)\right) c^2 dt^2 \\+\left(1+ \frac{2Gm}{c^2} \sum_{j=1}^2 \frac{\abs{a_j}^2}{\abs{\vec{x}-\vec{x}_j}} \mathrm{erf}\left(\frac{\abs{\vec{x}-\vec{x}_j}}{\sqrt{2} \sigma_j}\right)\right) d\vec{x}^2
\end{multline}
The extended past linearised metric is then
\begin{align}
    \underline{h}^-_{00} &= -\frac{2\Phi^-}{c^2}, \quad \underline{h}^-_{ij} = -\frac{2\Phi^-}{c^2} \delta_{ij}, \\
    \underline{h} &= -\underline{h}_{00}+\delta^{ij} \underline{h}_{ij} = -\frac{4 \Phi^-}{c^2}, \\
    \underline{\overline{h}}_{\mu\nu}^- &= -\frac{4 \Phi^-}{c^2} \delta^0_\mu \delta^0_\nu.
\end{align}
In particular these are $C^\infty$ in $\mathbb{M}$. 

\subsubsection{The shell stress-energy and mass}

We can then apply the results of Sec.~\ref{sec:measurements and divergence} here. The source restricted to the cone ($u=ct-r=0$, $v=ct+r=2r$) satisfies
\begin{align}
    \Delta T_{vv}^{(i)}(0,2r,\theta) &= \Delta T^{(i)}_{uv}(0,2r,\theta) \stackrel{\eqref{eq:vc-source-components}}{=} \frac{c^2}{4} \Delta \varrho_m^{(i)}(r\vec{\omega}), \\
    \Delta T_{vA}^{(i)}(0,2r,\theta) &\stackrel{\eqref{eq:vc-source-components}}{=} 0,
\end{align}
where $r=\abs{\vec{x}}$ and $\vec{\omega} = \vec{x}/r$. We can then insert these components into the characteristic constraints, which gives
\begin{equation}
    \mathfrak{b}^{(i)}_{vv}(2r,\theta) \stackrel{\eqref{eqn:characteristic bvv bvA}}{=} \frac{\kappa c^2 r}{4} \Delta \varrho^{(i)}_m(r\vec{\omega}),    
\end{equation}
Moreover,
\begin{equation}
    \mathfrak{b}_{vA}^{(i)}(2r,\theta) \stackrel{\eqref{eqn:vA solution}}{=} \frac{\kappa c^2}{2r^2} D_A \int_0^r s^3 \Delta \varrho_m^{(i)}(s \vec{\omega}) ds.
\end{equation}
Since
\begin{align}
    \int_0^v & D^A \mathfrak{b}_{vA}(s,\theta) ds \nonumber \\ &= \frac{\kappa c^2}{2} D^A D_A \int_0^v \frac{ds}{s^2} \int^s_0 (s')^3 \Delta \varrho_m^{(i)}(s' \vec{\omega}) ds' \\
    &=\kappa c^2 D^A D_A \int_0^r (s')^3 \Delta \varrho_m^{(i)}(s'\vec{\omega}) \left( \int_{s'}^r \frac{ds}{s^2}\right) ds' \\
    &=\kappa c^2 D^A D_A \int_0^r (s')^3 \Delta \varrho_m^{(i)}(s'\vec{\omega}) \left(\frac{1}{s'}-\frac{1}{r}\right) ds' \\
    &= \kappa c^2 D^A D_A \int_0^r \left(s^2 - \frac{s^3}{r}\right) \Delta \varrho_m^{(i)}(s \vec{\omega}) ds,
\end{align}
where we have integrated over the region $0 \leq s' \leq s \leq r$, we have
\begin{multline}
    \Omega^{AB} \mathfrak{b}_{AB}^{(i)}(2r,\theta)\stackrel{\eqref{eqn:AB solution}}{=} \kappa c^2 \Bigg(2 \int_0^r s^2 \Delta \varrho^{(i)}_m(s \vec{\omega}) ds \\ + D^A D_A \int_0^r \left(s^2 - \frac{s^3}{r}\right) \Delta \varrho_m^{(i)}(s \vec{\omega}) ds \Bigg) .
\end{multline}
Furthermore, \eqref{eq:BIc-coefficients} gives
\begin{align}
    \label{eqn:shell stress energy qubit example}
    \mu^{(i)}(2r,\theta) &= -\frac{c^2}{r^2} \Bigg[2\int_0^r s^2 \Delta \varrho_m^{(i)}(s\vec{\omega}) ds \nonumber \\
    &+ D^A D_A \int_0^r \left(s^2 - \frac{s^3}{r}\right) \Delta \varrho_m^{(i)}(s\vec{\omega}) ds\Bigg], \\
    j^{(i),A}(2r,\theta)&= \frac{c^2}{2r^4} D^A \int_0^r s^3 \Delta \varrho_m^{(i)}(s\vec{\omega}) ds, \\
    p^{(i)}(2r,\theta) &= -\frac{c^2 r}{4} \Delta \varrho_m^{(i)}(r\vec{\omega})
\end{align}
where in this normalisation ($k=\partial_v$, $U=c\partial_{ct}$, $-k\cdot U = \frac{c}{2}$) $T_{\mu\nu}^{\mathrm{sing},(i)} = 2 S^{(i)}_{\mu\nu} \delta(u)$ and the resulting shell stress-energy is given by \eqref{eq:BIc-S} and \eqref{eq:BIc-normalized-delta}. From Eqns.~\eqref{eq:BIc-energy}, \eqref{eqn:shell mass} and \eqref{eqn:difference stress energy}, the corresponding mass on the shell is
\begin{align}
    M^{(i)}_\mathrm{shell}(t) &= \frac{t^2}{2} \iint_{S^2} \mu(2ct,\theta) d\Omega \\ &= - \iiint_{\abs{\vec{x}}<ct} \Delta \varrho_m^{(i)}(t,\vec{x}) d^3 \vec{x}.
\end{align}
The bulk mass is
\begin{multline}
    M_\mathrm{bulk}^{(i)}(t) = \iiint_{\mathbb{R}^3} [\varrho_m^-(t,\vec{x}) + \chi_{\abs{\vec{x}} < ct}(\vec{x}) \Delta \varrho_m^{(i)}(t,\vec{x})] d^3 \vec{x} \\ = m - M^{(i)}_{\mathrm{shell}}(t).
\end{multline}
Consequently, $M_{\mathrm{bulk}}^{(i)}(t)+M_{\mathrm{shell}}^{(i)}(t) = m$ for each $t$. That is, the mass on the shell balances the change in bulk mass between $\M^-$ and $\M^+$.

\begin{figure*}
    \centering
    \includegraphics[scale=0.7]{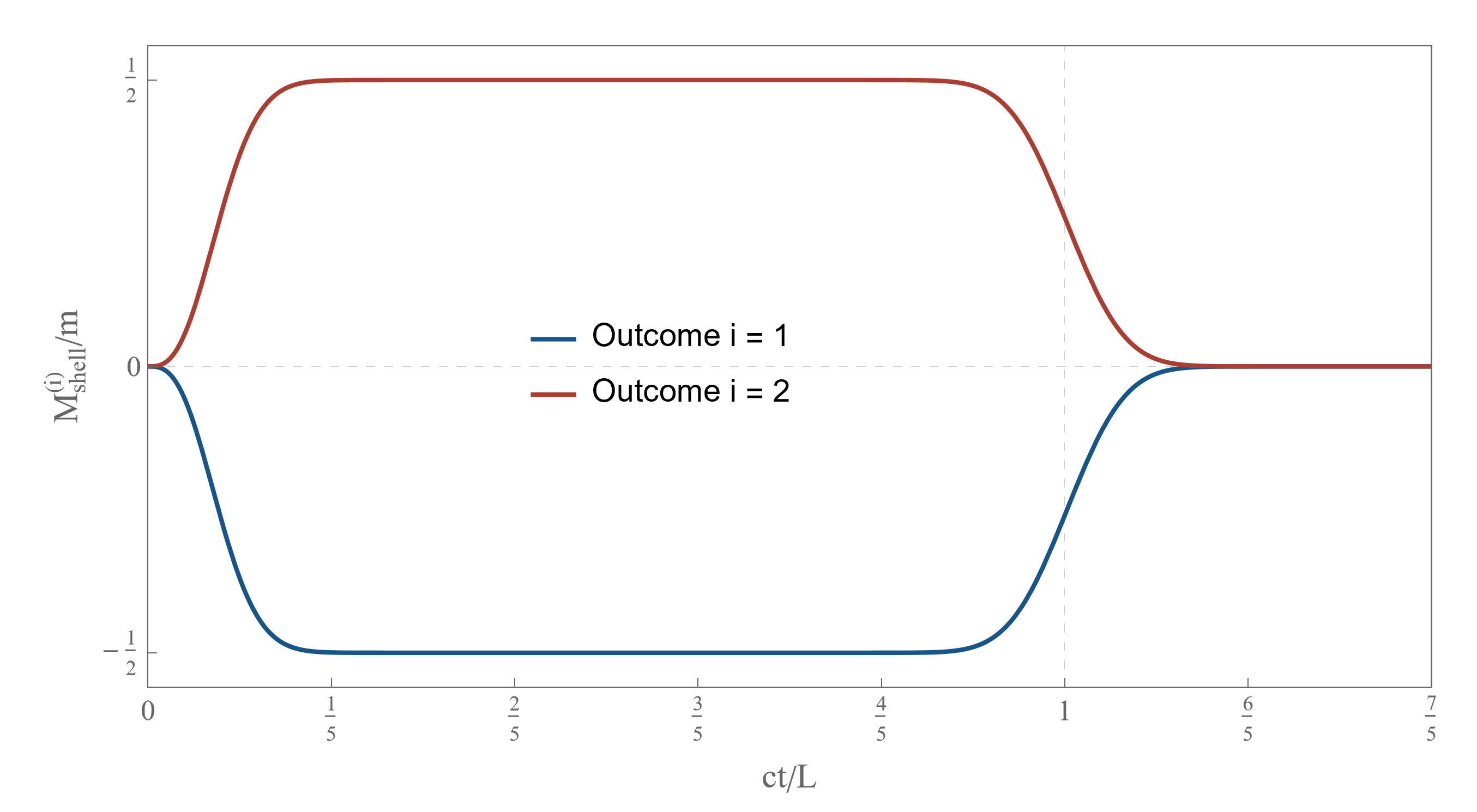}
    \caption{\justifying Gaussian shell-mass histories for equal branch weights $\abs{a_1}^2=\abs{a_2}^2 = \frac{1}{2}$, $c=1$, $L=1$ and equal wavepacket widths $\sigma_1=\sigma_2=\frac{L}{20}$. When $ct \approx L$, the contribution of the shell mass to the total mass starts to decrease: the other Gaussian is crossing the collapse hypersurface, compensating the mass loss. At late times, the shell mass is exponentially small.}
    \label{fig:shell mass history}
\end{figure*}

\begin{figure}
    \centering
    \begin{tikzpicture}[x=0.75pt,y=0.75pt,yscale=-1,xscale=1]

\draw  [fill={rgb, 255:red, 0; green, 0; blue, 0 }  ,fill opacity=1 ] (349.73,175.47) .. controls (349.73,174.1) and (350.84,173) .. (352.2,173) .. controls (353.56,173) and (354.67,174.1) .. (354.67,175.47) .. controls (354.67,176.83) and (353.56,177.93) .. (352.2,177.93) .. controls (350.84,177.93) and (349.73,176.83) .. (349.73,175.47) -- cycle ;
\draw    (205.33,4.4) -- (352.2,175.47) ;
\draw    (512.67,5.07) -- (352.2,175.47) ;
\draw    (306,161.73) -- (341.45,174.59) ;
\draw [shift={(343.33,175.27)}, rotate = 199.93] [color={rgb, 255:red, 0; green, 0; blue, 0 }  ][line width=0.75]    (10.93,-3.29) .. controls (6.95,-1.4) and (3.31,-0.3) .. (0,0) .. controls (3.31,0.3) and (6.95,1.4) .. (10.93,3.29)   ;
\draw    (315.33,209.73) .. controls (338.67,209.07) and (331.33,189.07) .. (344.67,189.73) .. controls (358,190.4) and (356.67,209.73) .. (375.33,210.4) ;
\draw    (398,211.73) .. controls (421.33,211.07) and (414,191.07) .. (427.33,191.73) .. controls (440.67,192.4) and (439.33,211.73) .. (458,212.4) ;
\draw    (311.33,219.07) .. controls (313.33,235.07) and (381.33,223.73) .. (390,223.73) .. controls (398.67,223.73) and (454.67,233.07) .. (456.67,220.4) ;
\draw    (390.67,251) -- (390.05,225.73) ;
\draw [shift={(390,223.73)}, rotate = 88.6] [color={rgb, 255:red, 0; green, 0; blue, 0 }  ][line width=0.75]    (10.93,-3.29) .. controls (6.95,-1.4) and (3.31,-0.3) .. (0,0) .. controls (3.31,0.3) and (6.95,1.4) .. (10.93,3.29)   ;
\draw    (426.67,186.33) -- (426.02,104.73) ;
\draw [shift={(426,101.73)}, rotate = 89.55] [fill={rgb, 255:red, 0; green, 0; blue, 0 }  ][line width=0.08]  [draw opacity=0] (8.93,-4.29) -- (0,0) -- (8.93,4.29) -- cycle    ;
\draw    (340,116.4) .. controls (348.67,116.4) and (344.67,99.73) .. (351.33,99.07) .. controls (358,98.4) and (354.67,115.73) .. (362.67,115.07) ;
\draw    (352,167) -- (350.77,132.73) ;
\draw [shift={(350.67,129.73)}, rotate = 87.95] [fill={rgb, 255:red, 0; green, 0; blue, 0 }  ][line width=0.08]  [draw opacity=0] (8.93,-4.29) -- (0,0) -- (8.93,4.29) -- cycle    ;

\draw (344.67,115.07) node [anchor=north west][inner sep=0.75pt]  [font=\scriptsize]  {$\psi_1$};
\draw (354.2,176) node [anchor=north west][inner sep=0.75pt]  [font=\scriptsize] [align=left] {$x$};
\draw (205.33,147.33) node [anchor=north west][inner sep=0.75pt]  [font=\scriptsize] [align=left] {Position measurement};
\draw (230.67,73.73) node [anchor=north west][inner sep=0.75pt]  [font=\scriptsize]  {$\Lambda ^{+}( x)$};
\draw (352.67,50.73) node [anchor=north west][inner sep=0.75pt]  [font=\scriptsize]  {$I^{+}( x)$};
\draw (219.33,188.73) node [anchor=north west][inner sep=0.75pt]  [font=\scriptsize]  {$\mathcal{M}\smallsetminus J^{+}(x)$};
\draw (355.33,249.07) node [anchor=north west][inner sep=0.75pt]  [font=\scriptsize]  {$\frac{1}{\sqrt{2}}\left(\psi_1 +\psi_2\right)$};
\draw (334.67,203.07) node [anchor=north west][inner sep=0.75pt]  [font=\scriptsize]  {$\psi_1$};
\draw (419.33,201.73) node [anchor=north west][inner sep=0.75pt]  [font=\scriptsize]  {$\psi_2$};
\draw (352.67,141.13) node [anchor=north west][inner sep=0.75pt]  [font=\tiny]  {$+\frac{m}{2}$};
\draw (412.67,78.47) node [anchor=north west][inner sep=0.75pt]  [font=\tiny]  {$-\frac{m}{2}$};
\draw (430.67,100.47) node [anchor=north west][inner sep=0.75pt]  [font=\scriptsize]  {$x_{2}$};
\draw (300.76,191.18) node [anchor=north west][inner sep=0.75pt]  [font=\scriptsize]  {$\frac{m}{2}$};
\draw (459.76,189.68) node [anchor=north west][inner sep=0.75pt]  [font=\scriptsize]  {$\frac{m}{2}$};
\draw (379.14,265.78) node [anchor=north west][inner sep=0.75pt]  [font=\scriptsize] [align=left] {Mass $m$};
\end{tikzpicture}
    \caption{\justifying Quantum measurement of a single particle of mass $m$ in a superposition $\frac{1}{\sqrt{2}}(\psi_1+\psi_2)$ of two sharply localised wavepackets $\psi_1$ and $\psi_2$ around $\vec{x}_1$ and $\vec{x}_2$ respectively in some foliation $\{\Sigma_t\}$. At $x \equiv (0,\vec{0})$, Alice performs a position measurement and finds outcome ``1''. In the neighbourhood of $x$ and within $\mathcal{M} \smallsetminus J^+(x)$, the mass density integrates to a mass of $m/2$. Shortly after (depending on the spread of the wavefunction), in the vicinity of $x$ in $I^+(x)$, the mass density now integrates to a mass of $m$. Loosely speaking, an additional mass $m/2$ ``appears'' in the vicinity of $x=(0,\vec{0})$ after $\Lambda^+(x)$. Conservation of mass density imposes that a negative mass density lies on $\Lambda^+(0,\vec{0})$ in the vicinity of the line segment between $(0,\vec{0})$ and $(t',\vec{x}_2)$ where $t'$ is the time at which $\psi_2$ crosses the future light cone of Alice's quantum measurement, at which point the mass of the $\psi_2$ wavepacket vanishes.}
    \label{fig:mass disappearance}
\end{figure}
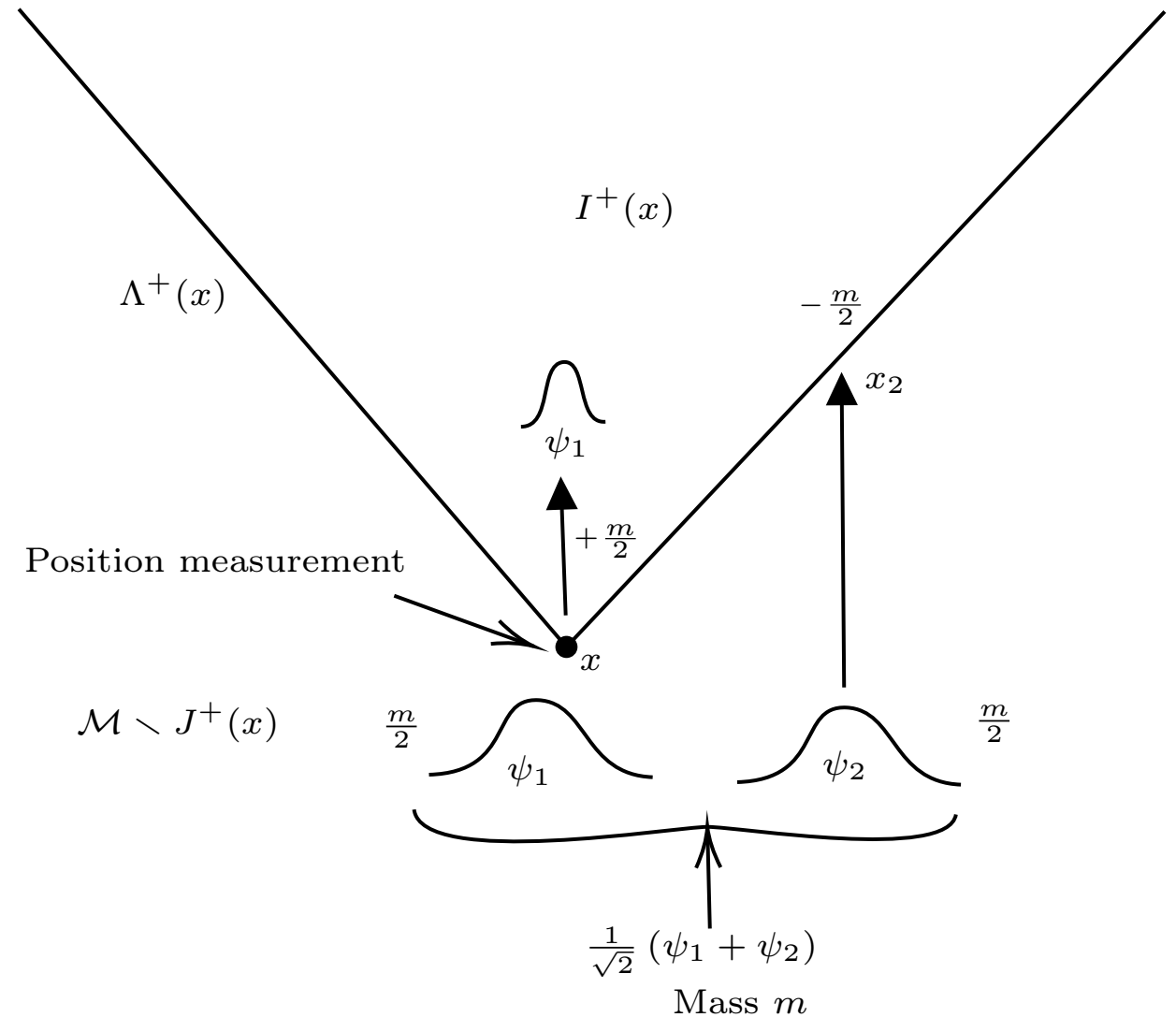

For example, if the wavepackets are given by \eqref{eqn:psi gaussian}, one verifies (see Appendix \ref{app:gaussian shell mass}) that
\begin{widetext}
    \begin{multline}
        \label{eqn:shell mass gaussian}
        M_{\mathrm{shell}}^{(i)}(t) = (-1)^i m \abs{a_{\bar{i}}}^2 \Bigg(\mathrm{erf}\left(\frac{ct}{\sqrt{2}\sigma_1}\right) - \sqrt{\frac{2}{\pi}} \frac{ct}{\sigma_1} \exp\left(\frac{-(ct)^2}{2\sigma_1^2}\right) -\frac{1}{2}\Bigg[\mathrm{erf}\left(\frac{ct+L}{\sqrt{2}\sigma_2}\right)+\mathrm{erf}\left(\frac{ct-L}{\sqrt{2}\sigma_2}\right)\Bigg] \\ -\frac{\sigma_2}{\sqrt{2\pi}L} \left[\exp\left(-\frac{(ct+L)^2}{2\sigma_2^2}\right)-\exp\left(-\frac{(ct-L)^2}{2\sigma_2^2}\right)\right]\Bigg) 
    \end{multline}
\end{widetext}
This mass history is displayed in Fig.~\eqref{fig:shell mass history}.
The shell starts continuously with zero mass. For example, a Taylor expansion at small $t$ gives
\begin{multline}
    M_\mathrm{shell}^{(1)}(t) = -\frac{4\pi}{3} m \abs{a_2}^2 (ct)^3 \left[\frac{1}{(2\pi \sigma_1^2)^{3/2}}-\frac{\exp\left(\frac{-L^2}{2\sigma_2^2}\right)}{(2\pi \sigma_2^2)^{3/2}}\right] \\ + \mathcal{O}((ct)^5)
\end{multline}
After $\psi_2$ has crossed, $M_\mathrm{shell}^{(1)}(t)$ and $M_\mathrm{shell}^{(2)}(t)$ tend to $0$, as also shown in Fig.~\ref{fig:shell mass history}. Closer to the wavepacket supports, the situation is interesting: mass is appearing or disappearing in places across the future light cone of the measurement event, though in a way consistent with conservation of total mass and energy, as is displayed in Fig. \ref{fig:mass disappearance}. 

Indeed, once both $ct \gg \sigma_1$ but $L-ct \gg \sigma_2$, the wavepacket $\psi_1$ has crossed $\Lambda^+(0)$ while $\psi_2$ has not. Consequently, as shown in Fig.~\ref{fig:shell mass history},
\begin{equation}
    M_\mathrm{shell}^{(1)}(t) \approx -m\abs{a_2}^2, \quad M_\mathrm{shell}^{(2)}(t) \approx +m\abs{a_1}^2.
\end{equation}
In some scenarios, there seems to be an interesting phenomenon: some \emph{negative mass} seems to be carried through the shell upon the quantum measurement when outcome ``1'' is obtained. This violates the null energy condition: for $N$ given as in Eqn.\eqref{eqn:N vector}, $S_{\mu\nu}^{(i)} N^\mu N^\nu = \mu^{(i)}$. But near the origin, $\Delta \varrho_m^{(1)} > 0$, so Eqn.~\eqref{eqn:shell stress energy qubit example} gives $\mu^{(i)} < 0$. 

In classical GR, shells with $\mu < 0$ are typically dismissed as unphysical as they require exotic matter. Here they arise from a quantum measurement, a process with no classical analogue. Whether the weaker averaged null energy conditions \cite{flanagan_does_1996,fewster_quantum_2005,graham_achronal_2007,balakrishnan_general_2019} are also violated in this model may be interesting to examine. Whether this should be regarded as a prediction of the theory (exotic null shells are produced by quantum measurements) or as evidence against the theory (violations of the null energy condition in these scenarios are unacceptable) is open to discussion.
It is not evident to us that it breaks other physical principles (for example, that it allows superluminal signalling or violates the second law of thermodynamics), nor that it implies the existence of wormholes (wormholes typically require violations of energy conditions \cite{kontou_wormhole_2024}, but the converse is not clear). Violations of the null energy condition also occur in QFT on fixed background spacetimes \cite{epstein_nonpositivity_1965}.  We thus see this as a feature of the model that deserves further investigation, but not an immediate reason to dismiss it.

\subsubsection{The future metric and gravitational field}

Let us now derive the future metric in this example. The stationary continuation of the past field is
\begin{equation}
    \underline{\overline{h}}^-_{\mu\nu} = -\frac{4 \Phi^-}{c^2} \delta^0_{\mu} \delta^0_\nu,
\end{equation}
where $\Phi^-$ is given in \eqref{eqn:retarded past gravitational field}. The previous construction gives
\begin{align}
    \overline{h}^{+,(i)} &= \overline{\underline{h
    }}^- + \overline{\mathfrak{H}}^{(i)} \nonumber \\ &+ 2 \kappa G_\mathrm{ret} * \left[\Theta(u) \Delta T^{(i)} - \kappa^{-1} \Theta(u) G^{(1)}[\mathfrak{H}^{(i)}]\right] \\ 
    &= \overline{h}^{\mathrm{ret},(i)} + \overline{\mathfrak{H}}^{(i)} - 2 G_\mathrm{ret} * \left[\Theta(u) G^{(1)}[\mathfrak{H}^{(i)}]\right],
        \label{eqn:h plus theta u G}
\end{align}
where $\overline{h}^{\mathrm{ret},(i)} = \underline{\overline{h}}^- + 2 \kappa G_{\mathrm{ret}} * \left[\Theta(u) \Delta T^{(i)}\right]$. At a fixed point $(ct,r) \in I^+(0)$, $ct-r>0$ and $ct+r>0$ so
\begin{equation}
    \frac{ct-r}{ct+r} > \epsilon \Longleftrightarrow ct > \frac{1+\epsilon}{1-\epsilon} r.
\end{equation}
When this inequality holds, $\chi_\epsilon(u/v) = 0$ on a neighbourhood of the point and thus $\mathfrak{H} = 0$ there, thus
\begin{equation}
    \overline{h}^{+,(i)}_{\mu\nu} = \overline{h}^{\mathrm{ret},(i)}_{\mu\nu} + \delta_{\mu\nu}^{(i)}.
\end{equation}
Since $2\kappa/(4\pi)=4G/c^4$ and
$\Delta T_{\mu\nu}^{(i)}=c^2\Delta\varrho_m^{(i)}\delta^0_\mu\delta^0_\nu$,
the two terms are explicitly
\begin{align}
 &\overline h^{\mathrm{ret},(i)}_{\mu\nu}(t,\vec x)
 =-\frac{4\Phi^-(\vec x)}{c^2}\delta^0_\mu\delta^0_\nu
   \nonumber \\ &+\frac{4G}{c^2}\delta^0_\mu\delta^0_\nu
       \int_{\mathbb R^3}
       \frac{\Theta(ct-|\vec x-\vec y|-|\vec y|)
                      \Delta\varrho_m^{(i)}(\vec y)}{|\vec x-\vec y|}\,d^3\vec y,
 \label{full:bulk-retarded}\\
 &\delta_{\mu\nu}^{(i)}(t,\vec x) \nonumber \\
 &=-\frac1{2\pi}\int_{\mathbb R^3}
       \frac{\Theta(ct-|\vec x-\vec y|-|\vec y|)}{|\vec x-\vec y|} \nonumber \\ & \qquad \qquad \qquad \cdot 
       G^{(1)}_{\mu\nu}[\mathfrak H^{(i)}]
       \left(t-\frac{|\vec x-\vec y|}{c},\vec y\right)d^3\vec y.
 \label{full:cutoff-correction}
\end{align}

Now note that 
\begin{equation}
    G^{(1)}[\Theta(u) \mathfrak{H}^{(i)}] = \Theta(u) G^{(1)}[\mathfrak{H}^{(i)}] + \kappa T^{\mathrm{sing},(i)},
\end{equation}
so solving this identity for $\Theta(u) G^{(1)}[\mathfrak{H}^{(i)}]$ and substituting into Eqn.~\eqref{eqn:h plus theta u G} yields
\begin{multline}
 \overline h^{+,(i)} =\overline h^{\mathrm{ret},(i)}+2\kappa G_{\mathrm{ret}}*T^{\mathrm{sing},(i)}
    \\ +\overline{\mathfrak H}^{(i)}
    -2G_{\mathrm{ret}}*G^{(1)}[\Theta(u)\mathfrak H^{(i)}].
 \label{full:split-shell-gauge}
\end{multline}
We have
\begin{align}
 2G^{(1)}_{\mu\nu}[\Theta(u)\mathfrak H^{(i)}]
 ={}&-\Box[\Theta(u)\overline{\mathfrak H}^{(i)}_{\mu\nu}]
 +\partial_\mu\partial^\rho[\Theta(u)\overline{\mathfrak H}^{(i)}_{\rho\nu}]
 \nonumber\\
 &+\partial_\nu\partial^\rho[\Theta(u)\overline{\mathfrak H}^{(i)}_{\rho\mu}]
 -\eta_{\mu\nu}\partial^\rho\partial^\sigma
                [\Theta(u)\overline{\mathfrak H}^{(i)}_{\rho\sigma}].
 \label{full:Einstein-identity}
\end{align}
But
\begin{equation}
     G_\mathrm{ret} * \partial_\alpha \partial_\beta[\Theta(u) \overline{\mathfrak{H}}^{(i)}_{\mu\nu}] = (\partial_\alpha \partial_\beta G_\mathrm{ret}) * [\Theta(u) \overline{\mathfrak{H}}_{\mu\nu}^{(i)}]
\end{equation}
so contracting with $\eta^{\alpha \beta}$ gives
\begin{multline}
    G_{\mathrm{ret}} * \Box [\Theta(u) \overline{\mathfrak{H}}^{(i)}_{\mu\nu}] = (\Box G_{\mathrm{ret}}) * [\Theta(u) \overline{\mathfrak{H}}^{(i)}_{\mu\nu}] \\ = - \delta^{(4)} * [\Theta(u) \overline{\mathfrak{H}}^{(i)}_{\mu\nu}] = - \Theta(u) \overline{\mathfrak{H}}^{(i)}_{\mu\nu}.
\end{multline}
Note that Eqn.~\eqref{full:split-shell-gauge} is being evaluated in $I^+(0)$, where
$\Theta(u)\overline{\mathfrak H}^{(i)}=\overline{\mathfrak H}^{(i)}$, so this term cancels the $\overline{\mathfrak{H}}^{(i)}$ term in Eqn.~\eqref{full:split-shell-gauge}. Thus,
\begin{align}
 \overline h^{+,(i)}_{\mu\nu}
 &-\left(\overline h^{\mathrm{ret},(i)}_{\mu\nu}
                   +2\kappa G_{\mathrm{ret}}*T^{\mathrm{sing},(i)}_{\mu\nu}\right)
 \nonumber\\
 ={}&-\partial_\mu G_{\mathrm{ret}}*
             \partial^\rho[\Theta(u)\overline{\mathfrak H}^{(i)}_{\rho\nu}]
     -\partial_\nu G_{\mathrm{ret}}*
             \partial^\rho[\Theta(u)\overline{\mathfrak H}^{(i)}_{\rho\mu}]
 \nonumber\\
 &\quad+\eta_{\mu\nu}\partial^\sigma G_{\mathrm{ret}}*
             \partial^\rho[\Theta(u)\overline{\mathfrak H}^{(i)}_{\rho\sigma}].
 \label{full:trace-gauge-difference}
\end{align}
Contracting this difference with $\eta^{\mu\nu}$ gives $2 \partial^\sigma G_\mathrm{ret} * \partial^\rho[\Theta(u) \overline{\mathfrak{H}}^{(i)}_{\rho \sigma}]$ (the first two terms on the right-hand side contribute $-2$ copies and the last contributes $4$ copies). Trace-reversal subtracts $1/2$ of this trace times $\eta_{\mu\nu}$ which cancels the last line. Writing $\xi_{\mu}^{(i)} := -G_\mathrm{ret} * \partial^\rho[\Theta(u) \overline{\mathfrak{H}}^{(i)}_{\rho\mu}]$, we thus have that
\begin{multline}
    h^{+,(i)}_{\mu\nu} - \left( h^{\mathrm{ret},(i)}_{\mu\nu}
                   +2\kappa G_{\mathrm{ret}}*\overline{T}^{\mathrm{sing},(i)}_{\mu\nu}\right) \\ = \partial_\mu \xi_\nu^{(i)} + \partial_\nu \xi_\mu^{(i)},
\end{multline}
i.e. they are equal up to a linearised diffeomorphism, so we can switch to a representative such that
\begin{equation}
    \label{eqn:h+ as hret and Tsing}
    \overline{h}^{+,(i)} = \overline{h}^{\mathrm{ret},(i)} + 2 \kappa G_{\mathrm{ret}} * T^{\mathrm{sing},(i)}.
\end{equation}
Since the total source is conserved
\begin{equation}
    \partial^\mu(\underline{T}^-_{\mu\nu} + \Theta(u) \Delta T_{\mu\nu}^{(i)} + T^{\mathrm{sing},(i)}_{\mu\nu})=0,
\end{equation}
its divergence is
\begin{equation}
    \partial^\mu \overline{h}^{+,(i)}_{\mu\nu} = 2 \kappa G_{\mathrm{ret}} * \partial^\mu(\underline{T}^-_{\mu\nu} + \Theta(u) \Delta T_{\mu\nu}^{(i)} + T^{\mathrm{sing},(i)}_{\mu\nu})=0,
\end{equation}
i.e. this \emph{local} representative is in Lorenz gauge. Another way to express this is that
\begin{equation}
    \lim_{\epsilon \downarrow 0} \delta_{\mu\nu}^{(i)} = 2 \kappa G_{\mathrm{ret}} * T^{\mathrm{sing},(i)},
\end{equation}
i.e.
\begin{equation}
    \lim_{\epsilon \downarrow 0} \overline{h}^{+,(i)} = \overline{h}^{\mathrm{ret},(i)} + 2 \kappa G_{\mathrm{ret}} * T^{\mathrm{sing},(i)}.
\end{equation}

Now note that since $\varrho_m^- + \Delta \varrho_m^{(i)} = m \abs{\psi_i}^2$, we have from Eqn.~\eqref{full:bulk-retarded} that
\begin{equation}
    \overline{h}^\mathrm{ret}_{\mu\nu} = -\frac{4\Phi^{+,(i)}}{c^2} \delta^0_{\mu} \delta^0_\nu
\end{equation}
where
\begin{equation}
    \Phi^{+,(i)}(t,\vec{x}) = - Gm\iiint_{\mathbb{R}^3} \frac{\abs{\psi_i(t,\vec{y})}^2}{\abs{\vec{x}-\vec{y}}} d^3 \vec{y}.
\end{equation}
We show in Appendix \ref{app:future metric qubit} that for sufficiently large $ct$,\footnote{We require $(ct,\vec{x})$ to be such that $\operatorname{supp } \psi_i(ct,\cdot) \mathrel{\underset{\approx}{\subset}} \{\vec{y} \mid \abs{\vec{y}} < \frac{ct-r}{2}\}$ where $r=\abs{\vec{x}}$, $i=1,2$. For sharply peaked Gaussians with $\operatorname{supp} \psi_i(t,\cdot) = \mathbb{R}^3$, this is still a good approximation: the tails are exponentially small.} the additional term sourced by $T^\mathrm{sing}$ is subleading in this regime: at fixed $\vec{x}$,
\begin{align}
    G_\mathrm{ret} * T_{00}^{\mathrm{sing},(i)} &= \mathcal{O}((ct)^{-3}), \\
    G_\mathrm{ret} * T_{0j}^{\mathrm{sing},(i)} &= \mathcal{O}((ct)^{-2}), \\
    G_\mathrm{ret} * T_{jk}^{\mathrm{sing},(i)} &= \mathcal{O}((ct)^{-1}),
\end{align}
so that to leading order
\begin{align}
    h^{+,(i)}_{00}(t,\vec{x}) &\approx -\frac{2\Phi^{+,(i)}}{c^2}, \\
    h^{+,(i)}_{0j}(t,\vec{x}) &\approx 0, \\
    h^{+,(i)}_{jk}(t,\vec{x}) &\approx -\frac{2 \Phi^{+,(i)} }{c^2} \delta_{jk}
\end{align}
Here, $g_{\mu\nu}^{+,(i)} = \eta_{\mu\nu} + h_{\mu\nu}^{+,(i)}$ at the retained order, and so locally in $I^+(0)$ and for sufficiently large $ct$ we have that
\begin{multline}
    ds_+^2 \approx -\left(1+\frac{2\Phi^{+,(i)}}{c^2}\right) c^2 dt^2 + \left(1-\frac{2\Phi^{+,(i)}}{c^2}\right)d\vec{x}^2 \\ + \mathcal{O}(h^2). 
\end{multline}
In the Gaussian example, 
\begin{equation}
    \Phi^{+,(i)}(t,\vec{x}) \approx - G m \frac{\mathrm{erf}\left(\frac{\abs{\vec{x}-\vec{x}_i}}{\sqrt{2}\sigma_i}\right)}{\abs{\vec{x}-\vec{x}_i}}
\end{equation}
Thus,
\begin{align}
    \overline{h}^{+,(i)}_{00}(t,\vec{x}) &\approx \frac{4Gm}{c^2} \frac{\mathrm{erf}\left(\frac{\abs{\vec{x}-\vec{x}_i}}{\sqrt{2}\sigma_i}\right)}{\abs{\vec{x}-\vec{x}_i}}, \\
    \overline{h}^{+,(i)}_{0j}(t,\vec{x}) &\approx \overline{h}^{+,(i)}_{jk}(t,\vec{x}) \approx 0,
\end{align}
and so
\begin{align}
    h^{+,(i)}_{00}(t,\vec{x}) &\approx \frac{2Gm}{c^2} \frac{\mathrm{erf}\left(\frac{\abs{\vec{x}-\vec{x}_i}}{\sqrt{2}\sigma_i}\right)}{\abs{\vec{x}-\vec{x}_i}}, \\
    h^{+,(i)}_{0j}(t,\vec{x}) &\approx 0, \\
    h^{+,(i)}_{jk}(t,\vec{x}) &\approx \frac{2Gm \delta_{jk}}{c^2} \frac{\mathrm{erf}\left(\frac{\abs{\vec{x}-\vec{x}_i}}{\sqrt{2}\sigma_i}\right)}{\abs{\vec{x}-\vec{x}_i}}
\end{align}
 Equivalently, we have that 
\begin{align}
    ds_+^2 \approx &-\left(1-\frac{2Gm}{c^2} \frac{\mathrm{erf}\left(\frac{\abs{\vec{x}-\vec{x}_i}}{\sqrt{2}\sigma_i}\right)}{\abs{\vec{x}-\vec{x}_i}}\right) c^2 dt^2 \nonumber \\
    &+ \left(1+\frac{2Gm}{c^2} \frac{\mathrm{erf}\left(\frac{\abs{\vec{x}-\vec{x}_i}}{\sqrt{2}\sigma_i}\right)}{\abs{\vec{x}-\vec{x}_i}}\right) d\vec{x}^2.
\end{align}    

\subsection{Entangled qudits and collapses over spacelike separations}

\label{subsec:mmt Bell state}

Although a single quantum state collapse (or a series of timelike-separated collapses) is, as we have seen, not contradictory with the distributional semiclassical Einstein field equations (at least in the weak-field limit), the case of quantum state collapses occurring over spacelike separations requires separate scrutiny due to crossing shells. 

Let us consider a simple example of two agents, Alice (A) and Bob (B), who have agreed in the distant past to each carry out a projective quantum measurement on their respective qudit of mass $m_A$ and $m_B$, once they are spacelike separated at a spatial distance $d$ in some foliation $\Sigma_t$. Again, we assume that the associated collapses occur at points. Their qudits are initially (at time $t=\tau_i$) in the state
\begin{multline}
    \label{eqn:qudit}
    \Psi(t,\vec{x}_A,\vec{x}_B) = \sum_{i,j=1}^n a_{ij} \psi_i^{(A)}(t,\vec{x}_A) \psi_j^{(B)}(t,\vec{x}_B) 
\end{multline}
where $\sum_{i,j=1}^n \abs{a_{ij}}^2 = 1$, and $\psi^A$ and $\psi^B$ can each be in $n$ sharply localised wavepackets $\psi^A_i$ and $\psi^B_j$, respectively, around $\vec{x}_{i,A}$ and $\vec{x}_{j,B}$, respectively, $i,j=1,\cdots,n$, in some foliation $\Sigma_t$. Each wavepacket is smooth, normalised, and rapidly decreasing, and their motion and spreading are neglected over the light-travel times considered, with the same assumptions as in Sec.~\ref{subsec:mmt qubit in superposition}. Here, $\abs{\vec{x}_{i,A}-\vec{x}_{j,A}} \ll d$ and $ \abs{\vec{x}_{i,B}-\vec{x}_{j,B}} \ll d$ for all $i,j=1,\cdots,n$.

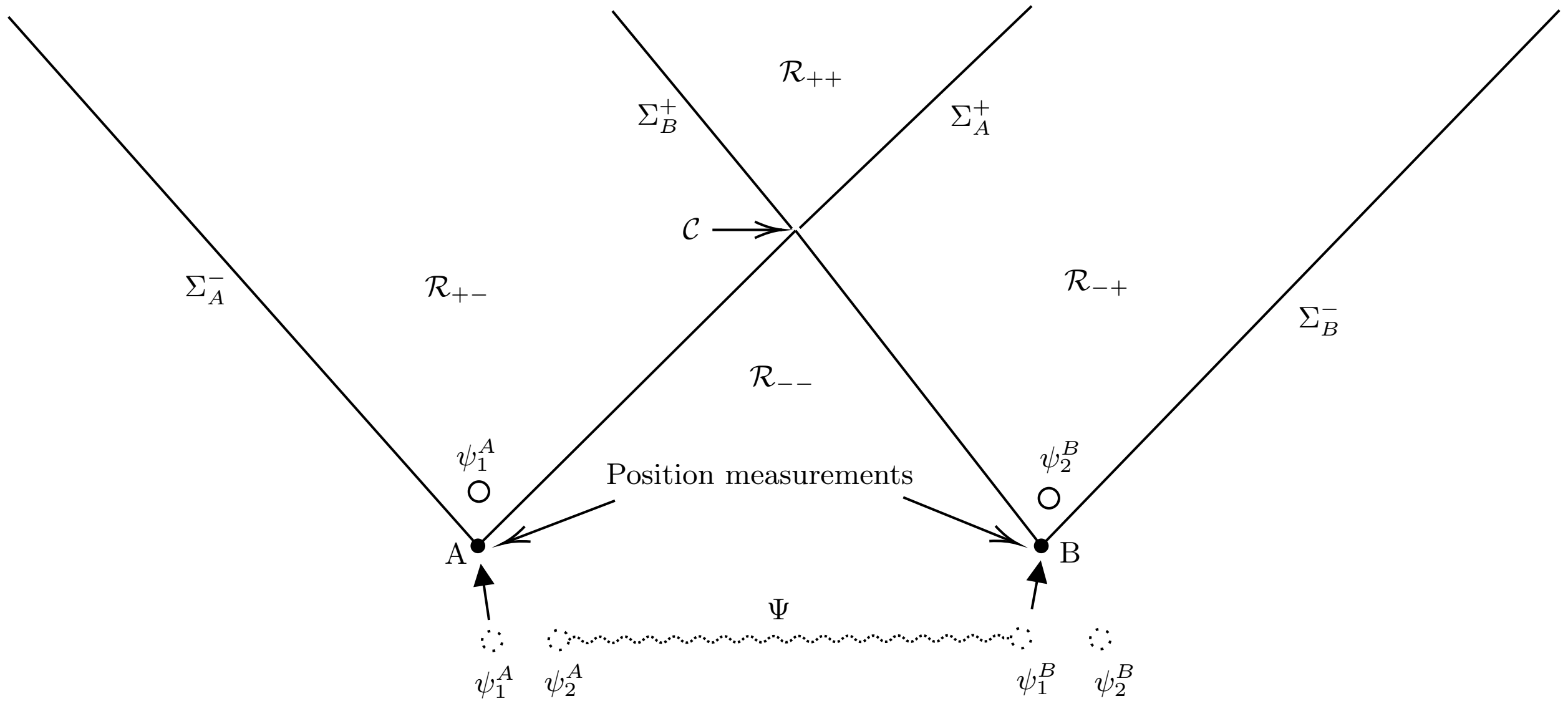
\begin{figure*}
    \centering
    \begin{tikzpicture}[x=0.75pt,y=0.75pt,yscale=-1,xscale=1]

\draw  [fill={rgb, 255:red, 0; green, 0; blue, 0 }  ,fill opacity=1 ] (233.07,230.47) .. controls (233.07,229.1) and (234.17,228) .. (235.53,228) .. controls (236.9,228) and (238,229.1) .. (238,230.47) .. controls (238,231.83) and (236.9,232.93) .. (235.53,232.93) .. controls (234.17,232.93) and (233.07,231.83) .. (233.07,230.47) -- cycle ;
\draw  [fill={rgb, 255:red, 0; green, 0; blue, 0 }  ,fill opacity=1 ] (459.73,230.47) .. controls (459.73,229.1) and (460.84,228) .. (462.2,228) .. controls (463.56,228) and (464.67,229.1) .. (464.67,230.47) .. controls (464.67,231.83) and (463.56,232.93) .. (462.2,232.93) .. controls (460.84,232.93) and (459.73,231.83) .. (459.73,230.47) -- cycle ;
\draw    (46.67,17.6) -- (235.53,230.47) ;
\draw    (363.33,103.6) -- (462.2,230.47) ;
\draw    (363.33,103.6) -- (235.53,230.47) ;
\draw    (670.67,14.93) -- (462.2,230.47) ;
\draw    (458.33,13.27) -- (365,102.6) ;
\draw  [dash pattern={on 0.75pt off 0.75pt}]  (272.11,268.35) .. controls (273.77,266.68) and (275.44,266.67) .. (277.11,268.33) .. controls (278.78,269.98) and (280.45,269.97) .. (282.11,268.3) .. controls (283.77,266.63) and (285.44,266.62) .. (287.11,268.28) .. controls (288.78,269.94) and (290.45,269.93) .. (292.11,268.26) .. controls (293.77,266.59) and (295.44,266.58) .. (297.11,268.24) .. controls (298.78,269.9) and (300.45,269.89) .. (302.11,268.22) .. controls (303.77,266.55) and (305.44,266.54) .. (307.11,268.2) .. controls (308.78,269.86) and (310.45,269.85) .. (312.11,268.18) .. controls (313.77,266.51) and (315.44,266.5) .. (317.11,268.16) .. controls (318.78,269.82) and (320.45,269.81) .. (322.11,268.14) .. controls (323.77,266.47) and (325.44,266.46) .. (327.11,268.12) .. controls (328.78,269.77) and (330.45,269.76) .. (332.11,268.09) .. controls (333.77,266.42) and (335.44,266.41) .. (337.11,268.07) .. controls (338.78,269.73) and (340.45,269.72) .. (342.11,268.05) .. controls (343.77,266.38) and (345.44,266.37) .. (347.11,268.03) .. controls (348.78,269.69) and (350.45,269.68) .. (352.11,268.01) .. controls (353.77,266.34) and (355.44,266.33) .. (357.11,267.99) .. controls (358.78,269.65) and (360.45,269.64) .. (362.11,267.97) .. controls (363.77,266.3) and (365.44,266.29) .. (367.11,267.95) .. controls (368.78,269.61) and (370.45,269.6) .. (372.11,267.93) .. controls (373.77,266.26) and (375.44,266.25) .. (377.11,267.91) .. controls (378.78,269.56) and (380.45,269.55) .. (382.11,267.88) .. controls (383.77,266.21) and (385.44,266.2) .. (387.11,267.86) .. controls (388.78,269.52) and (390.45,269.51) .. (392.11,267.84) .. controls (393.77,266.17) and (395.44,266.16) .. (397.11,267.82) .. controls (398.78,269.48) and (400.45,269.47) .. (402.11,267.8) .. controls (403.77,266.13) and (405.44,266.12) .. (407.11,267.78) .. controls (408.78,269.44) and (410.45,269.43) .. (412.11,267.76) .. controls (413.77,266.09) and (415.44,266.08) .. (417.11,267.74) .. controls (418.78,269.4) and (420.45,269.39) .. (422.11,267.72) .. controls (423.77,266.05) and (425.44,266.04) .. (427.11,267.7) .. controls (428.78,269.36) and (430.45,269.35) .. (432.11,267.68) .. controls (433.77,266.01) and (435.44,266) .. (437.11,267.65) .. controls (438.78,269.31) and (440.45,269.3) .. (442.11,267.63) .. controls (443.77,265.96) and (445.44,265.95) .. (447.11,267.61) -- (450,267.6) -- (450,267.6) ;
\draw  [dash pattern={on 0.84pt off 2.51pt}] (237.23,268.67) .. controls (237.23,266.43) and (239.05,264.62) .. (241.28,264.62) .. controls (243.52,264.62) and (245.33,266.43) .. (245.33,268.67) .. controls (245.33,270.9) and (243.52,272.72) .. (241.28,272.72) .. controls (239.05,272.72) and (237.23,270.9) .. (237.23,268.67) -- cycle ;
\draw    (240,260.27) -- (237.1,240.57) ;
\draw [shift={(236.67,237.6)}, rotate = 81.63] [fill={rgb, 255:red, 0; green, 0; blue, 0 }  ][line width=0.08]  [draw opacity=0] (8.93,-4.29) -- (0,0) -- (8.93,4.29) -- cycle    ;
\draw    (458.4,255.95) -- (461.46,239.22) ;
\draw [shift={(462,236.27)}, rotate = 100.37] [fill={rgb, 255:red, 0; green, 0; blue, 0 }  ][line width=0.08]  [draw opacity=0] (8.93,-4.29) -- (0,0) -- (8.93,4.29) -- cycle    ;
\draw    (290.67,212.27) -- (247.2,228.89) ;
\draw [shift={(245.33,229.6)}, rotate = 339.08] [color={rgb, 255:red, 0; green, 0; blue, 0 }  ][line width=0.75]    (10.93,-3.29) .. controls (6.95,-1.4) and (3.31,-0.3) .. (0,0) .. controls (3.31,0.3) and (6.95,1.4) .. (10.93,3.29)   ;
\draw    (406.67,210.93) -- (450.81,228.85) ;
\draw [shift={(452.67,229.6)}, rotate = 202.09] [color={rgb, 255:red, 0; green, 0; blue, 0 }  ][line width=0.75]    (10.93,-3.29) .. controls (6.95,-1.4) and (3.31,-0.3) .. (0,0) .. controls (3.31,0.3) and (6.95,1.4) .. (10.93,3.29)   ;
\draw    (289.67,15.27) -- (361.87,102.8) ;
\draw  [dash pattern={on 0.84pt off 2.51pt}] (264.01,268.35) .. controls (264.01,266.11) and (265.82,264.3) .. (268.06,264.3) .. controls (270.29,264.3) and (272.11,266.11) .. (272.11,268.35) .. controls (272.11,270.58) and (270.29,272.4) .. (268.06,272.4) .. controls (265.82,272.4) and (264.01,270.58) .. (264.01,268.35) -- cycle ;
\draw  [dash pattern={on 0.84pt off 2.51pt}] (450,267.6) .. controls (450,265.36) and (451.81,263.55) .. (454.05,263.55) .. controls (456.29,263.55) and (458.1,265.36) .. (458.1,267.6) .. controls (458.1,269.84) and (456.29,271.65) .. (454.05,271.65) .. controls (451.81,271.65) and (450,269.84) .. (450,267.6) -- cycle ;
\draw  [dash pattern={on 0.84pt off 2.51pt}] (481.9,267.95) .. controls (481.9,265.71) and (483.71,263.9) .. (485.95,263.9) .. controls (488.19,263.9) and (490,265.71) .. (490,267.95) .. controls (490,270.19) and (488.19,272) .. (485.95,272) .. controls (483.71,272) and (481.9,270.19) .. (481.9,267.95) -- cycle ;
\draw    (329.89,103.31) -- (355.89,103.31) ;
\draw [shift={(357.89,103.31)}, rotate = 180] [color={rgb, 255:red, 0; green, 0; blue, 0 }  ][line width=0.75]    (10.93,-3.29) .. controls (6.95,-1.4) and (3.31,-0.3) .. (0,0) .. controls (3.31,0.3) and (6.95,1.4) .. (10.93,3.29)   ;
\draw   (231.9,208.62) .. controls (231.9,206.38) and (233.71,204.57) .. (235.95,204.57) .. controls (238.19,204.57) and (240,206.38) .. (240,208.62) .. controls (240,210.85) and (238.19,212.67) .. (235.95,212.67) .. controls (233.71,212.67) and (231.9,210.85) .. (231.9,208.62) -- cycle ;
\draw   (461.23,211.28) .. controls (461.23,209.05) and (463.05,207.23) .. (465.28,207.23) .. controls (467.52,207.23) and (469.33,209.05) .. (469.33,211.28) .. controls (469.33,213.52) and (467.52,215.33) .. (465.28,215.33) .. controls (463.05,215.33) and (461.23,213.52) .. (461.23,211.28) -- cycle ;

\draw (350.33,250.4) node [anchor=north west][inner sep=0.75pt]  [font=\small]  {$\Psi$};
\draw (225.33,186.07) node [anchor=north west][inner sep=0.75pt]  [font=\small]  {$\psi _{1}^{A}$};
\draw (220.67,228) node [anchor=north west][inner sep=0.75pt]  [font=\small] [align=left] {A};
\draw (468,227.33) node [anchor=north west][inner sep=0.75pt]  [font=\small] [align=left] {B};
\draw (460,186.73) node [anchor=north west][inner sep=0.75pt]  [font=\small]  {$\psi _{2}^{B}$};
\draw (285.67,195.67) node [anchor=north west][inner sep=0.75pt]  [font=\small] [align=left] {Position measurements};
\draw (116,118.07) node [anchor=north west][inner sep=0.75pt]  [font=\small]  {$\Sigma _{A}^{-}$};
\draw (564,130.73) node [anchor=north west][inner sep=0.75pt]  [font=\small]  {$\Sigma _{B}^{-}$};
\draw (343.33,156.4) node [anchor=north west][inner sep=0.75pt]  [font=\small]  {$\mathcal{R}_{--}$};
\draw (212.67,120.4) node [anchor=north west][inner sep=0.75pt]  [font=\small]  {$\mathcal{R}_{+-}$};
\draw (470,117.73) node [anchor=north west][inner sep=0.75pt]  [font=\small]  {$\mathcal{R}_{-+}$};
\draw (355.33,33.73) node [anchor=north west][inner sep=0.75pt]  [font=\small]  {$\mathcal{R}_{++}$};
\draw (316.48,97.2) node [anchor=north west][inner sep=0.75pt]  [font=\small]  {$\mathcal{C}$};
\draw (232.67,277.4) node [anchor=north west][inner sep=0.75pt]  [font=\small]  {$\psi _{1}^{A}$};
\draw (260.67,276.73) node [anchor=north west][inner sep=0.75pt]  [font=\small]  {$\psi _{2}^{A}$};
\draw (482,276.73) node [anchor=north west][inner sep=0.75pt]  [font=\small]  {$\psi _{2}^{B}$};
\draw (450.67,276.07) node [anchor=north west][inner sep=0.75pt]  [font=\small]  {$\psi _{1}^{B}$};
\draw (298,49.07) node [anchor=north west][inner sep=0.75pt]  [font=\small]  {$\Sigma _{B}^{+}$};
\draw (424,49.73) node [anchor=north west][inner sep=0.75pt]  [font=\small]  {$\Sigma _{A}^{+}$};

\end{tikzpicture}
    \caption{\justifying Quantum measurements of two particles of respective mass $m_1$ and $m_2$ in an entangled qubit ($n=2$) state $\Psi = \sum_{i,j=1}^2 a_{ij} \psi_i^A \psi_j^B$, where $\psi^A$ and $\psi^B$ can each be in two sharply localised wavepackets $\psi^A_1$ and $\psi^A_2$, and $\psi^B_1$ and $\psi_2^B$, respectively, around $\vec{x}_{1,A}$ and $\vec{x}_{2,A}$, and $\vec{x}_{1,B}$ and $\vec{x}_{2,B}$, respectively, in some foliation $\Sigma_t$. At $x_A \approx (0,\vec{x}_{1,A})$, Alice performs a position measurement and finds outcome ``1'', while at $x_B \approx (0,\vec{x}_{1,B}) \perp x_a$, Bob performs a position measurement and finds outcome ``2''. The local stress-energies jump accordingly across $\Lambda^+(x_A)$ and $\Lambda^+(x_B)$, respectively. At $\mathcal{C} = \Lambda^+(x_A) \cap \Lambda^+(x_B)$, both stress-energies contribute independently.}
    \label{fig:spacelike measurements}
\end{figure*}

Each carry out a position measurement, with collapse events at spacelike separated spacetime points $x_A$ and $x_B$; for simplicity, choose a foliation in which $t_A = t_B = 0$ (this is without loss of generality). This is shown in Fig. \ref{fig:spacelike measurements}. Let us examine how spacetime backreacts from this scenario.

Write 
\begin{align}
    p_A^{(i)} &= \sum_{j=1}^n \abs{a_{ij}}^2,  \\
    p_B^{(j)} &= \sum_{i=1}^n \abs{a_{ij}}^2 .
\end{align}
For example, if $n=2$, $p_A^{(1)} = \abs{a_{11}}^2 + \abs{a_{12}}^2$, $p_A^{(2)} = \abs{a_{21}}^2 + \abs{a_{22}}^2$, $p_B^{(1)} = \abs{a_{11}}^2 + \abs{a_{21}}^2$ and $p_B^{(2)} = \abs{a_{12}}^2 + \abs{a_{22}}^2$.

Set
\begin{equation}
    u_A = ct - \abs{\vec{x}-\vec{x}_A}, \qquad u_B = ct - \abs{\vec{x}-\vec{x}_B}.
\end{equation}
The four spacetime regions and four null-shell segments of interest are
\begin{align}
    \mathcal{R}_{--} &= \{u_A < 0, u_B < 0\} =  \mathcal{M} \smallsetminus (J^+(x_A) \cup J^+(x_B)) \, , \\    
    \mathcal{R}_{++} &= \{u_A > 0, u_B > 0\} = I^+(x_A) \cap I^+(x_B) \, , \\
    \mathcal{R}_{-+} &= \{u_A < 0, u_B > 0\} = I^+(x_B) \smallsetminus J^+(x_A) \, , \\
    \mathcal{R}_{+-} &= \{u_A > 0, u_B < 0\} = I^+(x_A) \smallsetminus J^+(x_B) \, , \\
    \Sigma_A^- &= \{u_A = 0, u_B < 0\} = \Lambda^+(x_A) \smallsetminus J^+(x_B) \, , \\
    \Sigma_B^- &= \{u_B = 0, u_A < 0\} = \Lambda^+(x_B) \smallsetminus J^+(x_A) \, , \\
    \Sigma_A^+ &= \{u_A = 0,u_B > 0\} = \Lambda^+(x_A) \cap I^+(x_B) \, , \\
    \Sigma_B^+ &= \{u_B = 0,u_A > 0\} = \Lambda^+(x_B) \cap I^+(x_A) \, , \\
    \mathcal{C} &= \{u_A = u_B = 0\} = \Lambda^+(x_A) \cap \Lambda^+(x_B) \, .
\end{align}
The local quantum state prescription conditions on the outcomes whose measurement events lie in the causal past of the observation point.
Suppose Alice obtains outcome ``$\alpha$'' and Bob obtains outcome ``$\beta$''.  Applying Alice's
projector to Eqn.~\eqref{eqn:qudit} gives
\begin{equation}
 \Psi\longmapsto
 \psi_\alpha^{(A)}
       \frac{\sum_{j=1}^{n}a_{\alpha j}\psi_j^{(B)}}
            {\sqrt{p_A^{(\alpha)}}}
 \qquad\text{in }\mathcal R_{+-}.
 \label{eqn:alice-conditioned-state}
\end{equation}
Similarly, applying Bob's projector gives
\begin{equation}
 \Psi\longmapsto
 \frac{\sum_{i=1}^{n}a_{i\beta}\psi_i^{(A)}}
      {\sqrt{p_B^{(\beta)}}}\psi_\beta^{(B)}
 \qquad\text{in }\mathcal R_{-+}.
 \label{eqn:bob-conditioned-state}
\end{equation}
In $\mathcal R_{--}$ the state is $\Psi$ given in Eqn.~\eqref{eqn:qudit}, whereas in $\mathcal R_{++}$
it is $\psi_\alpha^{(A)}\psi_\beta^{(B)}$ (up to a global phase).

As in Sec.~\ref{subsec:mmt qubit in superposition}, the leading-order contribution in $v/c$ to the stress-energy is the $T_{00}$ component. The associated mass densities are
\begin{multline}
    \varrho_m^{--}(t,\vec x)
 =m_A\sum_{i=1}^{n}p_A^{(i)}|\psi_i^{(A)}(t,\vec x)|^2
   \\+m_B\sum_{j=1}^{n}p_B^{(j)}|\psi_j^{(B)}(t,\vec x)|^2,
 \label{eqn:rho-minusminus}
\end{multline}
in $\mathcal{R}_{--}$ (before any quantum collapse has occurred),
\begin{multline}
    \varrho_m^{+-,(\alpha)}(t,\vec x)
 =m_A|\psi_\alpha^{(A)}(t,\vec x)|^2
   \\+m_B\sum_{j=1}^{n}
       \frac{|a_{\alpha j}|^2}{p_A^{(\alpha)}}
                        |\psi_j^{(B)}(t,\vec x)|^2
 \label{eqn:rho-plusminus}
\end{multline}
in $\mathcal{R}_{+-}$ (in the future of Alice's quantum measurement but before $\Lambda^+(x_B)$ reaches her),
\begin{multline}
    \varrho_m^{-+,(\beta)}(t,\vec x)
 =m_A\sum_{i=1}^{n}
       \frac{|a_{i\beta}|^2}{p_B^{(\beta)}}
                         |\psi_i^{(A)}(t,\vec x)|^2
   \\+m_B|\psi_\beta^{(B)}(t,\vec x)|^2
 \label{eqn:rho-minusplus}
\end{multline}
in $\mathcal{R}_{-+}$ (in the future of Bob's quantum measurement but before $\Lambda^+(x_A)$ reaches him), and
\begin{equation}
    \varrho_m^{++,(\alpha,\beta)}(t,\vec x)
 =m_A|\psi_\alpha^{(A)}(t,\vec x)|^2
   +m_B|\psi_\beta^{(B)}(t,\vec x)|^2.
 \label{eqn:rho-plusplus}
\end{equation}
in $\mathcal{R}_{++}$ (both quantum states have collapsed). Here $T^{\mathrm{loc},(\alpha,\beta)}_{00,\pm\pm}=c^2\varrho_m^{\pm\pm,(\alpha,\beta)}$ and
$T^{\mathrm{loc},(\alpha,\beta)}_{0j,\pm\pm}=T^{\mathrm{loc},(\alpha,\beta)}_{jk,\pm\pm}\approx 0$ at
the retained order. Again, we assume that these stress-energies are conserved in their respective regions. 

There are now four different interfaces to consider: (i) $\mathcal{R}_{--} \leftrightarrow \mathcal{R}_{+-}$, (ii) $\mathcal{R}_{--} \leftrightarrow \mathcal{R}_{-+}$, (iii) $\mathcal{R}_{-+} \leftrightarrow \mathcal{R}_{++}$ and (iv) $\mathcal{R}_{+-} \leftrightarrow \mathcal{R}_{++}$, separated by null hypersurfaces respectively. For interface (i), the null hypersurface of interest which separates both regions is $\Sigma_{A}^-$. For interface (ii), the null hypersurface of interest which separates both regions is $\Sigma_{B}^-$. For interface (iii), the null hypersurface of interest which separates both regions is $\Sigma_{A}^+$. For interface (iv), the null hypersurface of interest which separates both regions is $\Sigma_{B}^+$. 

The changes in the individual laboratories' local mass densities after their respective quantum measurements is
\begin{widetext}
    \begin{align}
    \Delta \varrho_{m,A}^\alpha(t,\vec{x}) &= m_A \left[\abs{\psi_\alpha^{(A)}(t,\vec{x})}^2-\sum_{i=1}^n p_A^{(i)} \abs{\psi_i^{(A)}(t,\vec{x})}^2\right] + \chi_{u_B<0}(t,\vec{x}) m_B \sum_{j=1}^n \left(\frac{|a_{\alpha j}|^2}{p_A^{(\alpha)}}-p_B^{(j)}\right) \abs{\psi_j^{(B)}(t,\vec{x})}^2, \\
    \Delta \varrho_{m,B}^\beta(t,\vec{x}) &= m_B \left[\abs{\psi_\beta^{(B)}(t,\vec{x})}^2-\sum_{j=1}^n p_B^{(j)} \abs{\psi_j^{(B)}(t,\vec{x})}^2\right] + \chi_{u_A<0}(t,\vec{x}) m_A \sum_{i=1}^n \left(\frac{|a_{i \beta}|^2}{p_B^{(\beta)}}-p_A^{(i)}\right) \abs{\psi_i^{(A)}(t,\vec{x})}^2.
\end{align}
\end{widetext}
Then the bulk mass density is
\begin{multline}
    \varrho^{\alpha\beta}_\mathrm{bulk}(t,\vec{x}) = \varrho_m^{--}(t,\vec{x}) + \chi_{u_A>0}(t,\vec{x}) \Delta \varrho_{m,A}^\alpha(t,\vec x) \\ + \chi_{u_B>0}(t,\vec{x}) \Delta \varrho_{m,B}^{\beta}(t,\vec{x}).
\end{multline}
If Alice's (respectively Bob's) wavepackets are supported in her (respectively his) laboratory in the sense that
\begin{align}
    \operatorname{supp} \psi_i^{(A)}(t,\cdot) &\mathrel{\underset{\approx}{\subset}} \left\{\vec{x} \in \mathbb{R}^3 \mid \abs{\vec{x}-\vec{x}_A} < \abs{\vec{x}-\vec{x}_B}\right\}, \\
    \operatorname{supp} \psi_j^{(B)}(t,\cdot) &\mathrel{\underset{\approx}{\subset}} \left\{\vec{x} \in \mathbb{R}^3 \mid \abs{\vec{x}-\vec{x}_B} < \abs{\vec{x}-\vec{x}_A}\right\},
\end{align}
for all $i,j =1,\cdots,n$, then 
\begin{align}
    \chi_{u_A>0}(t,\vec{x})\chi_{u_B<0}(t,\vec{x}) \abs{\psi_j^{(B)}(t,\vec{x})}^2 &\approx 0,
    \\
    \chi_{u_B>0}(t,\vec{x})\chi_{u_A<0}(t,\vec{x}) \abs{\psi_i^{(A)}(t,\vec{x})}^2 &\approx 0,
\end{align}
i.e. the contributions from the conditional expectation values (approximately) vanish. Moreover, 
\begin{align}
    \Delta T_{\mu\nu,A}^\alpha(t,\vec{x}) &\approx c^2 \Delta \varrho_{m,A}^\alpha(t,\vec{x}) \delta^0_\mu \delta^0_\nu, \\
    \Delta T_{\mu\nu,B}^\beta(t,\vec{x}) &\approx c^2 \Delta \varrho_{m,B}^\beta(t,\vec{x}) \delta^0_\mu \delta^0_\nu,
\end{align}
with $\partial^\mu \Delta T_{\mu\nu,A}^\alpha = \partial^\mu \Delta T_{\mu\nu,B}^\beta = 0$ in $\mathcal{R}_{\pm\pm}$ and $\mathcal{R}_{\pm\mp}$. The global bulk stress-energy built from the local quantum states is then
\begin{equation}
    T_{\mu\nu}^{\mathrm{loc},\alpha\beta}(t,\vec{x}) \approx c^2 \varrho_\mathrm{bulk}^{\alpha \beta}(t,\vec{x}) \delta^0_\mu \delta^0_\nu.
\end{equation}

As in Sec.~\ref{subsec:mmt qubit in superposition}, we can then compute the metric in each of the regions apply the Barrabès-Israel formalism to find the singular stress-energy supported on $\Sigma_{A}^\pm$ and $\Sigma_B^\pm$. The computations follow analogously to the single qubit case, so we shall not detail them again here. Qualitatively, in each region, the gravitational field is given (up to linearised diffeomorphisms) in a retarded way sourced by the local mass densities, with corrections in the near future of the collapse hypersurfaces.

Note that the null hypersurfaces interfacing each of the four regions are not light cones anymore, yet the gluing of metrics into a globally defined locally Lipschitz (including at the crossing $\mathcal{C}$) metric is still possible. The interfaces $\Sigma_A^\pm$ and $\Sigma_B^\pm$ are portions of two intersecting null cones. \cite[Theorem A.4]{manzano_generalizing_2025} guarantees that the gluing of such metrics is locally Lipschitz, but only applies locally away from $\mathcal{C}$. In the linearised regime considered here, we instead construct the perturbation globally in a common Minkowski coordinate system as
\begin{equation}
    h^{\mathrm{tot}}= \underline{h}^-+ \Theta(u_A) \mathfrak{H}^\alpha_A + H^\alpha_A+\Theta(u_B) \mathfrak{H}^\beta_B + H^\beta_B,
\end{equation}
where $\mathfrak{H}^\alpha_A$, $H^\alpha_A$. $\mathfrak{H}^\beta_B$ and $H^\beta_B$ are the locally Lipschitz corrections supplied by the single collapse construction. Their sum is locally Lipschitz, including at $\mathcal C$. The linearity of $G^{(1)}$ then establishes the distributional field equation with the sum of the two shell sources. Thus compatibility at the crossing follows.

Indeed, note that 
\begin{multline}
    \partial_\rho h_{\mu\nu}^\mathrm{tot} = (\partial_\rho h_{\mu\nu}^{\mathrm{tot}})_{\mathrm{strong}} + n_{A\rho} \underbrace{[h_{\mu\nu}^{\mathrm{tot}}]_A}_{=0} \delta(u_A)\\+n_{B\rho} \underbrace{[h_{\mu\nu}^{\mathrm{tot}}]_B}_{=0} \delta(u_B),
\end{multline}
where $(\partial_\rho h_{\mu\nu}^\mathrm{tot})_\text{strong}$ is the standard (strong) derivative computed separately on each side of the hypersurfaces, $[h_{\mu\nu}^\mathrm{tot}]$ is the difference between $h_{\mu\nu}^\mathrm{tot}|_{\M^+}$ and $h_{\mu\nu}^\mathrm{tot}|_{\M^-}$ on $\mathcal{N}$, and $n_{Aa}$ and $n_{Ba}$ here mean the components of $d u_A$ and $d u_B$, respectively, in the coordinates used. The latter two terms vanish by continuity. However, the remaining piecewise derivative can have nonzero jumps across both cones:
\begin{multline}
    \partial_\sigma \partial_\rho h_{\mu\nu}^\mathrm{tot} = (\partial_\sigma \partial_\rho h_{\mu\nu}^{\mathrm{tot}})_{\mathrm{strong}} + n_{A\sigma} [\partial_\rho h_{\mu\nu}^{\mathrm{tot}}]_A \delta(u_A)\\+n_{B\sigma} [\partial_\rho h_{\mu\nu}^{\mathrm{tot}}]_B \delta(u_B).
\end{multline}
There is no $\delta(u_A) \delta(u_B)$ nor any $\delta'$ term here. Thus, the contributions to each shell are independent, and there is no additional edge term on $\mathcal{C}$. The distributional stress-energies and corresponding shell masses that ensure the conservation of stress-energy and mass, respectively, can therefore be computed in complete analogy with the single qubit case independently from the other shell, at least at this linear order. 

The crossing shells thus admit a consistent solution in this example.   We find it plausible that such crossing shells are not an issue more generally in the linearised regime, though we have no proof at present.  We also have no results in the nonlinear regime.   We keep these open questions for future work. 

In summary, two spacelike-separated quantum measurements can be consistently described in this model, at least in this weak-field regime. We showed this holds to leading order in $v/c$ (and in $G$).  We see no reason for the reasoning to fail at higher orders in $v/c$ as it relies on additivity from linearity rather than nonrelativistic effects.

\section{Relativistic readout devices}

\label{sec:relativistic readout devices}

As we have seen in Sec. \ref{sec:weak field limit}, away from quantum measurements, the linearised gravitational field backreacts causally from the expectation value of the stress-energy tensor with respect to the local state as in Eqn. \eqref{eqn: weak field gravitational field}. We now show that, in the presence of quantum measurements, the weak-field potential is still retarded.

We showed in Eqn.~\eqref{eqn:h+ as hret and Tsing} that the linearised future metric can, at least locally, be constructed from the retarded potential $h^\mathrm{ret}$ and from the convolution of the retarded Green's function with the singular stress-energy. $T^{\mathrm{sing}}$ is local in $\mu$, $j^A$ and $p$ (see Eqn.~\eqref{eq:BIc-cartesian}) where $\mu$, $p$ and $j^A$ are expressed from $\mathfrak{b}_{\mu\nu}$ as in Eqn.~\eqref{eq:BIc-coefficients}.

From \eqref{eqn:characteristic bvv bvA} we also see that $\mathfrak{b}_{vv}$ is given locally in terms of $\Delta T$ i.e. in terms of $T^{\mathrm{loc},\pm}$, and is thus only dependent on retarded data. Moreover, $\mathfrak{b}_{vA}$ and $\Omega^{AB} \mathfrak{b}_{AB}$ are given by integrals of local functionals of $\Delta T$ from the origin of the cone $x=(0,\vec{0})$ to the evaluation point on the cone (i.e. these lie in the causal past of the point, see Appendix \ref{app:future metric qubit} to see how the singular stress energies are evaluated at points both on the cone and in the causal past of the evaluation point), i.e. is again retarded.

Hence, $\overline{h}^{+,(i)}$ only depends on retarded data: quantum measurements do not change this. Readings of the gravitational field remain causal. This is also important for the discussion on integrability of Sec.~\ref{sec:integrability}: the argument presented there holds even in the presence of quantum measurements, and thus dynamics remain integrable after quantum measurements, as they should.

The classical gravitational field can be probed through a Cavendish experiment. This leads to the existence of (finite precision probability) readout devices \cite{kent_nonlinearity_2005,kent_testing_2021,kent_quantum_2021, fedida_mixture_2025, fedida_thermodynamics_2026}, which abstractly can be understood as black boxes which provide a classical description of quantum states without perturbing them.

More specifically, consider \cite{fedida_thermodynamics_2026} a single particle of mass $m$ in a superposition of $k$ approximately localised wavepackets of the form
\begin{equation}
    \ket{\Psi} \approx \sum_{i=1}^k a_i \ket{\vec{x}_i}
\end{equation}
in the nonrelativistic limit of the theory, where $\sum_{i=1}^k \abs{a_i}^2 = 1$. Then, from Poisson's equation \eqref{eqn:N-body Poisson}, we find that the gravitational field at position $\vec{y}_j$ is
\begin{equation}
    \Phi(\vec{y}_j) \approx -Gm \sum_{i=1}^k \frac{\abs{a_i}^2}{\abs{\vec{x}_i-\vec{y}_j}} \, .
\end{equation}
Repeating Cavendish experiments at positions $\vec{y}_j$ for $j \in \{1,\cdots,k-1\}$ suffices (by normalisation) to get an estimate of the $\abs{a_i}^2$, $i = 1,\cdots,k$, without disturbing the quantum state. Thus, this semiclassical theory of gravity allows one to construct a finite-precision probability readout device (FPPRD) \cite{kent_nonlinearity_2005,kent_testing_2021,kent_quantum_2021, fedida_mixture_2025, fedida_thermodynamics_2026}, that is, a black box which takes as input a quantum state, a basis and a precision parameter and outputs a classical description of the diagonal elements (Born probabilities) of the state in that basis:
\begin{widetext}
    \begin{equation}
    \label{Semiclassical gravity FPRD}
    l, \text{basis}, \ket{\Psi} \xlongrightarrow{input} \boxed{\text{Semiclassical gravity (FP)PRD}} \xlongrightarrow{output} \begin{cases}
        \abs{a_1}^2 = 0.n_1^{(1)}n_1^{(2)}...n_1^{(l)} \\
        \abs{a_2}^2 = 0.n_2^{(1)}n_2^{(2)}...n_2^{(l)} \\
        \vdots \\
        \abs{a_k}^2 = 0.n_k^{(1)}n_k^{(2)}...n_k^{(l)}
    \end{cases} \quad .
\end{equation}
\end{widetext}
The amplitudes (up to an overall phase) can be retrieved using quantum state tomography \cite{fedida_thermodynamics_2026}. One can use this classical information to implement quantum state cloning. It has been argued \cite{galley_any_2023, mielnik_mobility_1979} that this feature, which is quite generic of mean-field models, makes the state space of the theory classical, in the sense that all states are (at least in principle) perfectly distinguishable.

\begin{figure}
    \centering
    \begin{tikzpicture}[x=0.75pt,y=0.75pt,yscale=-1,xscale=1]

\draw   (267.22,249.78) .. controls (267.22,247.57) and (269.01,245.78) .. (271.22,245.78) .. controls (273.43,245.78) and (275.22,247.57) .. (275.22,249.78) .. controls (275.22,251.99) and (273.43,253.78) .. (271.22,253.78) .. controls (269.01,253.78) and (267.22,251.99) .. (267.22,249.78) -- cycle ;
\draw   (301.22,249.78) .. controls (301.22,247.57) and (303.01,245.78) .. (305.22,245.78) .. controls (307.43,245.78) and (309.22,247.57) .. (309.22,249.78) .. controls (309.22,251.99) and (307.43,253.78) .. (305.22,253.78) .. controls (303.01,253.78) and (301.22,251.99) .. (301.22,249.78) -- cycle ;
\draw   (370.22,250.78) .. controls (370.22,248.57) and (372.01,246.78) .. (374.22,246.78) .. controls (376.43,246.78) and (378.22,248.57) .. (378.22,250.78) .. controls (378.22,252.99) and (376.43,254.78) .. (374.22,254.78) .. controls (372.01,254.78) and (370.22,252.99) .. (370.22,250.78) -- cycle ;
\draw   (404.22,250.78) .. controls (404.22,248.57) and (406.01,246.78) .. (408.22,246.78) .. controls (410.43,246.78) and (412.22,248.57) .. (412.22,250.78) .. controls (412.22,252.99) and (410.43,254.78) .. (408.22,254.78) .. controls (406.01,254.78) and (404.22,252.99) .. (404.22,250.78) -- cycle ;
\draw    (271.22,249.78) .. controls (273.47,248.84) and (275.1,249.45) .. (276.13,251.62) .. controls (277.11,253.77) and (278.65,254.34) .. (280.75,253.34) .. controls (282.8,252.33) and (284.49,252.95) .. (285.81,255.22) .. controls (286.62,257.29) and (288.19,257.87) .. (290.54,256.95) .. controls (292.39,255.84) and (293.86,256.37) .. (294.95,258.53) .. controls (296,260.66) and (297.57,261.21) .. (299.66,260.16) .. controls (301.66,259.07) and (303.29,259.6) .. (304.55,261.75) .. controls (305.44,263.76) and (306.95,264.21) .. (309.08,263.09) .. controls (311.07,261.9) and (312.78,262.34) .. (314.19,264.39) .. controls (315.33,266.34) and (316.91,266.65) .. (318.94,265.32) .. controls (320.79,263.89) and (322.41,264.09) .. (323.82,265.9) .. controls (325.35,267.65) and (327.05,267.69) .. (328.9,266.02) .. controls (330.31,264.27) and (331.99,264.13) .. (333.93,265.6) .. controls (335.8,266.99) and (337.38,266.71) .. (338.67,264.74) .. controls (339.87,262.73) and (341.52,262.29) .. (343.63,263.42) .. controls (345.59,264.55) and (347.21,264.01) .. (348.48,261.8) .. controls (349.44,259.67) and (351.02,259.06) .. (353.23,259.99) .. controls (355.2,260.99) and (356.74,260.36) .. (357.85,258.09) .. controls (358.64,255.94) and (360.11,255.3) .. (362.28,256.17) .. controls (364.52,257) and (366.11,256.3) .. (367.04,254.05) .. controls (367.61,251.96) and (369.1,251.28) .. (371.52,252.01) -- (374.22,250.78) ;
\draw    (305.22,249.78) .. controls (306.55,247.42) and (308.24,246.71) .. (310.31,247.65) .. controls (312.34,248.61) and (313.93,247.95) .. (315.08,245.67) .. controls (315.74,243.59) and (317.22,242.98) .. (319.53,243.83) .. controls (321.77,244.72) and (323.16,244.16) .. (323.71,242.15) .. controls (325,239.86) and (326.67,239.21) .. (328.72,240.2) .. controls (330.69,241.25) and (332.24,240.68) .. (333.36,238.51) .. controls (334.46,236.38) and (336.05,235.85) .. (338.14,236.93) .. controls (340.11,238.09) and (341.61,237.67) .. (342.62,235.67) .. controls (343.97,233.64) and (345.66,233.27) .. (347.67,234.57) .. controls (349.51,235.98) and (351.12,235.77) .. (352.5,233.94) .. controls (354.31,232.18) and (356.02,232.13) .. (357.63,233.79) .. controls (359.09,235.54) and (360.82,235.67) .. (362.81,234.2) .. controls (364.73,232.81) and (366.38,233.11) .. (367.76,235.08) .. controls (368.78,237.04) and (370.38,237.45) .. (372.57,236.32) .. controls (374.6,235.19) and (376.16,235.69) .. (377.24,237.8) .. controls (378.31,239.95) and (379.81,240.49) .. (381.74,239.43) .. controls (384.12,238.56) and (385.74,239.2) .. (386.6,241.34) .. controls (387.48,243.5) and (389.03,244.14) .. (391.25,243.27) .. controls (393.54,242.44) and (394.99,243.07) .. (395.6,245.15) .. controls (396.67,247.44) and (398.23,248.13) .. (400.26,247.2) .. controls (402.81,246.51) and (404.46,247.26) .. (405.23,249.43) -- (408.22,250.78) ;
\draw    (473.36,56.78) -- (309.36,193.78) ;
\draw   (266,193.78) -- (309.36,193.78) -- (309.36,233.78) -- (266,233.78) -- cycle ;
\draw   (369,169.78) -- (412.36,169.78) -- (412.36,209.78) -- (369,209.78) -- cycle ;
\draw    (266,233.78) .. controls (266.36,183.78) and (316.36,202.78) .. (309.36,233.78) ;
\draw    (270.79,216.18) -- (288,233) ;
\draw [shift={(269.36,214.78)}, rotate = 44.35] [color={rgb, 255:red, 0; green, 0; blue, 0 }  ][line width=0.75]    (10.93,-3.29) .. controls (6.95,-1.4) and (3.31,-0.3) .. (0,0) .. controls (3.31,0.3) and (6.95,1.4) .. (10.93,3.29)   ;
\draw   (264.22,173.78) .. controls (264.22,171.57) and (266.01,169.78) .. (268.22,169.78) .. controls (270.43,169.78) and (272.22,171.57) .. (272.22,173.78) .. controls (272.22,175.99) and (270.43,177.78) .. (268.22,177.78) .. controls (266.01,177.78) and (264.22,175.99) .. (264.22,173.78) -- cycle ;
\draw    (390.36,170.78) -- (390.68,189.78) ;
\draw  [fill={rgb, 255:red, 0; green, 0; blue, 0 }  ,fill opacity=1 ] (385.29,195.17) .. controls (385.29,192.19) and (387.7,189.78) .. (390.68,189.78) .. controls (393.66,189.78) and (396.07,192.19) .. (396.07,195.17) .. controls (396.07,198.15) and (393.66,200.56) .. (390.68,200.56) .. controls (387.7,200.56) and (385.29,198.15) .. (385.29,195.17) -- cycle ;
\draw   (367,68.78) -- (410.36,68.78) -- (410.36,108.78) -- (367,108.78) -- cycle ;
\draw    (388.36,69.78) -- (374.68,95.17) ;
\draw  [fill={rgb, 255:red, 0; green, 0; blue, 0 }  ,fill opacity=1 ] (369.29,95.17) .. controls (369.29,92.19) and (371.7,89.78) .. (374.68,89.78) .. controls (377.66,89.78) and (380.07,92.19) .. (380.07,95.17) .. controls (380.07,98.15) and (377.66,100.56) .. (374.68,100.56) .. controls (371.7,100.56) and (369.29,98.15) .. (369.29,95.17) -- cycle ;
\draw    (393.36,234.78) -- (393.36,219.78) ;
\draw [shift={(393.36,216.78)}, rotate = 90] [fill={rgb, 255:red, 0; green, 0; blue, 0 }  ][line width=0.08]  [draw opacity=0] (8.93,-4.29) -- (0,0) -- (8.93,4.29) -- cycle    ;
\draw    (393.36,160.28) -- (393.36,136.78) ;
\draw [shift={(393.36,133.78)}, rotate = 90] [fill={rgb, 255:red, 0; green, 0; blue, 0 }  ][line width=0.08]  [draw opacity=0] (8.93,-4.29) -- (0,0) -- (8.93,4.29) -- cycle    ;
\draw    (142.36,283.78) -- (138.41,61.78) ;
\draw [shift={(138.36,58.78)}, rotate = 88.98] [fill={rgb, 255:red, 0; green, 0; blue, 0 }  ][line width=0.08]  [draw opacity=0] (10.72,-5.15) -- (0,0) -- (10.72,5.15) -- (7.12,0) -- cycle    ;
\draw    (142.36,283.78) -- (186.36,284.72) ;
\draw [shift={(189.36,284.78)}, rotate = 181.22] [fill={rgb, 255:red, 0; green, 0; blue, 0 }  ][line width=0.08]  [draw opacity=0] (10.72,-5.15) -- (0,0) -- (10.72,5.15) -- (7.12,0) -- cycle    ;
\draw    (312.36,290.78) -- (366.36,290.78) ;
\draw [shift={(369.36,290.78)}, rotate = 180] [fill={rgb, 255:red, 0; green, 0; blue, 0 }  ][line width=0.08]  [draw opacity=0] (10.72,-5.15) -- (0,0) -- (10.72,5.15) -- (7.12,0) -- cycle    ;
\draw [shift={(309.36,290.78)}, rotate = 0] [fill={rgb, 255:red, 0; green, 0; blue, 0 }  ][line width=0.08]  [draw opacity=0] (10.72,-5.15) -- (0,0) -- (10.72,5.15) -- (7.12,0) -- cycle    ;
\draw   (370.22,131.78) .. controls (370.22,129.57) and (372.01,127.78) .. (374.22,127.78) .. controls (376.43,127.78) and (378.22,129.57) .. (378.22,131.78) .. controls (378.22,133.99) and (376.43,135.78) .. (374.22,135.78) .. controls (372.01,135.78) and (370.22,133.99) .. (370.22,131.78) -- cycle ;
\draw    (374.22,124.78) -- (374.22,114.28) ;
\draw [shift={(374.22,111.28)}, rotate = 90] [fill={rgb, 255:red, 0; green, 0; blue, 0 }  ][line width=0.08]  [draw opacity=0] (8.93,-4.29) -- (0,0) -- (8.93,4.29) -- cycle    ;
\draw    (268.36,160.28) -- (268.36,136.78) ;
\draw [shift={(268.36,133.78)}, rotate = 90] [fill={rgb, 255:red, 0; green, 0; blue, 0 }  ][line width=0.08]  [draw opacity=0] (8.93,-4.29) -- (0,0) -- (8.93,4.29) -- cycle    ;
\draw    (268.22,173.78) .. controls (269.52,171.72) and (271.21,171.28) .. (273.3,172.45) .. controls (275.35,173.63) and (276.95,173.2) .. (278.1,171.17) .. controls (279.19,169.15) and (280.95,168.67) .. (283.36,169.73) .. controls (285.24,170.93) and (286.65,170.53) .. (287.59,168.53) .. controls (288.91,166.41) and (290.67,165.89) .. (292.86,166.98) .. controls (294.6,168.2) and (296.04,167.76) .. (297.17,165.67) .. controls (298.22,163.59) and (299.94,163.04) .. (302.31,164.02) .. controls (304.28,165.12) and (305.87,164.59) .. (307.08,162.42) .. controls (307.86,160.37) and (309.34,159.84) .. (311.51,160.83) .. controls (313.66,161.81) and (315.2,161.23) .. (316.11,159.09) .. controls (317.24,156.84) and (318.83,156.2) .. (320.88,157.16) .. controls (322.93,158.11) and (324.44,157.46) .. (325.42,155.21) .. controls (326.35,152.97) and (327.82,152.3) .. (329.84,153.2) .. controls (332.15,153.96) and (333.76,153.2) .. (334.65,150.93) .. controls (335.26,148.78) and (336.76,148.05) .. (339.14,148.74) .. controls (341.28,149.54) and (342.69,148.84) .. (343.38,146.64) .. controls (344.11,144.41) and (345.6,143.67) .. (347.86,144.41) .. controls (350.16,145.13) and (351.75,144.34) .. (352.64,142.05) .. controls (353.25,139.89) and (354.61,139.22) .. (356.73,140.04) .. controls (359.26,140.67) and (360.91,139.88) .. (361.68,137.65) .. controls (362.51,135.41) and (363.89,134.76) .. (365.81,135.69) .. controls (368.17,136.42) and (369.85,135.64) .. (370.86,133.33) -- (374.22,131.78) ;

\draw (328,243.4) node [anchor=north west][inner sep=0.75pt]  [font=\scriptsize]  {$\ket{\Phi ^{+}}$};
\draw (189,184) node [anchor=north west][inner sep=0.75pt]   [align=left] {\begin{minipage}[lt]{46.36pt}\setlength\topsep{0pt}
\begin{center}
{\scriptsize Quantum}\\{\scriptsize Measurement}
\end{center}

\end{minipage}};
\draw (417,164) node [anchor=north west][inner sep=0.75pt]   [align=left] {\begin{minipage}[lt]{38.82pt}\setlength\topsep{0pt}
\begin{center}
{\scriptsize Cavendish}\\{\scriptsize Experiment}
\end{center}

\end{minipage}};
\draw (301,57) node [anchor=north west][inner sep=0.75pt]   [align=left] {\begin{minipage}[lt]{38.82pt}\setlength\topsep{0pt}
\begin{center}
{\scriptsize Cavendish}\\{\scriptsize Experiment}
\end{center}

\end{minipage}};
\draw (213,211.4) node [anchor=north west][inner sep=0.75pt]  [font=\scriptsize]  {$t_{QM}$};
\draw (431,193.4) node [anchor=north west][inner sep=0.75pt]  [font=\scriptsize]  {$t_{CAV}^{( 1)}$};
\draw (133,43.4) node [anchor=north west][inner sep=0.75pt]  [font=\scriptsize]  {$t$};
    \draw (195,279.4) node [anchor=north west][inner sep=0.75pt]  [font=\scriptsize]  {$\vec{x}$};
\draw (321,84.4) node [anchor=north west][inner sep=0.75pt]  [font=\scriptsize]  {$t_{CAV}^{( 2)}$};
\draw (218,255.4) node [anchor=north west][inner sep=0.75pt]  [font=\scriptsize]  {$t_{0}$};
\draw (338,278.4) node [anchor=north west][inner sep=0.75pt]  [font=\scriptsize]  {$d$};
\draw (279.22,257.78) node [anchor=north west][inner sep=0.75pt]   [align=left] {{\scriptsize Alice}};
\draw (381.22,256.78) node [anchor=north west][inner sep=0.75pt]   [align=left] {{\scriptsize Bob}};
\draw (305,137.4) node [anchor=north west][inner sep=0.75pt]  [font=\scriptsize]  {$\ket{LL}$};

\end{tikzpicture}
    \caption{\justifying Torsion pendulums of Cavendish experiments as relativistic gravitational readout devices. At time $t=t_0$, Alice and Bob, who are separated by a distance $d \gg \text{dist}(L,R)$, share an entangled state $\ket{\Phi^+} = \frac{1}{\sqrt{2}}(\ket{LL} + \ket{RR})$. At time $t = t_\text{QM}$, Alice performs a quantum measurement and reads ``L''. At time $t \approx t_\text{CAV}^{(1)} < t_\text{QM} + \frac{d}{c}$, Bob performs a Cavendish experiment -- his torsion pendulum reads a gravitational field coming from the middle of ``L'' and ``R''. At time $t \approx t^{(2)}_\text{CAV} > t_\text{QM} +\frac{d}{c}$, after Alice's light cone (from the moment of her quantum measurement) reaches him, his torsion pendulum now shows ``L'': his local quantum state has updated from Alice's quantum measurement and is $\rho^\mathrm{loc}_{(B)} = \ketbra{L}$.}
    \label{fig:gravitational RDs}
\end{figure}
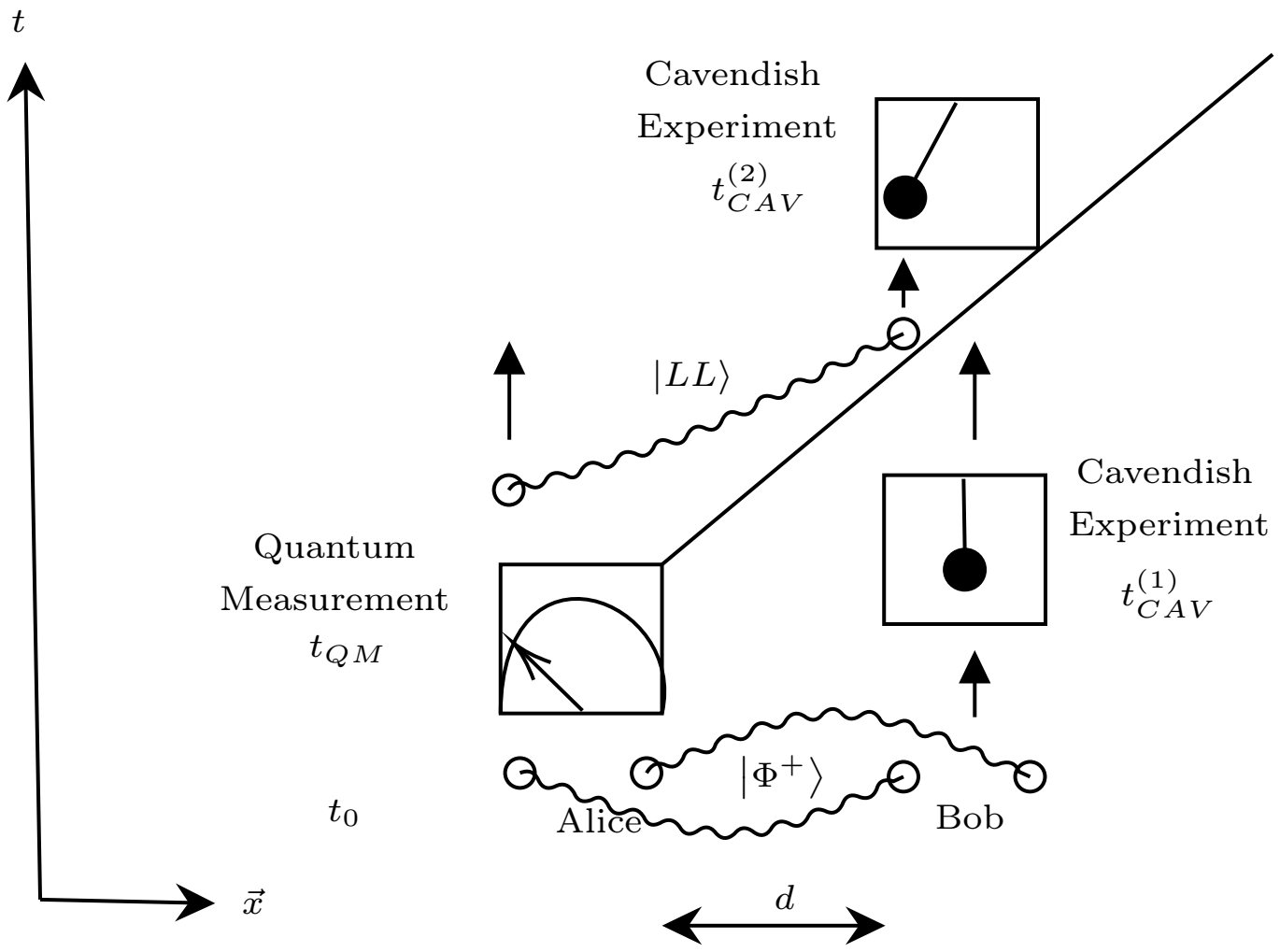

In this context however, this is consistent with the relativistic no-signalling principle, as was first shown in \cite{kent_nonlinearity_2005}. Consider the following thought experiment, depicted in Fig. \ref{fig:gravitational RDs}. Take a foliation whereby $t$ is a time parameter. At time $t=t_0$, Alice and Bob, who are two agents separated by a spatial distance $d$, share an entangled quantum state $\ket{\Phi^+} = \frac{1}{\sqrt{2}}(\ket{LL}+\ket{RR})$, where $\ket{L}_i$ (resp. $\ket{R}_i$) is a well-localised wavepacket on the left (resp. right) side of $i=\text{Alice, Bob}$. For simplicity, suppose $d \gg \text{dist}(L,R)$.

If either of them (or both) were to perform a Cavendish experiment on their own, they would each find that their respective torsion pendulum reads a gravitational field coming from the middle of their ``L'' and ``R'' wavepackets, since $\rho^\mathrm{loc}_i = \frac{1}{2}(\ketbra{L}+\ketbra{R})$.

For example, suppose Bob performs such an experiment at time $t=t_\text{CAV}^{(1)}$. Suppose that at time $t=t_\text{QM} > t_\text{CAV}^{(1)} - \frac{d}{c}$, Alice performs a quantum measurement in the $\{\ket{L},\ket{R}\}$ basis, and reads ``L'' on her pointer. Her local quantum state $\rho^\mathrm{loc}_{(A)}$ updates to $\ketbra{L}$\footnote{Note that it does not matter whether she decides to read the outcome or not; the gravitational field ``knows'' the outcome. In this sense, ``gravity selects'' even if one performs a nonselective update.}. Until the light cone from this quantum measurement reaches Bob, i.e. for $t<t_\text{QM} + \frac{d}{c}$, Bob's Cavendish experiment continues to show an average gravitational field between ``L'' and ``R''. At time $t=t_\text{QM} + \frac{d}{c}$, the information of Alice's outcome which travels at the speed of light (through the thin-shell of stress-energy of her light-cone), reaches Bob's torsion pendulum, which then starts to point to the left at $t \approx t_\text{CAV}^{(2)}$: his local quantum state updates to $\ketbra{L}$. 

This is a relativistic gravitational readout device.
Before Bob crosses the future light cone of Alice's measurement event, he can learn, without disturbing either his local quantum state nor the (global) polystate, his local quantum state in the position basis and thus infer that Alice's state is, to the best of his (local) knowledge, a proper mixture $\frac{1}{2} \ketbra{L} + \frac{1}{2} \ketbra{R}$. After he crosses the light cone, he can learn Alice's outcome and his updated local quantum state, again without disturbing his local quantum state nor the global state.

At any point, provided he can make use of quantum state tomography and apply arbitrary local unitaries, he may use his torsion pendulum to clone his local quantum state by encoding his classical information into a quantum state. However, neither this cloning nor the readout devices readings that enable it imply superluminal signalling.

\section{No-signalling principles}

\label{sec:nsp}

\subsection{The no-superluminal signalling principle}

In probabilistic theories such as quantum theory and its extensions, the statement of causality is a statement about probabilities of outcomes depending on agents' choices over spacelike separations. It is called the \emph{no-superluminal-signalling principle} (NSSP), and takes the following form. 
\begin{quote}
    \textbf{NSSP: } If Alice and Bob can perform measurements across spacelike-separated locations using input parameters $w_A$ and $z_B$ yielding outcomes $a$ and $b$, respectively, then the conditional probabilities for each agent should not depend on the other’s choice:
\begin{align}
    P(a \mid w_A,z_B) &= P(a \mid w_A) \\ P(b \mid w_A,z_B) &= P(b \mid z_B) \, .
\end{align}
\end{quote}
Over the years, many inequivalent versions of this principle have been proposed to formulate it in quantum theory, including microcausality, Einstein causality, the quantum no-signalling principle (discussed below), $C^*$-independence and several others. In the model discussed in this paper, we need an appropriate implementation of the impossibility of superluminal signalling through (quantum and non-quantum) measurements.

\subsection{The quantum no-signalling principle}

We also need to characterise other signalling principles that the model respects or violates.
There are subtle issues here.  Standard versions of principles discussed in the literature turn out, on close examination, to involve approximations or non-physical assumptions.  This is true even in models that do not involve gravity, and considering gravitational effects adds further nuances.
Consider for example a standard formulation of the quantum no-signalling principle (QNSP).
\begin{quote}
    \textbf{QNSP: } Given two \emph{non-interacting} systems (Alice and Bob), 
    the probability of a measurement outcome $a$ for a \emph{quantum} measurement $M^{(Q)}_A$ on Alice's subsystem is independent of whether or not a \emph{quantum} measurement $M^{(Q)}_B$ is performed on Bob's subsystem:
    \begin{equation}
    P(a \mid  M^{(Q)}_A,M^{(Q)}_B) = P(a \mid M^{(Q)}_A) \, .
\end{equation}
\end{quote}
This framing of the QNSP assumes a standard representation for
the local quantum measurements $M_A^{(Q)}, M_B^{(Q)}$, for example as POVMs, but not that the bipartite system is necessarily described by a quantum state or that the measurement probabilities are given by the Born rule.

The term ``non-interacting'' needs close examination.  
A standard scenario in quantum communication theory is that Alice and Bob are separated 
and control separate quantum subsystems whose states belong to Hilbert spaces ${H}_A, H_B$, which are often taken to be finite-dimensional.   Each may apply local unitaries $U_A, U_B$ and local quantum
measurements $M^{(Q)}_A, M^{(Q)}_B$.   Here ``non-interacting'' means that the scenario excludes 
global joint unitaries $U_{AB}$ and global joint measurements $M^{(Q)}_{AB}$, and also excludes classical and quantum communications between the parties.   In particular, this means that neither party communicates the outcomes of any quantum measurements, or the post-measurement states of their subsystem, to the other.

In quantum cryptography, this is generally formalised by assuming that Alice is working in a secure laboratory from which no classical or quantum information is transmitted except via communication channels that she controls, and that Bob is working in a similar separate (disjoint) laboratory.   If a protocol specifies no communications in either direction after a certain point, the QNSP can then be applied.

These assumptions need to be understood as characterising a domain of validity for the relevant models, rather than as absolute physical truths.   They model Alice's and Bob's subsystems within quantum theory rather than relativistic quantum field theory.    They assume the subsystems are pointlike, or at least perfectly localized within disjoint compact regions (for example, Alice's and Bob's laboratories). 
They neglect any electromagnetic signal that Alice's operations may generate and transmit to Bob, and vice versa; in the cryptographic context, this assumes that the laboratories have perfect shielding.
Similarly, they neglect any gravitational effect of Alice's operations on Bob's subsystem, and vice versa -- although assuming a laboratory that perfectly shields gravitational forces is not even theoretically coherent.
All of these approximations are often justified for practical purposes, and they underpin many key results in quantum information theory and quantum cryptography, with important practical applications.
Still, in principle, they are not quite true.
As a result, if ``non-interacting'' is taken as a precise condition, the QNSP only applies to spacelike-separated systems.
Discussions in the literature thus generally
implicitly weaken the statement of the QNSP, implicitly replacing ``non-interacting'' by
``approximately non-interacting'' and ``independent'' by ``for all practical purposes may be treated as 
independent'', 
without quantifying either term.   

We will follow these general conventions below.  
For example, in principle actions by Bob on his subsystem will generally alter its gravitational field at least slightly.   Bob's system's local gravitational field propagates causally and could in principle be detected by Alice after the signal reaches her.  We neglect this effect, assuming it is irrelevant for practical purposes, following the standard conventions. 

However, there is an additional nuance in our semiclassical model. 
Actions by Bob on his subsystem can also causally alter the local quantum state of Alice's subsystem.
Because the gravitational field of Alice's subsystem is sourced by its local quantum state, this is an effect that Alice may be able to measure directly.
This effect need {\it not} be negligible in practice, even when Alice and Bob are well separated and the gravitational field of each subsystem has negligible effect on the other.   It implies violations of the QNSP in contexts where it 
would otherwise hold, as we now show.   

Consider once again the situation depicted in Fig.~\ref{fig:gravitational RDs}. Suppose that, at time $t=t_0$ in some chosen foliation of spacetime, Alice and Bob are spacelike-separated (by a distance $d \gg \text{dist}(L,R)$) and share an entangled qubit $\ket{\Phi^+} = \frac{1}{\sqrt{2}}(\ket{LL} + \ket{RR})$. At time $t=t_\mathrm{QM}$, Alice performs a quantum measurement and obtains an outcome, say ``L''. As we have seen, at times $t < t_{\text{QM}} + \frac{d}{c}$, if Bob performs a Cavendish experiment, his torsion pendulum reads a gravitational field that is a classical superposition of fields from ``L'' and ``R''. However, after $t=t_{\text{QM}} + \frac{d}{c}$, Bob's gravitational field updates and his torsion pendulum now shows ``L'': his local quantum state has updated from Alice's quantum measurement and is $\rho^\mathrm{loc}(x_B) = \ketbra{L}$.
If Alice carries out no measurement, Bob's torsion pendulum continues to read a gravitational field that is a classical superposition of fields from ``L'' and ``R''.   Bob can thus distinguish the cases where Alice does and does not measure.   To do so via quantum measurements, he can use a quantum probe whose evolution in the relevant gravitational fields is measurably distinguishable, and measure the state of the probe after a suitable length of time.    

Alice's measurements are quantumly detectable by Bob even if Alice and Bob's quantum systems have no quantum interaction Hamiltonian. Alice's quantum measurement need not generate a physically mediated signal (e.g. photons or gravitational waves) that can be measured by Bob's possibly arbitrarily (finitely) precise gravitational or quantum measurements.  The update happens without an information carrier in the normal sense.   Nonetheless, Bob's subsystem statistics are influenced by Alice's operations in his causal past. That is, the QNSP (understood in the operational, approximate sense) is violated in local semiclassical gravity.   

In the approximate, operational understanding of ``non-interacting systems", our model also does not satisfy the post-quantum no-signalling principle (PQNSP) which imposes the same requirement on probabilities as the QNSP but now for both quantum \emph{and non-quantum measurement}, and takes the following form.
\begin{quote}
    \textbf{PQNSP: } Given 
    two \emph{non-interacting} systems (Alice and Bob),
    the probability of a measurement outcome $a$ for a measurement $M_A$ on Alice's subsystem is independent of whether or not a measurement $M_B$ is performed on Bob's subsystem:
    \begin{equation}
    P(a \mid M_A,M_B) = P(a \mid M_A) \, .
\end{equation}
\end{quote}
This principle is stronger than the QNSP. Since local semiclassical gravity violates the QNSP (in the operational sense), it violates the PQNSP (in the same operational sense).

It would be very interesting to verify whether other post-quantum theories (whose dynamical evolution laws are linear in the classical-quantum state) of classical gravity \cite{diosi_universal_1987,oppenheim_postquantum_2023,tilloy_general_2024} satisfy the PQNSP, but this is beyond our scope. (See \cite{diosi_classical-quantum_2024, grudka_renormalisation_2024,vaidhyanathan_lorentz_2025,oppenheim_covariant_2026} for recent developments and open problems in the formulation of relativistic classical-quantum theories of gravity).

\subsection{The local no-superluminal-signalling principle}

As we discussed previously, $\rho^\mathrm{loc}$ describes the state of maximal knowledge that is accessible to a local observer \cite{kent_nonlinearity_2005}. Hence, an appropriate definition of no-superluminal signalling (in contrast with no-signalling between \emph{independent} subsystems) in the context of the model discussed in this paper (i.e. one which contains quantum and non-quantum measurements) is one which involves local quantum states, in what we shall call the \emph{local no-superluminal-signalling principle} (LNSSP). Its statement is as follows.
\begin{quote}
    \textbf{LNSSP: } Given a bipartite system (Alice and Bob) in a spacetime $(\M,g)$, the probability of a measurement outcome $a$ for a measurement $M_A$ on Alice's subsystem with associated local quantum state of maximal knowledge $\rho^\mathrm{loc}(x_A)$ is independent of whether or not a measurement $M_B$ is performed on Bob's spacelike-separated subsystem:
\begin{multline}
    P(a | \rho^\mathrm{loc}(x_A),g(x_A); M_A, M_B) \\ = P(a | \rho^\mathrm{loc}(x_A),g(x_A); M_A) \, .
\end{multline}
\end{quote}
The local measurements here may be on Alice's local quantum state or on the metric in her local region, or both.   
One reason for considering the LNSSP rather than the (P)QNSP in our model is that the no-signalling principle of the theory under consideration should take into account the resources available to every observer. The global state is not available; the local state and the local gravitational field are. Likewise, we must impose no-signalling for both quantum and non-quantum (here, gravitational) measurements.

Our model satisfies the LNSSP.  Now $\rho^\mathrm{loc} (x_A) $ is independent of $M_B$ by definition, and $g(x_A)$ is independent of $M_B$, at least in the weak-field regime, because we use retarded potentials. Thus the model also satisfies the 
NSSP.

\subsection{Local beables and local causality}

\label{sec:local beables and local causality}

In local semiclassical gravity, $g_{ab}(x)$ is a beable \cite{bell_theory_2004} at $x \in \mathcal{M}$: it is a $c$-number assigned to every point of spacetime. This is the case notably because $g_{ab}$ is a (continuous) function $\mathcal{M} \to \mathbb{R}$. Whether $T_{ab}^\mathrm{tot}$, the Ricci, Weyl and Einstein tensors are also beables  is less clear. Indeed, these are generally distributions $\mathcal{D}(\mathcal{M}) \to \mathbb{R}$ and thus generate $c$-numbers depending on what probes them (i.e. what test function is evaluated against them). On the other hand, the metric is an element of reality that exists regardless of operational considerations.

It is then natural to ask whether metrics are necessarily Bell local in local semiclassical gravity. First, note that until local measurements around $x$ are conducted, $\rho^\mathrm{loc}(x)$ is independent of the outcome of any spacelike-separated quantum measurements, and $\hat{T}_{ab}(x)$ plays the role of an \emph{individual observable} in a polyperspective linguo, rather than that of a \emph{joint observable} (which would look like $\cdots \otimes \mathbb{1} \otimes \hat{T}_{ab}(x) \otimes \mathbb{1} \cdots$, including identity operators $\mathbb{1}_{x_\perp}$ at spacelike separated points $x_\perp \perp x$). Individual observables yield marginal probabilities, and are computed from local quantum states. This is not to say that there cannot be Bell correlations from the quantum stress-energy tensor: taking the expectation value of $\cdots \otimes \mathbb{1} \otimes \hat{T}_{ab}(x) \otimes \mathbb{1} \cdots$ with respect to a \emph{global} state $\Psi$ displays this. 

Thus, the gravitational field backreacts from a \emph{marginal probability} rather than from a conditional probability, and thus is not sensitive to the outcomes of measurements on spacelike separated systems \emph{unless the measurements are conducted in the vicinity of the gravitational field measurement itself}. We have that these beables are \emph{locally causal} in Bell's sense \emph{unless quantum measurements on bipartite systems are performed at both sites}: writing
\begin{equation}
    \mathscr{T}(x) := \left\{T_{ab}^\text{tot}(f) \mid \text{supp } f \subset J^-(x) \right\} 
\end{equation}
for the set of expectation values of the stress-energy tensor with respect to the local states (including quantum measurement updates) in the past light cone of $x$,
\begin{equation}
    \slashed{\mathscr{T}}(x) := \left\{T_{ab}^\text{tot}(f) \mid \text{supp } f \subset \mathcal{M} \smallsetminus J(x) \right\}
\end{equation}
for the set of expectation values of the stress-energy tensor with respect to the local states (including quantum measurement updates) in $\mathcal{M} \smallsetminus J(x)$,
\begin{equation}
    \slashed{g}(x) := \{g_{ab}(y) \mid y \notin J(x)\} 
\end{equation}
for the set of metrics at spacelike separations from $x$, and
\begin{equation}
    \mathscr{J}^-_g(x) := \{g_{ab}(y) \mid y \in J^-(x)\} 
\end{equation}
for the set of metrics in the past light cone of $x$, we have that $\forall x \in \mathcal{M}$, 
\begin{multline}
    \label{eqn:Bell locality condition}
    P\left(g_{ab}(x) \mid \mathscr{T}(x), \mathscr{J}^-_g(x), \slashed{\mathscr{T}}(x), \slashed{g}(x) \right) \\ = P\left(g_{ab}(x) \mid \mathscr{T}(x), \mathscr{J}^-_g(x) \right) \, ,
\end{multline}
whenever no \emph{quantum} measurement is conducted in the vicinity of $x$.  Unless \emph{quantum} measurements are conducted in the vicinity of $x$, the gravitational field at $x$ is insensitive (in the sense of Bell) to any measurement performed over spacelike separations. 

However, this ceases to be true if one allows for quantum measurements in the vicinity of the gravitational field measurements. For example, suppose Alice (at $x$) and Bob (at $y \perp x$) share a Bell state $\ket{\Phi^+} = \frac{1}{\sqrt{2}}(\ket{00} + \ket{11})$ in some spin basis (which can be translated to a position basis through a Stern-Gerlach apparatus). They perform a spin (quantum) measurement at $x_\text{meas} \in I^+(x) \cap (\mathcal{M} \smallsetminus J^+(y_\text{meas}))$ and $y_\text{meas} \in I^+(y) \cap (\mathcal{M} \smallsetminus J^+(x_\text{meas}))$, respectively, in the $\{\ket{0},\ket{1}\}$ basis. Suppose Bob measures outcome ``0''. The relevant polystate before either Alice or Bob measure is $\rho_\text{poly} = \frac{\mathbb{1}_A}{2} \oplus \frac{\mathbb{1}_B}{2} \oplus \ketbra{\Phi^+}$. The local quantum state of Alice after she measures (whether or not Bob has measured in the relevant foliation) is then $\rho_\text{loc}(x_\text{meas})=\ketbra{0}$. The gravitational field $g_{ab}^{(0)}$ on Alice's side after she measures will thus locally react from this outcome, which is Bell-correlated to Bob's outcome and thus breaks the Bell locality condition of Eqn. \eqref{eqn:Bell locality condition}. 

Thus, if both Alice and Bob share an entangled state and
\begin{enumerate}
    \item only Bob performs a quantum measurement on the entangled state, then Alice's gravitational field satisfies the Bell locality condition \eqref{eqn:Bell locality condition}, since her gravitational field still backreacts on $\mathbb{1}_A/2$ until she performs a quantum measurement.
    \item Alice performs a \emph{quantum} measurement on the entangled state, then Alice's gravitational field does not satisfy the Bell locality condition \eqref{eqn:Bell locality condition} since her gravitational field backreacts from the Bell-correlated outcome $\ketbra{0}$.
\end{enumerate}

We thus see that the gravitational sector of the theory is locally causal (Bell local) in the sense of Bell only if quantum measurements are not performed on spacelike separated entangled states. On the other hand, the quantum mechanical sector is not Bell local (as in the non-gravitational theory), and neither is the gravitational field if one allows for quantum measurements to be performed in the vicinity of the gravitational field. An immediate corollary is that experimentally demonstrating the Bell nonlocality of the gravitational field would not, by itself, be sufficient evidence for the quantisation of gravity. (See \cite{kent_proposed_2009,kent_semi-quantum_2018,das_how_2026,kent_2027_nonlocality} for experimental proposals to test the Bell (non-)locality of the gravitational field.)

\section{Falsifiability}

\label{sec:falsifiability}

We now explore ways to test the model developed in this paper. As discussed in the introduction, the Page-Geilker experiment \cite{page_indirect_1981} only falsifies Everettian approaches to semiclassical gravity. If one considers that after a quantum measurement the gravitational field backreacts only through the collapsed state, then there is no contradiction with experiments. More specifically, in the context of our model, the mass-selection and subsequent macroscopic amplification in the Page-Geilker setup lie in the balance's past light cone during the measurement. Hence, taking either an objective collapse approach to state updates, or an agnostic single-world view on selective updates, $\rho^{\mathrm{loc}}$ at the balance point already reflects the \textit{actual} branch, so the model predicts the observed (non-averaged) torque. Thus, the standard Page-Geilker setup does not falsify this model. Instead, one may look at other experimental setups to probe signatures of the semiclassicality of gravity. 

\subsection{GIE experiments}

Gravitationally induced entanglement (GIE) experiments consist in comparing the predictions between low-energy quantum gravity models, and those models of classical gravity which couple to quantum matter. The idea is to find experimental evidence for gravitationally mediated entanglement between two quantum particles: this would constitute some evidence for the quantum nature of gravity since the LOCC theorem ensures that local (in the quantum information theoretic sense) interactions and classical communication cannot, on average, generate entanglement.

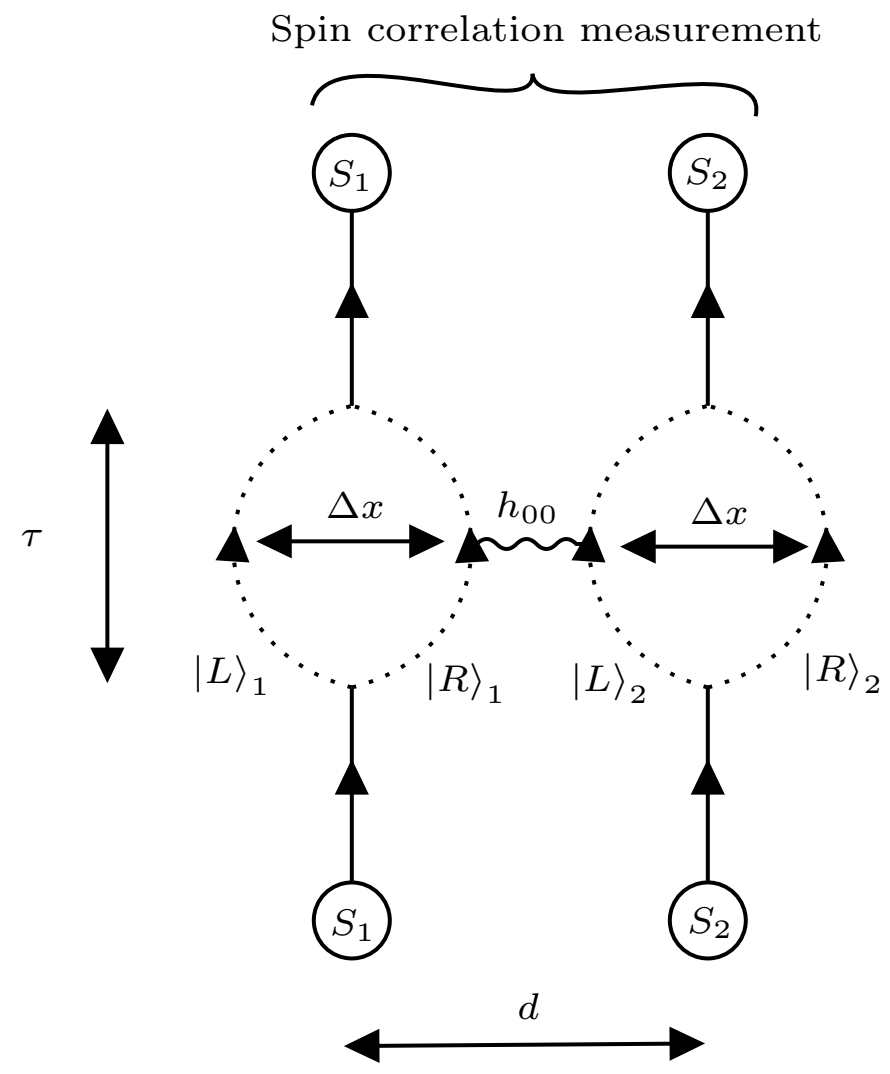
\begin{figure}
    \centering
    \begin{tikzpicture}[x=0.75pt,y=0.75pt,yscale=-1,xscale=1]

\draw    (370,191.07) -- (370,240) ;
\draw [shift={(370,209.03)}, rotate = 90] [fill={rgb, 255:red, 0; green, 0; blue, 0 }  ][line width=0.08]  [draw opacity=0] (8.93,-4.29) -- (0,0) -- (8.93,4.29) -- cycle    ;
\draw  [dash pattern={on 0.84pt off 2.51pt}]  (370,191.07) .. controls (410,181.07) and (410,130.4) .. (370,120.4) ;
\draw [shift={(399.64,150.74)}, rotate = 86.6] [fill={rgb, 255:red, 0; green, 0; blue, 0 }  ][line width=0.08]  [draw opacity=0] (8.93,-4.29) -- (0,0) -- (8.93,4.29) -- cycle    ;
\draw  [dash pattern={on 0.84pt off 2.51pt}]  (370,191.07) .. controls (330,180.4) and (330.67,130.4) .. (370,120.4) ;
\draw [shift={(340.64,150.52)}, rotate = 93.71] [fill={rgb, 255:red, 0; green, 0; blue, 0 }  ][line width=0.08]  [draw opacity=0] (8.93,-4.29) -- (0,0) -- (8.93,4.29) -- cycle    ;
\draw    (370,71.47) -- (370,120.4) ;
\draw [shift={(370,89.43)}, rotate = 90] [fill={rgb, 255:red, 0; green, 0; blue, 0 }  ][line width=0.08]  [draw opacity=0] (8.93,-4.29) -- (0,0) -- (8.93,4.29) -- cycle    ;
\draw    (280.67,191.07) -- (280.67,240) ;
\draw [shift={(280.67,209.03)}, rotate = 90] [fill={rgb, 255:red, 0; green, 0; blue, 0 }  ][line width=0.08]  [draw opacity=0] (8.93,-4.29) -- (0,0) -- (8.93,4.29) -- cycle    ;
\draw  [dash pattern={on 0.84pt off 2.51pt}]  (280.67,191.07) .. controls (320.67,181.07) and (320.67,130.4) .. (280.67,120.4) ;
\draw [shift={(310.3,150.74)}, rotate = 86.6] [fill={rgb, 255:red, 0; green, 0; blue, 0 }  ][line width=0.08]  [draw opacity=0] (8.93,-4.29) -- (0,0) -- (8.93,4.29) -- cycle    ;
\draw  [dash pattern={on 0.84pt off 2.51pt}]  (280.67,191.07) .. controls (240.67,180.4) and (241.33,130.4) .. (280.67,120.4) ;
\draw [shift={(251.3,150.52)}, rotate = 93.71] [fill={rgb, 255:red, 0; green, 0; blue, 0 }  ][line width=0.08]  [draw opacity=0] (8.93,-4.29) -- (0,0) -- (8.93,4.29) -- cycle    ;
\draw    (280.67,71.47) -- (280.67,120.4) ;
\draw [shift={(280.67,89.43)}, rotate = 90] [fill={rgb, 255:red, 0; green, 0; blue, 0 }  ][line width=0.08]  [draw opacity=0] (8.93,-4.29) -- (0,0) -- (8.93,4.29) -- cycle    ;
\draw    (312,155.07) .. controls (313.67,153.4) and (315.33,153.4) .. (317,155.07) .. controls (318.67,156.74) and (320.33,156.74) .. (322,155.07) .. controls (323.67,153.4) and (325.33,153.4) .. (327,155.07) .. controls (328.67,156.74) and (330.33,156.74) .. (332,155.07) .. controls (333.67,153.4) and (335.33,153.4) .. (337,155.07) -- (339.33,155.07) -- (339.33,155.07) ;
\draw   (271.13,249.53) .. controls (271.13,244.27) and (275.4,240) .. (280.67,240) .. controls (285.93,240) and (290.2,244.27) .. (290.2,249.53) .. controls (290.2,254.8) and (285.93,259.07) .. (280.67,259.07) .. controls (275.4,259.07) and (271.13,254.8) .. (271.13,249.53) -- cycle ;
\draw   (360.47,249.53) .. controls (360.47,244.27) and (364.73,240) .. (370,240) .. controls (375.27,240) and (379.53,244.27) .. (379.53,249.53) .. controls (379.53,254.8) and (375.27,259.07) .. (370,259.07) .. controls (364.73,259.07) and (360.47,254.8) .. (360.47,249.53) -- cycle ;
\draw   (271.13,61.93) .. controls (271.13,56.67) and (275.4,52.4) .. (280.67,52.4) .. controls (285.93,52.4) and (290.2,56.67) .. (290.2,61.93) .. controls (290.2,67.2) and (285.93,71.47) .. (280.67,71.47) .. controls (275.4,71.47) and (271.13,67.2) .. (271.13,61.93) -- cycle ;
\draw   (360.47,61.93) .. controls (360.47,56.67) and (364.73,52.4) .. (370,52.4) .. controls (375.27,52.4) and (379.53,56.67) .. (379.53,61.93) .. controls (379.53,67.2) and (375.27,71.47) .. (370,71.47) .. controls (364.73,71.47) and (360.47,67.2) .. (360.47,61.93) -- cycle ;
\draw    (281.67,281.04) -- (366.33,280.42) ;
\draw [shift={(369.33,280.4)}, rotate = 179.58] [fill={rgb, 255:red, 0; green, 0; blue, 0 }  ][line width=0.08]  [draw opacity=0] (8.93,-4.29) -- (0,0) -- (8.93,4.29) -- cycle    ;
\draw [shift={(278.67,281.07)}, rotate = 359.58] [fill={rgb, 255:red, 0; green, 0; blue, 0 }  ][line width=0.08]  [draw opacity=0] (8.93,-4.29) -- (0,0) -- (8.93,4.29) -- cycle    ;
\draw    (270.67,45) .. controls (274.67,26.4) and (327.33,50.4) .. (326,37.07) ;
\draw    (382,47.73) .. controls (386,29.13) and (327.33,50.4) .. (326,37.07) ;
\draw    (219.33,124.07) -- (219.33,187) ;
\draw [shift={(219.33,190)}, rotate = 270] [fill={rgb, 255:red, 0; green, 0; blue, 0 }  ][line width=0.08]  [draw opacity=0] (8.93,-4.29) -- (0,0) -- (8.93,4.29) -- cycle    ;
\draw [shift={(219.33,121.07)}, rotate = 90] [fill={rgb, 255:red, 0; green, 0; blue, 0 }  ][line width=0.08]  [draw opacity=0] (8.93,-4.29) -- (0,0) -- (8.93,4.29) -- cycle    ;
\draw    (392.33,155.73) -- (351,155.73) ;
\draw [shift={(348,155.73)}, rotate = 360] [fill={rgb, 255:red, 0; green, 0; blue, 0 }  ][line width=0.08]  [draw opacity=0] (8.93,-4.29) -- (0,0) -- (8.93,4.29) -- cycle    ;
\draw [shift={(395.33,155.73)}, rotate = 180] [fill={rgb, 255:red, 0; green, 0; blue, 0 }  ][line width=0.08]  [draw opacity=0] (8.93,-4.29) -- (0,0) -- (8.93,4.29) -- cycle    ;
\draw    (301,154.4) -- (259.67,154.4) ;
\draw [shift={(256.67,154.4)}, rotate = 360] [fill={rgb, 255:red, 0; green, 0; blue, 0 }  ][line width=0.08]  [draw opacity=0] (8.93,-4.29) -- (0,0) -- (8.93,4.29) -- cycle    ;
\draw [shift={(304,154.4)}, rotate = 180] [fill={rgb, 255:red, 0; green, 0; blue, 0 }  ][line width=0.08]  [draw opacity=0] (8.93,-4.29) -- (0,0) -- (8.93,4.29) -- cycle    ;

\draw (273.67,245.4) node [anchor=north west][inner sep=0.75pt]  [font=\scriptsize]  {$S_1$};
\draw (273,57.4) node [anchor=north west][inner sep=0.75pt]  [font=\scriptsize]  {$S_1$};
\draw (363.33,244.73) node [anchor=north west][inner sep=0.75pt]  [font=\scriptsize]  {$S_2$};
\draw (362.67,56.73) node [anchor=north west][inner sep=0.75pt]  [font=\scriptsize]  {$S_2$};
\draw (315.33,140.4) node [anchor=north west][inner sep=0.75pt]  [font=\scriptsize]  {$h_{00}$};
\draw (239,181.4) node [anchor=north west][inner sep=0.75pt]  [font=\scriptsize]  {$\ket{L}_{1}$};
\draw (334,183.4) node [anchor=north west][inner sep=0.75pt]  [font=\scriptsize]  {$\ket{L}_{2}$};
\draw (297.33,183.4) node [anchor=north west][inner sep=0.75pt]  [font=\scriptsize]  {$\ket{R}_{1}$};
\draw (392,181.07) node [anchor=north west][inner sep=0.75pt]  [font=\scriptsize]  {$\ket{R}_{2}$};
\draw (320.67,266.4) node [anchor=north west][inner sep=0.75pt]  [font=\scriptsize]  {$d$};
\draw (258.67,21) node [anchor=north west][inner sep=0.75pt]  [font=\scriptsize] [align=left] {Spin correlation measurement};
\draw (196,150.07) node [anchor=north west][inner sep=0.75pt]  [font=\scriptsize]  {$\tau $};
\draw (364,142.07) node [anchor=north west][inner sep=0.75pt]  [font=\scriptsize]  {$\Delta x$};
\draw (272.67,140.73) node [anchor=north west][inner sep=0.75pt]  [font=\scriptsize]  {$\Delta x$};

\end{tikzpicture}

    \caption{\justifying The Bose-et.al.-Marletto-Vedral gravitationally induced entanglement experiment \cite{bose_spin_2017, marletto_gravitationally_2017}. Two particles $S_1$ and $S_2$ are each placed in a superposition of wavepackets $\ket{L}+\ket{R}$. During at duration $\tau$, the wavepackets $\ket{R}_1$ and $\ket{L}_2$ are sufficiently close that, if gravity is quantum mechanical, the gravitational interaction between these branches generates a phase difference compared to the other branches, and thus generates entanglement which can, in principle, be measured. On the other hand, in (local) semiclassical gravity, no entanglement is generated.}
    \label{fig:BMV}
\end{figure}

One of the many proposals in the search for signatures of the quantum nature of gravity is the Bose-et.al-Marletto-Vedral (BMV) experiment \cite{bose_spin_2017,marletto_gravitationally_2017}, depicted in Fig. \ref{fig:BMV}, which we now review.

Consider two quantum particles $S_i$ with spin $1/2$, $i=1,2$, of mass $m_i$, initially in well-localised wavepackets of width $r_i$, separated by a distance $d \gg r_i$. Each $S_i$ individually is sent into a Stern-Gerlach experiment at time $t_0$ so that each $S_i$ is in a superposition $\frac{1}{\sqrt{2}}(\ket{L}_i + \ket{R}_i)$ just after $t=t_0$. These branches are separated by a distance $\Delta x \gg r_i$, and are centred around the initial trajectory prior to the beam splitting so that the distance between $\ket{R}_1$ and $\ket{L}_2$ is $d-\Delta x$ while the distance between $\ket{L}_1$ and $\ket{R}_2$ is $d+\Delta x$ and the distance between $\ket{L}_1$ and $\ket{L}_2$, and between $\ket{R}_1$ and $\ket{R}_2$, is $d$. We assume that $\braket{L}{R} \approx 0$ throughout.

At time $t = t_0 + \tau$ the wavepackets of each individual $S_1$ are recombined. Shortly after, a spin correlation measurement on the final state of $S_1+S_2$ is undertaken with spin entanglement witness
\begin{equation}
    W : \psi \mapsto \abs{\expval{\sigma_x^{(1)} \otimes \sigma_z^{(2)}}_\psi + \expval{\sigma_y^{(1)} \otimes \sigma_y^{(2)}}_\psi} \, .
\end{equation}
For a separable state $\rho$ we have $W(\rho) \leq 1$, while some entangled states give $W(\rho) > 1$, i.e. $W$ is indeed an entanglement witness.

Shortly after $t=t_0$, the quantum state of $S_1+S_2$ in both the quantum and semiclassical gravitational cases is\footnote{For simplicity, we ignore decoherence effects and interactions from non-gravitational sources including  Casimir effects in Eqns.~\eqref{eqn:state qg}-\eqref{eqn:witness scg}}. $\ket{\Psi(t_0)}_{12} = \frac{1}{2}(\ket{L}_1 + \ket{R}_1) \otimes (\ket{L}_2 + \ket{R}_2)$. Let us compare the predictions made by quantum gravity and (local) semiclassical gravity, respectively.

\paragraph*{Quantum gravity.} In the quantum gravitational case, each branch interacts gravitationally in a different fashion, which generates phase differences in between branches. The state at time $t=t_0+\tau$ is \cite[Equation 2]{bose_spin_2017}
\begin{multline}
    \label{eqn:state qg}
    \ket{\Psi(t_0 + \tau)}_{12} = \frac{e^{i \phi}}{2}(\ket{LL} + \ket{RR} \\ + e^{i\Delta \phi_{LR}} \ket{LR} + e^{i \Delta \phi_{RL}}\ket{RL})
\end{multline}
where $\phi = \frac{Gm_1m_2 \tau}{\hbar d}$, $\phi_{LR} = \frac{Gm_1m_2 \tau}{\hbar (d+\Delta x)}$, $\phi_{RL} = \frac{Gm_1m_2 \tau}{\hbar (d-\Delta x)}$, $\Delta \phi_{LR} = \phi_{LR}-\phi$, $\Delta \phi_{RL} = \phi_{RL}-\phi$. This is just evolving each branch unitarily using the standard Schrödinger equation with respect to their respective gravitational interactions.

The resulting value of the spin entanglement witness shortly after recombination is
\begin{multline}
    W(\Psi(t_0+\tau)) = \frac{1}{2}\Big|-1+\cos(\Delta \phi_{RL}) \\ -\cos(\Delta \phi_{LR}) + \cos(\Delta \phi_{LR}-\Delta\phi_{RL})\Big| \, .
\end{multline}
This is greater than $1$ whenever 
\begin{equation}
    \cos(\Delta \phi_{RL}) -\cos(\Delta \phi_{LR}) + \cos(\Delta \phi_{LR}-\Delta\phi_{RL}) < -1
\end{equation}
which is the case e.g. when $\Delta \phi_{RL} \approx \pi$ while $\Delta \phi_{LR} \approx 0$.

\paragraph*{Semiclassical gravity.} In the semiclassical gravitational case, each branch interacts gravitationally through the mean gravitational field. Writing $\psi_{ia}(\vec{x}_i) := \braket{\vec{x}_i}{a}$ for $i=1,2$ and $a=L,R$, we have
\begin{multline}
    \Psi(t_0;\vec{x}_1,\vec{x}_2) = \frac{1}{2} \left(\psi_{1L}(\vec{x}_1) + \psi_{1R}(\vec{x}_1)\right)\\ \cdot \left(\psi_{2L}(\vec{x}_2)+\psi_{2R}(\vec{x}_2)\right),
\end{multline}
where each wavepacket is normalised. Assuming negligible overlap with $\braket{L}{R} \approx 0$,
\begin{equation}
    \abs{\Psi(t_0;\vec{x}_1,\vec{x}_2)}^2 \approx \frac{1}{4} \sum_{a,b=L,R} \abs{\psi_{1a}(\vec{x}_1)}^2 \abs{\psi_{2b}(\vec{x}_2)}^2.
\end{equation}
Note that integrating over particle $1$ yields
\begin{align}
    \iiint_{\mathbb{R}^3} & \abs{\Psi(t_0;\vec{x}_1,\vec{x}_2)} d^3 \vec{x}_1  \nonumber \\&= \frac{1}{4} \sum_{a,b=L,R} \underbrace{\iiint_{\mathbb{R}^3} \abs{\psi_{1a}(\vec{x}_1)}^2 d^3 \vec{x}_1}_{=1} \abs{\psi_{2b}(\vec{x}_2)}^2 \\
    &= \frac{1}{2} \left(\abs{\psi_{2L}(\vec{x}_2)}^2 + \abs{\psi_{2R}(\vec{x}_2)}^2\right),
\end{align}
and likewise if one integrates over particle $2$. The mass density associated with $\Psi$ is
\begin{multline}
    \hat{\varrho}_\Psi(t,\vec{x}) = m_1 \iiint_{\mathbb{R}^3} \abs{\Psi(t;\vec{x},\vec{x}_2)}^2 d^3 \vec{x}_2 \\ + m_2 \iiint_{\mathbb{R}^3} \abs{\Psi(t;\vec{x}_1,\vec{x})}^2 d^3 \vec{x}_1\\ \approx \frac{m_1}{2}\left(\abs{\psi_{1L}(\vec{x})}^2 + \abs{\psi_{1R}(\vec{x})}^2\right) \\ + \frac{m_2}{2}\left(\abs{\psi_{2L}(\vec{x})}^2 + \abs{\psi_{2R}(\vec{x})}^2\right).
\end{multline}
Here, we assume that the mass density is roughly constant over the timescales considered. The gravitational potential satisfies
\begin{align}
    \Phi_\Psi(t,\vec{x}) &= - G \iiint_{\mathbb{R}^3} \frac{\hat{\varrho}_\Psi(t,\vec{x}')}{\abs{\vec{x}-\vec{x}'}} d^3 \vec{x}'\\ 
    &\approx -\frac{G}{2} \sum_{j=1}^2 m_j \sum_{b=L,R} \iiint_{\mathbb{R}^3} \frac{\abs{\psi_{jb}(\vec{x}')}^2}{\abs{\vec{x}-\vec{x}'}} d^3 \vec{x}'.
\end{align}
For $i,j = 1,2$ and $a,b = L,R$, define
\begin{equation}
    I_{ia,jb}[\Psi] := \int_{\mathbb{R}^3 \times \mathbb{R}^3} \frac{\abs{\psi_{ia}(\vec{x})}^2\abs{\psi_{jb}(\vec{x}')}^2}{\abs{\vec{x}-\vec{x}'}} d^3 \vec{x} d^3 \vec{x}'.
\end{equation}
The gravitational energy assigned to branch $a$ of particle $i$ is the matrix element
\begin{equation}
    E_{ia}[\Psi](t) := m_i \iiint_{\mathbb{R}^3} \abs{\psi_{ia}(\vec{x})}^2 \Phi_\Psi(t,\vec{x}) d^3 \vec{x}
\end{equation}
which is roughly constant in the setup we consider. For the full field,
\begin{align}
    E_{ia}[\Psi] &= -\frac{Gm_i}{2} \sum_{j=1}^2 m_j \sum_{b=L,R} I_{ia,jb}[\Psi] \\
    &= \underbrace{-\frac{G m_i^2}{2} (I_{ia,iL}[\Psi] + I_{ia,iR}[\Psi])}_{\text{self-interaction}} \\
    &\underbrace{- \frac{G m_i m_{\bar{i}}}{2} (I_{ia,\bar{i}L}[\Psi] + I_{ia,\bar{i}R}[\Psi])}_{\text{mutual interaction}}.
\end{align}
The joint branch $\ket{ab}$, $a,b=L,R$ has energy $E_{1a}[\Psi] + E_{2b}[\Psi]$. Moreover , we can write \cite{struyve_absence_2026}
\begin{equation}
    \ket{\Psi(t_0+\tau)} = \operatorname{T\exp}\left(-\frac{i}{\hbar} \int_{t_0}^{t_0+\tau} H[\Psi](t') dt'\right) \ket{\Psi(t_0)}
\end{equation}
for the state-dependent total Hamiltonian $H[\Psi]$. But the Schrödinger-Newton (gravitational) Hamiltonian is a multiplication operator in the position basis,
\begin{multline}
    (H_{SN}[\Psi] \Psi)(t;\vec{x}_1,\vec{x}_2) = (m_1 \Phi_\Psi(t,\vec{x}_1) \\ + m_2 \Phi_\Psi(t,\vec{x}_2))\Psi(t;\vec{x}_1,\vec{x}_2).
\end{multline}
Thus, writing $\psi_{ia}(\vec{x}) = \braket{\vec{x}}{a}_i$ for $i=1,2$ and $a=L,R$, we have
\begin{multline}
    \bra{ab} H_{SN}[\Psi]\ket{cd} = \int_{\mathbb{R}^3 \times \mathbb{R}^3} d^3 \vec{x}_1 d^3 \vec{x}_2 \psi_{1a}(\vec{x}_1)^* \psi_{2b}(\vec{x}_2)^* \\ \times (m_1 \Phi_\Psi(\vec{x}_1) + m_2 \Phi_\Psi(\vec{x}_2)) \\ \times \psi_{1c}(\vec{x}_1) \psi_{2d}(\vec{x}_2) \\ = V_{ac}^{(1)}[\Psi] \delta_{bd} + \delta_{ac} V^{(2)}_{bd}[\Psi]
\end{multline}
where 
\begin{equation}
    V_{ac}^{(i)}[\Psi] := m_i \iiint_{\mathbb{R}^3} d^3 \vec{x} \psi_{ia}(\vec{x})^* \Phi_\Psi(\vec{x}) \psi_{ic}(\vec{x})
\end{equation}
and we have assumed that the branches are orthonormal $\braket{a}{c}_i = \delta_{ac}$. Note that 
\begin{equation}
    V_{aa}^{(i)}[\Psi] = m_i \iiint_{\mathbb{R}^3}  \abs{\psi_{ia}(\vec{x})}^2 \Phi_\Psi(\vec{x})d^3 \vec{x} = E_{ia}[\Psi].
\end{equation}
Suppose that the $L$ and $R$ wavepackets have disjoint spatial supports, i.e. $\psi_{iL}(\vec{x})^* \psi_{iR}(\vec{x}) \approx 0$. Then
\begin{equation}
    V_{LR}^{(i)}[\Psi] = V_{RL}^{(i)}[\Psi] \approx 0.
\end{equation}
Then
\begin{equation}
    \bra{ab} H_{SN}[\Psi]\ket{cd} \approx (E_{1a}[\Psi] + E_{2b}[\Psi]) \delta_{ac} \delta_{bd}.
\end{equation}
Thus,
\begin{align}
    H_{SN}[\Psi] &\approx \sum_{a,b=L,R} (E_{1a}[\Psi] + E_{2b}[\Psi])\ketbra{ab} \\
    &= \sum_{a=L,R} E_{1a} \ketbra{a}_1 \otimes \mathbb{1} + \mathbb{1} \otimes \sum_{b=L,R} E_{2b}[\Psi] \ketbra{b}_2.
\end{align}
Thus, we have branch-wise state-dependent evolution of the form (ignoring other non-gravitational kinetic contributions)
\begin{multline}
    \ket{\Psi(t_0+\tau)} \approx \frac{1}{2} \left(e^{-iE_{1L} \tau/\hbar} \ket{L}_1 + e^{-iE_{1R} \tau/\hbar} \ket{R}_1 \right) \\ \otimes \left(e^{-iE_{2L} \tau/\hbar} \ket{L}_2 + e^{-iE_{2R} \tau/\hbar} \ket{R}_2\right).
\end{multline}
This is a product state\footnote{It would still be a product state if we were to consider non-constant gravitational fields over the timescales under consideration. Indeed, the exponentials of the gravitational energies would simply become time-ordered exponentials of the time-dependent gravitational energies.} and therefore
\begin{equation}
    \label{eqn:witness scg}
    W(\Psi(t_0+\tau)) \leq 1 \, .
\end{equation}
We can thus, in principle, distinguish the quantum gravitational case from the semiclassical gravitational case.

Several roadblocks affect the undertaking of this experiment to test the model presented in this paper. One of them is environmental decoherence. Another is the Casimir interaction between the $\ket{R}_1$ and $\ket{L}_2$ branches: these cannot be too close to each other, otherwise there may be non-gravitational sources of entanglement generation. Such GIE setups test quantum gravity against a wide class of alternatives, including semiclassical models as well as stochastic gravitational models. However, there are  simpler experiments comparing quantum gravity directly against local semiclassical gravity. We now examine some of these.

\subsection{Violations of the MEP} 

We can also consider a setup that provides a falsifiable prediction against low-energy quantum gravity within a thought experiment similar to that of Fig. \ref{fig:gravitational RDs} using violations of the mixture equivalence principle \cite{kent_quantum_2021,kent_testing_2021,fedida_mixture_2025}. Consider two small inertial labs $L$ and $R$ of characteristic size $r$, with synchronised clocks, separated by a distance $d$. Let $t=0$ at $L$ be the time at which a collapse-inducing measurement is performed there. Any point $x$ in $R$ is outside $J^+(L)$ until $t \approx d/c$ provided $r \ll d$. Let us compare ``before'' ($t<d/c -r/c$) and ``after'' ($t>d/c + r/c$) light-cone arrival regimes.

At $R$, prepare a mesoscopic point-like mass $m$ in a spatial superposition of two well-localised wavepackets $\ket{\vec{x}_1}, \ket{\vec{x}_2}$ separated by $\Delta = \abs{\vec{x}_1-\vec{x}_2}$. In this regime, the superposition principle holds to a very good approximation, and so prepare the joint state with $L$ as
\begin{equation}
    \ket{\psi}_{LR} \approx \frac{1}{\sqrt{2}}(\ket{0}_L \ket{\vec{x}_1}_R + \ket{1}_L \ket{\vec{x}_2}_R)
\end{equation}
where $\ket{0},\ket{1}$ are orthogonal internal states at $L$ (e.g. spin or polarisation). This can be created by a standard entangling gate.

At $t=0$ ($L$'s clock), perform a rapid, macroscopically amplified projective measurement on the $L$ qubit in the $\{\ket{0},\ket{1}\}$ basis so that wavefunction reduction is completed within a negligible time $T_r$ compared to $d/c$.

Continuously record (with time intervals $\delta t \ll d/c$). the probe signal at $R$ beginning well before $t=0$, through $t \approx d/c$, and after. 

Let us first focus on the semiclassical predictions. For a single mass in the two-packet state, the potential near the probe before any collapse signals have arrived is the Born-weighted average
\begin{equation}
    \Phi_{SC}(\vec{y},t<d/c) \approx -Gm \left(\frac{\abs{a_1}^2}{\abs{\vec{y}-\vec{x}_1}} + \frac{\abs{a_2}^2}{\abs{\vec{y}-\vec{x}_2}}\right)
\end{equation}
where $\abs{a_1} = \abs{a_2} = \frac{1}{\sqrt{2}}$. Once collapse information from $L$ is in $J^-(\vec{y},t)$ (i.e. $t > d/c)$, the local state is pure $\ket{\vec{x}_i}\bra{\vec{x}_i}$ and
\begin{equation}
    \Phi_{SC}(\vec{y},t>d/c) \approx \frac{-Gm}{\abs{\vec{y}-\vec{x}_i}}
\end{equation}
where $i=1$ or $i=2$. There is thus a swift (though continuous) change in the local value of the gravitational field between $t<d/c$ and $t>d/c$. This is very much analogous to the setup for a one-shot violation of the gravitational mixture equivalence principle \cite{fedida_mixture_2025}: a Cavendish-style measurement typically distinguishes an improper mixture (pre-collapse, averaged field) from a proper mixture (post-collapse, definite field).  On the other hand, in perturbative quantum gravity, the local gravitational field would only be sourced by a single source in each branch of the superposition. This yields a different prediction for the gravitational field compared to our semiclassical model.

It is sufficient for a test of local semiclassical gravity of this type to record whether or not there is a jump at $t=d/c$. One may simply compare whether the mean value of the signal, averaged over experimental noise, jumps at that time, or whether it does not.

\subsection{Testing local semiclassical gravity near measurement events} 

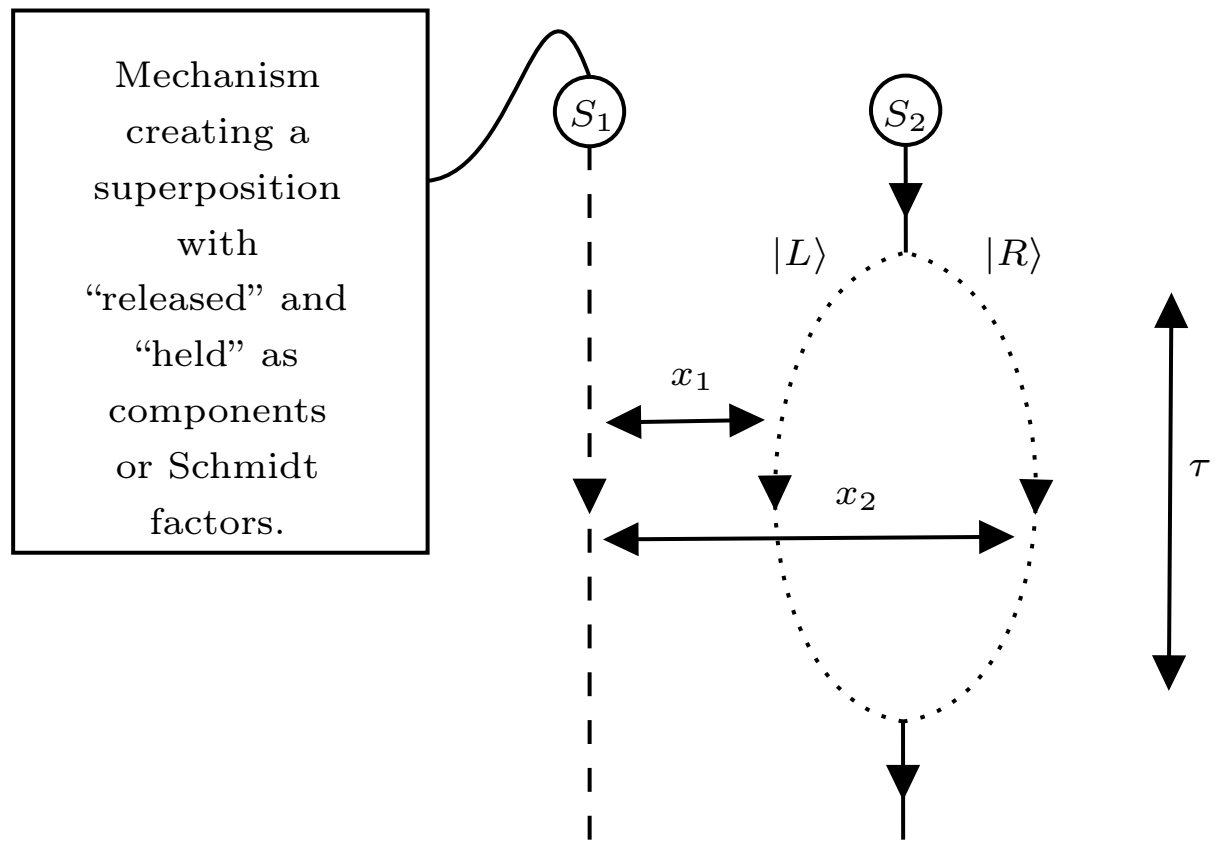
\begin{figure}
    \centering
    \begin{tikzpicture}[x=0.75pt,y=0.75pt,yscale=-1,xscale=1]

\draw   (306.6,55.03) .. controls (306.6,50.23) and (310.5,46.33) .. (315.3,46.33) .. controls (320.1,46.33) and (324,50.23) .. (324,55.03) .. controls (324,59.84) and (320.1,63.73) .. (315.3,63.73) .. controls (310.5,63.73) and (306.6,59.84) .. (306.6,55.03) -- cycle ;
\draw   (385.93,54.7) .. controls (385.93,49.9) and (389.83,46) .. (394.63,46) .. controls (399.44,46) and (403.33,49.9) .. (403.33,54.7) .. controls (403.33,59.5) and (399.44,63.4) .. (394.63,63.4) .. controls (389.83,63.4) and (385.93,59.5) .. (385.93,54.7) -- cycle ;
\draw    (394.63,63.4) -- (394.67,90.4) ;
\draw [shift={(394.66,81.9)}, rotate = 269.93] [fill={rgb, 255:red, 0; green, 0; blue, 0 }  ][line width=0.08]  [draw opacity=0] (8.93,-4.29) -- (0,0) -- (8.93,4.29) -- cycle    ;
\draw  [dash pattern={on 4.5pt off 4.5pt}]  (315.3,63.73) -- (315.33,238.4) ;
\draw [shift={(315.32,156.07)}, rotate = 269.99] [fill={rgb, 255:red, 0; green, 0; blue, 0 }  ][line width=0.08]  [draw opacity=0] (8.93,-4.29) -- (0,0) -- (8.93,4.29) -- cycle    ;
\draw    (393.97,208.07) -- (394,237.73) ;
\draw [shift={(393.99,227.9)}, rotate = 269.94] [fill={rgb, 255:red, 0; green, 0; blue, 0 }  ][line width=0.08]  [draw opacity=0] (8.93,-4.29) -- (0,0) -- (8.93,4.29) -- cycle    ;
\draw  [dash pattern={on 0.84pt off 2.51pt}]  (393.97,208.07) .. controls (350.67,201.07) and (351.33,100.4) .. (394.67,90.4) ;
\draw [shift={(361.92,155.26)}, rotate = 269.22] [fill={rgb, 255:red, 0; green, 0; blue, 0 }  ][line width=0.08]  [draw opacity=0] (8.93,-4.29) -- (0,0) -- (8.93,4.29) -- cycle    ;
\draw  [dash pattern={on 0.84pt off 2.51pt}]  (393.97,208.07) .. controls (437.33,202.4) and (439.33,101.73) .. (394.67,90.4) ;
\draw [shift={(427.15,156.26)}, rotate = 271.63] [fill={rgb, 255:red, 0; green, 0; blue, 0 }  ][line width=0.08]  [draw opacity=0] (8.93,-4.29) -- (0,0) -- (8.93,4.29) -- cycle    ;
\draw   (170.67,29.67) -- (274.67,29.67) -- (274.67,165.73) -- (170.67,165.73) -- cycle ;
\draw    (461.31,103.4) -- (460.69,197.4) ;
\draw [shift={(460.67,200.4)}, rotate = 270.38] [fill={rgb, 255:red, 0; green, 0; blue, 0 }  ][line width=0.08]  [draw opacity=0] (8.93,-4.29) -- (0,0) -- (8.93,4.29) -- cycle    ;
\draw [shift={(461.33,100.4)}, rotate = 90.38] [fill={rgb, 255:red, 0; green, 0; blue, 0 }  ][line width=0.08]  [draw opacity=0] (8.93,-4.29) -- (0,0) -- (8.93,4.29) -- cycle    ;
\draw    (274.67,72.4) .. controls (298,70.4) and (302.67,10.4) .. (315.3,46.33) ;
\draw    (322.33,133.02) -- (356.33,132.45) ;
\draw [shift={(359.33,132.4)}, rotate = 179.05] [fill={rgb, 255:red, 0; green, 0; blue, 0 }  ][line width=0.08]  [draw opacity=0] (8.93,-4.29) -- (0,0) -- (8.93,4.29) -- cycle    ;
\draw [shift={(319.33,133.07)}, rotate = 359.05] [fill={rgb, 255:red, 0; green, 0; blue, 0 }  ][line width=0.08]  [draw opacity=0] (8.93,-4.29) -- (0,0) -- (8.93,4.29) -- cycle    ;
\draw    (321.67,162.38) -- (419,161.75) ;
\draw [shift={(422,161.73)}, rotate = 179.63] [fill={rgb, 255:red, 0; green, 0; blue, 0 }  ][line width=0.08]  [draw opacity=0] (8.93,-4.29) -- (0,0) -- (8.93,4.29) -- cycle    ;
\draw [shift={(318.67,162.4)}, rotate = 359.63] [fill={rgb, 255:red, 0; green, 0; blue, 0 }  ][line width=0.08]  [draw opacity=0] (8.93,-4.29) -- (0,0) -- (8.93,4.29) -- cycle    ;

\draw (308.67,50.73) node [anchor=north west][inner sep=0.75pt]  [font=\scriptsize]  {$S_1$};
\draw (387.33,50.73) node [anchor=north west][inner sep=0.75pt]  [font=\scriptsize]  {$S_2$};
\draw (186.67,41) node [anchor=north west][inner sep=0.75pt]   [align=left] {\begin{minipage}[lt]{50.72pt}\setlength\topsep{0pt}
\begin{center}
{\scriptsize Mechanism creating a}\\{\scriptsize superposition  }\\{\scriptsize with ``released'' and ``held'' as components or Schmidt factors.}
\end{center}

\end{minipage}};
\draw (359.33,85.07) node [anchor=north west][inner sep=0.75pt]  [font=\scriptsize]  {$\ket{L}$};
\draw (412.67,85.07) node [anchor=north west][inner sep=0.75pt]  [font=\scriptsize]  {$\ket{R}$};
\draw (464,141.07) node [anchor=north west][inner sep=0.75pt]  [font=\scriptsize]  {$\tau $};
\draw (334,118.07) node [anchor=north west][inner sep=0.75pt]  [font=\scriptsize]  {$x_{1}$};
\draw (375.33,148.07) node [anchor=north west][inner sep=0.75pt]  [font=\scriptsize]  {$x_{2}$};

\end{tikzpicture}

    \caption{\justifying Testing quantum gravity near measurement events \cite{kent_testing_2021-1}.   A quantum particle (a sphere $S_1$) is put in a superposition of two states (``released'' or ``held''), or into part of a larger entangled system in which these states are Schmidt factors.   The experimental parameters are such that the superposition or entanglement is maintained through the experiment, until the final observation. Another quantum particle $S_2$ travelling parallel to the ``released'' path of $S_1$ is placed into a superposition of a left branch $\ket{L}$, close to the path of $S_1$, and a right branch $\ket{R}$, farther away. If local semiclassical gravity describes the world, there will be a local phase generated by the gravitational interaction between $S_1$ and $\ket{L}$ whether or not the outcome obtained from the indeterministic quantum experiment, actuated after this whole setup is undertaken, is ``released'' or ``held''. On the other hand, for low energy quantum gravity, one expects a local phase to be generated only in the ``released'' outcome, and so after post-selection on the ``held'' outcome one can verify which of these hypotheses (if any) is verified.}
    \label{fig:testing qg near mmts}
\end{figure}

GIE experiments are very challenging to implement in the lab because they require a very low temperature, very high quality vacuum, and separations between the mesoscopic masses that are constrained not to be too small by the Casimir-Polder effect. Instead, one may consider specific tests of mean-field models directly. We here recall a proposal first laid out in \cite{kent_testing_2021}, whose setup is shown in Fig. \ref{fig:testing qg near mmts}, which we adapt to our model.

Consider two spheres $S_i$ ($i=1,2$) of radius $r_i$ and mass $m_i$, of the same density $\rho$ i.e. $m_i = \frac{4}{3} \pi r_i^3 \rho$. Consider, in particular, the case $m_1 \geq m_2$. 

First, $S_1$ is prepared in a superposition state: one outcome is associated to its ``release'' at time $t=0$, after which it falls under the Earth's gravity; the other to ``holding'' it in place so that it does not fall.    In principle this may be a pure superposition state, or $S_1$ may be entangled with another mesoscopic system used to prepare the state and/or with environmental and/or other degrees of freedom.   
In any case, the assumption here is that the experimental parameters are such that no intrinsic collapse affecting the state of $S_1$ takes place during the experiment so that the superposition or entanglement of $S_1$ persists until the final observation.   This quantitative constraint on the experiment 
depends on the model of collapse or measurement being considered together with local semiclassical gravity.   It requires that the superposition/entanglement is not amplified ``too much'' (as parametrised by the relevant model) by interactions with the environment or experimental apparatus during the experiment, until the final observation.  It implies that $S_1$ remains in a superposition or improper mixture during the experiment, so that, according to local semiclassical gravity, its gravitational field is the weighted average of the ``released'' and ``held'' states.   

At time $t=0$, $S_2$ is then released in parallel to the trajectory of the ``released'' state of $S_1$ at a distance $d$. At time $t=t_0$, $S_2$ goes into a beam splitter (or a Stern-Gerlach apparatus) with one branch $\ket{L}$ held close (at distance $x_1$) to the ``released'' trajectory of $S_1$, while the other branch $\ket{R}$ is farther away (at distance $x_2$), so that initially $\ket{\Psi(t_0)}_{S_2} = \frac{1}{\sqrt{2}}(\ket{L} + \ket{R})$ (up to a global phase). At time $t=t_0 + \tau$, the branches recombine (through another beam splitter or a Stern-Gerlach apparatus). Let us compare the state of $S_2$ at time $t_1=t_0 + \tau - \epsilon$ for $\epsilon \ll \tau$ just before recombination depending on whether quantum gravity or semiclassical gravity governs nature.

\paragraph*{Quantum gravity.} In quantum gravity, the state of $S_2$ at time $t_1$ is (up to a global phase)
\begin{equation}
    \label{eqn:qg prediction held}
    \ket{\Psi(t_1)}_{S_2,\text{held},\text{qg}} \approx \frac{1}{\sqrt{2}}(\ket{L}+\ket{R}) \, ,
\end{equation}
in the branch where $S_1$ is held, while it is
\begin{equation}
    \label{eqn:qg prediction released}
    \ket{\Psi(t_1)}_{S_2,\text{released},\text{qg}} \approx \frac{e^{i \phi_L \tau}}{\sqrt{2}}(\ket{L}+e^{i(\phi_R-\phi_L)\tau}\ket{R})
\end{equation}
in the branch where $S_1$ is released, where
\begin{equation}
    \phi_L = \frac{G m_1 m_2}{\hbar x_1} , \quad \phi_R = \frac{G m_1 m_2}{\hbar x_2} \, .
\end{equation}

\paragraph*{Semiclassical gravity.} In (local) semiclassical gravity, the gravitational potential take the same (mean) value in both branches of $S_1$, and so the state of $S_2$ at time $t_1$ is
\begin{equation}
    \label{eqn:scg prediction}
    \ket{\Psi(t_1)0}_{S_2,\text{scg}} \approx \frac{e^{i \phi_L \tau/2}}{\sqrt{2}}(\ket{L}+e^{i(\phi_R-\phi_L)\tau/2}\ket{R}) \, .
\end{equation}
If
\begin{equation}
    (\phi_R - \phi_L) \tau \approx 1 \mod 2\pi \, ,
\end{equation}
we can distinguish Eqns. \eqref{eqn:qg prediction held}, \eqref{eqn:qg prediction released} and \eqref{eqn:scg prediction} by a measurement in the $\left\{\frac{\ket{L} \pm \ket{R}}{\sqrt{2}}\right\}$ basis.

The real advantage of this experimental proposal comes now: after the final recombination, the experimenter measures whether $S_1$ was held or released.  They discard the runs where it was found that the ``released'' outcome was actualised, and keep only the ``held'' runs. 
They then repeat the experiment in a different setup, in which $S_1$ is always prepared in the ``held'' state.  In this version of the experiment, both local semiclassical gravity and quantum gravity yield the same prediction \eqref{eqn:qg prediction held}.   They then, compare the results of this experiment with those of the subensemble in which the ``$S_1$ was held'' outcome was actualised: for quantum gravity, these are identical statistics; for local semiclassical gravity, these are different statistics. 

This is useful for implementing an experiment for the following reason. When $S_1$ is held, then so long as the initial distance $d$ between $S_1$ and $S_2$ are chosen so that their Casimir-Polder (CP) interactions \cite{casimir_influence_1948} are negligible, the
CP interactions can be neglected throughout any run of
the experiment (in either version) in which $S_1$ is not
released. These interactions are governed by quantum
electrodynamics, not by a semiclassical theory. Thus, $x_1$ can be very small compared to $d$ and the Casimir-Polder interaction between $S_1$ and $\ket{L}$ considerable when $S_1$ is released, but this is insignificant in the cases where $S_1$ is held. Even for runs where $S_1$ is released, this experiment
does not necessarily require the CP potential to be
smaller than the gravitational potential, so long as the CP
potential’s contribution to the overall phase can be precisely
estimated. This is one of the reasons such a proposal is more easily implementable than other GIE proposals \cite{bose_spin_2017,marletto_gravitationally_2017}: we can work with $m_2 \ll m_1$, i.e. the interferometry of $S_2$ can be done with small masses so long as the mass of $S_1$ is larger; and the interferometry can also last for a shorter amount of time. 
A further advantage is that interactions between the superposition state of $S_1$ and the environment can be allowed, even though they decohere the superposition, so long as they do not amplify the superposition to an extent that the relevant collapse or measurement model predicts that a collapse or measurement probably takes place.

A new challenge comes in that one needs to couple a decohered state that has not collapsed to a mesoscopic mass.   However, closely related microscopic-to-mechanical and mesoscopic mechanical quantum-control ingredients have already been demonstrated experimentally \cite{oconnell_quantum_2010,bild_schrodinger_2023}.

\subsection{Gravitational waves and gravitons} 

Another avenue for testing mean-field models is through gravitational waves. Ford \cite{ford_gravitational_1982} compared the classical gravitational wave energy flux obtained from the linearized semiclassical Einstein equation to the graviton-emission flux in quantised gravity. They agree for coherent states but disagree macroscopically for number states, thermal states, and superpositions: semiclassical radiation can vanish when the quantum gravitational prediction is nonzero, because the product of averages is not the average of products: $\expval{\hat{T}\hat{T}} \neq \expval{\hat{T}}\expval{\hat{T}}$ in general, which still holds true when the expectation value is taken with respect to the local state. This is a theoretical disagreement which, to our knowledge, has not yet been experimentally adjudicated. See \cite{tobar_detecting_2024} for a discussion on the experimental verifiability of this and related effects.

More recently, proposals to test some quantum aspects of gravity involve the direct detection of gravitons \cite{tobar_detecting_2024,loughlin_wave-particle_2025,tobar_gravito-phononic_2025}. However, it has been argued (e.g. see \cite{carney_graviton_2024,carney_comments_2025}) that direct graviton detection via current gravitational wave interferometers would not constitute evidence in favour of quantum gravity, since semiclassical gravity can replicate the expected signal through classical gravitational waves. It has been argued that \cite{carney_graviton_2024} ``a source of nonclassical (e.g., squeezed or entangled) gravitational radiation has to be produced, and then detected with a gravitational detector with high efficiency'' in order to differentiate between semiclassical and quantum gravity predictions. This is beyond current technology.

\section{Conclusions and Outlook}

\label{sec:conclusion}

``Naïve'' M{\o}ller-Rosenfeld semiclassical gravity is riddled by many issues, including 1) superluminal signalling, 2) likely inconsistency, 3) empirical falsification if consistent, 4) foliation dependence in the gravitational sector, 5) lack of integrability in the quantum dynamics. We have presented a theory that addresses these problems and gives a consistent picture, at least in the linearised regime in the examples we have analyzed. We showed that this ``cured'' semiclassical theory still recovers the Schrödinger-Newton system in the nonrelativistic limit, and so can be considered as a relativistic completion of this nonlinear quantum mechanical model.

The construction of local quantum states yields a gravitational field that only depends on past data. Hence, provided quantum measurements induce a collapse in the future light cone, no superluminal signalling is possible, either in the gravitational or in the quantum matter sector. We showed that although the gravitational field is classical in local semiclassical gravity, it is not Bell local in general. Local semiclassical gravity makes distinct predictions which
may perhaps seem exotic but have not been falsified to date.

We finished by discussing proposals for some tabletop experimental tests. 

Open questions remain. For example, it would be interesting to show the independence of the choice of the sequence $\{\Sigma_n\} \to \Lambda^-(x)$ on $\rho^\mathrm{loc}$ in the appropriate topology of quantum states. A related question is whether the local quantum state defines a normal algebraic state (i.e. a density operator), and whether it is Hadamard. It would also be interesting to understand the precise setting in which the three constructions of the notion of local quantum state coincide. We leave this for future work.

It also remains to be shown that the model is consistent, even in the linearised regime, for more than two measurements in arbitrary spacetime configurations.

In our model, the classical null energy condition can be violated in the future light cone of the region of a quantum measurement. One open question is to see whether averaged or quantum energy conditions are also violated. Another interesting question is to understand whether this can be exploited to generate exotic and possibly problematic behaviours, i.e. whether the theory is endangered by such violations.

A related open question is whether the distributional ``collapse energy'' only interacts gravitationally: there is no direct contact term between the Barrabès-Israel stress-energy and quantum matter or quantum forces.  If so, the shell can only be probed operationally through gravitational experiments.
If not, other effects (of unknown form and magnitude) should in principle be observable when two independent masses are placed in separate superpositions, the position of one is observed, and the future light cone of this observation crosses the other while it is still superposed.    

Even more speculatively, it would be interesting to understand whether this ``collapse energy'' could possibly be related to dark energy. We showed in Sec.~\ref{subsec:mmt qubit in superposition} that ``collapse energy'' can violate the classical null energy condition depending on the outcome of quantum state collapses and thus generate negative pressure, hence potentially altering the cosmic expansion rate over large scales. Note that in the stationary wavepackets approximation we have considered in Sec.~\ref{sec:examples},
\begin{equation}
    \sum_{i} p_i T_{\mu\nu}^{+,(i)} \approx \underline{T}_{\mu\nu}^- \Rightarrow \sum_{i} p_i \Delta T_{\mu\nu}^{(i)} \approx 0,
\end{equation}
so since the singular stress-energy in linear in the source jump (see Eqns.~\eqref{eq:BIc-p-source}--
 \eqref{eq:BIc-mu-source-intermediate}), we have
\begin{equation}
    \sum_{i} p_i T_{\mu\nu}^{\mathrm{sing},(i)} \approx 0.
\end{equation}
Thus the existence of positive and negative conditional shells does not itself demonstrate a nonzero mean cosmological source. This could nonetheless stem, for example, from random fluctuations in the outcome distribution of quantum state collapses, or from looking at more general scenarios beyond the stationary wavepackets approximation.  We presently have no estimate of the magnitude or even the sign of this potential effect in cosmology, however.

We have shown that, in the linearised regime, there is no issue arising from crossing shells when collapses occur over spacelike separations. An important open question is whether this also holds when post-Newtonian gravitational corrections are included.

The study of renormalisation should also be undertaken in future work. In the gravitational sector, this issue is related to the well-posedness of the semiclassical field equations given sufficiently regular initial data and allowing quantum state collapses. If the gravitational field suffers from finite-time blowups (perhaps through the formation of a black hole singularity), the addition of $R^2$ terms in the field equations may be worthy of consideration. This would lead to a form of semiclassical quadratic gravity.

We have presented several ways to probe the phenomenology of this theory in tabletop experiments. It would be very interesting to explore other ways to test this model against low energy perturbative quantum gravity and other alternatives to quantum gravity, e.g. \cite{oppenheim_postquantum_2023}. 

In summary, although important open questions remain, we do not presently see a strong reason not to consider the model presented in this paper as a credible alternative to quantum gravity, whether on experimental, technical or conceptual grounds.

We finish with three open questions that deserve further exploration.

\subsection{Is there a black hole information paradox in our model?}

\label{subsec:BHIP}

More work is required to explore the black hole information paradox in the context of the model presented in this paper. In semiclassical gravity, quantum information in the sense of the von Neumann entropy is not the appropriate thermodynamic notion of information \cite{fedida_thermodynamics_2026}, and indeed von Neumann entropy is not necessarily conserved under nonlinear time evolution. To see this, consider the following example \cite{peres_quantum_1993,fedida_thermodynamics_2026}: let $\ket{\alpha(t)}$ and $\ket{\beta(t)}$ be two pure states, and
\begin{equation}
    \rho(t) = \epsilon \ketbra{\alpha(t)} + (1-\epsilon) \ketbra{\beta(t)}
\end{equation}
be a \emph{proper} mixture with $0 < \epsilon < 1$ and nonvanishing eigenvalues
\begin{equation}
    \lambda_{1,2}(t) = \frac{1}{2} \pm \sqrt{\frac{1}{4} - \epsilon(1-\epsilon)[1-x(t)]}
\end{equation}
where $x(t) = \abs{\braket{\alpha(t)}{\beta(t)}}^2$. The von Neumann entropy of this \emph{proper} mixture $S_{VN}(t) = - \sum_{i} \lambda_i(t) \log(\lambda_i(t))$ satisfies $\frac{dS_{VN}}{dx(t)}<0$ for all $\epsilon$ and $t$. Thus,
\begin{equation}
    \forall t, \, \frac{dS_{VN}}{dt}(t) \geq 0 \Leftrightarrow \forall t, \, \frac{dx(t)}{dt} \leq 0
\end{equation}
hence $\forall t \geq 0$, $\abs{\braket{\alpha(t)}{\beta(t)}}^2 \leq \abs{\braket{\alpha(0)}{\beta(0)}}^2$. Consider a complete orthogonal set $\{\ket{\alpha_k(t)}\}$ where, for every $\ket{\beta(t)}$,
\begin{equation}
    \sum_k \abs{\braket{\alpha_k(t)}{\beta(t)}}^2 = 1 \, .
\end{equation}
Then if there exists an $m$ such that $\abs{\braket{\alpha_m(t)}{\beta(t)}}^2 < \abs{\braket{\alpha_m(0)}{\beta(0)}}^2$ then there must be some $n$ such that $\abs{\braket{\alpha_n(t)}{\beta(t)}}^2 > \abs{\braket{\alpha_n(0)}{\beta(0)}}^2$ which makes the von Neumann entropy decrease unless $\abs{\braket{\alpha_k(t)}{\beta(t)}}^2 = \abs{\braket{\alpha_k(0)}{\beta(0)}}^2$ for every $\ket{\alpha_k(t)}$ and $\ket{\beta(t)}$ and all $t$, which from Wigner's theorem implies that time evolution is unitary (antiunitarity being excluded by continuity). Thus, since time evolution is non-unitary, we have that the von Neumann entropy of a proper mixture can decrease through time evolution, and so \emph{quantum} information can fluctuate in the Schrödinger-Newton equation. Note however that there is no violation of the second law for an appropriately generalised notion of entropy \cite{fedida_thermodynamics_2026}.

The unitarity of the quantum state evolution, alongside the conservation of quantum information, is often considered to be the other major challenge in explaining the emergence of a mixed Hawking-Gibbs state from a pure quantum state. Here, the time evolution is nonlinear in the quantum state and does not conserve von Neumann entropy. Is this sufficient to explain a state evolution from approximately pure to mixed of the form $\ket{\psi}\bra{\psi} \to \rho$? The answer is not straightforwardly positive. To see this, consider the nonrelativistic limit of local semiclassical gravity, i.e. the Schrödinger-Newton system, and how a \emph{pure} state $\rho(t) = \ketbra{\Psi(t)}$ evolves. We have
\begin{equation}
    \frac{d\rho}{dt} = \left(\frac{\partial}{\partial t} \ket{\Psi(t)}\right) \bra{\Psi(t)} + \ket{\Psi(t)} \left(\frac{\partial}{\partial t} \bra{\Psi(t)}\right) \, .
\end{equation}
Note that the Schrödinger-Newton Hamiltonian is Hermitian, so for pure states
\begin{equation}
    \frac{d\rho}{dt} = \frac{1}{i\hbar} \comm{H_{SN}[\Psi]}{\rho} \, .
\end{equation}
Thus,
\begin{align}
    \frac{d}{dt}\Tr[\rho^2] &= 2 \Tr\left[\rho \frac{d\rho}{dt}\right] \\
    &= \frac{2}{i\hbar} \left(\Tr[\rho \left(H_{SN}[\Psi] \rho - \rho H_{SN}[\Psi]\right)]\right) \\
    &= \frac{2}{i\hbar} \left(\Tr[\rho^2 H_{SN}[\Psi]] - \Tr[\rho^2 H_{SN}[\Psi]]\right) \\
    &= 0 \, .
\end{align}
Hence, the Schrödinger-Newton system maps pure quantum states \emph{of the full system} to pure quantum states \emph{of the full system}. Thus, a consistent coupling between classical gravity and quantum matter need \emph{not} be diffusive \emph{on all quantum states}: the gravitational field does not decohere a pure quantum state of the full system in local semiclassical gravity. 

If this were the whole story, we would still be faced with an issue: a Hawking-Gibbs state cannot be obtained from a pure quantum state through a Schrödinger-Newton evolution. This however ignores the fact that the MEP is violated because of nonlinear evolution \cite{fedida_mixture_2025}. Further note that the ``amount'' of entanglement shared by multipartite pure states evolves through the Schrödinger-Newton equation \eqref{eqn:SNE}. For a pure total state $\ket{\Psi_{AB}}$, the purity of a reduced state $\rho_A$ is a measure of entanglement: if entanglement increases, purity decreases and \emph{vice-versa}. Hence, the purity $\Tr[\rho_A^2]$ can fluctuate.

Thus, a step towards a full understanding of this black hole scenario is that we do not have $\ketbra{\psi} \to \rho$ but rather $\rho_\text{region} \to \rho$, where $\rho_\text{region} \in \mathcal{A}(\mathscr{H})_{\mathrm{phys}}$ is a polystate which, even if it is initially \emph{almost} pure, might credibly evolve to a Gibbs state over the timescales of black hole evaporation. Note however that this theory cannot explain a transition between a quantum state of the full system which is \emph{exactly} pure and one which is mixed, even over an arbitrarily long timescale. If the initial state is indeed exactly pure, the paradox might perhaps still be solved by assuming it is a subsystem interacting gravitationally with other systems. In such cases, there seems to be no fundamental issue surrounding black hole evaporation, though a precise analysis of black holes would be very interesting to pursue in this model.

We here recall that there exists a difference between this semiclassical theory and the semiclassical limit of quantum gravity in the context of black hole spacetimes. The gravitational backreaction of the Hawking radiation from a quantum gravitational black hole in the semiclassical limit is empirically distinguishable to that of a fundamentally semiclassical black hole. This stems from the difference between gravitational backreactions sourced from proper and improper mixtures \cite{fedida_mixture_2025}. We thus encourage further exploration of the behaviour of black holes in this theory where spacetime is fundamentally classical and coupled to quantum matter in this local fashion. 

\subsection{Is there an action principle for local semiclassical gravity?}

It would be interesting to derive the present theory from an action principle, though the discussion of the model presented in this paper does not require one. An action exists for M{\o}ller-Rosenfeld semiclassical gravity, but becomes much trickier to formulate for the version of local semiclassical gravity we have examined.  Møller-Rosenfeld semiclassical gravity, which sources the gravitational field with the expectation value of the stress-energy tensor with respect to the \emph{global} state $\psi$, can be recovered from an action principle as \cite{kibble_non-linear_1980, kibble_is_1981}
\begin{multline}
    S[\psi,\alpha] = \int_{\mathbb{R}} \Big(\Im \braket{\dot{\psi}(t)}{\psi(t)} - \expval{\hat{H}}_{\psi(t)
    } \\ + \alpha(t)[\braket{\psi(t)} - 1]\Big) dt
\end{multline}
where $\alpha$ works as a Lagrange multiplier, and
\begin{equation}
    S[g] = \frac{1}{2 \kappa c} \int_{\mathcal{M}} R \sqrt{-g} \, d^4 x
\end{equation}
is the classical action for Einstein gravity, such that
\begin{equation}
    S[g,\psi,\alpha] = S[g] + S[\psi,\alpha] \, .
\end{equation}
In M{\o}ller-Rosenfeld semiclassical gravity, it is possible to work with certain renormalisation schemes in semiclassical theories \cite{randjbar-daemi_renormalization_1980} using such an action. However, it is unclear whether one can find a fully consistent renormalisation scheme with no ghosts unless one considers the theory to be an effective field theory: $R_{\mu \nu}R^{\mu \nu}$ renormalisation terms yield ghosts with a Hamiltonian unbounded from below (although whether such ghosts generally arise in field theories of classical gravity is unclear \cite{grudka_renormalisation_2024}.) For local semiclassical gravity, the discussion of renormalisability and ghosts may depend on the local nature of the theory we develop. 

\subsection{Is local semiclassical gravity underdetermined?}

Another interesting discussion in action-based formulations of semiclassical gravity which has been explored in the literature \cite{duff_inconsistency_1980} as an argument against such models, is that of a possible underdetermination which affects for semiclassical theories which include quantum fields. Though this argument has already been replied to by Kibble \cite{kibble_is_1981}, the point is interesting to highlight for completeness and pedagogy. The idea is the following: in ``standard'' QFT, field redefinitions at the classical level yields quantumly equivalent theories on shell. Consider a classical theory described by an action $S[g,\{\phi\}]$, with equations of motion
\begin{equation}
    \frac{\delta S}{\delta g_{ab}} = 0 \, , \qquad \frac{\delta S}{\delta \phi^i} = 0 \, .
\end{equation}
Consider another action $S'[g',\{\phi'\}]$ obtained from $S[g,\{\phi\}]$ through field redefinitions
\begin{align}
    g'_{ab} &= g_{ab} + f_g(g,\{\phi\}) \\
    \phi'^i &= \phi^i + f_i(g,\{\phi\})
\end{align}
where $f_g$ and $f_i$ are such that
\begin{equation}
    S'[g',\{\phi'\}] = S[g,\{\phi\}] \, .
\end{equation}
If these changes of variables do not involve derivatives of the fields, and both variables coincide in the weak-field limit (i.e. excluding changes of the form $\phi' \sim \phi^k$ where $k \neq 1$), then classically $S$ and $S'$ describe the same theory, with boundary conditions of the primed theory fully determined from those in the unprimed theory. $S$ and $S'$ also describe the same theory on-shell if all the fields are quantised\footnote{This can be seen by the invariance of the $S$-matrix under field redefinitions.}, but do \emph{not} describe the same theory (even on-shell) if only some fields are quantised. A concrete example showing inequivalent one-loop counterterms to on-shell divergences in the unprimed and primed theory of a real scalar field is given in \cite{duff_inconsistency_1980}.

Renormalisability also depends on classical field redefinitions: some classical field redefinitions lead to renormalisable semiclassical theories, others do not. In Duff's words \cite{duff_inconsistency_1980}, 
\begin{quote}
    ``[...] one must single out one choice of field variable for preferential treatment in the quantum theory, from among the infinitely many equivalent choices in the classical theory''
\end{quote}

This seems to lead to a notion of classical underdetermination. Finding a Grand Unified Theory for the matter sector is already difficult and usually driven by principles such as symmetry requirements, unitarity, renormalisability and so on, some of which need to be relaxed in this theory. It might thus be unclear why one would expect these methods to have worked in the first place (when gravity is not involved). 

Arguably this is not really an issue in that it ``simply'' gives physicists more work to do to find an accurate description of nature. Furthermore, this difficulty is arguably just a manifestation of the special essence of the gravitational field and of spacetime in contrast to quantum matter. In the words of Kibble \cite{kibble_is_1981},
\begin{quote}
    ``A semiclassical theory is necessarily based on the concept that the gravitational field has a distinguished place. In that context, it is not unreasonable that we must forbid transformations of the metric that are dependent on other fields.''
\end{quote}

We agree with this view: if gravity is indeed an ``arena'' (in the classical sense) in which quantum matter is of a different essence than spacetime and should thus be treated on a different footing, then the metric has a special role, and one cannot freely mix classical and quantum degrees of freedom together whilst hoping to end up with the same final theory.

\begin{acknowledgments}
We acknowledge financial support from UK-Canada Quantum for Science research collaboration grant OPP640. 
S.F. is funded by a studentship from the Engineering and Physical Sciences Research Council, grant number 2882481.  A.K. was supported in part by Perimeter Institute for
Theoretical Physics. Research at Perimeter Institute is supported by
the Government of Canada through the Department of Innovation, Science
and Economic Development and by the Province
of Ontario through the Ministry of Research, Innovation and Science.
Portions of Sections \ref{sec:weak field solutions}, \ref{subsec:late time}, \ref{sec:uniqueness future metric}, \ref{subsec:shell mass} and \ref{subsec:mmt qubit in superposition}, and of Appendices \ref{app:limit sequences Minkowski}, \ref{app:future metric}, \ref{app:distributional conservation} and \ref{app:computations qubit in superposition}, benefited from suggestions from Claude (Anthropic) and ChatGPT (OpenAI). These were also used to review the pre-submission draft and correct typos. All statements were independently drafted and verified by the authors.

\end{acknowledgments}

\appendix

\section{Limiting sequences in Minkowski spacetime}

\label{app:limit sequences Minkowski}

\begin{figure*}
    \centering
    \begin{subfigure}[t]{0.4\textwidth}
    \includegraphics[scale=0.31]{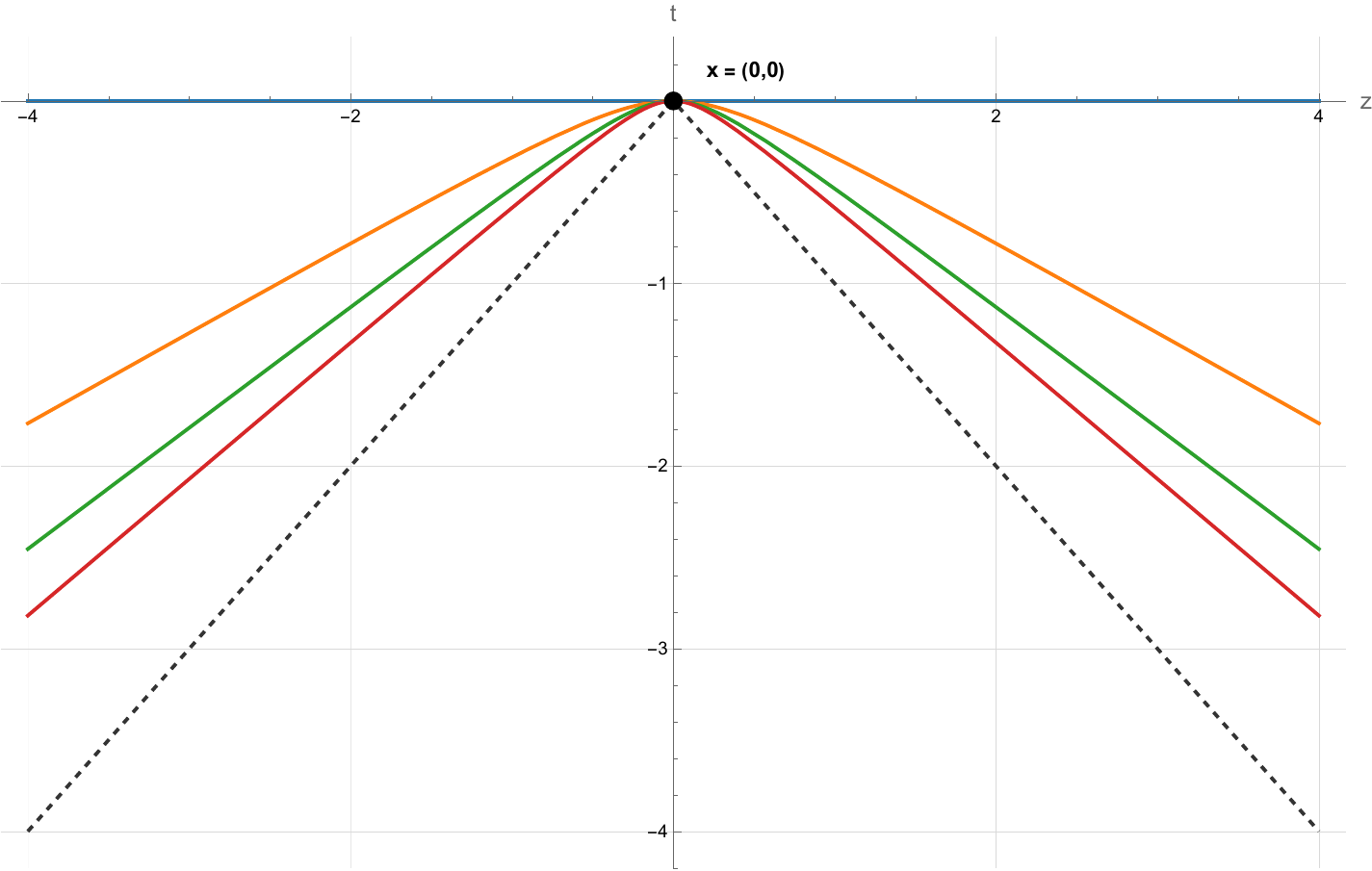}
    \caption{\justifying The sequence of spacelike Cauchy hypersurfaces $\Sigma_n$ in $1+1$ dimensions. $\Sigma_1$ is the hypersurface $t=0$ (shown in blue). As $n$ increases, they tend towards $\Lambda^-(0)$ shown in black dashed lines.}
    \label{fig:convergence 2d}
    \end{subfigure}
    \qquad
    \begin{subfigure}[t]{0.4\textwidth}
        \includegraphics[scale=0.4]{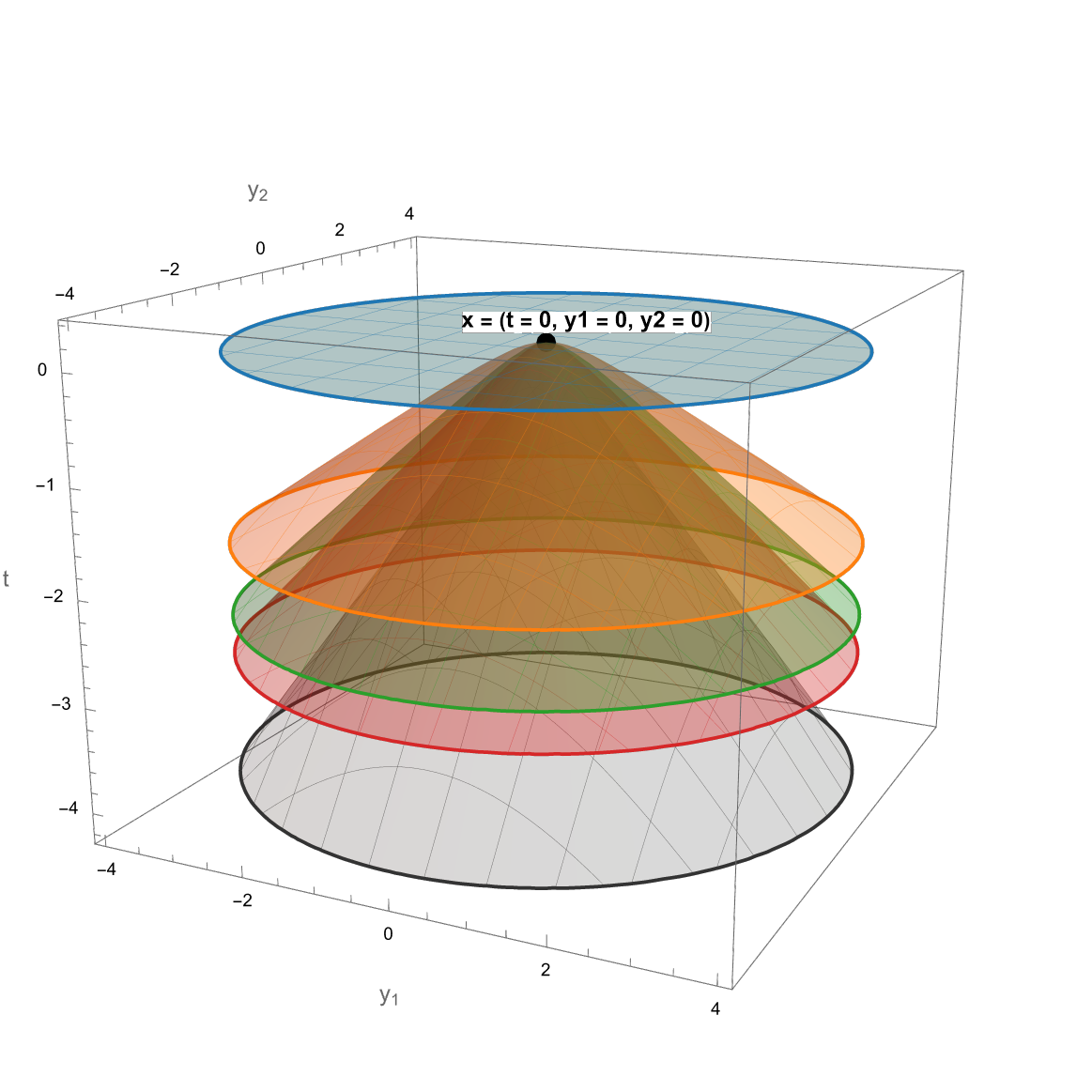}
    \caption{\justifying The sequence of spacelike Cauchy hypersurfaces $\Sigma_n$ in $1+2$ dimensions. $\Sigma_1$ is the hypersurface $t=0$ (shown in blue). As $n$ increases, they tend towards $\Lambda^-(0)$ shown in black.}
    \label{fig:convergence 3d}
    \end{subfigure}
    \caption{\justifying The first few terms of the sequence $\Sigma_n = \left\{(ct,\vec{y}) \mid ct=-\left(1-\frac{1}{n}\right)\left(\sqrt{\abs{\vec{y}}^2 + \frac{1}{n^2}}-\frac{1}{n}\right)\right\}$ for $c=1$, in $1+1$ (left) and $1+2$ (right) dimensional Minkowski spacetime.}
    \label{fig:convergence}
\end{figure*}

Choose global inertial Cartesian coordinates of Minkowski such that $x = (ct,\vec{x}) \in \mathbb{M}$. We construct a sequence of smooth spacelike Cauchy hypersurfaces $(\Sigma_n)$ such that
\begin{enumerate}
    \item $0 = (t=0, \vec{0}) \in \Sigma_n$ for all $n \in \mathbb{N}$,
    \item $\Sigma_n \to \Lambda^-(0)$ in $(\mathrm{CL}(\M),\tau_\mathrm{CF})$.
\end{enumerate}
Given a function $f: \mathbb{R}^3 \to \mathbb{R}$, we write
\begin{align}
    \mathrm{graph}(f) &:= \{(f(\vec{y}),\vec{y}) \mid \vec{y} \in \mathbb{R}^{3}\} \\
    &= \{(ct,\vec{y}) \in \mathbb{M} \mid ct = f(\vec{y})\}.
\end{align}
Let $f(\vec{y}) = - \abs{\vec{y}}$. Then
\begin{equation}
    \mathrm{graph}(f) = \{(-\abs{\vec{y}},\vec{y}) \mid \vec{y} \in \mathbb{R}^3\} = \{(ct,\vec{y}) \mid ct = -\abs{\vec{y}}\}.
\end{equation}
Note
\begin{equation}
    J^-(0) = \{(ct,\vec{y}) \mid ct \leq - \abs{\vec{y}}\},
\end{equation}
so
\begin{equation}
    \Lambda^-(0) = \partial J^-(0) = \{(ct,\vec{y}) \mid ct = -\abs{\vec{y}}\} = \mathrm{graph}(f).
\end{equation}
For all $n \geq 1$, let
\begin{equation}
    f_n(\vec{y}) = -\left(1-\frac{1}{n}\right)\left(\sqrt{\abs{\vec{y}}^2 + \frac{1}{n^2}}-\frac{1}{n}\right).
\end{equation}
Then let
\begin{align}
    \Sigma_n &:= \mathrm{graph}(f_n) \\
    &= \left\{(ct,\vec{y}) \mid ct=-\left(1-\frac{1}{n}\right)\left(\sqrt{\abs{\vec{y}}^2 + \frac{1}{n^2}}-\frac{1}{n}\right)\right\}.
\end{align}
The first four $\Sigma_n$ are shown alongside $\Lambda^-(0)$ in $1+1$ and $1+2$-dimensional Minkowski spacetime (for simplicity of visualisation) in Fig.\ref{fig:convergence}.

We need to show that
\begin{enumerate}
    \item Each $\Sigma_n$ is smooth,
    \item Each $\Sigma_n$ is spacelike,
    \item Each $\Sigma_n$ is a Cauchy hypersurface,
    \item $\forall p \in \Lambda^-(0)$, $\exists p_n \in \Sigma_n$ such that $p_n \to p$,
    \item For $n_j \uparrow \infty$, $p_j \in \Sigma_{n_j}$, $p_j \to p \Rightarrow p \in \Lambda^-(0)$.
\end{enumerate}

For 1., since $\vec{x} \mapsto \sqrt{\abs{\vec{x}}^2 + \alpha}$ is smooth on all of $\mathbb{R}^3$ including at $\vec{x}=\vec{0}$ if and only if $\alpha>0$, we have that for any fixed $n \geq 1$, $f_n$ is smooth everywhere including at $\vec{y}=\vec{0}$. The map
\begin{equation}
    F_n : \mathbb{R}^3 \to \mathbb{M}, \qquad F_n(\vec{y}) = (f_n(\vec{y}),\vec{y})
\end{equation}
is a smooth embedding, thus $\Sigma_n$ is a smooth hypersurface.

For 2., we have that
\begin{equation}
    \nabla f_n(\vec{y}) = -\left(1-\frac{1}{n}\right) \frac{\vec{y}}{\sqrt{\abs{\vec{y}}^2+\frac{1}{n^2}}},
\end{equation}
and thus $\abs{\nabla f_n(\vec{y})} \leq (1-\frac{1}{n}) < 1$. A tangent vector at $p = (f_n(\vec{y}),\vec{y}) \in \Sigma_n$ is the velocity of a smooth curve lying in $\Sigma_n$ and passing through $p$, which can be written $\lambda \mapsto (f_n(\vec{y}(\lambda)),\vec{y}(\lambda))$ with $\vec{y}(0) = \vec{y}$. Let $\vec{v} = \frac{d \vec{y}(\lambda)}{d\lambda}|_{\lambda = 0}$. The chain rule gives $\frac{d}{d\lambda} f_n(\vec{y}(\lambda))|_{\lambda=0} = \nabla f_n(\vec{y}) \cdot \vec{v}$, i.e. tangent vectors are of the form $(\nabla f_n(\vec{y}) \cdot \vec{v},\vec{v})$ for $\vec{v} \in \mathbb{R}^3$. Equivalently, 
\begin{equation}
    T_p \Sigma_n = \{(\nabla f_n(\vec{y}) \cdot \vec{v},\vec{v}) \mid \vec{v} \in \mathbb{R}^3\}.
\end{equation}
For any tangent vector $V \in T_p \Sigma_n$ with $\vec{v} \neq \vec{0}$,
\begin{multline}
    \eta(V,V) = \abs{\vec{v}}^2 - (\nabla f_n(\vec{y}) \cdot \vec{v})^2 \\ \geq \left(1-\left(1-\frac{1}{n}\right)^2\right) \abs{\vec{v}}^2 \\ = \frac{2n-1}{n^2} \abs{\vec{v}}^2 > 0.
\end{multline}
Thus every nonzero tangent vector is spacelike: the induced metric on $\Sigma_n$ is positive definite, i.e. $\Sigma_n$ is a spacelike hypersurface.

To prove Condition 3., we must establish the existence of an intersection between $\Sigma_n$ and every inextendible causal curve. Take an arbitrary inextendible future-directed causal curve. We parametrise the curve by inertial time $\tau = ct$. Along the future direction of the curve, $\tau$ increases. We write the curve as $\gamma(\tau) = (\tau,\vec{y}(\tau))$, where $\tau$ initially ranges over some interval $I \subseteq \mathbb{R}$. Note that since this is a causal curve and $\tau=ct$, 
\begin{equation}
    \abs{\frac{d\vec{y}}{d\tau}} \leq 1 \qquad \text{ almost everywhere},
\end{equation}
with, for any $\tau_1 < \tau_2 \in I$,
\begin{equation}
    \label{eqn:lipschitz causal curve}
    \abs{\vec{y}(\tau_2) - \vec{y}(\tau_1)} \leq \tau_2 - \tau_1.
\end{equation}
We now show that inextendibility forces the time range to be all of $\mathbb{R}$. Suppose, for contradiction, that the time coordinate has a finite upper endpoint $b$. As $\tau_1,\tau_2 \uparrow b$, we get
\begin{equation}
    \abs{\vec{y}(\tau_2) - \vec{y}(\tau_1)} \leq \abs{\tau_2-\tau_1} \rightarrow 0.
\end{equation}
Thus, the spatial position is Cauchy as $\tau \uparrow b$. Cauchy completeness of $\mathbb{R}^3$ gives a limit $\vec{y}(\tau) \to \vec{y}_b$. The curve consequently approaches the spacetime point $(b,\vec{y}_b)$. We could now continue it by remaining at the fixed spatial position $\vec{y}_b$:
\begin{equation}
    \tilde{\gamma}(\tau) := \begin{cases}
        (\tau,\vec{y}(\tau)) \qquad & \tau \in I, \\
        (\tau,\vec{y}_b) & \tau \geq b.
    \end{cases}
\end{equation}
This is a valid causal extension since taking the limit $\tau_2 \uparrow b$ gives
\begin{equation}
    \abs{\vec{y}_b - \vec{y}(\tau_1)} \leq b - \tau_1,
\end{equation}
so the spatial bound remains valid across the joining point and the appended portion is timelike and remains locally Lipschitz. This contradicts future inextendibility. The same argument applies for past inextendibility, i.e. the time coordinate is unbounded from above and below, as expected, i.e. $I=\mathbb{R}$ for any inextendible causal curve in Minkowski spacetime.

We use this to prove the existence of the crossing of all such inextendible causal curves with $\Sigma_n$. To do so, we measure the curve's position relative to the graph. For any $\tau \in \mathbb{R}$,
\begin{align}
    d_n(\tau) &:= \tau - f_n(\vec{y}(\tau)) \\
    &= \tau + \left(1-\frac{1}{n}\right)\left(\sqrt{\abs{\vec{y}(\tau)}^2 + \frac{1}{n^2}}-\frac{1}{n}\right).
\end{align}
This is continuous. Since at the spatial position $\vec{y}(\tau)$, the hypersurface has time coordinate $f_n(\vec{y}(\tau))$, we have that
\begin{itemize}
    \item $d_n(\tau) = 0$ means precisely that $\gamma(\tau) \in \Sigma_n$,
    \item $d_n(\tau) > 0$ means that the curve lies in the timelike future of the graph at that spatial position,
    \item $d_n(\tau) <0$ means that the curve lies in the timelike past of the graph at that spatial position.
\end{itemize}
This is because given a curve $\gamma(\tau) = (\tau,\vec{y}(\tau))$, the unique point on $\Sigma_n$ with spatial coordinates $\vec{y}(\tau)$ is $(f_n(\vec{y}(\tau)),\vec{y}(\tau))$. Thus, $\gamma(\tau) \in \Sigma_n$ if and only if $\tau = f_n(\vec{y}(\tau))$ if and only if $d_n(\tau)=0$. We must thus show that $d_n$ has exactly one zero.

Since $\abs{\nabla f_n(\vec{w})} \leq 1-\frac{1}{n}$ for any  $\vec{w} \in \mathbb{R}^3$,
\begin{multline}
    \label{eqn:bound difference fn's}
    \abs{f_n(\vec{z}) - f_n(\vec{y})} = \abs{\int_0^1 \nabla f_n(\vec{y}+\lambda(\vec{z}-\vec{y})) \cdot (\vec{z}-\vec{y}) d\lambda}  \\
     \leq \left(\int_0^1 \abs{\nabla f_n(\vec{y}+\lambda(\vec{z}-\vec{y}))} d\lambda\right) \times \abs{\vec{z}-\vec{y}}\\ \leq \left(1-\frac{1}{n}\right) \abs{\vec{z}-\vec{y}}
\end{multline}
by the fundamental theorem of calculus, and so we have, for $\tau_2 > \tau_1$, 
\begin{align}
    d_n(\tau_2) - d_n(\tau_1) &= (\tau_2-\tau_1) - [f_n(\vec{y}(\tau_2))-f_n(\vec{y}(\tau_1))] \\
    &\stackrel{\eqref{eqn:bound difference fn's}}{\geq} (\tau_2-\tau_1)- \left(1-\frac{1}{n}\right)\abs{\vec{y}(\tau_2)-\vec{y}(\tau_1)} \\
    &\stackrel{\eqref{eqn:lipschitz causal curve}}{\geq} (\tau_2-\tau_1) - \left(1-\frac{1}{n}\right)(\tau_2-\tau_1) \\ &= \frac{\tau_2-\tau_1}{n} > 0.
\end{align}
Fix any $\tau_0 \in \mathbb{R}$. For $\tau \geq \tau_0$, $d_n(\tau) \geq d_n(\tau_0) + \frac{\tau-\tau_0}{n}$, i.e.
\begin{equation}
    \lim_{\tau \to +\infty} d_n(\tau) = +\infty.
\end{equation}
For $\tau \leq \tau_0$, $d_n(\tau) \leq d_n(\tau_0) + \frac{\tau-\tau_0}{n}$, i.e.
\begin{equation}
    \lim_{\tau \to -\infty} d_n(\tau) =- \infty.
\end{equation}
Since $d_n$ is continuous, the intermediate value theorem implies a time $\tau_*$ such that $d_n(\tau_*)=0$. Since $d_n$ is strictly increasing, this intersection is unique. Thus, every inextendible causal curve intersects $\Sigma_n$ exactly once. Hence, every $\Sigma_n$ is a smooth spacelike Cauchy hypersurface.

For 4., given
\begin{equation}
    p = (-\abs{\vec{y}},\vec{y}) \in \Lambda^-(0),
\end{equation}
take
\begin{equation}
    p_n = (f_n(\vec{y}), \vec{y}) \in \Sigma_n.
\end{equation}
Then $p_n \to p$, which proves 4. 

Condition 5.~says ``take any strictly increasing sequence $n_j$ and any points $p_j \in \Sigma_{n_j}$. If $p_j$ converges to a spacetime point $p$, then $p$ must lie on $\Lambda^-(0)$''. Set $\epsilon_n = \frac{1}{n}$, $r = \abs{\vec{y}}$ so that
\begin{equation}
    f_n(\vec{y}) = -(1-\epsilon_n)\left(\sqrt{r^2+\epsilon_n^2}-\epsilon_n\right), \quad f(\vec{y})=-r.
\end{equation}
For $r \geq 0$, $r \leq \sqrt{r^2 + \epsilon_n^2} \leq r+\epsilon_n$. Hence,
\begin{align}
    f_n(\vec{y}) - f(\vec{y}) &= r - (1-\epsilon_n)\left(\sqrt{r^2 + \epsilon_n^2} - \epsilon_n\right) \\
    &= \underbrace{r - \left(\sqrt{r^2+\epsilon_n^2}-\epsilon_n\right)}_{\in [0,\epsilon_n]} + \epsilon_n\underbrace{\left(\sqrt{r^2+\epsilon_n^2}-\epsilon_n\right)}_{\in[0,r]} .
\end{align}
Hence,
\begin{equation}
    \label{eqn:bound fny - fy}
    0 \leq f_n(\vec{y}) - f(\vec{y}) \leq \epsilon_n + \epsilon_n r = \frac{1+\abs{\vec{y}}}{n}.
\end{equation}

Let $n_j \uparrow \infty$ and suppose that
\begin{equation}
    p_j = (f_{n_j}(\vec{y}_j),\vec{y}_j) \in \Sigma_{n}, \qquad p_j \longrightarrow p = (ct,\vec{y}).
\end{equation}
Componentwise convergence gives $\vec{y}_j \to \vec{y}$ and $f_{n_j}(\vec{y}_j) \to ct$. Combining Eqn.~\eqref{eqn:bound fny - fy} with Eqn.~\eqref{eqn:bound difference fn's} gives
\begin{align}
    \abs{f_{n_j}(\vec{y}_j)-f(\vec{y})} &\leq \abs{f_{n_j}(\vec{y}_j) - f_{n_j}(\vec{y})} + \abs{f_{n_j}(\vec{y}) - f(\vec{y})} \\ &\leq \left(1-\frac{1}{n_j}\right) \abs{\vec{y}_j - \vec{y}} + \frac{1+\abs{\vec{y}}}{n_j} \\
    &\longrightarrow 0.
\end{align}
Hence,
\begin{equation}
    ct = f(\vec{y}) = - \abs{\vec{y}}.
\end{equation}
Thus $p \in \Lambda^-(0)$. Thus, no sequence of points on the sequence $(\Sigma_n)$ can converge to a point away from the light cone.

\section{Convergence of metrics in spacetime}

\label{app:convergence metrics}

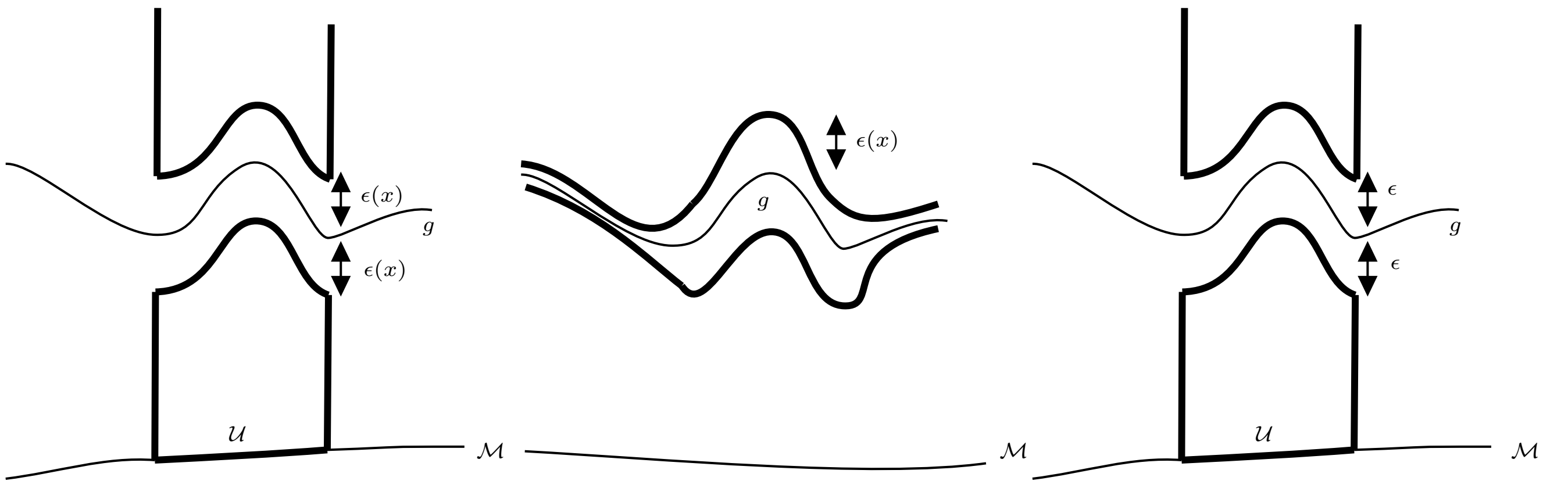
\begin{figure*}
    \centering
    \begin{subfigure}[t]{0.32\textwidth}
        \begin{tikzpicture}[x=0.75pt,y=0.75pt,yscale=-1,xscale=1]

\draw [line width=2.25]    (302.78,237.85) .. controls (328.4,236.53) and (355.07,235.2) .. (377.22,233.42) ;
\draw    (238.4,245.87) .. controls (261.08,243.1) and (280.99,236.19) .. (302.78,237.85) ;
\draw    (377.22,233.42) .. controls (417.04,232.04) and (395.47,232.04) .. (436.4,232.04) ;
\draw [line width=2.25]    (303.07,165.2) -- (302.78,237.85) ;
\draw [line width=2.25]    (377.73,166.53) -- (377.22,233.42) ;
\draw    (238.4,109.87) .. controls (251.73,109.2) and (282.45,140.03) .. (303.07,140.53) .. controls (323.68,141.03) and (321.07,122.53) .. (339.73,111.2) .. controls (358.4,99.87) and (370.11,142.34) .. (377.73,141.87) .. controls (385.36,141.39) and (407.07,127.2) .. (422.4,129.87) ;
\draw [line width=2.25]    (303.07,165.2) .. controls (330.4,164.53) and (330.4,134.53) .. (346.4,134.53) .. controls (362.4,134.53) and (363.07,161.87) .. (377.73,166.53) ;
\draw [line width=2.25]    (303.73,115.2) .. controls (331.07,114.53) and (331.07,84.53) .. (347.07,84.53) .. controls (363.07,84.53) and (363.73,111.87) .. (378.4,116.53) ;
\draw [line width=2.25]    (304.02,42.55) -- (303.73,115.2) ;
\draw [line width=2.25]    (378.91,49.65) -- (378.4,116.53) ;
\draw    (383.07,146.2) -- (383.07,164.2) ;
\draw [shift={(383.07,167.2)}, rotate = 270] [fill={rgb, 255:red, 0; green, 0; blue, 0 }  ][line width=0.08]  [draw opacity=0] (8.93,-4.29) -- (0,0) -- (8.93,4.29) -- cycle    ;
\draw [shift={(383.07,143.2)}, rotate = 90] [fill={rgb, 255:red, 0; green, 0; blue, 0 }  ][line width=0.08]  [draw opacity=0] (8.93,-4.29) -- (0,0) -- (8.93,4.29) -- cycle    ;
\draw    (383.07,116.2) -- (383.07,134.2) ;
\draw [shift={(383.07,137.2)}, rotate = 270] [fill={rgb, 255:red, 0; green, 0; blue, 0 }  ][line width=0.08]  [draw opacity=0] (8.93,-4.29) -- (0,0) -- (8.93,4.29) -- cycle    ;
\draw [shift={(383.07,113.2)}, rotate = 90] [fill={rgb, 255:red, 0; green, 0; blue, 0 }  ][line width=0.08]  [draw opacity=0] (8.93,-4.29) -- (0,0) -- (8.93,4.29) -- cycle    ;

\draw (440,229.07) node [anchor=north west][inner sep=0.75pt]  [font=\scriptsize]  {$\mathcal{M}$};
\draw (333.33,221.73) node [anchor=north west][inner sep=0.75pt]  [font=\scriptsize]  {$\mathcal{U}$};
\draw (416.67,133.4) node [anchor=north west][inner sep=0.75pt]  [font=\scriptsize]  {$g$};
\draw (391.33,149.73) node [anchor=north west][inner sep=0.75pt]  [font=\scriptsize]  {$\epsilon ( x)$};
\draw (390,117.73) node [anchor=north west][inner sep=0.75pt]  [font=\scriptsize]  {$\epsilon ( x)$};

\end{tikzpicture}
    \caption{A neighbourhood $B(\U,\epsilon,g)$ of the metric $g$ in the compact-open topology consists of all metrics which lie within $\epsilon(x)$ of $g$ over the compact set $\U$ of $\mathcal{M}$.}
    \label{fig:compact open topology}
    \end{subfigure}
    \begin{subfigure}[t]{0.32\textwidth}
        \begin{tikzpicture}[x=0.75pt,y=0.75pt,yscale=-1,xscale=1]

\draw    (220.67,233.67) .. controls (278.4,236.87) and (371.73,245.53) .. (419.73,238.87) ;
\draw    (219.07,114.2) .. controls (232.4,113.53) and (263.12,144.37) .. (283.73,144.87) .. controls (304.35,145.37) and (301.73,126.87) .. (320.4,115.53) .. controls (339.07,104.2) and (350.78,146.68) .. (358.4,146.2) .. controls (366.02,145.72) and (387.73,131.53) .. (403.07,134.2) ;
\draw [line width=2.25]    (288.4,162.2) .. controls (298.4,177.53) and (311.07,138.87) .. (327.07,138.87) .. controls (343.07,138.87) and (341.73,170.2) .. (358.4,170.87) .. controls (375.07,171.53) and (355.07,146.87) .. (399.07,137.53) ;
\draw [line width=2.25]    (293.07,126.87) .. controls (303.73,118.2) and (309.73,88.2) .. (325.73,88.2) .. controls (341.73,88.2) and (341.73,113.53) .. (352.4,124.2) .. controls (363.07,134.87) and (370.4,136.2) .. (399.07,126.87) ;
\draw [line width=2.25]    (219.07,109.53) .. controls (249.07,111.53) and (269.07,157.53) .. (293.07,126.87) ;
\draw [line width=2.25]    (221.07,119.53) .. controls (253.07,129.53) and (274.4,151.53) .. (288.4,162.2) ;
\draw    (355.07,91.2) -- (355.07,109.2) ;
\draw [shift={(355.07,112.2)}, rotate = 270] [fill={rgb, 255:red, 0; green, 0; blue, 0 }  ][line width=0.08]  [draw opacity=0] (8.93,-4.29) -- (0,0) -- (8.93,4.29) -- cycle    ;
\draw [shift={(355.07,88.2)}, rotate = 90] [fill={rgb, 255:red, 0; green, 0; blue, 0 }  ][line width=0.08]  [draw opacity=0] (8.93,-4.29) -- (0,0) -- (8.93,4.29) -- cycle    ;

\draw (424,228.73) node [anchor=north west][inner sep=0.75pt]  [font=\scriptsize]  {$\mathcal{M}$};
\draw (362,93.4) node [anchor=north west][inner sep=0.75pt]  [font=\scriptsize]  {$\epsilon ( x)$};
\draw (319.33,122.4) node [anchor=north west][inner sep=0.75pt]  [font=\scriptsize]  {$g$};

\end{tikzpicture}
    \caption{A neighbourhood $B(\U,\epsilon,g)$ of the metric $g$ in the open topology consists of all metrics which lie within $\epsilon(x)$ of $g$ over the set $U$ of $\mathcal{M}$. Note that U can equal $\mathcal{M}$, and that $\epsilon$ can go to zero at infinity.}
    \label{fig:open topology}
    \end{subfigure}
    \begin{subfigure}[t]{0.32\textwidth}
        \begin{tikzpicture}[x=0.75pt,y=0.75pt,yscale=-1,xscale=1]

\draw [line width=2.25]    (291.04,244.18) .. controls (316.67,242.87) and (343.33,241.53) .. (365.49,239.75) ;
\draw    (226.67,252.2) .. controls (249.34,249.43) and (269.25,242.52) .. (291.04,244.18) ;
\draw    (365.49,239.75) .. controls (405.31,238.37) and (383.74,238.37) .. (424.67,238.37) ;
\draw [line width=2.25]    (291.33,171.53) -- (291.04,244.18) ;
\draw [line width=2.25]    (366,172.87) -- (365.49,239.75) ;
\draw    (226.67,116.2) .. controls (240,115.53) and (270.72,146.37) .. (291.33,146.87) .. controls (311.95,147.37) and (309.33,128.87) .. (328,117.53) .. controls (346.67,106.2) and (358.38,148.68) .. (366,148.2) .. controls (373.62,147.72) and (395.33,133.53) .. (410.67,136.2) ;
\draw [line width=2.25]    (291.33,171.53) .. controls (318.67,170.87) and (318.67,140.87) .. (334.67,140.87) .. controls (350.67,140.87) and (351.33,168.2) .. (366,172.87) ;
\draw [line width=2.25]    (292,121.53) .. controls (319.33,120.87) and (319.33,90.87) .. (335.33,90.87) .. controls (351.33,90.87) and (352,118.2) .. (366.67,122.87) ;
\draw [line width=2.25]    (292.29,48.89) -- (292,121.53) ;
\draw [line width=2.25]    (367.18,55.98) -- (366.67,122.87) ;
\draw    (371.33,152.53) -- (371.33,170.53) ;
\draw [shift={(371.33,173.53)}, rotate = 270] [fill={rgb, 255:red, 0; green, 0; blue, 0 }  ][line width=0.08]  [draw opacity=0] (8.93,-4.29) -- (0,0) -- (8.93,4.29) -- cycle    ;
\draw [shift={(371.33,149.53)}, rotate = 90] [fill={rgb, 255:red, 0; green, 0; blue, 0 }  ][line width=0.08]  [draw opacity=0] (8.93,-4.29) -- (0,0) -- (8.93,4.29) -- cycle    ;
\draw    (371.33,122.53) -- (371.33,140.53) ;
\draw [shift={(371.33,143.53)}, rotate = 270] [fill={rgb, 255:red, 0; green, 0; blue, 0 }  ][line width=0.08]  [draw opacity=0] (8.93,-4.29) -- (0,0) -- (8.93,4.29) -- cycle    ;
\draw [shift={(371.33,119.53)}, rotate = 90] [fill={rgb, 255:red, 0; green, 0; blue, 0 }  ][line width=0.08]  [draw opacity=0] (8.93,-4.29) -- (0,0) -- (8.93,4.29) -- cycle    ;

\draw (431.6,235.4) node [anchor=north west][inner sep=0.75pt]  [font=\scriptsize]  {$\mathcal{M}$};
\draw (321.6,228.07) node [anchor=north west][inner sep=0.75pt]  [font=\scriptsize]  {$\mathcal{U}$};
\draw (404.93,139.73) node [anchor=north west][inner sep=0.75pt]  [font=\scriptsize]  {$g$};
\draw (379.6,156.07) node [anchor=north west][inner sep=0.75pt]  [font=\scriptsize]  {$\epsilon $};
\draw (378.27,124.07) node [anchor=north west][inner sep=0.75pt]  [font=\scriptsize]  {$\epsilon $};

\end{tikzpicture}
    \caption{A neighbourhood $B(\epsilon,g)$ of the metric $g$ in the fine topology consists of all the metrics which lie within $\epsilon$ of $g$ over any compact set $\U$ and coincide with the metric $g$ outside $\U$.}
    \label{fig:fine}
    \end{subfigure}
    \caption{The $C^0$ compact-open, open and fine topologies on the space of $C^0$ Lorentzian metrics on $\mathcal{M}$ \cite[Figures 1, 2 and 3]{hawking_stable_1971}.}
    \label{fig:topologies}
\end{figure*}

Here, we briefly review three natural topologies on $C^0$ spacetimes. We start by discussing those on the space of $C^0$ Lorentzian metrics on $\mathcal{M}$. We closely follow \cite{hawking_stable_1971}, and reproduce its Figures 1, 2 and 3 in Fig.~\ref{fig:topologies}.

Consider the space $L_0(\mathcal{M})$ of all $C^0$ Lorentzian metrics $g$ on a four-dimensional manifold $\mathcal{M}$. There are then three main possibilities\footnote{There also exists Sobolev $W^0$ topologies, which are coarser than their $C^0$ counterparts, whereby instead of requiring the difference between two nearby metrics to be pointwise small, they require the integrals of the squares of these differences to be small. These topologies are relevant for the initial value problem in General Relativity \cite{hawking_stable_1971} and may prove to be meaningful in the distributional formulation of semiclassical gravity. We keep such a discussion for future work.} for understanding how ``close'' two metrics are \cite{hawking_stable_1971,minguzzi_widening_2009}:
\begin{enumerate}
    \item The $C^0$ compact-open topology: the metrics are required to be ``close'' only on compact regions of the manifold, with no control at infinity (see Fig.~\ref{fig:compact open topology}). If $g$ is a Lorentzian metric, $\U$ a compact set of $\mathcal{M}$ and $\epsilon$ a continuous positive function on $\mathcal{M}$, the neighbourhood $B(\U,\epsilon,g)$ of $g$ can be defined as the set of all Lorentzian metrics that differ from $g$ by less than $\epsilon$ on $\U$.
    The set of all such neighbourhoods forms a sub-basis for the $C^0$ compact-open topology for $L_0(\mathcal{M})$.
    \item The $C^0$ open topology: the metrics are required to be ``close'' on any $\U \subseteq \mathcal{M}$. This means that ``nearby'' metrics must be nearby everywhere and must have limiting behaviour at infinity (see Fig.~\ref{fig:open topology}).
    \item The $C^0$ fine topology: For any compact set $\U$ of $\mathcal{M}$, define the set $F(\U,\epsilon,g)$ of all Lorentzian metrics that differ from $g$ by less than $\epsilon$ and that coincide with $g$ outside of $\U$.
    The neighbourhood $B(\epsilon,g)$ is defined as the union of the $F(\U,\epsilon,g)$ for all compact sets $\U$ (see Fig.~\ref{fig:fine}). The neighbourhoods $B(\epsilon,g)$ forms a sub-basis for the fine topology of $L_0(\mathcal{M})$.
\end{enumerate}
The fine topology is finer than the open topology which in turn is finer than the compact-open topology. These are topologies relevant for the convergence of \emph{metrics}.

In the construction of local quantum states of Sec.~\ref{sec:local quantum states Kent}, we argue that it makes most sense to use the $C^0$ open topology in $L_0(\M)$.   The $C^0$ compact-open topology does not control what happens at infinity (very early times near the past light cone).   In contrast, the $C^0$ fine topology, in the sense of Fig.~\ref{fig:fine}, requires two metrics to actually coincide outside some compact set.  This requires an exact gravitational measurement that is not operationally accessible.  
Our intuition is that the $C^0$ open topology is more natural and useful because the gravitational field will be sourced by the local quantum state, and the $C^0$ open topology both controls the tails at infinity and allows for small errors outside any operationally accessible gravitational measurement.

\section{Barrabès-Israel null-shell formalism}

\label{app:BI}

We here follow the exposition of \cite{poisson_reformulation_2002,poisson_hypersurfaces_2004}. A null hypersurface $\mathcal{N}$ partitions the spacetime $\mathcal{M}$ into two regions $\mathcal{M}^\pm$, in which the metric is $g^\pm_{\alpha \beta}$ when expressed in coordinates $x^\alpha_{\pm}$ -- the coordinate systems on each side of the hypersurface may differ. We take $\mathcal{M}^-$ to be the ``past spacetime'' and $\mathcal{M}^+$ to be the ``future spacetime''.

Consider the coordinates $y^a = (\lambda,\theta^A)$ on $\mathcal{N}$, where $a=1,2,3$ and $A = 2,3$, which are taken to be the same on both sides of $\mathcal{N}$. Take $\lambda$ to be an arbitrary parameter on the null generators of the hypersurface, and use $\theta^A$ to label the generators: to each generator we assign two coordinate $\theta^A$. Changing $\lambda$ while keeping $\theta^A$ constant produces a displacement along a single generator; changing $\theta^A$ while keeping $\lambda$ constant produces a displacement across generators. Hence, $\mathcal{N}$ as seen from $\mathcal{M}^\pm$ is described by a set of parametric relations $x^\alpha_\pm(y^a)$, which can be used to define tangent vectors $e^\alpha_{\pm a} = \partial x^\alpha_\pm/\partial y^a$ on each side of the hypersurface. The null vector
\begin{equation}
    k_\pm^\alpha = \left(\frac{\partial x_\pm^\alpha}{\partial \lambda}\right)_{\theta^A} \equiv e^\alpha_{\pm \lambda}
\end{equation}
is proportional to the null generator, while the two spacelike vectors
\begin{equation}
    e^\alpha_{\pm A} = \left(\frac{\partial x_\pm^\alpha}{\partial \theta^A}\right)_\lambda
\end{equation}
are orthogonal to the null generator.  By construction,
\begin{equation}
    k_{\pm \alpha } k^{\alpha}_\pm = 0 = k_{\pm \alpha} e^{\alpha}_{\pm A} \, .
\end{equation}
The remaining inner products, that we write
\begin{equation}
    \sigma_{AB}^\pm(\lambda,\theta^C) = g_{\alpha \beta}^{\pm} e^{\alpha}_{\pm A} e^\beta_{\pm B}
\end{equation}
do not vanish, though we assume by construction that they are the same on both sides of $\mathcal{N}$:
\begin{equation}
    [\sigma_{AB}] = 0
\end{equation}
where we introduced the notation
\begin{equation}
    [A] \equiv A(\mathcal{M}^+)|_{\mathcal{N}} - A(\mathcal{M}^-)|_{\mathcal{N}} \, .
\end{equation}
A vanishing $[A]$ means that $A$ is continuous at $\mathcal{N}$, while a non-zero $[A]$ means that $A$ is discontinuous and $[A]$ gives the jump of $A$ across $\mathcal{N}$.

$\sigma_{AB}^\pm$ acts as a metric (with inverse $\sigma^{AB}_\pm$) on $\mathcal{N}$: a displacement on the hypersurface is given by $dx_\pm^\alpha = k_\pm^\alpha d\lambda + e^\alpha_{\pm A} d\theta^A$, so on both sides of $\mathcal{N}$ the line element is
\begin{equation}
    ds^2_{\mathcal{N}} = \sigma_{AB}^\pm d\theta^A d\theta^B \, .
\end{equation}
Thus, the condition $[\sigma_{AB}] = 0$ ensures that the hypersurface possesses a well-defined intrinsic geometry, which is required for the null-shell formalism. A violation of this condition would require tools beyond distributional techniques; thus, we will take it as a reasonable assumption for the time being, and will need to check that it indeed holds in the physical scenarios of interest. In this sense, we require similar assumptions to the approaches of \cite{juarez-aubry_initial_2023,mucino_fully_2025}, for which the collapse hypersurfaces need a suitable intrinsic geometry.

The tangent vectors $k_\pm^\alpha$ and $e_{\pm A}^\alpha$ provide on a partial basis for the decomposition of vector fields on either side of $\mathcal{N}$. We can complete the basis by adding a transverse null vector $N_\pm^\alpha$ satisfying
\begin{equation}
    \label{eqn:N vector}
    N_{\pm \alpha} N^\alpha_\pm = 0, \quad N_{\pm \alpha} k^\alpha_\pm = -1, \quad N_{\pm \alpha} e^{\alpha}_{\pm A} = 0 \, .
\end{equation}
These equations specify $N^\alpha_\pm$ uniquely. In this basis,
\begin{equation}
    g^{\alpha \beta}_{\pm} = - k_\pm^\alpha N_\pm^\beta - N^\alpha_\pm k^\beta_\pm + \sigma^{AB} e^\alpha_{\pm A} e^\beta_{\pm B} \, .
\end{equation}

Although $\sigma_{AB}$ is the same on both sides of the light cone, the metric's transverse derivative will be discontinuous across it. The information about the metric's transverse derivative is encoded in the transverse curvature $C^\pm_{ab}$ defined by
\begin{equation}
    C^\pm_{ab} = -N_{\pm \alpha} e^{\alpha}_{\pm a;\beta} e^\beta_{\pm b} = C_{ba}^\pm \, .
\end{equation}
In general, $[C_{ab}] \neq 0$. The jump in the transverse curvature is directly related to the stress-energy tensor on $\mathcal{N}$. We define
\begin{align}
    \mu &= -\frac{c^4}{8\pi G} \sigma^{AB}[C_{AB}] \, , \\
    j^A &= \frac{c^4}{8\pi G} \sigma^{AB}[C_{\lambda B}] \, , \\
    p &= -\frac{c^4}{8\pi G} [C_{\lambda \lambda}] \, .
\end{align}
It can be shown \cite{poisson_reformulation_2002,poisson_hypersurfaces_2004} that the surface stress-energy flux on either sides of $\mathcal{N}$ is given by
\begin{equation}
    \label{eqn:singular S}
    S^{\alpha \beta}_\pm = \mu k^\alpha_\pm k^\beta_\pm + j^A(k^\alpha_\pm e^\beta_{\pm A} + e^{\alpha}_{\pm A} k^\beta_\pm) + p \sigma^{AB} e^\alpha_{\pm A} e^{\beta}_{\pm B} \, .
\end{equation}
The first term represents a flow of matter along the null generators of the hypersurface, i.e. $\mu$ represents a mass density; the second term represents a flow of matter in the directions transverse to the generators, i.e. $j^A$ represents a current density; the last term expresses the fact that $p$ is an isotropic pressure. As a distribution, the stress-energy tensor is
\begin{equation}
    \label{eqn:stress-energy shell}
    T^{ab}_{\mathcal{N}} = (-k \cdot U)^{-1} S^{ab} \delta(\tau)
\end{equation}
where
\begin{itemize}
    \item $S^{\alpha \beta}_\pm = S^{ab} e^{\alpha}_{\pm a} e^{\beta}_{\pm b}$, where $S^{ab}$ is intrinsic to $\mathcal{N}$. It can be shown \cite{poisson_reformulation_2002,poisson_hypersurfaces_2004} that $[C_{ab}]$ and $\sigma_{AB}$ are sufficient to reconstruct $S_{ab}$ as an intrinsic quantity.
    \item $U$ is a timelike geodesic such that at proper time $\tau$, the geodesic intersects $\mathcal{N}$, i.e. for $\tau < 0$ it is in $\mathcal{M}^-$ and for $\tau > 0$ it is in $\mathcal{M}^+$.
\end{itemize}

To see this, introduce a future-directed timelike vector $U^a$, transverse to $\Lambda^+(x)$, such that $U^a$ is tangent to an arbitrary congruence of timelike geodesics parametrized by
proper time $\tau$ chosen so that $\tau=0$ on $\Lambda^+(x)$ (the congruence represents a family of observers making measurements on the shell) \cite{poisson_hypersurfaces_2004}. One then writes
\begin{equation}
\label{eq:null shell distribution}
    T_{\Lambda^+(x)}^{ab} = (-k_cU^c(\tau))^{-1} S^{ab}(k) \delta(\tau),
\end{equation}
where $k=(k^a)$ is a future-directed null vector proportional to the null generator of
$\Lambda^+(x)$ (defined below), and $S^{ab}$ is the surface stress-energy
tensor associated with the chosen generators (see Eqn. \eqref{eqn:singular S}).

As Poisson \cite{poisson_hypersurfaces_2004} puts it ($\Sigma$, for us, is $\Lambda^+(x)$ or $\mathcal{N}$ more generally): 

\begin{quote}
    To complete the geometric setup we introduce a congruence of geodesics that cross the hypersurface. In [the case] in which $\Sigma$ was either timelike or spacelike, the congruence was selected by demanding that the geodesics intersect the hypersurface orthogonally: the vector field $U^\alpha_\pm$ tangent to the congruence was set equal (on $\Sigma$) to the normal vector $N_\pm^\alpha$. When the hypersurface is null, however, this requirement does not produce a unique congruence, because a vector orthogonal to $k_\alpha$ can still possess an arbitrary component along $k_\alpha$. We shall have to give up on the idea of adopting a unique congruence. An important aspect of our description of null shells is therefore that it involves an arbitrary congruence of timelike geodesics intersecting $\Sigma$. This arbitrariness comes with the lightlike nature of the singular hypersurface, and it cannot be removed. It can, however, be physically motivated: the arbitrary vector field $U^\alpha_\pm$ that enters our description of null shells can be identified with the four-velocity of a family of observers making measurements on the shell; since many different families of observers could be introduced to make such measurements, there is no reason to demand that the vector field be uniquely specified. We therefore introduce a congruence of timelike geodesics $\gamma$ that arbitrarily intersect the hypersurface. The geodesics are parametrized by proper time $\tau$, which is adjusted so that $\tau$ = 0 when a geodesic crosses $\Sigma$; thus, $\tau < 0$  in $V^-$ and $\tau > 0$ in $V^+$. 
    
    The vector tangent to the geodesics is $U^\alpha_\pm$, and a displacement along a member of the congruence is described by $dx^\alpha = U^\alpha d\tau$. To ensure that the congruence is smooth at the hypersurface, we demand that $U^\alpha_\pm$ be ``the same'' on both sides of $\Sigma$. This means that $U_\alpha e^\alpha_a$, the tangential projections of the vector field, must be equal when evaluated on either side of the hypersurface: $[-U^\alpha k_\alpha] = 0 = [U_\alpha e^\alpha_A]$. If, for example, $U^\alpha_-$ is specified in $V^-$, then the[se] three conditions are sufficient (together with the geodesic equation) to determine the three independent components of $U^\alpha_+$ in $V^+$. 
    
    We note that $-U^\alpha N_\alpha$, the transverse projection of the four-velocity, is allowed to be discontinuous at $\Sigma$. The proper-time parameter on the timelike geodesics can be viewed as a scalar field $\tau(x^\alpha_\pm)$ defined in a neighbourhood of $\Sigma$: Select a point $x^\alpha_\pm$ off the hypersurface and locate the unique geodesic $\gamma$ that connects this point to $\Sigma$; the value of the scalar field at $x^\alpha_\pm$ is equal to the proper-time parameter of this geodesic at that point. 
    
    The hypersurface $\Sigma$ can then be described by the statement $\tau(x^\alpha) = 0$, and its normal vector $k^\pm_\alpha$ will be proportional to the gradient of $\tau(x^\alpha_\pm)$ evaluated at $\Sigma$. 
    
    It is easy to check that the expression $k_\alpha = -(k_\mu U^\mu) \frac{\partial \tau}{\partial x^\alpha}$ is compatible with $dx^\alpha = U^\alpha d\tau$. We recall that the factor $-k_\mu U^\mu$ is continuous across $\Sigma$.
\end{quote}

The appearance of $U^a$ in Eqn. \eqref{eq:null shell distribution} does not
mean that the shell stress-energy depends on $U^a$. The vector $U^a$ is
only used to keep track of an observer's proper-time delta function $\delta(\tau)$. To see this explicitly, introduce coordinates $(\tau,\lambda,\theta^A)$ near $\Lambda^+(x)$, with
\begin{equation}
    \left.\frac{\partial}{\partial\tau}\right|_{\Lambda^+(x)}=U,
    \qquad
    \left.\frac{\partial}{\partial\lambda}\right|_{\Lambda^+(x)}=k,
    \qquad
    e_A^a=\frac{\partial x^a}{\partial\theta^A}.
\end{equation}
Let $\sigma_{AB}=g_{ab}e_A^ae_B^b$ be the metric on the cone, such that $\sqrt{\abs{g}} = (-k \cdot U) \sqrt{\sigma}$ on $\Lambda^+(x)$. Then, on $\Lambda^+(x)$,
\begin{equation}
    \sqrt{\abs{g}}\,d\tau\,d\lambda\,d^2\theta = (-k\cdot U)\sqrt{\sigma}\,
    d\tau\,d\lambda\,d^2\theta .
\end{equation}
Therefore, for any (smooth compactly supported) test function $f_{ab}$,
\begin{multline}
    \int_\mathcal{M} f_{ab} (-k\cdot U)^{-1} S^{ab}\delta(\tau)
    \,\sqrt{\abs{g}} d\tau\,d\lambda\,d^2\theta  \\= \int_{\mathcal{M}} f_{ab} (-k\cdot U)^{-1} S^{ab} \delta(\tau)
    (-k\cdot U)\sqrt{\sigma}\,
    d\tau\,d\lambda\,d^2\theta\\
    =
    \int_{\Lambda^+(x)}
    f_{ab}S^{ab}
    \sqrt{\sigma}\,d\lambda\,d^2\theta .
\end{multline}
The right-hand side contains no $U^a$. Thus changing the auxiliary timelike
congruence $U^a$, and hence changing the proper-time coordinate $\tau$,
changes only the representation of the same distribution. It does not change
$T_{\Lambda^+(x)}^{ab}$.

The role of $k^a$ is different. The null hypersurface $\Lambda^+(x)$ has a
canonical null direction at each point. Choose any parameter $\lambda$ along the null generators and define
\begin{equation}
    k^a = \left(\frac{\partial x^a}{\partial\lambda}\right)_{\theta^A}.
\end{equation}
A general change of parameter $\lambda\mapsto\lambda'(\lambda,\theta^A)$ gives a changed $k \mapsto k'$ as well as a changed $S_{ab} \mapsto S_{ab}'$. However, it can be shown \cite{poisson_reformulation_2002,poisson_hypersurfaces_2004} that this reparametrisation keeps $T^{ab}_{\Lambda^+(x)}$ invariant, that is,
\begin{equation}
    \frac{S^{ab}}{-k\cdot U}\delta(\tau) = \frac{S'^{ab}}{-k'\cdot U}\delta(\tau).
\end{equation}
Thus, the surface stress-energy tensor $S^{ab}(k)$ indeed depends on this
normalisation choice, while the intrinsic stress-energy tensor $T^{ab}_{\Lambda^+(x)}$ of $\Lambda^+(x)$ does not. The physical shell stress-energy is intrinsic to the null hypersurface $\Lambda^+(x)$. It is not observer-dependent and it is not dependent on an arbitrary parametrisation of those generators. Only the intermediate surface quantities depend on the chosen generator normalization.

We can also examine the distributional nature of the Einstein tensor. Let us assume that $[g_{\alpha \beta}] = [k^\alpha] = [e^\alpha_A] = [N^\alpha] = [u^\alpha] = 0$ in the coordinate system we work with\footnote{See \cite{poisson_reformulation_2002,poisson_hypersurfaces_2004} for a detailed discussion of these choices of coordinates and assumptions.}. We then have $g_{\alpha \beta} = g^+_{\alpha \beta} \Theta(\tau) + g_{\alpha \beta}^- (1-\Theta(\tau))$, where $\Theta(\tau)$ is the Heaviside step function, and with $g$ being continuous across $\tau = 0$, and so the Riemann tensor is given by
\begin{equation}
    \label{eqn: Riemann tensor BI detail}
    R^{\alpha}_{ \; \beta \gamma \delta} = R^{\; \alpha}_{+ \; \beta \gamma \delta} \Theta(\tau) + R^{\; \alpha}_{- \; \beta \gamma \delta} (1-\Theta(\tau)) + R^{\; \alpha}_{\mathcal{N} \; \beta \gamma \delta}
\end{equation}
where
\begin{equation}
    \label{eqn:singular Riemann}
    R^{\; \alpha}_{\mathcal{N} \; \beta \gamma \delta} = -(-k \cdot u)^{-1} ([\Gamma^\alpha_{\; \beta \delta}]k_\gamma - [\Gamma^{\alpha}_{\; \beta \gamma} k_\delta]) \delta(\tau) \, .
\end{equation}
Here,
\begin{equation}
    [\Gamma^\alpha_{\; \beta \delta}] = -\frac{1}{2}(\gamma^\alpha_{\; \beta} k_\delta + \gamma^\alpha_{\; \delta} k_\beta - \gamma_{\beta \delta} k^\alpha ) 
\end{equation}
denotes the jump in the affine connection across $\mathcal{N}$, and
\begin{equation}
    \gamma_{\alpha \beta} = [g_{\alpha \beta,\delta}] N^\delta \, .
\end{equation}
Note that the discontinuity in the derivatives of the metric is purely along $k_\delta$:
\begin{equation}
    [g_{\alpha \beta,\delta}] = - \gamma_{\alpha \beta} k_\delta
\end{equation}
whilst the tangential derivatives of the metric are continuous:
\begin{equation}
    [g_{\alpha \beta,\delta}] k^\delta = 0 \, .
\end{equation}
In turn, Eqns. \eqref{eqn:singular S} and \eqref{eqn:singular Riemann} imply that for the Einstein field equations to hold distributionally, it is necessary and sufficient that \begin{align}
    \mu &= -\frac{c^4}{16 \pi G} \sigma^{AB} \gamma_{\alpha \beta} e^{\alpha}_A e^{\beta}_B \, , \\
    j^A &= \frac{c^4}{16 \pi G} \sigma^{AB} \gamma_{\alpha \beta} e^\alpha_A k^\beta \, , \\
    p &= -\frac{c^4}{16\pi G} \gamma_{\alpha \beta} k^\alpha k^\beta \, .
\end{align}

Note that both the Weyl and Ricci parts of the Riemann tensor are singular. The singularity for the Ricci part can be associated with a material stress-energy tensor -- a thin shell or surface layer. However, the singularity for the Weyl tensor can be associated to an independent impulsive tidal wave. It would be interesting to understand physically how this translates to this quantum measurement setting, and whether there are any observable physical consequences to such discontinuities.

\section{Future metric solution for the linearised Einstein field equations}

\label{app:future metric}

First, we review angular and null-coordinate geometry. Let $\theta^A, A=1,2$ be local angular coordinates. The unit-sphere metric is denoted by $\Omega_{AB}$ and given, in standard angular coordinates $(\vartheta,\varphi)$, by
\begin{equation}
    \Omega_{AB} d\theta^A d\theta^B = d\vartheta^2 + \sin^2 \vartheta d\varphi^2.
\end{equation}
Its Levi-Civita connection is given by
\begin{equation}
    {}^\Omega \Gamma^C_{\; AB} = \frac{1}{2} \Omega^{CD}(\partial_A \Omega_{BD} + \partial_B \Omega_{AD} - \partial_D \Omega_{AB}).
\end{equation}
The associated angular derivative $D_A$ compares tensor components at neighbouring points on a curved sphere. On a scalar $f$,
\begin{equation}
    D_A f = \partial_A f,
\end{equation}
while for a covector $X_B$,
\begin{equation}
    D_A X_B = \partial_A X_B - {}^\Omega \Gamma^C_{\; AB} X_C
\end{equation}
and for a vector $X^B$,
\begin{equation}
    D_A X^B = \partial_A X^B + {}^\Omega \Gamma^B_{\; AC} X^C.
\end{equation}
More generally, on tensors,
\begin{multline}
    D_A T^{B_1\cdots B_p}{}_{C_1\cdots C_q} = \partial_A T^{B_1\cdots B_p}{}_{C_1\cdots C_q} \\
    + \sum_{i=1}^p {}^\Omega \Gamma^{B_i}{}_{AD} T^{B_1\cdots D \cdots B_p}{}_{C_1\cdots C_q} \\
    - \sum_{j=1}^q {}^\Omega \Gamma^{D}{}_{A C_j} T^{B_1 \cdots B_p}{}_{C_1\cdots D \cdots C_q}.
\end{multline}

We use $D^A:= \Omega^{AB} D_B$. For any point $y \in \mathbb{M}$, let $r = \abs{\vec{y}}$ and $u = ct-r$, $v=ct+r$. In null coordinates, the Minkowski metric becomes
\begin{equation}
    \eta = -dudv + r^2 \Omega_{AB} d\theta^A d\theta^B
\end{equation}
with
\begin{align}
    \eta_{uv} &= \eta_{vu} = - \frac{1}{2},\\
    \eta^{uv} &= \eta^{vu} = -2, \\
    \eta^{AB} &= r^{-2} \Omega^{AB}.
\end{align}
Then
\begin{equation}
    n_u = 1, \quad  n_v = n_A = 0, \quad n^\mu = \eta^{\mu\nu}n_\nu = -2\delta^\mu_{\; v}.
\end{equation}
The nonzero Christoffel symbols, apart from symmetry in the lower indices, are
\begin{align}
    \Gamma^u_{\; AB} = r \Omega_{AB},& \qquad \Gamma^v_{AB} = -r \Omega_{AB}, \\ \Gamma^{A}_{\; uB} = -\frac{1}{2r} \delta^A_B,& \qquad \Gamma^A_{\; vB} = \frac{1}{2r} \delta^A_B,
\end{align}
and
\begin{equation}
    \Gamma^{A}_{\; BC} = {}^\Omega\Gamma^A_{\; BC}.
\end{equation}
We have 
\begin{equation}
    \nabla_\alpha \nabla_\beta u = \nabla_\alpha n_\beta, \qquad \nabla_A \nabla_B u = -r\Omega_{AB},
\end{equation}
and
\begin{equation}
    \Box u \equiv \nabla^\alpha \nabla_\alpha u = - \frac{2}{r}.
\end{equation}

\subsection{Computing $\mathfrak{H}$}

Note that $n^\mu G^{(1)}_{\mu\nu}[\mathfrak{H}] \propto G^{(1)}_{v\nu}[\mathfrak{H}]$, so those components are the ones that describe the characteristic constraints on $\mathcal{N}$. This is what we now derive.

\subsubsection{The constraints on $\Lambda^+(0)$}

We start the analysis on $\mathcal{N}_\mathrm{reg}$ since the consideration of the vertex requires more care; we will examine it later. Pick any point $x_\mathrm{reg} \in \mathcal{N}_\mathrm{reg}$, for which $r > 0$ (and $ct = r$) and so $v > 0$ in $\M^+$, let $0 < a < 1$ and
\begin{equation}
    \mathcal{C}^+_{a} := \{(u,v,\theta) : v>0, 0 \leq u < av, \theta \in S^2\}
\end{equation}
be a one-sided conical collar lying inside the future causal cone. For a fixed $(v,\theta)$ and $u < av$, the segment $s \mapsto (su,v,\theta)$ lies inside $\mathcal{C}^+_{a}$ for $0 \leq s \leq 1$. Since $\mathfrak{H}|_\mathcal{N} = 0$, Taylor's theorem to first order with an integral remainder of second order in $\mathcal{C}^+_a$ for fixed $(v,\theta)$ gives
\begin{equation}
    \mathfrak{H}_{\mu\nu} = u \mathfrak{b}_{\mu\nu} + u^2 \mathfrak{s}_{\mu\nu} 
\end{equation}
away from the vertex of the future light cone, where
\begin{equation}
    \mathfrak{b}_{\mu\nu}(v,\theta) := \partial_u \mathfrak{H}_{\mu\nu}|_{\mathcal{N},+},
\end{equation}
and
\begin{equation}
    \mathfrak{s}_{\mu\nu}(u,v,\theta) = \int_0^1 (1-\lambda) \partial^2_u \mathfrak{H}_{\mu\nu}(\lambda u,v,\theta) d\lambda .
\end{equation}
We abuse notation and write
\begin{equation}
    \mathfrak{b}_{\mu\nu}(u,v,\theta) \stackrel{!}{=} \mathfrak{b}_{\mu\nu}(0,v,\theta) := \mathfrak{b}_{\mu\nu}(v,\theta) 
\end{equation}
such that $\partial_u \mathfrak{b}_{\mu\nu} = 0$.

We want to check that the characteristic components $G^{(1)}_{v \nu}|_\mathcal{N}$ only depend on the first transverse jet $\mathfrak{b}_{\mu\nu}$, not on $\mathfrak{s}_{\mu\nu}$: that is, that $\partial \mathfrak{H}_{\mu\nu}|_{\mathcal{N}}$ fully determines the contribution of $\mathfrak{H}$ to the characteristic components linearised Einstein tensor on $\mathcal{N}$. This is not trivial because the linearised Einstein tensor is second order in derivatives of $u$ so could in principle contain a $u^2 \mathfrak{s}$ contribution. Let $\overline{\mathfrak{s}}_{\mu\nu} = \mathfrak{s}_{\mu\nu} - \frac{1}{2}\eta_{\mu\nu} \eta^{\rho\sigma} \mathfrak{s}_{\rho\sigma}$ be its trace reversal.

At $u=0$, 
\begin{align}
    \nabla_\alpha \nabla_\beta(u^2 \overline{\mathfrak{s}})_{\mu\nu} &= \nabla_\alpha (2 u n_\beta \overline{\mathfrak{s}}_{\mu\nu}) + \nabla_\alpha (u^2 \nabla_\beta \overline{\mathfrak{s}}_{\mu\nu}) \\
    &= 2 n_\alpha n_\beta \overline{\mathfrak{s}}_{\mu\nu} + 2 u (\nabla_\alpha n_\beta) \overline{\mathfrak{s}}_{\mu\nu} \nonumber \\
    &+ 2u n_\beta \nabla_\alpha \overline{\mathfrak{s}}_{\mu\nu} + 2u n_\alpha \nabla_\beta \overline{\mathfrak{s}}_{\mu\nu} \nonumber \\ &+ u^2 \nabla_\alpha \nabla_\beta \overline{\mathfrak{s}}_{\mu\nu}.
\end{align}
Every term except the first contains at least one explicit factor of $u$, i.e. on $\mathcal{N}_\mathrm{reg}$ we have
\begin{equation}
    \nabla_\alpha \nabla_\beta(u^2 \overline{\mathfrak{s}})_{\mu\nu}|_{\mathcal{N}} = 2 n_\alpha n_\beta \overline{\mathfrak{s}}_{\mu\nu}.
\end{equation}
Define $C_\nu[u^2 \mathfrak{s}_{\mu\nu}] = \nabla^\rho \overline{u^2 \mathfrak{s}_{\mu\nu}}_{\rho \nu}$. Then
\begin{multline}
    \Box (u^2 \overline{\mathfrak{s}})_{\mu\nu}|_\mathcal{N} = \eta^{\alpha\beta} \nabla_\alpha \nabla_\beta (u^2 \overline{\mathfrak{s}})_{\mu\nu}|_{\mathcal{N}} = 2 \eta^{\alpha \beta} n_\alpha n_\beta \overline{\mathfrak{s}}_{\mu\nu}|_{\mathcal{N}} \\ = 2 n^2 \overline{\mathfrak{s}}_{\mu\nu}|_{\mathcal{N}} = 0
\end{multline}
since $n = du$ is null. Moreover, 
\begin{equation}
    C_\nu[u^2 \mathfrak{s}]|_{\mathcal{N}} = 2 u n^\rho \overline{\mathfrak{s}}_{\rho\nu}|_{\mathcal{N}} + u^2 \nabla^\rho \overline{\mathfrak{s}}_{\rho \nu}|_{\mathcal{N}} = 0.
\end{equation}
Furthermore,
\begin{equation}
    \nabla_\mu C_\nu[u^2 \mathfrak{s}]|_{\mathcal{N}} = 2n_\mu n^\rho  \overline{\mathfrak{s}}_{\rho \nu},
\end{equation}
and
\begin{equation}
    \nabla_\rho C^\rho[u^2 \mathfrak{s}]|_{\mathcal{N}} = 2 n^\rho n^\sigma \overline{\mathfrak{s}}_{\rho \sigma}.
\end{equation}
Substitution into $G^{(1)}_{\mu\nu}[u^2 \mathfrak{s}_{\mu\nu}]$ (see \eqref{eqn:linearised Einstein tensor}) yields
\begin{equation}
    G^{(1)}_{\mu\nu}[u^2 \mathfrak{s}]|_\mathcal{N} = n_\mu n^\rho \overline{\mathfrak{s}}_{\rho \nu}+ n_{\nu} n^\rho \overline{\mathfrak{s}}_{\rho \mu} - \eta_{\mu\nu} n^\rho n^\sigma \overline{\mathfrak{s}}_{\rho \sigma}.
\end{equation}
Using $n_v=n_A=0$ and $\eta_{vv}=\eta_{vA}=0$, we get
\begin{equation}
    G^{(1)}_{vv}[u^2 \mathfrak{s}]|_{\mathcal{N}} = G^{(1)}_{vA}[u^2 \mathfrak{s}]|_{\mathcal{N}} = 0.
\end{equation}
For the $uv$ component,
\begin{equation}
    G^{(1)}_{uv}[u^2 \mathfrak{s}]|_{\mathcal{N}} = 1 \cdot (-2 \overline{\mathfrak{s}}_{vv}) + 0 - \left(-\frac{1}{2}\right)(4 \overline{\mathfrak{s}}_{vv}) = 0.
\end{equation}
Thus the four characteristic components $G^{(1)}_{v\nu}[\mathfrak{H}]|_{\mathcal{N}}$ are fully determined by $\mathfrak{b}_{\mu\nu}$.

Let us now compute $G^{(1)}_{v\nu}[u\mathfrak{b}]$. Let $\overline{\mathfrak{b}}_{\mu\nu} := \mathfrak{b}_{\mu\nu} - \frac{1}{2} \eta_{\mu\nu} \eta^{\rho\sigma} \mathfrak{b}_{\rho\sigma}$. Note that $\overline{u\mathfrak{b}}_{\mu\nu} = u \overline{\mathfrak{b}}_{\mu\nu}$. We have
\begin{align}
    \label{eqn:box u b}
    \Box(u\overline{\mathfrak{b}})_{\mu\nu}|_{\mathcal{N}} &= (\Box u) \overline{\mathfrak{b}}_{\mu\nu}|_{\mathcal{N}} + 2 n^\alpha \nabla_\alpha \overline{\mathfrak{b}}_{\mu\nu}|_{\mathcal{N}} \\
    &= -\frac{2}{r} \overline{\mathfrak{b}}_{\mu\nu} - 4\nabla_v \overline{\mathfrak{b}}_{\mu\nu}|_{\mathcal{N}}.
\end{align}
Moreover,
\begin{equation}
    C_\nu[u\mathfrak{b}] = n^\rho\overline{\mathfrak{b}}_{\rho\nu} + u \nabla^\rho \overline{\mathfrak{b}}_{\rho \nu}.
\end{equation}
Since $\nabla_A n^\rho = \Gamma^\rho_{\; A \lambda} n^\lambda = - 2 \Gamma^\rho_{\; A v} = -\frac{1}{r}\delta^\rho_A$, differentiating then restricting to $u=0$ gives
\begin{align}
    \nabla_v C_\nu[u\mathfrak{b}]|_{\mathcal{N}} &= -2\nabla_v \overline{\mathfrak{b}}_{v\nu}|_{\mathcal{N}}, \\
    \nabla_A C_\nu[u\mathfrak{b}]|_{\mathcal{N}} &= - 2 \nabla_A \overline{\mathfrak{b}}_{v\nu}|_{\mathcal{N}} - \frac{1}{r} \overline{\mathfrak{b}}_{A\nu}|_{\mathcal{N}}, \\
    \nabla_u C_\nu[u\mathfrak{b}]|_{\mathcal{N}} &= - 2 \nabla_u \overline{\mathfrak{b}}_{v\nu}|_{\mathcal{N}} + \nabla^\rho \overline{\mathfrak{b}}_{\rho\nu}|_{\mathcal{N}} \\
    \nabla_\rho C^\rho[u\mathfrak{b}]|_{\mathcal{N}} &= -\frac{\Omega^{AB}\overline{\mathfrak{b}}_{AB}}{r^3} - 4 \nabla^\rho \overline{\mathfrak{b}}_{\rho v}|_{\mathcal{N}}
\end{align}

\paragraph*{The $vv$ components.} We have
\begin{align}
    \Box(u \overline{\mathfrak{b}})_{vv}|_{\mathcal{N}} - 2 \nabla_v C_v[u\mathfrak{b}]|_{\mathcal{N}} = -\frac{2}{r} \overline{\mathfrak{b}}_{vv}.
\end{align}
Since $\eta_{vv}=0$, we have
\begin{multline}
    G_{vv}^{(1)}[\mathfrak{H}]|_{\mathcal{N}} = G_{vv}^{(1)}[u\mathfrak{b}]|_{\mathcal{N}} = \frac{1}{r} \overline{\mathfrak{b}}_{vv} = \frac{1}{r} \mathfrak{b}_{vv} \\ = \frac{1}{r}\partial_u \mathfrak{H}_{vv}|_{\mathcal{N}}.
\end{multline}

\paragraph*{The $vA$ components.} We have
\begin{align}
    \Box(u \overline{\mathfrak{b}})_{vA}|_{\mathcal{N}} &- \nabla_v C_A[u\mathfrak{b}] - \nabla_A C_v[u\mathfrak{b}] \nonumber \\
    &= -\frac{2}{r} \overline{\mathfrak{b}}_{vA} - 4\nabla_v \overline{\mathfrak{b}}_{vA}|_\mathcal{N} \nonumber \\ &+2\nabla_v \overline{\mathfrak{b}}_{vA}|_{\mathcal{N}} + 2 \nabla_A \overline{\mathfrak{b}}_{vv}|_{\mathcal{N}} +\frac{1}{r} \overline{\mathfrak{b}}_{vA}  \\
    &= -\frac{1}{r}\overline{\mathfrak{b}}_{vA}-2\nabla_v \overline{\mathfrak{b}}_{vA}|_{\mathcal{N}} + 2 \nabla_A \overline{\mathfrak{b}}_{vv}|_{\mathcal{N}}.
\end{align}
Since
\begin{equation}
    \nabla_v \overline{\mathfrak{b}}_{vA} = \partial_v \overline{\mathfrak{b}}_{vA} - \frac{1}{2r}\overline{\mathfrak{b}}_{vA},
\end{equation}
and
\begin{equation}
    \nabla_A \overline{\mathfrak{b}}_{vv} = D_A \overline{\mathfrak{b}}_{vv} - \frac{1}{r} \overline{\mathfrak{b}}_{vA},
\end{equation}
and trace-reversal leaves the $vv$ and $vA$ components of $\mathfrak{b}_{\mu\nu}$ unchanged, we therefore have
\begin{multline}
    G^{(1)}_{vA}[ \mathfrak{H}]|_{\mathcal{N}}=G^{(1)}_{vA}[u \mathfrak{b}]|_{\mathcal{N}} \\ = \partial_v \mathfrak{b}_{vA}|_{\mathcal{N}}+\frac{2}{v}\mathfrak{b}_{vA}-D_A \mathfrak{b}_{vv}|_{\mathcal{N}} \\
    = \partial_u \partial_v \mathfrak{H}_{vA}|_{\mathcal{N}}+\frac{2}{v}\partial_u \mathfrak{H}_{vA}|_{\mathcal{N}} - D_A \partial_u \mathfrak{H}_{vv}|_{\mathcal{N}}.
\end{multline}

\paragraph*{The $uv$ components.} Since
\begin{equation}
    \eta^{\rho\sigma} \mathfrak{b}_{\rho \sigma} = -4 \mathfrak{b}_{uv} + \frac{1}{r^2} \Omega^{AB} \mathfrak{b}_{AB},
\end{equation}
it follows that
\begin{equation}
    \overline{\mathfrak{b}}_{uv}=\frac{\Omega^{AB} \mathfrak{b}_{AB}}{4r^2}, \qquad \Omega^{AB}\overline{\mathfrak{b}}_{AB} = 4r^2 \mathfrak{b}_{uv}.
\end{equation}
Moreover, since $v-u=2r$, $\partial_v r = \frac{1}{2}$ at fixed $u$ and so
\begin{equation}
    \label{eqn:buv dbuv dv omega bab}
    \frac{2}{r}\overline{\mathfrak{b}}_{uv}|_{\mathcal{N}} + 2 \partial_v \overline{\mathfrak{b}}_{uv}|_{\mathcal{N}} = \frac{1}{2r^2} \partial_v \Omega^{AB}\mathfrak{b}_{AB}|_{\mathcal{N}}.
\end{equation}

Moreover,
\begin{align}
    \label{eqn:nabla rho overline b rho v}
    \nabla^\rho \overline{\mathfrak{b}}_{\rho v}|_{\mathcal{N}} &= - 2 \nabla_u \overline{\mathfrak{b}}_{vv}|_{\mathcal{N}} - 2 \nabla_v \overline{\mathfrak{b}}_{uv}|_{\mathcal{N}} \nonumber \\ & \qquad + \frac{1}{r^2} \Omega^{AB} \nabla_A \overline{\mathfrak{b}}_{Bv}|_{\mathcal{N}}, \\
    \label{eqn:Omega AB nabla A b Bv}
    \Omega^{AB} \nabla_A \overline{\mathfrak{b}}_{Bv}|_{\mathcal{N}}&= \Omega^{AB} \Big(\partial_A \overline{\mathfrak{b}}_{Bv} - \Gamma^C{}_{AB}\overline{\mathfrak{b}}_{Cv} \nonumber \\ &-\Gamma^{u}{}_{AB}\overline{\mathfrak{b}}_{uv} - \Gamma^v{}_{AB}\overline{\mathfrak{b}}_{vv} - \Gamma^C{}_{Av} \overline{\mathfrak{b}}_{BC} \Big)|_{\mathcal{N}} \\
    &= D^A \mathfrak{b}_{vA}|_{\mathcal{N}} - 2r \overline{\mathfrak{b}}_{uv}|_{\mathcal{N}}+2r \mathfrak{b}_{vv}|_{\mathcal{N}} \nonumber \\ & \qquad - \frac{1}{2r} \Omega^{AB} \overline{\mathfrak{b}}_{AB}|_{\mathcal{N}}.
\end{align}
Thus, since $r=\frac{v}{2}$ on $\mathcal{N}$,
\begin{align}
    G^{(1)}_{uv}[ \mathfrak{H}]|_{\mathcal{N}}&=G^{(1)}_{uv}[u \mathfrak{b}]|_{\mathcal{N}} \\ 
    &= -\frac{1}{2} \Box(u \overline{\mathfrak{b}}_{uv})|_{\mathcal{N}} + \frac{1}{2}(\nabla_u C_v[u\mathfrak{b}] \nonumber \\ &\qquad+\nabla_v C_u[u\mathfrak{b}])|_{\mathcal{N}} - \frac{1}{2} \eta_{uv} \nabla_\rho C^\rho[u\mathfrak{b}]|_{\mathcal{N}} \\
    &= \frac{1}{r} \overline{\mathfrak{b}}_{uv}|_{\mathcal{N}} + \nabla_v \overline{\mathfrak{b}}_{uv}|_{\mathcal{N}} - \nabla_u \overline{\mathfrak{b}}_{vv}|_{\mathcal{N}} \nonumber \\ &\qquad \qquad  - \frac{1}{2}\nabla^\rho \overline{\mathfrak{b}}_{\rho v}|_{\mathcal{N}} - \frac{\Omega^{AB}\overline{\mathfrak{b}}_{AB}}{4r^3}|_{\mathcal{N}} \\
    &\stackrel{\eqref{eqn:nabla rho overline b rho v}}{=} \frac{1}{r} \overline{\mathfrak{b}}_{uv}|_{\mathcal{N}} + 2 \nabla_v \overline{\mathfrak{b}}_{uv}|_{\mathcal{N}}  \nonumber \\ & \quad  - \frac{1}{2r^2} \Omega^{AB} \nabla_A \overline{\mathfrak{b}}_{Bv}|_{\mathcal{N}} - \frac{\Omega^{AB}\overline{\mathfrak{b}}_{AB}}{4r^3}|_{\mathcal{N}} \\
    &\stackrel{\eqref{eqn:Omega AB nabla A b Bv}}{=} \frac{2}{r} \overline{\mathfrak{b}}_{uv}|_{\mathcal{N}} + 2 \partial_v \overline{\mathfrak{b}}_{uv}|_{\mathcal{N}} \nonumber \\ &  \qquad - \frac{1}{r} \mathfrak{b}_{vv}|_{\mathcal{N}} - \frac{1}{2r^2} D^A \mathfrak{b}_{vA}|_{\mathcal{N}} \\
    &\stackrel{\eqref{eqn:buv dbuv dv omega bab}}{=} -\frac{1}{r} \mathfrak{b}_{vv}|_{\mathcal{N}} + \frac{1}{2r^2}\Big(\partial_v \Omega^{AB} \mathfrak{b}_{AB} \nonumber \\ & \qquad \qquad - D^A \mathfrak{b}_{vA}\Big)|_{\mathcal{N}} \\
    &= -\frac{2}{v^2}(v \mathfrak{b}_{vv} + D^A \mathfrak{b}_{vA} - \partial_v \Omega^{AB}\mathfrak{b}_{AB})|_{\mathcal{N}} \\ 
    &= -\frac{2}{v^2}(v \partial_u \mathfrak{H}_{vv} + D^A \partial_u \mathfrak{H}_{vA} \nonumber \\ & \qquad \qquad  - \partial_v  \partial_u \Omega^{AB} \mathfrak{H}_{AB})|_{\mathcal{N}}.
\end{align}

Thus, the characteristic constraint equations
\begin{equation}
    G_{v\nu}^{(1)}[\mathfrak{H}]_{\mathcal{N}} = \kappa \Delta T_{v\nu}|_{\mathcal{N}}
\end{equation}
reduce to the system of equations
\begin{align}
    \label{eqn:Tvv EFE}
    \frac{2}{v} \mathfrak{b}_{vv} &= \kappa \Delta T_{vv}, \\
    \label{eqn:TvA EFE}
    \left(\partial_v + \frac{2}{v}\right) \mathfrak{b}_{vA} - D_A \mathfrak{b}_{vv} &= \kappa \Delta T_{vA}, \\
    \label{eqn:Tuv EFE}
    -\frac{2}{v^2}(v \mathfrak{b}_{vv}+D^A \mathfrak{b}_{vA} - \partial_{v} \Omega^{AB} \mathfrak{b}_{AB}) &= \kappa \Delta T_{uv},
\end{align}
where every source component is evaluated at $(u,v,\theta) = (0,v,\theta)$ on $\mathcal{N}$. We thus have
\begin{equation}
    \label{eqn:vv solution}
    \partial_u \mathfrak{H}_{vv}|_{\mathcal{N}} = \mathfrak{b}_{vv}(v,\theta) = \frac{\kappa v}{2}\Delta T_{vv}(0,v,\theta),
\end{equation}
from which we find
\begin{multline}
    \label{eqn:vA solution}
    \partial_u \mathfrak{H}_{vA}|_{\mathcal{N}} = \mathfrak{b}_{vA}(v,\theta) \\ = \frac{\kappa}{v^2} \int_0^v s^2 \left(\Delta T_{vA} + \frac{s}{2} D_A \Delta T_{vv} \right)(0,s,\theta) ds,
\end{multline}
and so
\begin{multline}
    \label{eqn:AB solution}
    \partial_u \Omega^{AB} \mathfrak{H}_{AB}|_{\mathcal{N}} = \Omega^{AB} \mathfrak{b}_{AB}(v,\theta) = \int^v_0 \Big(D^A \mathfrak{b}_{vA}(s,\theta) \\ +
    \frac{\kappa s^2}{2} (\Delta T_{vv} + \Delta T_{uv})(0,s,\theta)\Big) ds
\end{multline}
Note that there is homogeneous freedom (constants of integration) for some of the components. We want solutions that are regular at the vertex; the above are then fixed by vertex regularity, as we now see.

\subsubsection{Vertex regularity}

\label{app:vertex regularity}

We focused on $\mathcal{N}_\mathrm{reg}$, but we want to understand whether $\mathfrak{H}$ is also regular at the vertex. The equations for $\mathfrak b_{vv}$, $\mathfrak b_{vA}$ and $B=\Omega^{AB}\mathfrak b_{AB}$ use components in the coordinates $(u,v,\theta^A)$. Here $A$ takes two values corresponding to the usual polar angles $(\theta,\phi)$ wherever that particular angular chart is regular. The assertion that the source is regular at the vertex, however, is an assertion about its components in the ordinary Cartesian coordinates $X^a=(ct,x,y,z)$. Lower-case Latin indices $a,b$ range over $0,1,2,3$, with $X^0=ct$; spatial indices $i,j$ range over $1,2,3$. The precise smoothness hypothesis used below is that the Cartesian components of $\Delta T$ extend smoothly to a neighbourhood of the closed future cone. In particular, we assume their Cartesian derivatives of every specified finite order are bounded on each compact portion of that closed cone. 

Write $\omega^i(\theta)$ for the unit vector specifying a direction on the unit sphere, so that $\omega^i\omega_i=1$, and put $e_A^i=\partial_A\omega^i$. The identity $\omega_i e_A^i=0$ follows by differentiating
$\omega^i\omega_i=1$. The unit-sphere metric and its inverse are
\begin{equation}
 \Omega_{AB}=e_A^i e_B^j\delta_{ij},\qquad
 \Omega^{AC}\Omega_{CB}=\delta^A_B.
\end{equation}
The coordinate transformation and its derivatives are
\begin{align}
 X^0&=\frac{u+v}{2}, \quad
 X^i=\frac{v-u}{2}\omega^i(\theta),\\
 J_u^a&:=\frac{\partial X^a}{\partial u}
   =\frac12(1,-\omega^i), \\
 J_v^a&:=\frac{\partial X^a}{\partial v}
   =\frac12(1,\omega^i),\\
 J_A^a&:=\frac{\partial X^a}{\partial\theta^A}
   =(0,r e_A^i). 
 \label{eq:vc-Jacobians}
\end{align}
On the future cone $u=0$ we have $ct=r=v/2$, and hence
$J_A^a=(0,v e_A^i/2)$. Thus the vectors $J_u$ and $J_v$ have bounded Cartesian components as $v\downarrow0$, while each angular coordinate vector $J_A$ has Cartesian size proportional to $v$. Moreover,
\begin{equation}
 T_{\mu\nu}=J_\mu^a J_\nu^b T^{\rm Cart}_{ab}.
 \label{eq:vc-tensor-transform}
\end{equation}
Define the Cartesian source sampled along the cone ($ct=r \Rightarrow ct=v/2, r=v/2$) by
\begin{equation}
 \mathcal{T}_{ab}(v,\theta) :=\Delta T^{\rm Cart}_{ab}\left(\frac{v}{2},\frac{v}{2}\vec{\omega}(\theta)\right).
 \label{eq:vc-S}
\end{equation}
Direct substitution in \eqref{eq:vc-tensor-transform} gives every group of null components:
\begin{align}
 \Delta T_{vv}(0,v,\theta)
 &=\frac14\left(\mathcal{T}_{00}+2\omega^i\mathcal{T}_{0i}
                              +\omega^i\omega^j\mathcal{T}_{ij}\right),\\
 \Delta T_{uu}(0,v,\theta)
 &=\frac14\left(\mathcal{T}_{00}-2\omega^i\mathcal{T}_{0i}
                              +\omega^i\omega^j\mathcal{T}_{ij}\right),\\
 \Delta T_{uv}(0,v,\theta)
 &=\frac14\left(\mathcal{T}_{00}-\omega^i\omega^j\mathcal{T}_{ij}\right),\\
 \Delta T_{vA}(0,v,\theta)
 &=\frac v4 e_A^i\left(\mathcal{T}_{0i}+\omega^j\mathcal{T}_{ji}\right),\\
 \Delta T_{uA}(0,v,\theta)
 &=\frac v4 e_A^i\left(\mathcal{T}_{0i}-\omega^j\mathcal{T}_{ji}\right),\\
 \Delta T_{AB}(0,v,\theta)
 &=\frac{v^2}{4}e_A^i e_B^j\mathcal{T}_{ij}.
 \label{eq:vc-source-components}
\end{align}
These are the components to insert into the constraint integrals. In particular, an expression such as $\Delta T_{vA}(0,s,\theta)$ means the \emph{null-coordinate component} at the source point with $u=0$, $v=s$.

For $f(v,\theta)$, the notation $f\sim\mathcal{O}(v^p)$ as $v\downarrow0$ means that there are constants $C$ and $v_*>0$ such that $|f(v,\theta)|\leq Cv^p$ for $0<v\leq v_*$. Saying that the estimate is \emph{uniform in angle} means that $C$ does not depend on the point $\theta$ under consideration. For angular tensors one uses the norm defined by the unit-sphere metric. For example,
\begin{equation}
 |a|_\Omega^2=\Omega^{AB}a_Aa_B,
 \qquad
 |t|_\Omega^2=\Omega^{AC}\Omega^{BD}t_{AB}t_{CD}.
\end{equation}
We then see that
\begin{align}
 \Delta T_{vv}&\sim\mathcal{O}(1),\quad
 \Delta T_{uv}\sim\mathcal{O}(1),\\
 \Delta T_{vA}&\sim\mathcal{O}(v),\quad
 \Delta T_{AB}\sim\mathcal{O}(v^2),
 \label{eq:vc-source-orders}
\end{align}
with the same powers after any prescribed number of angular derivatives. Changing from ordinary angular derivatives to $D$ only adds smooth bounded sphere coefficients and lower angular derivatives, so it preserves these estimates.

The vertex-selected particular solutions of the constraint equations are
\begin{align}
 \mathfrak b_{vv}(v,\theta)
 &=\frac{\kappa v}{2}\Delta T_{vv}(0,v,\theta),\\
 \mathfrak b_{vA}(v,\theta)
 &=\frac\kappa{v^2}\int_0^v s^2
   \Big[\Delta T_{vA}(0,s,\theta) \nonumber \\
               &\qquad \qquad +\frac s2D_A\Delta T_{vv}(0,s,\theta)\Big]ds,\\
 \Omega^{AB} \mathfrak{b}_{AB}(v,\theta)
 &=\int_0^v\Big[D^A\mathfrak b_{vA}(s,\theta) \nonumber \\ & \qquad \qquad
       +\frac{\kappa s^2}{2}
             (\Delta T_{vv}+\Delta T_{uv})(0,s,\theta)\Big]ds.
 \label{eq:vc-integrals}
\end{align}

But
\begin{multline}
 \abs{f(s,\theta)}\leq C s^p,\quad p>-1
 \\ \Longrightarrow\quad
 \left|\int_0^v f(s,\theta)\,ds\right|
 \leq C\int_0^v s^p ds
 =\frac{C}{p+1}v^{p+1}.
 \label{eq:vc-integration-rule}
\end{multline}
The condition $p>-1$ ensures convergence at the lower endpoint. Uniformity of the original bound in $\theta$ makes the resulting bound uniform as well.

First, multiplication of $\Delta T_{vv}\sim\mathcal{O}(1)$ by $v$ immediately gives $\mathfrak b_{vv}\sim\mathcal{O}(v)$. Second, $\Delta T_{vA}\sim\mathcal{O}(s)$ and $D_A\Delta T_{vv}\sim\mathcal{O}(1)$ imply
\begin{align}
 \Delta T_{vA}+\frac s2D_A\Delta T_{vv}&\sim\mathcal{O}(s),\\
 s^2\left[\Delta T_{vA}+\frac s2D_A\Delta T_{vv}\right]&\sim\mathcal{O}(s^3).
\end{align}
Integrating $O(s^3)$ from $0$ to $v$ gives $\mathcal{O}(v^4)$, and dividing by
$v^2$ gives $\mathfrak b_{vA}\sim\mathcal{O}(v^2)$.  Third, the angular metric and its angular derivative coefficients do not depend on $s$. Consequently,
$D^A\mathfrak b_{vA}\sim\mathcal{O}(v^2)$. The integrand defining $\Omega^{AB} \mathfrak{b}_{AB}$ is therefore
$O(s^2)+s^2\mathcal{O}(1)\sim\mathcal{O}(s^2)$. Its integral is $\mathcal{O}(v^3)$. We have established
\begin{equation}
 \mathfrak b_{vv}\sim\mathcal{O}(v),\qquad
 \mathfrak b_{vA}\sim\mathcal{O}(v^2),\qquad \Omega^{AB} \mathfrak{b}_{AB}\sim\mathcal{O}(v^3),
 \label{eq:vc-b-orders}
\end{equation}
with the same powers after any fixed number of angular derivatives.

Suppose two solutions of the same constraint equations are compared. Their difference has zero source. The $vv$ equation \eqref{eqn:Tvv EFE} gives $\delta\mathfrak b_{vv}=0$. The $vA$ equation \eqref{eqn:TvA EFE} then becomes
\begin{equation}
 \partial_v\delta\mathfrak b_{vA} +\frac2v\delta\mathfrak b_{vA}=0, \qquad \partial_v\left(v^2\delta\mathfrak b_{vA}\right)=0.
\end{equation}
For each fixed angle, the quantity inside the last derivative is constant along $v$. Calling it $C_A(\theta)$ gives $\delta\mathfrak b_{vA}=C_A/v^2$. The remaining equation \eqref{eqn:Tuv EFE} reads $\partial_v\delta \Omega^{AB} \mathfrak{b}_{AB}=D^A\delta\mathfrak b_{vA}$. Hence
\begin{align}
 \delta\mathfrak b_{vv}&=0,\\
 \delta\mathfrak b_{vA}&=\frac{C_A(\theta)}{v^2},\\
 \delta \Omega^{AB} \mathfrak{b}_{AB}&=-\frac{D^AC_A(\theta)}v+C(\theta).
 \label{eq:vc-homogeneous}
\end{align}
Indeed, $\partial_v[-(D^AC_A)/v]=(D^AC_A)/v^2$. $C_A$ and $C$ are functions of $\theta$, not constants that must be the same in all directions.

These terms should be excluded since we require that
\begin{equation}
 \mathfrak{b}_{vv}\sim \mathcal{O}(v),\quad
 \mathfrak{b}_{vA}\sim\mathcal{O}(v^2),\quad
 \Omega^{AB}\mathfrak{b}_{AB}\sim\mathcal{O}(v^3).
 \label{eq:vc-minimal-orders}
\end{equation}
Thus, at each fixed angle,
\begin{equation}
 \frac{\abs{C_A(\theta)}_\Omega}{v^2}\leq Mv^2
\end{equation}
for some constant $M$. Letting $v\downarrow0$ forces $C_A(\theta)=0$. This holds at every angle. With $C_A=0$, its divergence also vanishes and $\delta \Omega^{AB} \mathfrak{b}_{AB}=C(\theta) = 0$ as well. Thus vertex regularity eliminates all the homogeneous terms in \eqref{eq:vc-homogeneous}.

\subsubsection{An explicit expression for $\mathfrak{H}$}

The preceding conditions select three parts of the first transverse derivative: these constraints only determine $\mathfrak{b}_{vv}$, $\mathfrak{b}_{vA}$ and $\Omega^{AB} \mathfrak{b}_{AB}$. One chooses (non-uniquely) the remaining components to be 
\begin{equation}
    \mathfrak{b}_{uu} = \mathfrak{b}_{uv} = \mathfrak{b}_{uA} = 0
\end{equation}
and
\begin{equation}
    \label{eqn:angular bAB}
    \mathfrak{b}_{AB} = \frac{1}{2}\Omega_{AB} \Omega^{CD} \mathfrak{b}_{CD}.
\end{equation}
We now construct an actual $\mathfrak{H}$ having those parts. Its role is to correct the source flux at the cone before the retarded wave equation is solved. Choose a fixed $0<\epsilon<1$. A smooth cutoff function equal to $1$ for $z\leq \epsilon/2$ and zero for $z\geq \epsilon$ can be specified: define
\begin{align}
 e(s)&=\begin{cases}\exp(-1/s),&s>0,\\0,&s\leq0,\end{cases}
 \\
 \chi_\epsilon(z)&=\frac{e(\epsilon-z)}{e(\epsilon-z)+e\left(z-\frac{\epsilon}{2}\right)}.
 \label{eq:vc-cutoff}
\end{align}
For $s>0$, each derivative of $e(s)$ is a polynomial in $s^{-1}$ times $\exp(-1/s)$. Every such derivative tends to zero as $s\downarrow0$, because exponential decay beats every power. Thus $e$ is smooth across zero (it is a bump function), with all derivatives zero there. The denominator in \eqref{eq:vc-cutoff} is positive at every real $z$: if its first term were zero then $z\geq \epsilon>\epsilon/2$, making its second term positive, and conversely. This proves all the stated properties of $\chi_\epsilon$, including smooth transitions to its constant values.

For $ct>r>0$ and $0<\epsilon<1$, set
\begin{multline}
 \mathfrak{H}(u,v,\theta)
 =u\chi_\epsilon(u/v)\Big[
    \mathfrak b_{vv}(v,\theta)\,dv^2
   \\+2\mathfrak b_{vA}(v,\theta)\,dv\,d\theta^A
   \\+\frac{\Omega^{CD} \mathfrak{b}_{CD}(v,\theta)}2\Omega_{AB}\,d\theta^A d\theta^B
                   \Big].
 \label{eq:vc-collar}
\end{multline}
The notation denotes a symmetric tensor: the coefficient of
$dv\,d\theta^A$ is twice its $vA$ component. Set the field to zero where $u/v\geq \epsilon$, including the axis $r=0$, $ct>0$. This definition agrees smoothly with \eqref{eq:vc-collar}. On the cone, $u=0$ and $\chi_\epsilon(0)=1$, so the field vanishes. Its first $u$-derivative there has
exactly the specified components, with the auxiliary choices
\begin{equation}
 \mathfrak b_{uu}=\mathfrak b_{uv}=\mathfrak b_{uA}=0,
 \qquad \mathfrak b_{AB}=\frac{\Omega^{CD} \mathfrak{b}_{CD}}{2}\Omega_{AB}.
\end{equation}
These assignments define this particular $\mathfrak{H}$; they do not impose additional equations on the source. This $\mathfrak{H}$ is easily seen to be $C^\infty(I^+(0))$ and locally Lipschitz across $\mathcal{N}$. 

\subsection{Computing $H$}

We finish with the remaining $H$ term, which we require to satisfy the linearised Einstein field equation in Lorenz gauge sourced by 
\begin{equation}
    \tau_{\mu\nu}(u,v,\theta) := \Theta(u)\left(\Delta T_{\mu\nu}(u,v,\theta) - \kappa ^{-1}G^{(1)}_{\mu\nu}[\mathfrak{H}](u,v,\theta)\right)
\end{equation}
as in \eqref{eqn:H source}, i.e.
\begin{equation}
    \Box \overline{H}_{\mu\nu}(ct,\vec{x}) = -2\kappa \tau_{\mu\nu}(ct,\vec{x}), \qquad C_\nu[H](ct,\vec{x})=0.
\end{equation}
Then
\begin{align}
    \overline{H}_{\mu\nu}(ct,\vec{x}) &= 2\kappa \int_{\mathbb{M}} G_{\mathrm{ret}}(x,x') \tau_{\mu\nu}(x') d^4x' \\
    &= 2 \kappa \int_{J^-(x) \cap \M^+} G_\mathrm{ret}(x,x') \nonumber \\ & \qquad (\Delta T_{\mu\nu} - \kappa^{-1}G^{(1)}_{\mu\nu}[\mathfrak{H}])(x') d^4x' \\
    &= \frac{4G}{c^4} \iiint_{\mathbb{R}^3} \frac{\tau_{\mu\nu}(ct-\abs{\vec{x}-\vec{x}'},\vec{x}')}{\abs{\vec{x}-\vec{x}'}} d^3 \vec{x}'
\end{align}
for $x=(ct,\vec{x})$ in $\M^+$. This indeed satisfies the Lorenz gauge:
\begin{equation}
    \Box(\partial^\mu \overline{H}_{\mu\nu}) = - 2 \kappa \partial^\mu \tau_{\mu\nu}=0,
\end{equation}
and $\partial^\mu \overline{H}_{\mu\nu}$ vanishes in the past of $\mathrm{supp} \; \tau$, so by uniqueness of the retarded solution \cite[Corollary 4.2]{bar_green-hyperbolic_2015} it vanishes identically. Hence,
\begin{equation}
    G^{(1)}[H]=-\frac{1}{2} \Box \overline{H} = \kappa \tau \quad \text{ in } \M^+,
\end{equation}
and so
\begin{equation}
    G^{(1)}[\mathfrak{h}^+] = G^{(1)}[\mathfrak{H}] + G^{(1)}[H] = \kappa \Delta T \quad \text{ in } \M^+.
\end{equation}
Moreover, the jump $\partial_u \mathfrak{h}^+|_\mathcal{N}$, and thus $T^\mathrm{sing}$, are carried by $\mathfrak{b}$ i.e. by $\mathfrak{H}$. Indeed, since $\overline{H}|_{\mathcal{N}}=0$, we write $a_{\mu\nu} := \partial_u \overline{H}_{\mu\nu}|_{\mathcal{N}}$, and so Eqn.~\eqref{eqn:box u b} applied to $\Box \overline{H} = -2\kappa \tau$ yields
\begin{equation}
    \left(\nabla_{v} + \frac{1}{v}\right) a_{\mu\nu} = \frac{\kappa}{2}\tau_{\mu\nu}|_{\mathcal{N},+}
\end{equation}
hence
\begin{align}
    \frac{1}{v} \partial_v(v a_{vv}) &= \frac{\kappa}{2} \tau_{vv}|_{\mathcal{N},+} \\
    \partial_v a_{vA} &= \frac{\kappa}{2} \tau_{vA}|_{\mathcal{N},+},\\
    \frac{1}{v}\partial_v(v a_{uv}) &= \frac{\kappa}{2} \tau_{uv}|_{\mathcal{N},+}.
\end{align}
The constraint on $\mathfrak{H}$ is exactly $\tau_{v\nu}|_{\mathcal{N}} = 0$, so the vertex-regular branch gives $a_{v\nu}=0$ i.e. the $H$-jump has vanishing $vv$,$vA$ and $\Omega^{AB}$-trace components. Since $\mathfrak{H}_{\mu\nu}$ is $C^\infty$ in $\M^+$, so is $G^{(1)}_{\mu\nu}[\mathfrak{H}]$ in $\M^+$. Thus, since $\Delta T_{\mu\nu}$ is $C^\infty$ in $\M^+$, so is $\tau_{\mu\nu}$ in $\M^+$. $\overline{H}_{\mu\nu}$ is sourced by $G_\mathrm{ret} * \tau_{\mu\nu}$ at points $J^-(x) \cap \M^+ \subset \M^+$ so $\overline{H}_{\mu\nu}$ is $C^\infty$ in $\M^+$. Thus,
\begin{equation}
    G^{(1)}[h^+] = G^{(1)}[\underline{h}^-] + G^{(1)}[\mathfrak{h}^+] = \kappa T^+ \quad \text{ in } \M^+,
\end{equation}
and $h^+ \in C^\infty(I^+(0))$ and locally Lipschitz across $\Lambda^+(0)$.

\subsection{Computing the Barrabès-Israel stress-energy}

\label{app:BI linearised}

We now compute the Barrabès-Israel stress-energy corresponding to $h$.

\subsubsection{$H$ contribution to $T^\mathrm{sing}$}

 Both $\mathfrak{H}$ and the global retarded field $H$ vanish on $\mathcal{N}$. Their transverse derivatives need not vanish. The smooth continuation $\underline{h}^-$ has no derivative jump. Thus, at every regular cone point,
\begin{equation}
 [\partial_\alpha h_{\mu\nu}]
 =n_\alpha\left(
    \mathfrak b_{\mu\nu}
    +\left.\partial_uH_{\mu\nu}\right|_{\mathcal N,+}
                 \right).
 \label{eq:BIc-derivative-jump}
\end{equation}
Here $[\cdot]$ means future minus past, as in  Appendix \ref{app:BI}. Let
\begin{equation}
 \mathfrak{a}_{\mu\nu}(v,\theta)
 :=\left.\partial_u\overline H_{\mu\nu}\right|_{\mathcal N,+}.
\end{equation}
Since $\overline{H}|_{\mathcal N}=0$, its covariant first derivative on the cone is
\begin{equation}
 \nabla_\alpha\overline H_{\mu\nu}|_{\mathcal N,+}
 =n_\alpha \mathfrak{a}_{\mu\nu}.
\end{equation}
Indeed, the tangential derivatives vanish and the connection terms multiply the zero boundary value. Taking the continuous one-sided limit of the interior Lorenz condition therefore gives
\begin{equation}
 0=\nabla^\mu\overline H_{\mu\nu}|_{\mathcal N,+}
 =n^\mu \mathfrak{a}_{\mu\nu}=-2\mathfrak{a}_{v\nu}.
 \label{eq:BIc-a-vanishes}
\end{equation}
We want to convert this trace-reversed statement to a statement about the unbarred metric. Differentiating trace reversal at the cone does not produce extra terms from differentiating the background metric: all such terms multiply $H|_{\mathcal N}=0$. Also,
\begin{equation}
    \eta^{\rho\sigma}\partial_uH_{\rho\sigma}|_{\mathcal N,+}
 =-4\partial_uH_{uv}|_{\mathcal N,+}
 +r^{-2}\Omega^{AB}
           \partial_uH_{AB}|_{\mathcal N,+}.
\end{equation}
It follows, component by component, that
\begin{align}
 \mathfrak{a}_{vv}&=\partial_uH_{vv}|_{\mathcal N,+},\\
 \mathfrak{a}_{vA}&=\partial_uH_{vA}|_{\mathcal N,+},\\
 \mathfrak{a}_{uv}&=\partial_uH_{uv}|_{\mathcal N,+}
 -\partial_uH_{uv}|_{\mathcal N,+} \nonumber \\
   &+\frac{1}{4}r^{-2}\Omega^{AB}\partial_uH_{AB}|_{\mathcal N,+} \\
          &=\frac1{4r^2}\Omega^{AB}
                    \partial_uH_{AB}|_{\mathcal N,+}.
\end{align}
Together with \eqref{eq:BIc-a-vanishes}, these identities give exactly
the projections needed below:
\begin{align}
 \partial_uH_{vv}|_{\mathcal N,+}&=0,\\
 \quad \partial_uH_{vA}|_{\mathcal N,+}&=0,\\
 \quad \Omega^{AB}\partial_uH_{AB}|_{\mathcal N,+}&=0.
 \label{eq:BIc-H-projections}
\end{align}
They do not imply that every component of $\partial_uH$ vanishes. In particular, the angular trace-free part $\partial_u H_{AB}|_{\mathcal{N},+}$ can be nonzero.

\subsubsection{The singular stress-energy}

We have
\begin{equation}
 C_{ab}=-N_\rho e_a^\mu\nabla_\mu e_b^\rho,
 \qquad a,b\in\{v,A\}.
 \label{eq:BIc-C-definition}
\end{equation}
The lower-case Latin indices label the three tangent vectors; Greek indices label spacetime components. The transverse curvature measures the part of the change of a tangent vector that is detected by the transverse null vector $N$.

In the null-coordinate chart, the coordinate components of $e_v$ and $e_A$ are constant: $e_v = \partial_v$, $e_A = \partial_A$, so $e^\mu_v = \delta^\mu_v$ and $e^\mu_A = \delta^\mu_A$. Hence their ordinary coordinate derivatives vanish, while their covariant derivatives contain the Christoffel symbols: $\nabla_\mu e^\rho_b = \Gamma^\rho_{\; \mu b} = \Gamma^\rho_{\; \mu \nu} e^\nu_b$ for $b \in \{v,A\}$. Thus,
\begin{equation}
 [C_{ab}]=-N_\rho[\Gamma^\rho_{\mu\nu}]e_a^\mu e_b^\nu.
\end{equation}
The connection jump follows by taking the jump of the Christoffel formula. Since the metric is continuous, the inverse metric is continuous. The inverse metric expansion is $g^{\rho \sigma} = \eta^{\rho \sigma} - \eta^{\rho \alpha} \eta^{\sigma \beta} h_{\alpha \beta} + \mathcal{O}(h^2)$; thus, replacing its value by $\eta^{-1}$ changes the answer by $\mathcal{O}(h[\partial h])$ i.e. at second order. Thus
\begin{align}
 [\Gamma^\rho_{\mu\nu}] = \frac{1}{2} \eta^{\rho \sigma}\Big[&[\partial_\mu h_{\nu\sigma}] + [\partial_\nu h_{\mu \sigma}] - [\partial_\sigma h_{\mu\nu}]\Big] + \mathcal{O}(h^2)\\
 =\frac12\eta^{\rho\sigma}\Big[&
 n_\mu\left(\mathfrak b_{\nu\sigma}
       +\left.\partial_uH_{\nu\sigma}\right|_{\mathcal N,+}\right) \nonumber \\
 &+n_\nu\left(\mathfrak b_{\mu\sigma}
       +\left.\partial_uH_{\mu\sigma}\right|_{\mathcal N,+}\right)
 \nonumber\\[-2pt]
 &-n_\sigma\left(\mathfrak b_{\mu\nu}
       +\left.\partial_uH_{\mu\nu}\right|_{\mathcal N,+}\right)
 \Big].
 \label{eq:BIc-connection}
\end{align}
Here, $n_\alpha = \partial_\alpha u = \delta^u_\alpha$ i.e. $[\partial_\alpha h_{\mu\nu}] = n_\alpha [\partial_u h_{\mu\nu}]$. Since $n_\mu e^\mu_a=0$, the first two terms vanish after contraction with $e^\mu_a e^\nu_b$, and $n_\rho N^\rho=2$ so
\begin{equation}
 [C_{ab}]=\left(\mathfrak b_{\mu\nu}
       +\left.\partial_uH_{\mu\nu}\right|_{\mathcal N,+}\right)
                          e_a^\mu e_b^\nu.
 \label{eq:BIc-C-jump}
\end{equation}
In particular,
\begin{align}
 [C_{vv}]&=\mathfrak b_{vv},\\
 [C_{vA}]&=\mathfrak b_{vA},\\
 [C_{AB}]&=\mathfrak b_{AB}
                +\left.\partial_uH_{AB}\right|_{\mathcal N,+},\\
 \Omega^{AB}[C_{AB}]&=\Omega^{AB} \mathfrak{b}_{AB}.
 \label{eq:BIc-C-components}
\end{align}
As a check using the notation of Appendix \ref{app:BI},
\begin{equation}
 \gamma_{\mu\nu}
 =[\partial_\rho h_{\mu\nu}]N^\rho
 =2\left(\mathfrak b_{\mu\nu}
       +\left.\partial_uH_{\mu\nu}\right|_{\mathcal N,+}\right).
\end{equation}
Then $[\partial_\rho h_{\mu\nu}]=-\gamma_{\mu\nu}k_\rho$ where $k=\partial_v$, and $[C_{ab}]=\gamma_{\mu\nu}e_a^\mu e_b^\nu/2$, in agreement with \eqref{eq:BIc-derivative-jump} and \eqref{eq:BIc-C-jump}. The Barrabès-Israel relations reviewed earlier give, with $\lambda=v$,
\begin{align}
 \mu&=-\frac1\kappa\sigma^{AB}[C_{AB}],\\
 j^A&=\frac1\kappa\sigma^{AB}[C_{vB}],\\
 p&=-\frac1\kappa[C_{vv}].
 \label{eq:BIc-BI-relations}
\end{align}
On a cross-section of the cone, $u=0$ and $v$ is fixed, i.e. $du=dv=0$, and the background metric restricted to that cross-section is $\sigma_{AB} = r^2 \Omega_{AB}$, i.e. $\sigma^{AB} = \frac{1}{r^2} \Omega^{AB}$. Substituting \eqref{eq:BIc-C-components} gives
\begin{align}
 \mu(v,\theta)&=-\frac{\Omega^{AB} \mathfrak{b}_{AB}(v,\theta)}{\kappa r^2},\\
 j^A(v,\theta)&=\frac{\Omega^{AB}\mathfrak b_{vB}(v,\theta)}{\kappa r^2},\\
 \qquad p(v,\theta)&=-\frac{\mathfrak b_{vv}(v,\theta)}\kappa,
 \label{eq:BIc-coefficients}
\end{align}
where $r=\frac v2$ on $u=0$. The surface stress-energy tensor is
\begin{equation}
 S^{\mu\nu}=\mu k^\mu k^\nu
 +j^A(k^\mu e_A^\nu+e_A^\mu k^\nu)
 +p\sigma^{AB}e_A^\mu e_B^\nu.
 \label{eq:BIc-S}
\end{equation}
Because $k^v=1$ and $e_A^B=\delta_A^B$, every group of contravariant null-coordinate components is
\begin{align}
 S^{vv}&=\mu,\qquad S^{vA}=j^A,\\
 S^{AB}&=\frac{p}{r^2}\Omega^{AB},\qquad S^{u\nu}=0.
 \label{eq:BIc-S-up}
\end{align}
Lowering both indices with the background metric gives
\begin{align}
 S_{uu}&=\eta_{uv}^2 S^{vv}=\frac\mu4=-\frac{\Omega^{AB} \mathfrak{b}_{AB}}{4\kappa r^2},\\
 S_{uA}&=\eta_{uv}\,r^2\Omega_{AB}S^{vB}
       =-\frac{\mathfrak b_{vA}}{2\kappa},\\
 S_{AB}&=r^4\Omega_{AC}\Omega_{BD}S^{CD}
       =-\frac{r^2}{\kappa}\Omega_{AB}\mathfrak b_{vv},\\
 S_{v\nu}&=0.
 \label{eq:BIc-S-down}
\end{align}
The distributional stress-energy on $\mathcal{N}$ is thus
\begin{equation}
    \label{eqn:distributional stress energy raw}
 T^{\mathrm{sing}}_{\mu\nu}
 =\frac{S_{\mu\nu}}{-k\cdot U}\delta(\tau).
\end{equation}
Here the scalar $\tau$ denotes the observer's proper-time (not the tensor $\tau_{\mu\nu}$ sourcing $H$). Note that $U^\mu=dX^\mu/d\tau$ has norm $-c^2$ in coordinates $X^0=ct$. For inertial observers at fixed spatial position,
\begin{align}
 U&=c\partial_ {ct},\qquad \tau=t-r/c=u/c,
 \\ -k\cdot U&=\frac c2,
 \qquad \delta(\tau)=c\delta(u).
\end{align}
Substitution gives $2S\delta(u)$ for the right-hand side of Eqn.~\eqref{eqn:distributional stress energy raw}. For an arbitrary crossing observer, the same cancellation follows from
\begin{align}
 \frac{du}{d\tau}&=n_\mu U^\mu=-2k\cdot U>0,
 \\ \delta(\tau)&=(-2k\cdot U)\delta(u)
 \quad\text{on the cone}.
\end{align}
Thus,
\begin{equation}
 T^{\mathrm{sing}}_{\mu\nu}=2S_{\mu\nu}\delta(u).
 \label{eq:BIc-normalized-delta}
\end{equation}
We have therefore determined every covariant component of the shell:
\begin{equation}
 \begin{aligned}
 T^{\mathrm{sing}}_{uu}
   &=-\frac{\Omega^{AB} \mathfrak{b}_{AB}(v,\theta)}{2\kappa r^2}\delta(u),\\
 T^{\mathrm{sing}}_{uA}=T^{\mathrm{sing}}_{Au}  
   &=-\frac{\mathfrak b_{vA}(v,\theta)}\kappa\delta(u),\\
 T^{\mathrm{sing}}_{AB}
   &=-\frac{2r^2}{\kappa}\Omega_{AB}
                  \mathfrak b_{vv}(v,\theta)\delta(u),\\
 T^{\mathrm{sing}}_{v\nu}&=0.
 \end{aligned}
 \label{eq:BIc-final-components}
\end{equation}

We therefore see that the jump in $\partial_u H_{AB}|_{\mathcal{N},+}$ can only contribute to the Weyl part of the distributional Riemann curvature tensor, which describes an impulsive tidal field, but does not contribute to the $\delta$ term in the distributional Einstein tensor.  Thus, only $\mathfrak{H}$ contributes to the Barrabès-Israel stress-energy.

\subsubsection{Explicit dependence on the source jump}

Inserting the constraint integrals into \eqref{eq:BIc-coefficients} gives
\begin{align}
 p(v,\theta)
 &=-\frac v2\Delta T_{vv}(0,v,\theta),
 \label{eq:BIc-p-source}\\
 j^A(v,\theta)
 &=\frac4{v^4}\int_0^v s^2
 \left[\Omega^{AB}\Delta T_{vB}
       +\frac s2D^A\Delta T_{vv}\right](0,s,\theta)\,ds,
 \label{eq:BIc-j-source}\\
 \mu(v,\theta)
 &=-\frac4{\kappa v^2}\int_0^v
 \Big[D^A\mathfrak b_{vA}(s,\theta) \nonumber \\
       &+\frac{\kappa s^2}{2}
           (\Delta T_{vv}+\Delta T_{uv})(0,s,\theta)
           \Big]ds.
 \label{eq:BIc-mu-source-intermediate}
\end{align}

To eliminate $\mathfrak b$ from $\mu$, recall Eqn.~\eqref{eq:vc-integrals}. The angular geometry does not depend on $v$, so
\begin{align}
 D^A\mathfrak b_{vA}(s,\theta)
 =\frac\kappa{s^2}&\int_0^s (s')^2
 \Big[D^A\Delta T_{vA} \nonumber \\
       &+\frac{s'}2D^A D_A\Delta T_{vv}
 \Big](0,s',\theta)\,ds'.
\end{align}
The prime on $s'$ only distinguishes the inner integration variable. The bracket is $\mathcal{O}(s')$ at the vertex, thus the double integral is absolutely convergent and its order may be interchanged. For fixed $s'$, the outer variable ranges
from $s'$ to $v$, and
\begin{equation}
 (s')^2\int_{s'}^v\frac{ds}{s^2}
 =(s')^2\left(\frac1{s'}-\frac1v\right)
 =s'-\frac{(s')^2}{v}.
\end{equation}
After renaming the integration variable, the result is
\begin{equation}
 \begin{aligned}
 \mu(v,\theta)={}&-\frac2{v^2}\int_0^v
        \Bigg[s^2(\Delta T_{vv}+\Delta T_{uv})(0,s,\theta)\\ &+ 2 \big(s-\frac{s^2}{v}\big)
 \Big[D^A\Delta T_{vA} \nonumber \\
       &+\frac s2D^AD_A\Delta T_{vv}\Big](0,s,\theta) \Bigg]\,ds.
 \end{aligned}
 \label{eq:BIc-mu-source}
\end{equation}

\section{Distributional conservation and the cone vertex}

\label{app:distributional conservation}

Although each ordinary branch is separately conserved, the step in the ordinary source has divergence
\begin{multline}
 \nabla^\mu\big(\underline T^-_{\mu\nu}
                         +\Theta(u)\Delta T_{\mu\nu}\big)
 =n^\mu\Delta T_{\mu\nu}|_{\mathcal N}\delta(u)\\
 =-2\Delta T_{v\nu}|_{\mathcal N}\delta(u).
 \label{eq:BIc-bulk-div}
\end{multline}
We verify directly that the shell cancels this term. To perform a calculation near a regular cone point, extend $\Omega^{AB} \mathfrak{b}_{AB}$ and $\mathfrak b_{v\nu}$ as functions of $(v,\theta)$ and use \eqref{eq:BIc-S-down} with $r=(v-u)/2$ off the cone as well. This is only an auxiliary extension: multiplying by $\delta(u)$ retains its boundary values alone. This choice has $S_{v\nu}=0$ throughout that neighbourhood, so
\begin{equation}
 n^\mu S_{\mu\nu}=-2S_{v\nu}=0.
\end{equation}
It follows that differentiating the delta function contributes no $\delta'(u)$:
\begin{equation}
 \nabla^\mu T^{\mathrm{sing}}_{\mu\nu}
 =2(\nabla^\mu S_{\mu\nu})|_{\mathcal N}\delta(u).
 \label{eq:BIc-shell-div-rule}
\end{equation}

Here the covariant derivatives, including the null-coordinate Christoffel terms, are essential. Using the Christoffel symbols listed in Appendix \ref{app:future metric} gives
\begin{align}
 \nabla^\mu S_{\mu v}
 &=\frac{\mathfrak b_{vv}}{\kappa r},
 \label{eq:BIc-div-v}\\
 \nabla^\mu S_{\mu A}
 &=\frac1\kappa\left(
       \partial_v\mathfrak b_{vA}+\frac{\mathfrak b_{vA}}r
                              -D_A\mathfrak b_{vv}\right),
 \label{eq:BIc-div-A}\\
 \nabla^\mu S_{\mu u}
 &=\frac{\Omega^{AB} \partial_v\mathfrak{b}_{AB}-D^A\mathfrak b_{vA}}{2\kappa r^2}
                          -\frac{\mathfrak b_{vv}}{\kappa r}.
 \label{eq:BIc-div-u}
\end{align}
The contractions can be seen explicitly. For the $v$ equation only the angular connection term remains:
\begin{equation}
 r^{-2}\Omega^{AB}\nabla_A S_{Bv}
 =-\frac1{2r^3}\Omega^{AB}S_{AB}
 =\frac{\mathfrak b_{vv}}{\kappa r}.
\end{equation}
For the $u$ equation the two nonzero contributions are
\begin{align}
 -2\nabla_v S_{uu}
 &=\frac{\Omega^{AB} \partial_v \mathfrak{b}_{AB}}{2\kappa r^2}
                            -\frac{\Omega^{AB} \mathfrak{b}_{AB}}{2\kappa r^3},\\
 r^{-2}\Omega^{AB}\nabla_A S_{Bu}
 &=-\frac{D^A\mathfrak b_{vA}}{2\kappa r^2}
   +\frac{\Omega^{AB} \mathfrak{b}_{AB}}{2\kappa r^3}-\frac{\mathfrak b_{vv}}{\kappa r}.
\end{align}
Their $\Omega^{AB} \mathfrak{b}_{AB}/r^3$ terms cancel. Moreover, 
\begin{align}
 -2\nabla_vS_{uA}
 &=\frac1\kappa\partial_v\mathfrak b_{vA}
                      -\frac{\mathfrak b_{vA}}{2\kappa r},\\
 r^{-2}\Omega^{BC}\nabla_BS_{CA}
 &=-\frac1\kappa D_A\mathfrak b_{vv}
                      +\frac{3\mathfrak b_{vA}}{2\kappa r}.
\end{align}
These sums give precisely \eqref{eq:BIc-div-v}--\eqref{eq:BIc-div-u}. On the cone, $r=v/2$, and the constraint equations read
\begin{align}
 \frac{\mathfrak b_{vv}}r&=\kappa\Delta T_{vv},\\
 \partial_v\mathfrak b_{vA}+\frac2v\mathfrak b_{vA}
                   -D_A\mathfrak b_{vv}
 &=\kappa\Delta T_{vA},\\
 \partial_v\Omega^{AB} \mathfrak{b}_{AB}-D^A\mathfrak b_{vA}
 &=\frac{\kappa v^2}{2}(\Delta T_{vv}+\Delta T_{uv}).
\end{align}
Substitution gives $\nabla^\mu S_{\mu\nu}|_{\mathcal N}
=\Delta T_{v\nu}|_{\mathcal N}$ for all four values of $\nu$.
Equations \eqref{eq:BIc-bulk-div} and
\eqref{eq:BIc-shell-div-rule} therefore cancel, proving
\begin{equation}
 \nabla^\mu\left(
 T^-_{\mu\nu}\chi_{\mathcal M^-}
 +T^+_{\mu\nu}\chi_{\mathcal M^+}
 +T^{\mathrm{sing}}_{\mu\nu}\right)=0
 \label{eq:BIc-conservation}
\end{equation}
on the punctured neighbourhood. The shell is generally not conserved by itself; it exchanges energy and momentum with the ordinary source.

We need to answer three remaining questions in the linearised regime:
\begin{enumerate}
    \item Does the shell stress-energy define a meaningful distribution near the vertex?
    \item Could the Einstein tensor contain an additional term concentrated only at the vertex, missed by the calculation elsewhere?
    \item Could stress-energy conservation fail by a term concentrated only at the vertex?
\end{enumerate}

First, we need to understand how to integrate the shell stress-energy near the vertex: the factor $\delta(ct-r)$ is present in the shell components. For example,
\begin{equation}
    \label{eqn:t00 sing}
    T_{00}^\mathrm{sing}(ct,\vec{x}) = \frac{\mu(2r,\theta)}{2}\delta(ct-r)
\end{equation}
for $r=\abs{\vec{x}} > 0$. This delta expresses concentration of the stress-energy on the cone. To decide whether this expression makes sense near its vertex, we must check that integrating it against a smooth weight remains finite there.

Let $f(ct,\vec x)$ be any smooth compactly supported test function in $\mathcal{D}(\mathbb{M})$ with support in a bounded closed region in a neighbourhood of the vertex. Its
value at the vertex need not vanish. The distribution in the vicinity of the vertex is specified by its action on these functions; we write this action as $\langle T^{\mathrm{sing}}_{\mu\nu},f\rangle$. This is the precise meaning of the formal spacetime integral of
$T^{\mathrm{sing}}_{\mu\nu}f$.

Initially keep $\vec x\ne0$, where the shell formula is already known. For each fixed such $\vec x$, integrate over the time coordinate first. The one-dimensional delta identity gives
\begin{equation}
 \int_{\mathbb R} f(ct,\vec x)\delta(ct-|\vec x|)\,d(ct)
 =f(|\vec x|,\vec x).
 \label{eq:vertex-time-integration}
\end{equation}
Thus, after this time integral, the only spacetime point sampled for a given nonzero $\vec x$ is $(|\vec x|,\vec x)$, which lies on the cone away from its vertex. For these points alone, set
\begin{equation}
 s=v=2r>0,\qquad
 \vec\omega=\frac{\vec x}{r},\qquad
 (ct,\vec x)=\left(\frac s2,\frac s2\vec\omega\right).
 \label{eq:vertex-punctured-parameters}
\end{equation}
At the vertex $s=v=0$, and $\vec\omega$ is undefined. We do not use these parameters at that point. The inclusion of the vertex is handled by a convergent limit. In particular, $s$ is the null coordinate $v$ along the cone.

To implement this limit, first exclude the spatial ball $|\vec x|\le\epsilon$, where $\epsilon>0$. By \eqref{eq:vertex-time-integration}, the contribution of \eqref{eqn:t00 sing} from $|\vec x|>\epsilon$ is
\begin{align}
 &\int_{|\vec x|>\epsilon}
       \frac{\mu(2|\vec x|,\theta)}{2}
       f(|\vec x|,\vec x)\,d^3\vec x\nonumber\\
 &\qquad=\frac1{16}\int_{2\epsilon}^{\infty}s^2\,ds
       \iint_{S^2}\mu(s,\theta)
       f\left(\frac s2,\frac s2\vec\omega\right)d\Omega.
 \label{eq:vertex-truncated-shell}
\end{align}
Here the spatial volume element is $r^2 drd\Omega=\frac{s^2}{8}\,ds\,d\Omega$: the sphere has area element $(s/2)^2d\Omega$, and the radial differential is $dr=ds/2$. The remaining factor $1/2$ comes from the coefficient $\mu/2$ in \eqref{eqn:t00 sing}. The other Cartesian components of $T_{\mu\nu}^\mathrm{sing}$ are integrated in exactly the same way, with their respective coefficients in place of $\mu/2$.

Use the previously obtained bounds $\mathfrak b_{vv} \sim \mathcal{O}(v) \sim \mathcal{O}(r)$, $\mathfrak b_{vA}\sim\mathcal{O}(v^2) \sim \mathcal{O}(r^2)$, and $\Omega^{AB} \mathfrak{b}_{AB}\sim\mathcal{O}(v^3) \sim \mathcal{O}(r^3)$. They imply
\begin{equation}
 \mu\sim \mathcal{O}(r),\qquad p\sim\mathcal{O}(r),\qquad
 j^A=\mathcal{O}(1).
\end{equation}
Note that the Cartesian stress-energy \eqref{eq:BIc-cartesian} contains the combination $rj^A \partial_A \omega^i \sim \mathcal{O}(r)$. Thus every Cartesian coefficient multiplying $\delta(ct-r)$ in $T^{\mathrm{sing}}_{\mu\nu}$ is of order $\mathcal{O}(r)$ near the vertex, i.e. is uniformly bounded by $Cr$ for some $C > 0$.

To check convergence, use the bound $Cr=Cs/2$ for any one of those coefficients. For the part of the cone with $0<s<2\epsilon$, the absolute value of its weighted integral is at most
\begin{multline}
\label{eq:vertex-shell-bound}
\abs{\int_{0<\abs{\vec{x}}<\epsilon}
       2 S_{\mu\nu}(\abs{\vec{x}},\vec{x})
       f(|\vec x|,\vec x)\,d^3\vec x} \\ \leq 
 \frac18\int_0^{2\epsilon}s^2\,ds
       \iint_{S^2}\frac{Cs}{2}\,\sup|f|\,d\Omega \\
 =\frac{C\pi}{4}\sup|f|\int_0^{2\epsilon}s^3\,ds\\
 =C\pi\epsilon^4\sup|f|\longrightarrow0.
\end{multline}
We used $\iint_{S^2}d\Omega=4\pi$ and
$\int_0^{2\epsilon}s^3ds=(2\epsilon)^4/4$.
Consequently the limit of \eqref{eq:vertex-truncated-shell} exists as $\epsilon\downarrow0$. This defines the shell on test functions whose support includes the vertex: an integral with lower endpoint $s=0$ here means this limit; it never requires a direction at $s=0$.

Each Cartesian component therefore defines a locally finite signed measure: its absolute integral is finite in every bounded closed region under consideration. In particular it is a distribution.

For use below, introduce a neighbourhood specified in Cartesian coordinates,
\begin{equation}
 \mathcal U_\epsilon
 =\{(ct,\vec x):(ct)^2+|\vec x|^2<\epsilon^2\}.
 \label{eq:vertex-cartesian-ball}
\end{equation}
It is simply a Euclidean ball in the four coordinate variables, used to specify closeness to the vertex; it is not a Lorentzian proper-distance ball. Its four-dimensional volume is proportional to $\epsilon^4$. Its intersection with the cone is contained in $0<s<2\epsilon$, so \eqref{eq:vertex-shell-bound} also bounds the absolute shell measure in $\mathcal U_\epsilon$ by $C\epsilon^4$. Throughout the argument $C$ denotes a finite constant independent of sufficiently small $\epsilon$; its value may change between estimates. Since the absolute measure of these shrinking neighbourhoods tends to zero, the shell measure assigns zero weight to the vertex itself. Thus the first of our three questions is settled: the shell stress-energy does indeed define a meaningful distribution near the vertex.

This does not yet settle the second question. The shell measure has just been extended through the vertex, but the calculation of the Einstein tensor away from that point could, in principle, have missed an additional distribution supported only there. We must compare the Einstein tensor with the proposed source on every test function, including those that are nonzero at the vertex.

For this comparison, all derivatives are Cartesian derivatives, and
\begin{equation}
 d^4x=d(ct)\,d^3\vec x,\qquad
 \partial^\mu=\eta^{\mu\alpha}\partial_\alpha.
\end{equation}
The covariant derivative in the conservation law is the connection of the Minkowski background. It is therefore an ordinary partial derivative in these coordinates.

Choose a smooth function $\zeta : \mathbb{M} \to [0,1]$ of four real variables, with values between zero and one, equal to zero inside the unit ball and equal to one outside the ball of radius two. Define
\begin{equation}
 \zeta_\epsilon(ct,\vec x)
 =\zeta\left(\frac{ct}{\epsilon},\frac{\vec x}{\epsilon}\right).
 \label{eq:vertex-window}
\end{equation}
This is a smooth window: it is zero in $\mathcal U_\epsilon$, one outside $\mathcal U_{2\epsilon}$, and changes between them. We use a smooth window because derivatives will be taken. 

The chain rule can be used to check the size of its derivatives:
\begin{align}
 \partial_\alpha\zeta_\epsilon(ct,\vec x)
 &=\frac1\epsilon(\partial_\alpha\zeta)
       \left(\frac{ct}{\epsilon},\frac{\vec x}{\epsilon}\right),\\
 \partial_\alpha\partial_\beta\zeta_\epsilon(ct,\vec x)
 &=\frac1{\epsilon^2}(\partial_\alpha\partial_\beta\zeta)
       \left(\frac{ct}{\epsilon},\frac{\vec x}{\epsilon}\right).
\end{align}
The derivatives of the fixed function $\zeta$ are bounded. Hence the first derivatives of $\zeta_\epsilon$ are bounded by $C \epsilon^{-1}$ and its second derivatives by $C \epsilon^{-2}$. These derivatives vanish outside the transition region. Although their bounds grow as that region shrinks, its four-dimensional volume decreases as $\epsilon^4$.

For any fixed test function we have the exact decomposition
\begin{equation}
 f=\zeta_\epsilon f+(1-\zeta_\epsilon)f.
 \label{eq:vertex-split-test}
\end{equation}
The first term vanishes near the vertex, so the equations already proved away from the vertex apply to it. The second term is supported in the closed ball of radius $2\epsilon$. It is the only part still to be checked. What we will show is that its contribution to the relevant integrals becomes small.

The potentially delicate part of the metric is the zero extension $\Theta(u)\mathfrak H$. The previously established Cartesian bound $\mathfrak H=O((ct)^2)$ on the future side implies
\begin{equation}
 |\Theta(u)\mathfrak H_{\alpha\beta}|
 \le C\epsilon^2\quad\hbox{in }\mathcal U_{2\epsilon}.
 \label{eq:vertex-collar-size}
\end{equation}
Indeed, there $|ct|=r<2\epsilon$, and outside the future side the zero extension vanishes. The value assigned to the field at the single vertex does not affect any ordinary spacetime integral.

We do not assume classical second derivatives at the vertex. Distributional derivatives are defined by integration by parts:
\begin{equation}
 \left\langle
 \partial_\alpha\partial_\beta
       (\Theta(u)\mathfrak H_{\rho\sigma}),f
 \right\rangle
 =\int\Theta(u)\mathfrak H_{\rho\sigma}
       \partial_\alpha\partial_\beta f\,d^4x.
 \label{eq:vertex-derivative-definition}
\end{equation}
In Cartesian coordinates, $G^{(1)}$ is a finite linear combination of second derivatives of metric components, with constant coefficients formed from $\eta$. Therefore it suffices to estimate the right-hand side of \eqref{eq:vertex-derivative-definition}, replacing $f$ by
$(1-\zeta_\epsilon)f$. The product rule gives
\begin{align}
 \partial_\alpha\partial_\beta((1-\zeta_\epsilon)f)
 ={}&(1-\zeta_\epsilon)\partial_\alpha\partial_\beta f
       -(\partial_\alpha\zeta_\epsilon)\partial_\beta f\nonumber\\
    &-(\partial_\beta\zeta_\epsilon)\partial_\alpha f
       -(\partial_\alpha\partial_\beta\zeta_\epsilon)f.
 \label{eq:vertex-test-product}
\end{align}
Every term is supported in the closed ball of radius $2\epsilon$. Multiply each by the field bound $C\epsilon^2$ and integrate over a volume bounded by $C\epsilon^4$. The term with no derivative on the window gives $C\epsilon^6\sup|\partial^2 f|$. Each term with one window derivative gives $C\epsilon^2\epsilon^{-1}\epsilon^4\sup|\partial f| =C\epsilon^5\sup|\partial f|$. The term with two window derivatives gives $C\epsilon^2\epsilon^{-2}\epsilon^4\sup|f| =C\epsilon^4\sup|f|$. Thus
\begin{align}
 &\left|\left\langle G^{(1)}_{\mu\nu}[\Theta(u)\mathfrak H],
                  (1-\zeta_\epsilon)f\right\rangle\right|\nonumber\\
 &\qquad\le C\left(\epsilon^6\sup|\partial^2 f|
             +\epsilon^5\sup|\partial f|
             +\epsilon^4\sup|f|\right)\longrightarrow0.
 \label{eq:vertex-Einstein-bound}
\end{align}
Here the derivative suprema are finite and independent of $\epsilon$, since $f$ is fixed.

The equation already obtained away from the vertex is
\begin{equation}
 G^{(1)}_{\mu\nu}[\Theta(u)\mathfrak H]
 =\Theta(u)G^{(1)}_{\mu\nu}[\mathfrak H]
       +\kappa T^{\mathrm{sing}}_{\mu\nu}.
 \label{eq:vertex-collar-equation}
\end{equation}
Its regular term is locally bounded, using the previously
established boundedness of $\Delta T$ and $\tau$ and the relation
\begin{equation}
 \Theta(u)G^{(1)}_{\mu\nu}[\mathfrak H]
 =\kappa\Theta(u)\Delta T_{\mu\nu}-\kappa\tau_{\mu\nu}.
 \label{eq:vertex-ordinary-source}
\end{equation}
A locally bounded regular term integrated over the ball of radius
$2\epsilon$ contributes at most $C\epsilon^4\sup|f|$.
The shell term has the same bound by
\eqref{eq:vertex-shell-bound}. Hence both sides of
\eqref{eq:vertex-collar-equation} have contributions tending to zero
on $(1-\zeta_\epsilon)f$.

For every $\epsilon>0$ the two sides agree on $\zeta_\epsilon f$,
because that test function vanishes near the vertex. Using
\eqref{eq:vertex-split-test} and then taking $\epsilon\downarrow0$
shows that they agree on $f$ itself. Since $f$ was arbitrary,
\eqref{eq:vertex-collar-equation} holds as a distributional equation
through the vertex. No additional vertex-supported distribution is
left over. This excludes both a point source, whose action would be
proportional to $f(0,\vec0)$, and derivatives of a point source,
whose actions would involve derivatives of $f$ there.

For the retarded correction $H$, the already established global identity $G^{(1)}[H]=\kappa\tau$ has a locally bounded right-hand side. It therefore contains no point-supported term. The continuation $\underline h^-$ is smooth through the vertex. Adding their equations to \eqref{eq:vertex-collar-equation}, and using $h=\underline h^-+\Theta(u)\mathfrak H+H$, gives
\begin{equation}
 G^{(1)}_{\mu\nu}[h]
 =\kappa\left(\underline T^-_{\mu\nu}
       +\Theta(u)\Delta T_{\mu\nu}
       +T^{\mathrm{sing}}_{\mu\nu}\right)
 \label{eq:vertex-Einstein-global}
\end{equation}
on a neighbourhood including the vertex. This answers the second question.

Finally, we check conservation through the vertex directly. Denote the total source appearing in \eqref{eq:vertex-Einstein-global} by
\begin{equation}
 T^{\mathrm{tot}}_{\mu\nu}
 :=\underline T^-_{\mu\nu}+\Theta(u)\Delta T_{\mu\nu}
       +T^{\mathrm{sing}}_{\mu\nu}.
 \label{eq:vertex-total-source}
\end{equation}
Its regular part is locally bounded, and its shell part satisfies \eqref{eq:vertex-shell-bound}. Consequently the absolute measure of each component in $\mathcal U_{2\epsilon}$ is at most $C\epsilon^4$.

The conservation law proved on the punctured neighbourhood gives
\begin{equation}
 \left\langle\partial^\mu T^{\mathrm{tot}}_{\mu\nu},
                       \zeta_\epsilon f\right\rangle=0.
\end{equation}
The decomposition \eqref{eq:vertex-split-test} therefore leaves only $(1-\zeta_\epsilon)f$. Transferring the single derivative to this test function and expanding the product gives
\begin{align}
 \left\langle\partial^\mu T^{\mathrm{tot}}_{\mu\nu},f\right\rangle
 ={}&-\left\langle T^{\mathrm{tot}}_{\mu\nu},
              (1-\zeta_\epsilon)\partial^\mu f\right\rangle\nonumber\\
    &+\left\langle T^{\mathrm{tot}}_{\mu\nu},
              (\partial^\mu\zeta_\epsilon)f\right\rangle.
 \label{eq:vertex-conservation-test}
\end{align}
The first term is bounded by $C\epsilon^4\sup|\partial f|$. The second includes one window derivative, bounded by $C\epsilon^{-1}$, and is therefore bounded by $C\epsilon^{-1}\epsilon^4\sup|f|=C\epsilon^3\sup|f|$. Thus
\begin{equation}
 \left|\left\langle\partial^\mu T^{\mathrm{tot}}_{\mu\nu},f
              \right\rangle\right|
 \le C\epsilon^4\sup|\partial f|
       +C\epsilon^3\sup|f|\longrightarrow0.
 \label{eq:vertex-conservation-bound}
\end{equation}
The left-hand side is independent of $\epsilon$. Since it is bounded by quantities tending to zero, it is zero. This holds for every $f$ and every $\nu$, proving
\begin{equation}
 \partial^\mu T^{\mathrm{tot}}_{\mu\nu}=0
 \label{eq:vertex-conservation-global}
\end{equation}
as a distributional identity including the vertex. There is no additional vertex-supported failure of conservation. Thus the linearised Einstein field equations and \eqref{eq:BIc-conservation} extend through the vertex. 

\section{Computations: a qubit in superposition}

\label{app:computations qubit in superposition}

\subsection{The semiclassical Newtonian potential of a Gaussian wavefunction}

\label{app:semiclassical newtonian potential}

We compute the gravitational potential of a Gaussian wavepacket. First, note that for a fixed source radius $s$ and writing $z=\cos(\theta)$,
\begin{align}
    \iint_{S^2}&\frac{d\Omega}{\sqrt{\abs{\vec{x}-\vec{x}_i}^2 + s^2 - 2s \abs{\vec{x}-\vec{x}_i}\cos(\theta)}} \\
    &= 2\pi \int_{-1}^1 \frac{dz}{\sqrt{\abs{\vec{x}-\vec{x}_i}^2 + s^2 - 2s \abs{\vec{x}-\vec{x}_i}z}} \\
    &= \frac{2\pi}{s\abs{\vec{x}-\vec{x}_i}} (\abs{\vec{x}-\vec{x}_i}+s - \abs{\abs{\vec{x}-\vec{x}_i}-s}) \\
    &= \begin{cases}
        \frac{4\pi}{\abs{\vec{x}-\vec{x}_i}}, \qquad &0 < s < \abs{\vec{x}-\vec{x}_i}, \\
        \frac{4\pi}{s}, \qquad & s > \abs{\vec{x}-\vec{x}_i}.
    \end{cases}
\end{align}
The potential of a single wavepacket separates into two integrals:
\begin{multline}
    -Gm \iiint_{\mathbb{R}^3} \frac{\abs{\psi_i(\vec{y})}^2}{\abs{\vec{x}-\vec{y}}} d^3 \vec{y}  \\ = - \frac{4\pi G m}{(2\pi \sigma_i^2)^{3/2}} \Bigg[\frac{1}{\abs{\vec{x}-\vec{x}_i}} \int_0^{\abs{\vec{x}-\vec{x}_i}} s^2 \exp\left(\frac{-s^2}{2\sigma_i^2}\right) ds \\ + \underbrace{\int_{\abs{\vec{x}-\vec{x}_i}}^\infty s \exp\left(\frac{-s^2}{2\sigma_i^2}\right)ds}_{=\sigma_i^2 \exp\left(\frac{-\abs{\vec{x}-\vec{x}_i}^2}{2\sigma_i^2}\right)} \Bigg] 
\end{multline}
The first integral can be solved through integration by parts by writing $s^2 e^{-s^2/(2\sigma_i^2)}=-\sigma_i^2 s d(e^{-s^2/(2\sigma_i^2)})/ds$, with
\begin{multline}
    \int_0^{\abs{\vec{x}-\vec{x}_i}} s^2 \exp\left(\frac{-s^2}{2\sigma_i^2}\right) ds = -\sigma_i^2 \abs{\vec{x}-\vec{x}_i} \exp\left(\frac{-\abs{\vec{x}-\vec{x}_i}^2}{(2\sigma_i^2)}\right) \\ + \sigma_i^2 \int_0^{\abs{\vec{x}-\vec{x}_i}} \exp\left(\frac{-s^2}{2\sigma_i^2}\right) ds \\
    = -\sigma_i^2 \abs{\vec{x}-\vec{x}_i} \exp\left(\frac{-\abs{\vec{x}-\vec{x}_i}^2}{(2\sigma_i^2)}\right) \\ +\sigma_i^3 \sqrt{\frac{\pi}{2}} \mathrm{erf}\left(\frac{\abs{\vec{x}-\vec{x}_i}}{\sqrt{2}\sigma_i}\right).
\end{multline}
Simplifying thus yields
\begin{equation}
    \label{eqn:gravitational potential gaussian}
    -Gm \iiint_{\mathbb{R}^3} \frac{\abs{\psi_i(\vec{y})}^2}{\abs{\vec{x}-\vec{y}}} d^3 \vec{y} = -\frac{Gm}{\abs{\vec{x}-\vec{x}_i}}\mathrm{erf}\left(\frac{\abs{\vec{x}-\vec{x}_i}}{\sqrt{2}\sigma_i}\right).
\end{equation}
Note that this is finite at $\vec{x}=\vec{x}_i$:
\begin{equation}
    \frac{\mathrm{erf}\left(\frac{\abs{\vec{x}-\vec{x}_i}}{\sqrt{2}\sigma_i}\right)}{\abs{\vec{x}-\vec{x}_i}} = \sqrt{\frac{2}{\pi \sigma_i^2}} \left(1-\frac{\abs{\vec{x}-\vec{x}_i}^2}{6 \sigma_i^2} + \mathcal{O}\left(\frac{\abs{\vec{x}-\vec{x}_i}^4}{ \sigma_i^4}\right)\right),
\end{equation}
i.e. at $\vec{x}=\vec{x}_i$ the value of the potential is $-\sqrt{\frac{2}{\pi}} \frac{Gm}{\sigma_i}$. In the Newtonian regime, a linear superposition of such Gaussian wavepackets simply gives a linear combination of such potentials. More generally, in the relativistic weak-field linearised regime, integrals are retarded but the same idea applies.

\subsection{The time interval over which the stationary source is a good approximation}

\label{app:interval stationary approximation}

Over the timescales under consideration, we want the gravitational potential given in Eqn.\eqref{eqn:gravitational potential gaussian} to remain a good approximation, and so for the stationary approximation to hold. Two different changes in the wavepackets must be small: wavepacket spreading and motion of the wavepacket by self-gravity. A negligible overlap between Gaussian wavepackets in superposition is not sufficient alone to ensure these conditions.

Note that one may also want the wavepacket spreading and gravitational self-attraction to be much smaller than $L$. However, we will require that they are much smaller than $\sigma_i$, and $\sigma_i \ll L$ by assumption, so this is automatically satisfied. 

\subsubsection{Wavepacket spreading}

It has been shown numerically in \cite[Section 4]{oseka-lenart_quantifying_2025} that wavepacket spreading is slower in the Schrödinger-Newton system than in the standard unitary Schrödinger equation. Physically this can be understood as the wavepacket staying sharply localised under gravitational self-attraction and thus resisting dispersion. It is thus sufficient for the following to place an upper bound on the validity of this stationary approximation from dispersion by looking at dispersion timescales in unitary quantum theory. 

The Gaussian propagation kernel for unitary quantum mechanics gives, for $\tau > 0$, \cite{shankar_simple_1994}
\begin{equation}
    \psi_i(t+\tau,\vec{x}) = \frac{\exp\left(-\frac{\abs{\vec{x}-\vec{x}_i}^2}{4\sigma_i^2 (1+i\hbar \tau/(2m\sigma_i^2))}\right)}{(2\pi \sigma_i^2)^{3/4}(1+i\hbar \tau/(2m \sigma_i^2))^{3/2}}
\end{equation}
Thus, at the wavepacket's centre $\vec{x}=\vec{x}_i$, its density relative to the initial central density is
\begin{equation}
    \frac{\abs{\psi_i(t+\tau,\vec{x}_i)}^2}{\abs{\psi_i(t,\vec{x}_i)}^2} = \left(1+\frac{\hbar^2 \tau^2}{4m^2 \sigma_i^4}\right)^{-3/2}
\end{equation}
This is approximately equal to $1$ if
\begin{equation}
    \frac{\hbar^2 \tau^2}{4m^2 \sigma_i^4} \ll 1 \Longleftrightarrow \tau \ll \frac{2m \sigma_i^2}{\hbar} =: \tau_\mathrm{spread} \qquad (\tau > 0).
\end{equation}

\subsubsection{Gravitational self-attraction}

We also want the wavepackets' centres to move a distance much less than $\sigma_i$. The gravitational acceleration between branches of mass $\abs{a_i}^2 m$ is $a_G \sim \frac{Gm}{L^2}$. The approximate displacement induced by gravity during a window of total time $\tau$ is $\Delta X_G \sim \frac{1}{2} a_G \tau^2$. Thus, the wavepackets' centres do not shift beyond a standard deviation of their Gaussian profile over such timescales if
\begin{equation}
    \Delta X_G \ll \sigma \Longleftrightarrow \tau \ll \sqrt{\frac{2\sigma_i L^2}{Gm}} =: \tau_\mathrm{self-grav} \, .
\end{equation} 

Likewise, after some time, the wavepackets merge from gravitational self-interactions and the wavefunction \eqref{eqn:qubit} becomes one big wavepacket centred somewhere in between $\psi_1$ and $\psi_2$. However, the timescales for this process are several orders of magnitude greater than those considered here. Indeed, there is no ``merging of the wavefunction'' over such timescales if
\begin{equation}
    \Delta X_G \ll L \Longleftrightarrow \tau \ll \sqrt{\frac{2L^3}{Gm}} =: \tau_\mathrm{grav-merge} \, .
\end{equation} 
This timescale is longer than $\tau_\mathrm{self-grav}$ since $L \gg \sigma_i$.

\subsection{Gaussian shell mass}

\label{app:gaussian shell mass}

We want to compute the mass on the collapse hypersurface.

We start by examining the contribution to this mass from $\psi_1$.  We have
\begin{multline}
 \iiint_{\abs{\vec x}<r}\abs{\psi_1(t,\vec x)}^2d^3\vec x  \\
 =\frac{4\pi}{(2\pi\sigma_1^2)^{3/2}}
       \int_0^r s^2e^{-s^2/(2\sigma_1^2)}ds\\
 =\frac{4\pi}{(2\pi\sigma_1^2)^{3/2}}
 \left[-\sigma_1^2r e^{-r^2/(2\sigma_1^2)}
 +\sigma_1^3\sqrt{\frac\pi2}
       \mathrm{erf}\left(\frac r{\sqrt2\sigma_1}\right)\right]
 \\
 =\mathrm{erf}\!\left(\frac r{\sqrt2\sigma_1}\right)
 -\sqrt{\frac2\pi}\frac r{\sigma_1}e^{-r^2/(2\sigma_1^2)}.
 \label{full:ball-centered}
\end{multline}
Let us now compute the contribution from $\psi_2$. Choose the polar axis along $\vec x_2$ and put $z=\cos\theta$. Then
$\abs{s\vec\omega-\vec x_2}^2=s^2+L^2-2sLz$, and
\begin{align}
 &\iiint_{\abs{\vec x}<r}|\psi_2(t,\vec x)|^2d^3\vec x \nonumber \\
 &=\frac{2\pi}{(2\pi\sigma_2^2)^{3/2}}
       \int_0^r s^2 e^{-(s^2+L^2)/(2\sigma_2^2)}
       \left(\int_{-1}^1 e^{sLz/\sigma_2^2}dz\right)ds,\\
 &\int_{-1}^1 e^{sLz/\sigma_2^2}dz
 =\left[\frac{\sigma_2^2}{sL}e^{sLz/\sigma_2^2}\right]_{-1}^{1} \\ & \qquad \qquad \qquad
 =\frac{\sigma_2^2}{sL}
       \left(e^{sL/\sigma_2^2}-e^{-sL/\sigma_2^2}\right).
\end{align}
At $s=0$, the last expression has its continuous value $2$.
Combining the exponents and cancelling one factor of $s$ gives
\begin{multline}
 \iiint_{\abs{\vec x}<r}|\psi_2(t,\vec x)|^2d^3\vec x
 \\ =\frac1{\sqrt{2\pi}\sigma_2L}
 \int_0^r s\left[e^{-(s-L)^2/(2\sigma_2^2)}
                  -e^{-(s+L)^2/(2\sigma_2^2)}\right]ds.
 \label{full:ball-displaced-radial}
\end{multline}
In the first integral write $s=(s-L)+L$; in the second write $s=(s+L)-L$. Their antiderivatives are, respectively,
\begin{align}
 \int& s e^{-(s-L)^2/(2\sigma_2^2)}ds \nonumber \\
 &=-\sigma_2^2 e^{-(s-L)^2/(2\sigma_2^2)}
 +L\sigma_2\sqrt{\frac\pi2}
       \mathrm{erf}\!\left(\frac{s-L}{\sqrt2\sigma_2}\right),\\
 \int& s e^{-(s+L)^2/(2\sigma_2^2)}ds \nonumber \\
 &=-\sigma_2^2 e^{-(s+L)^2/(2\sigma_2^2)}
 -L\sigma_2\sqrt{\frac\pi2}
       \mathrm{erf}\!\left(\frac{s+L}{\sqrt2\sigma_2}\right).
\end{align}
Subtracting these primitives, the exponential terms cancel at $s=0$,
as do the error functions by oddness. Evaluation at $s=r$ therefore gives
\begin{multline}
 \iiint_{\abs{\vec x}<r}|\psi_2(t,\vec x)|^2d^3\vec x
 \\=\frac12\left[
       \mathrm{erf}\!\left(\frac{r-L}{\sqrt2\sigma_2}\right)
      +\mathrm{erf}\!\left(\frac{r+L}{\sqrt2\sigma_2}\right)\right]
 \\
 +\frac{\sigma_2}{\sqrt{2\pi}L}
       \left[e^{-(r+L)^2/(2\sigma_2^2)}
             -e^{-(r-L)^2/(2\sigma_2^2)}\right].
 \label{full:ball-displaced}
\end{multline}

For outcome $i=1$,
$\Delta\varrho_m^{(1)}=m\abs{a_2}^2(\abs{\psi_1}^2-\abs{\psi_2}^2)$; for outcome $i=2$, $\Delta\varrho_m^{(2)}=-m\abs{a_1}^2(\abs{\psi_1}^2-\abs{\psi_2}^2)$. Combining Eqns.~\eqref{full:ball-centered} and \eqref{full:ball-displaced} with $r=ct$ gives Eqn.~\eqref{eqn:shell mass gaussian}.

\subsection{The future metric}

\label{app:future metric qubit}

Our starting point is
\begin{equation}
    \overline{h}^{+,(i)}_{\mu\nu} = \overline{h}^{\mathrm{ret},(i)}_{\mu\nu} + 2 \kappa G_{\mathrm{ret}} * T^{\mathrm{sing},(i)}_{\mu\nu}.
\end{equation}
We here want to compute the second term $2 \kappa G_{\mathrm{ret}} * T^{\mathrm{sing},(i)}_{\mu\nu}$ and show it is subleading in $ct$.

Write the source position as $\vec{y} = s \vec{\omega}$, $s = \abs{\vec{y}}$, $\abs{\vec{\omega}}=1$. The outgoing shell is at $y^0 = s$. Consequently,
\begin{equation}
    T_{\mu\nu}^\mathrm{sing,(i)}(y^0,s) = 2 S_{\mu\nu}^{(i)} \delta(y^0-s)
\end{equation}
where
\begin{equation}
    S^{\mu\nu,(i)} = \mu^{(i)} k^\mu k^\nu + j^{A,(i)}(k^\mu e_A^\nu + e_A^\mu k^\nu) + \frac{p^{(i)}}{s^2} \Omega^{AB} e^\mu_A e^\nu_B.
\end{equation}
The vectors appearing here have Cartesian components
\begin{align}
    k^0&=\frac{1}{2}=-k_0, \qquad k^j = \frac{\omega_j}{2}, \\
    e^0_A &= 0, \qquad e^j_A = \frac{\partial(s \omega_j)}{\partial \theta^A} = sD_A \omega_j.
\end{align}
Let us compute each component. For the energy component,
\begin{equation}
    2 S^{(i)}_{00} = 2 \mu^{(i)} k_0 k_0 = \frac{\mu^{(i)}}{2}.
\end{equation}
For the mixed (current-like) component,
\begin{multline}
    2 S_{0j}^{(i)} = 2\mu^{(i)} k_0 k_j + 2 j^{A,(i)} k_0 e_{Aj} \\ = -\frac{\mu^{(i)}}{2}\omega_j - s j^{A,(i)} D_A \omega_j. 
\end{multline}
For the spatial (pressure-like) components,
\begin{multline}
    2 S_{jk}^{(i)} = \frac{\mu^{(i)}}{2} \omega_j \omega_k + s j^{A,(i)} (\omega_j D_A \omega_k + \omega_k D_A \omega_j) \\ +2 p^{(i)} \Omega^{AB} D_A \omega_j D_B \omega_k.
\end{multline}
But the last contraction is the projection onto the tangent plane of the unit sphere:
\begin{equation}
    \Omega^{AB} D_A \omega_j D_B \omega_k = \delta_{jk} - \omega_j \omega_k.
\end{equation}
Thus,
\begin{align}
    T_{00}^{\mathrm{sing},(i)} &= \frac{\mu^{(i)}}{2} \delta(y^0-s), \\
    T_{0j}^{\mathrm{sing},(i)} &= - \left(\frac{\mu^{(i)}}{2} \omega_j+ s j^{A,(i)} D_A \omega_j\right) \delta(y^0-s), \\
    T_{jk}^{\mathrm{sing},(i)} &= \Big[\frac{\mu^{(i)}}{2} \omega_j \omega_k + s j^{A,(i)} (\omega_j D_A \omega_k + \omega_k D_A \omega_j) \nonumber \\ & \qquad +2 p^{(i)}(\delta_{jk} - \omega_j \omega_k) \Big] \delta(y^0-s).
\end{align}
The coefficients $\mu^{(i)}, j^{A,(i)}$ and $p^{(i)}$ are evaluated on the shell. Since $v=y^0 + s$, the shell condition $y^0 = s$ means that their arguments are $(2s,\theta)$.

We want to integrate these with the retarded kernel in $G_{\mathrm{ret}}$:
\begin{multline}
    2 \kappa G_{\mathrm{ret}} * T_{\mu\nu}^{\mathrm{sing},(i)} = \frac{2\kappa}{4\pi} \iiint_{\mathbb{R}^3} d^3\vec{y} \int_\mathbb{R} dy^0  \\ \left(\frac{\delta(ct-y^0-\abs{\vec{x}-\vec{y}})}{\abs{\vec{x}-\vec{y}}} T_{\mu\nu}^{\mathrm{sing},(i)}(y^0,y)\right).
\end{multline}
There are two delta functions, with
\begin{multline}
\int_{\mathbb R}dy^{0}\,
\delta\!\left(ct-y^{0}-|\vec x-s\vec\omega|\right)
\delta(y^{0}-s) \\ =
\delta\!\left(ct-s-|\vec x-s\vec\omega|\right).
\end{multline}

This remaining delta function requires the source point $(y^{0},\vec y)=(s,s\vec\omega)$
to lie both on the outgoing shell from the measurement event and on the past light cone of the observation event \((ct,\vec x)\).

We also change the spatial volume element $d^{3}\vec y=s^{2}ds\,d\Omega$. For example, the \(00\) convolution is now
\begin{multline}
    \label{eqn:T00 convolution s2 mu}
2\kappa G_{\mathrm{ret}}*T^{\mathrm{sing},(i)}_{00}
=
\frac{4G}{c^{4}}
\int_{S^{2}}d\Omega\int_0^\infty ds\\ \left(
\frac{
\delta\!\left(ct-s-|\vec x-s\vec\omega|\right)}
{|\vec x-s\vec\omega|}
\frac{s^{2}\mu^{(i)}(2s,\theta)}{2}\right).
\end{multline}
We see that there is here a coefficient $s^2 \mu$ that needs to be integrated for the $00$ contribution. For the $0j$ contribution, there is a coefficient $s^2 \mu$ and a coefficient $s^3 j^A$ that need to be integrated. For the $jk$ contribution, there is a coefficient $s^2 \mu$, a coefficient $s^3 j^A$, and a coefficient $s^2 p$, that need to be integrated.

We want to perform the \(s\)-integration in \eqref{eqn:T00 convolution s2 mu}. Its delta function requires $ct-s-|\vec x-s\vec\omega|=0$, i.e. $ct-s=|\vec x-s\vec\omega|$. Squaring gives
\begin{equation}
    (ct)^2-2cts+s^2
=
r^2-2s\vec x\cdot\vec\omega+s^2,
\qquad r=|\vec x|
\end{equation}
Cancel \(s^2\) and collect the terms containing \(s\):
\begin{equation}
    (ct)^2-r^2
=
2s(ct-\vec x\cdot\vec\omega).
\end{equation}
Therefore,
\begin{equation}
    s=\frac{(ct)^2-r^2}
{2(ct-\vec x\cdot\vec\omega)}.
\end{equation}

To integrate the radial delta function, we change the radial
variable locally from \(s\) to the argument of that delta
function,
\begin{equation}
    z=ct-s-|\vec x-s\vec\omega|.
\end{equation}
The required Jacobian is therefore the derivative \(dz/ds\). First differentiate the distance:
\begin{align}
    \frac{d}{ds}|\vec x-s\vec\omega|
    &=
    \frac{d}{ds}
    \left(r^2-2s\vec x\cdot\vec\omega+s^2\right)^{1/2}
    \nonumber\\
    &=
    \frac{-2\vec x\cdot\vec\omega+2s}
         {2\left(r^2-2s\vec x\cdot\vec\omega+s^2\right)^{1/2}}
    \nonumber\\
    &=
    \frac{s-\vec x\cdot\vec\omega}
         {|\vec x-s\vec\omega|}.
\end{align}
It follows that
\begin{align}
    \frac{dz}{ds}
    &=-1-\frac{s-\vec x\cdot\vec\omega}
                  {|\vec x-s\vec\omega|}
    \nonumber\\
    &=-\frac{
        |\vec x-s\vec\omega|+s-\vec x\cdot\vec\omega}
        {|\vec x-s\vec\omega|}.
\end{align}
At the root, \( |\vec x-s\vec\omega|=ct-s\), so
\begin{align}
    \left.\frac{dz}{ds}\right|_{|\vec x-s\vec\omega|=ct-s}
    &=
    -\left.
    \frac{(ct-s)+s-\vec x\cdot\vec\omega}
         {|\vec x-s\vec\omega|}
    \right|_{|\vec x-s\vec\omega|=ct-s}
    \nonumber\\
    &=
    -\left.
    \frac{ct-\vec x\cdot\vec\omega}
         {|\vec x-s\vec\omega|}
    \right|_{|\vec x-s\vec\omega|=ct-s}.
\end{align}
Both the numerator and denominator are positive. In particular,
the derivative is nonzero, and the local change of variable
is valid.

Under this change of variable, the delta function is
\(\delta(z)\), while the integration measure contributes
\(|ds/dz|\,dz\). Since \(\delta(z)\) evaluates the remaining
integrand at \(z=0\), the radial integral acquires the factor
\begin{equation}
    \left|\frac{ds}{dz}\right|_{z=0}
    =
    \left.
    \frac{|\vec x-s\vec\omega|}
         {ct-\vec x\cdot\vec\omega}
    \right|_{|\vec x-s\vec\omega|=ct-s}.
\end{equation}
Applying this to the radial integral in
\eqref{eqn:T00 convolution s2 mu} gives
\begin{align}
    &\int_0^\infty ds\,
    \frac{s^2\mu^{(i)}(2s,\theta)}
         {2|\vec x-s\vec\omega|}
    \delta(ct-s-|\vec x-s\vec\omega|)
    \nonumber\\
    &\quad=
    \left.
    \left[
    \frac{s^2\mu^{(i)}(2s,\theta)}
         {2|\vec x-s\vec\omega|}
    \frac{|\vec x-s\vec\omega|}
         {ct-\vec x\cdot\vec\omega}
    \right]\right|_{|\vec x-s\vec\omega|=ct-s}
    \nonumber\\
    &\quad=
    \left.
    \frac{s^2\mu^{(i)}(2s,\theta)}
         {2(ct-\vec x\cdot\vec\omega)}
    \right|_{|\vec x-s\vec\omega|=ct-s}.
\end{align}
Thus the cancellation of the spatial denominator occurs
between two factors evaluated at the same retarded root.

The three shell-field components are consequently
\begin{align}
    2\kappa G_{\mathrm{ret}}*
    T^{\mathrm{sing},(i)}_{00}
    &=
    \frac{2G}{c^4}\int_{S^2}
    \left.
    \frac{s^2\mu^{(i)}}
         {ct-\vec x\cdot\vec\omega}
    \right|_{|\vec x-s\vec\omega|=ct-s}d\Omega,
    \label{eq:shell-decay-exact-00}\\
    2\kappa G_{\mathrm{ret}}*
    T^{\mathrm{sing},(i)}_{0j} &= -\frac{4G}{c^4}\int_{S^2} d\Omega \nonumber \\
    &
    \left.
    \frac{s^2\mu^{(i)}\omega_j/2
          +s^3j^{A,(i)}D_A\omega_j}
         {ct-\vec x\cdot\vec\omega}
    \right|_{|\vec x-s\vec\omega|=ct-s}.
    \label{eq:shell-decay-exact-0j} \\
    2\kappa G_{\mathrm{ret}}*
    T^{\mathrm{sing},(i)}_{jk}
    ={}&\frac{4G}{c^4}\int_{S^2}
    \frac{d\Omega}{ct-\vec x\cdot\vec\omega}
    \nonumber\\
    &\times
    \Bigl[
       s^2\mu^{(i)}\omega_j\omega_k/2
       \nonumber\\
    &\qquad
       +s^3j^{A,(i)}
        (\omega_jD_A\omega_k+\omega_kD_A\omega_j)
       \nonumber\\
    &\qquad
       +2s^2p^{(i)}
        (\delta_{jk}-\omega_j\omega_k)
    \Bigr]_{|\vec x-s\vec\omega|=ct-s}.
    \label{eq:shell-decay-exact-jk}
\end{align}
In these equations, the coefficients have arguments \((2s,\theta)\).

We now take \(ct\) sufficiently large so that the supports of the wavepackets approximately satisfy
\begin{equation}
    \operatorname{supp}\psi_1\cup\operatorname{supp}\psi_2
   \mathrel{\underset{\approx}{\subset}}
    \left\{\vec y:|\vec y|<\frac{ct-r}{2}\right\}.
    \label{eq:shell-decay-late-region}
\end{equation}
Since
\begin{equation}
    \frac{ct-r}{2}
    \leq s=
    \frac{(ct)^2-r^2}
         {2(ct-\vec x\cdot\vec\omega)}
    \leq
    \frac{ct+r}{2}
    <ct,
\end{equation}
every root then lies outside both wavepacket supports. Hence
\begin{equation}
    \Delta\varrho_m^{(i)}(s\vec\omega)|_{|\vec x-s\vec\omega|=ct-s}=0,
    \qquad
    p^{(i)}(2s,\theta)|_{|\vec x-s\vec\omega|=ct-s}=0,
\end{equation}
and for $n=2,3$,
\begin{equation}
    \int_0^s(s')^n
    \Delta\varrho_m^{(i)}(s'\vec\omega)\,ds'
    =
    \int_0^\infty(s')^n
    \Delta\varrho_m^{(i)}(s'\vec\omega)\,ds'
\end{equation}
since the density difference of the wavepackets (approximately) vanishes throughout the radial interval $[s,\infty)$ for every direction. 

This is useful for the computation of Eqn.~\eqref{eqn:shell stress energy qubit example}, which contributes to Eqn.~\eqref{eq:shell-decay-exact-00}. To simplify the $00$ coefficient, first use linearity of the integral:
\begin{align}
    &\int_0^\infty
    \left((s')^2-\frac{(s')^3}{s}\right)
    \Delta\varrho_m^{(i)}(s'\vec\omega)\,ds'
    \nonumber\\
    &\quad=
    \int_0^\infty(s')^2
    \Delta\varrho_m^{(i)}(s'\vec\omega)\,ds'
    -\frac1s\int_0^\infty(s')^3
    \Delta\varrho_m^{(i)}(s'\vec\omega)\,ds'.
\end{align}
Since the angular derivatives in the shell coefficient are
taken at fixed \(s\), the factor \(1/s\) is constant under
those derivatives. Multiplying the coefficient formulas \eqref{eqn:shell stress energy qubit example} by
the powers of \(s\) present in
\eqref{eq:shell-decay-exact-00}--\eqref{eq:shell-decay-exact-jk}
therefore gives
\begin{align}
    s^2\mu^{(i)}
    ={}&-c^2(D^AD_A+2)
       \int_0^\infty(s')^2
       \Delta\varrho_m^{(i)}(s'\vec\omega)\,ds'
       \nonumber\\
    &+\frac{c^2}{s}D^AD_A
       \int_0^\infty(s')^3
       \Delta\varrho_m^{(i)}(s'\vec\omega)\,ds',\\
    s^3j^{A,(i)}
    ={}&\frac{c^2}{2s}D^A
       \int_0^\infty(s')^3
       \Delta\varrho_m^{(i)}(s'\vec\omega)\,ds'.
    \label{eq:shell-decay-late-coefficients}
\end{align}

The third-moment terms in the metric contain the factor
\(1/[s(ct-\vec x\cdot\vec\omega)]\). Substitution of the root
gives
\begin{align}
    \frac1{s(ct-\vec x\cdot\vec\omega)}
    &=
    \frac{1}{
      \displaystyle
      \frac{(ct)^2-r^2}
           {2(ct-\vec x\cdot\vec\omega)}
      (ct-\vec x\cdot\vec\omega)}
      \nonumber\\
    &=\frac{2}{(ct)^2-r^2}.
    \label{eq:shell-decay-third-prefactor}
\end{align}
This factor is independent of the integration direction. Substitution of \eqref{eq:shell-decay-late-coefficients} into \eqref{eq:shell-decay-exact-00} gives
\begin{widetext}
    \begin{equation}
        2\kappa G_{\mathrm{ret}}*
    T^{\mathrm{sing},(i)}_{00}
    =-\frac{2G}{c^2}\int_{S^2}
    \frac{
       (D^AD_A+2)\displaystyle\int_0^\infty(s')^2
       \Delta\varrho_m^{(i)}(s'\vec\omega)\,ds'}
       {ct-\vec x\cdot\vec\omega}\,d\Omega
    +\frac{2G}{c^2}\int_{S^2}
    \frac{
       D^AD_A\displaystyle\int_0^\infty(s')^3
       \Delta\varrho_m^{(i)}(s'\vec\omega)\,ds'}
       {s(ct-\vec x\cdot\vec\omega)}\,d\Omega.
    \end{equation}
\end{widetext}
Using \eqref{eq:shell-decay-third-prefactor}, the second term is
\begin{multline}
    \frac{4G}{c^2((ct)^2-r^2)}
    \int_{S^2}D^AD_A
    \left[
       \int_0^\infty(s')^3
       \Delta\varrho_m^{(i)}(s'\vec\omega)\,ds'
    \right]d\Omega
    \\
    =0
\end{multline}
by the divergence theorem on the sphere.

For the $0j$ component, the numerator is
\begin{align}
    \frac{s^2\mu^{(i)}}2\omega_j
    &+s^3j^{A,(i)}D_A\omega_j
    = \nonumber \\
    &-\frac{c^2}{2}\omega_j(D^AD_A+2)
       \int_0^\infty(s')^2
       \Delta\varrho_m^{(i)}(s'\vec\omega)\,ds'
       \nonumber\\
    &+\frac{c^2}{2s}\omega_jD^AD_A
       \int_0^\infty(s')^3
       \Delta\varrho_m^{(i)}(s'\vec\omega)\,ds'
       \nonumber\\
    &+\frac{c^2}{2s}(D_A\omega_j)D^A
       \int_0^\infty(s')^3
       \Delta\varrho_m^{(i)}(s'\vec\omega)\,ds'.
\end{align}

Thus the second-moment term acquires the coefficient
\[
    -\frac{4G}{c^4}\left(-\frac{c^2}{2}\right)
    =\frac{2G}{c^2}.
\]
The two third-moment terms instead have the combined
coefficient
\[
    -\frac{4G}{c^4}\frac{c^2}{2}
    \frac1{s(ct-\vec x\cdot\vec\omega)}
    =
    -\frac{4G}{c^2((ct)^2-r^2)}.
\]
Their angular numerator is a single divergence, because
\begin{align}
    &\omega_jD^AD_A
      \left[\int_0^\infty(s')^3
      \Delta\varrho_m^{(i)}(s'\vec\omega)\,ds'\right]
      \nonumber\\
    &\quad+
      (D_A\omega_j)D^A
      \left[\int_0^\infty(s')^3
      \Delta\varrho_m^{(i)}(s'\vec\omega)\,ds'\right]
      \nonumber\\
    &=
      D_A\left\{
      \omega_jD^A
      \left[\int_0^\infty(s')^3
      \Delta\varrho_m^{(i)}(s'\vec\omega)\,ds'\right]
      \right\}.
\end{align}
Consequently, their contribution is exactly
\begin{multline}
    -\frac{4G}{c^2((ct)^2-r^2)}
    \int_{S^2}d \Omega \\ D_A\left\{
      \omega_jD^A
      \left[\int_0^\infty(s')^3
      \Delta\varrho_m^{(i)}(s'\vec\omega)\,ds'\right]
    \right\} \\
    =0.
\end{multline}

Finally, let us examine the $jk$ components. The pressure term is zero in the late region. The remaining
numerator is
\begin{align}
    &\frac{s^2\mu^{(i)}}2\omega_j\omega_k
    +s^3j^{A,(i)}
      (\omega_jD_A\omega_k+\omega_kD_A\omega_j)
    \nonumber\\
    &\quad=
    -\frac{c^2}{2}\omega_j\omega_k(D^AD_A+2)
      \int_0^\infty(s')^2
      \Delta\varrho_m^{(i)}(s'\vec\omega)\,ds'
      \nonumber\\
    &\qquad+
    \frac{c^2}{2s}\omega_j\omega_kD^AD_A
      \int_0^\infty(s')^3
      \Delta\varrho_m^{(i)}(s'\vec\omega)\,ds'
      \nonumber\\
    &\qquad+
    \frac{c^2}{2s}
      (\omega_jD_A\omega_k+\omega_kD_A\omega_j)D^A
      \int_0^\infty(s')^3
      \Delta\varrho_m^{(i)}(s'\vec\omega)\,ds'.
\end{align}
Multiplying the first term by \(4G/c^4\) gives
\(-2G/c^2\). The third-moment terms have the common coefficient
\[
    \frac{4G}{c^4}\frac{c^2}{2}
    \frac1{s(ct-\vec x\cdot\vec\omega)}
    =
    \frac{4G}{c^2((ct)^2-r^2)}.
\]
The product rule gives
\begin{equation}
    D_A(\omega_j\omega_k)
    =\omega_jD_A\omega_k+\omega_kD_A\omega_j.
\end{equation}
Hence their angular numerator is
\begin{align}
    &\omega_j\omega_kD^AD_A
      \left[\int_0^\infty(s')^3
      \Delta\varrho_m^{(i)}(s'\vec\omega)\,ds'\right]
      \nonumber\\
    &\quad+
      D_A(\omega_j\omega_k)D^A
      \left[\int_0^\infty(s')^3
      \Delta\varrho_m^{(i)}(s'\vec\omega)\,ds'\right]
      \nonumber\\
    &=
      D_A\left\{
      \omega_j\omega_kD^A
      \left[\int_0^\infty(s')^3
      \Delta\varrho_m^{(i)}(s'\vec\omega)\,ds'\right]
      \right\}.
\end{align}
Its spherical integral is thus zero. Thus the third radial moment cancels in all three types of component.

After these cancellations, the shell contribution is
\begin{widetext}
    \begin{align}
    2\kappa G_{\mathrm{ret}}*
    T^{\mathrm{sing},(i)}_{00}
    &=
    -\frac{2G}{c^2}\int_{S^2}
    \frac{
    (D^AD_A+2)\displaystyle\int_0^\infty
    s^2\Delta\varrho_m^{(i)}(s\vec\omega)\,ds}
    {ct-\vec x\cdot\vec\omega}\,d\Omega,
    \label{eq:shell-decay-late-00}\\
    2\kappa G_{\mathrm{ret}}*
    T^{\mathrm{sing},(i)}_{0j}
    &=
    \frac{2G}{c^2}\int_{S^2}
    \frac{
    \omega_j(D^AD_A+2)\displaystyle\int_0^\infty
    s^2\Delta\varrho_m^{(i)}(s\vec\omega)\,ds}
    {ct-\vec x\cdot\vec\omega}\,d\Omega,
    \label{eq:shell-decay-late-0j}\\
    2\kappa G_{\mathrm{ret}}*
    T^{\mathrm{sing},(i)}_{jk}
    &=
    -\frac{2G}{c^2}\int_{S^2}
    \frac{
    \omega_j\omega_k(D^AD_A+2)\displaystyle\int_0^\infty
    s^2\Delta\varrho_m^{(i)}(s\vec\omega)\,ds}
    {ct-\vec x\cdot\vec\omega}\,d\Omega.
    \label{eq:shell-decay-late-jk}
\end{align}
\end{widetext}

We will later expand the denominator $1/(ct-\vec{x}\cdot \vec{\omega})$. We next determine which terms in the denominator expansion survive angular integration. Using the divergence theorem,
\begin{align}
    &\int_{S^2}(D^AD_A+2)
    \left[\int_0^\infty s^2
    \Delta\varrho_m^{(i)}(s\vec\omega)\,ds\right]d\Omega
    \nonumber\\
    &=
    \int_{S^2}D_A\left\{
    D^A\left[\int_0^\infty s^2
    \Delta\varrho_m^{(i)}(s\vec\omega)\,ds\right]
    \right\}d\Omega
    \nonumber\\
    &\qquad+
    2\int_{S^2}d\Omega\int_0^\infty s^2
    \Delta\varrho_m^{(i)}(s\vec\omega)\,ds
    \nonumber\\
    &=0+
    2\int_{\mathbb R^3}
    \Delta\varrho_m^{(i)}(\vec y)\,d^3\vec y
    =0.
    \label{eq:shell-decay-monopole}
\end{align}
The change from the radial and angular integrals to the spatial
integral uses \(d^3\vec y=s^2ds\,d\Omega\).

The Cartesian coordinate functions on the unit sphere obey
\(D^AD_A\omega_j=-2\omega_j\). For example, with
\(\omega_3=\cos\theta\),
\begin{align}
    D^AD_A\omega_3
    &=
    \frac1{\sin\theta}
    \frac{\partial}{\partial\theta}
    \left(\sin\theta
    \frac{\partial\cos\theta}{\partial\theta}\right)
    \nonumber\\
    &=
    \frac1{\sin\theta}
    \frac{\partial}{\partial\theta}(-\sin^2\theta)
    =-2\cos\theta.
\end{align}
Rotational invariance gives the same identity for \(j=1,2\). Integrating by parts twice on the sphere gives
\begin{align}
    &\int_{S^2}\omega_jD^AD_A
    \left[\int_0^\infty s^2
    \Delta\varrho_m^{(i)}(s\vec\omega)\,ds\right]d\Omega
    \nonumber\\
    &=
    -\int_{S^2}(D_A\omega_j)D^A
    \left[\int_0^\infty s^2
    \Delta\varrho_m^{(i)}(s\vec\omega)\,ds\right]d\Omega
    \nonumber\\
    &=
    \int_{S^2}(D^AD_A\omega_j)
    \left[\int_0^\infty s^2
    \Delta\varrho_m^{(i)}(s\vec\omega)\,ds\right]d\Omega
    \nonumber\\
    &=
    -2\int_{S^2}\omega_j
    \left[\int_0^\infty s^2
    \Delta\varrho_m^{(i)}(s\vec\omega)\,ds\right]d\Omega.
\end{align}
Adding the \(+2\) part of the angular operator therefore yields
\begin{equation}
    \int_{S^2}\omega_j(D^AD_A+2)
    \left[\int_0^\infty s^2
    \Delta\varrho_m^{(i)}(s\vec\omega)\,ds\right]d\Omega
    =0.
    \label{eq:shell-decay-dipole}
\end{equation}

For \(ct>r\),
\begin{align}
    \frac1{ct-\vec x\cdot\vec\omega}-\frac1{ct}
    &=
    \frac{ct-(ct-\vec x\cdot\vec\omega)}
         {ct(ct-\vec x\cdot\vec\omega)}
    \nonumber\\
    &=
    \frac{\vec x\cdot\vec\omega}
         {ct(ct-\vec x\cdot\vec\omega)}.
    \label{eq:shell-decay-first-remainder}
\end{align}
But
\begin{align}
    &\frac1{ct-\vec x\cdot\vec\omega}
     -\frac1{ct}
     -\frac{\vec x\cdot\vec\omega}{(ct)^2}
    \nonumber\\
    &\quad=
    \frac{\vec x\cdot\vec\omega}{ct}
    \left(
      \frac1{ct-\vec x\cdot\vec\omega}-\frac1{ct}
    \right)
    \nonumber\\
    &\quad=
    \frac{(\vec x\cdot\vec\omega)^2}
         {(ct)^2(ct-\vec x\cdot\vec\omega)}.
\end{align}
Thus
\begin{equation}
    \frac1{ct-\vec x\cdot\vec\omega}
    =
    \frac1{ct}
    +\frac{\vec x\cdot\vec\omega}{(ct)^2}
    +\frac{(\vec x\cdot\vec\omega)^2}
          {(ct)^2(ct-\vec x\cdot\vec\omega)}.
    \label{eq:shell-decay-exact-expansion}
\end{equation}

Insert \eqref{eq:shell-decay-exact-expansion} into
\eqref{eq:shell-decay-late-00}. The \(1/(ct)\) coefficient
vanishes by \eqref{eq:shell-decay-monopole}. The
\(1/(ct)^2\) coefficient also vanishes, because
\begin{align}
    &\int_{S^2}(\vec x\cdot\vec\omega)(D^AD_A+2)
    \left[\int_0^\infty s^2
    \Delta\varrho_m^{(i)}(s\vec\omega)\,ds\right]d\Omega
    \nonumber\\
    &\quad=
    \sum_{a=1}^3 x_a
    \int_{S^2}\omega_a(D^AD_A+2)
    \left[\int_0^\infty s^2
    \Delta\varrho_m^{(i)}(s\vec\omega)\,ds\right]d\Omega
    =0.
\end{align}
Consequently, the \(00\) field is exactly the remainder term:
\begin{align}
    2\kappa G_{\mathrm{ret}}*
    T^{\mathrm{sing},(i)}_{00}
    ={}&-\frac{2G}{c^2}
    \int_{S^2}
    \frac{(\vec x\cdot\vec\omega)^2}
         {(ct)^2(ct-\vec x\cdot\vec\omega)}
    \nonumber\\
    \times
    (D^AD_A+2)&
    \left[\int_0^\infty s^2
    \Delta\varrho_m^{(i)}(s\vec\omega)\,ds\right]d\Omega.
\end{align}
Since $1/(ct-r) \sim 1/(ct)$ for fixed $\vec{x}$, taking absolute values gives
\begin{widetext}
    \begin{multline}
        \left|
    2\kappa G_{\mathrm{ret}}*
    T^{\mathrm{sing},(i)}_{00}
    \right|
    \leq{}
    \frac{2Gr^2}{c^2(ct)^2(ct-r)}
    \int_{S^2}\left|
    (D^AD_A+2)
    \int_0^\infty s^2
    \Delta\varrho_m^{(i)}(s\vec\omega)\,ds
    \right|d\Omega
    \nonumber\\
    \lesssim{}
    \frac{4Gr^2}{c^2(ct)^3}
    \int_{S^2}\left|
    (D^AD_A+2)
    \int_0^\infty s^2
    \Delta\varrho_m^{(i)}(s\vec\omega)\,ds
    \right|d\Omega.
    \end{multline}
\end{widetext}
The final angular integral is finite and independent of \(t\).
This proves the claimed \(\mathcal O((ct)^{-3})\) estimate.

For the $0j$ term, in \eqref{eq:shell-decay-late-0j}, the coefficient of \(1/(ct)\)
vanishes directly by \eqref{eq:shell-decay-dipole}. We may
therefore subtract \(1/(ct)\) from the kernel and use
\eqref{eq:shell-decay-first-remainder}:
\begin{align}
    2\kappa G_{\mathrm{ret}}*
    T^{\mathrm{sing},(i)}_{0j}
    ={}&\frac{2G}{c^2}
    \int_{S^2}
    \frac{\omega_j(\vec x\cdot\vec\omega)}
         {ct(ct-\vec x\cdot\vec\omega)}
    \nonumber\\
    \times
    (D^AD_A+2)&
    \left[\int_0^\infty s^2
    \Delta\varrho_m^{(i)}(s\vec\omega)\,ds\right]d\Omega.
\end{align}
Since \(|\omega_j|\leq1\),
\begin{widetext}
    \begin{multline}
        \left|
    2\kappa G_{\mathrm{ret}}*
    T^{\mathrm{sing},(i)}_{0j}
    \right|
    \leq{}
    \frac{2Gr}{c^2ct(ct-r)}\int_{S^2}\left|
    (D^AD_A+2)
    \int_0^\infty s^2
    \Delta\varrho_m^{(i)}(s\vec\omega)\,ds
    \right|d\Omega
    \\
    \lesssim
    \frac{4Gr}{c^2(ct)^2}
    \int_{S^2}\left|
    (D^AD_A+2)
    \int_0^\infty s^2
    \Delta\varrho_m^{(i)}(s\vec\omega)\,ds
    \right|d\Omega.
    \end{multline}
\end{widetext}
This proves the \(\mathcal O((ct)^{-2})\) estimate.

For the $jk$ term, in \eqref{eq:shell-decay-late-jk}, no angular cancellation
is needed to prove the stated bound. Since
\(|\omega_j\omega_k|\leq1\),
\begin{widetext}
    \begin{multline}
         \left|
    2\kappa G_{\mathrm{ret}}*
    T^{\mathrm{sing},(i)}_{jk}
    \right|
    \leq{}
    \frac{2G}{c^2(ct-r)} \int_{S^2}\left|
    (D^AD_A+2)
    \int_0^\infty s^2
    \Delta\varrho_m^{(i)}(s\vec\omega)\,ds
    \right|d\Omega\\
    \lesssim
    \frac{4G}{c^2ct}
    \int_{S^2}\left|
    (D^AD_A+2)
    \int_0^\infty s^2
    \Delta\varrho_m^{(i)}(s\vec\omega)\,ds
    \right|d\Omega.
    \end{multline}
\end{widetext}
This proves the \(\mathcal O((ct)^{-1})\) estimate.

We have therefore established
\begin{align}
        2\kappa G_{\mathrm{ret}}*
        T^{\mathrm{sing},(i)}_{00}
        &=\mathcal O((ct)^{-3}),\\
        2\kappa G_{\mathrm{ret}}*
        T^{\mathrm{sing},(i)}_{0j}
        &=\mathcal O((ct)^{-2}),\\
        2\kappa G_{\mathrm{ret}}*
        T^{\mathrm{sing},(i)}_{jk}
        &=\mathcal O((ct)^{-1}).
\end{align}

\bibliographystyle{unsrt}
\bibliography{references,references-2}

\end{document}